\documentclass[]{tADP2e}
\usepackage{rotate,fancybox,epsfig,graphics}
\usepackage{mathbbol,amsmath,amsfonts,amssymb,bm,color}
\usepackage{wasysym}
\usepackage{graphicx,psfrag}

\def\dual{\ \stackrel{\Phi_\d}{\longrightarrow}\ }
\def\idual{\ \stackrel{\Phi^{-1}_\d}{\longrightarrow}\ }
\def\qdual{\ \stackrel{\Phi_\d (?)}{\longrightarrow}\ }

\def\d{{{\sf d}}}
\def\r{{\bm{r}}}
\def\x{{\bm{x}}}
\def\i{{\bm{e_1}}}
\def\j{{\bm{e_2}}}
\def\k{{\bm{e_3}}}
\def\Z{{\mathbb{Z}}}

\def\tr{{{\sf Tr}\ }}

\def\pf{\mathcal{Z}}

\newcommand{\hh}{\begin{picture}(13,9)(-2,2)
    \put (0,0) {\line (1,0) {8}}
    \put (8,8) {\line (-1,0) {8}}
    \put (0,0) {\circle*{3}}
    \put (0,8) {\circle*{3}}
    \put (8,0) {\circle*{3}}
    \put (8,8) {\circle*{3}}
    \end{picture}
}
\newcommand{\vv}{\begin{picture}(13,9)(-2,2)
    \put (0,0) {\line (0,1) {8}}
    \put (8,8) {\line (0,-1) {8}}
    \put (0,0) {\circle*{3}}
    \put (0,8) {\circle*{3}}
    \put (8,0) {\circle*{3}}
    \put (8,8) {\circle*{3}}
    \end{picture}
}

\begin{document}
\doi{10.1080/0001873YYxxxxxxxx}
\issn{1460-6976}
\issnp{0001-8732}  \jvol{00} \jnum{00} \jyear{2011} \jmonth{June}

\markboth{E. Cobanera, G. Ortiz, and Z. Nussinov}{Advances in Physics}

\title{The Bond-Algebraic Approach to Dualities}

\author{Emilio Cobanera$^{\rm 1}$ $^{\ast}$\thanks{$^\ast$Corresponding author. 
Email: ecobaner@indiana.edu
\vspace{6pt}}, Gerardo Ortiz$^{\rm 1}$, and Zohar Nussinov$^{\rm 2}$ 
\\\vspace{6pt} 
$^{\rm 1}${\em{Department of Physics, Indiana University, Bloomington,
IN 47405, USA}} \\ 
$^{\rm 2}${\em{Department of Physics, Washington University, St.
Louis, MO 63160, USA}}
\\\vspace{6pt}\received{received 2011}
}

\maketitle

\begin{abstract}

An algebraic theory of dualities is developed based on the notion of 
bond algebras. It deals  with classical and quantum dualities in a
unified fashion explaining  the precise connection between quantum
dualities  and the low temperature (strong-coupling)/high temperature 
(weak-coupling) dualities of classical statistical mechanics (or
(Euclidean)  path integrals). Its range of applications includes
discrete lattice, continuum field, and  gauge theories. Dualities are
revealed to be local, structure-preserving mappings between
model-specific  bond algebras that can be implemented as unitary
transformations,  or partial isometries if gauge symmetries are
involved.  This characterization permits to search systematically  for
dualities and self-dualities in quantum  models of arbitrary system
size, dimensionality and complexity, and any classical model admitting a
transfer matrix or operator representation. In particular, special
dualities like exact dimensional reduction,  emergent, and
gauge-reducing dualities that solve  gauge constraints can be easily
understood in terms of mappings of  bond algebras. As a new example, we
show that the \(\mathbb{Z}_2\) Higgs model is dual to the extended toric
code model {\it in any number of dimensions}. Non-local transformations
like dual variables and Jordan-Wigner dictionaries are  algorithmically
derived from the local mappings of bond algebras. This permits to
establish a precise connection between quantum dual and  classical
disorder variables. Our bond-algebraic approach goes beyond the standard
approach to classical dualities, and could help resolve the long
standing problem of obtaining duality transformations for lattice
non-Abelian models. As an illustration, we present new dualities in any
spatial  dimension for the quantum Heisenberg model.  Finally, we
discuss various applications including location of phase boundaries, 
spectral behavior and, notably, we show how bond-algebraic dualities
help  constrain and realize fermionization in an arbitrary number of
spatial  dimensions. 

\bigskip

\bigskip

\newpage

\vspace*{0.2cm}

\centerline{\bfseries CONTENTS }
\bigskip

\hbox to \textwidth{\hsize\textwidth\vbox{\hsize32pc
\hspace*{-12pt} {1.}      Introduction: The power of dualities\\ \\ 
\hspace*{-2pt} {2.}       Traditional approaches to dualities\\    
\hspace*{7pt} {2.1.}      Dualities in perspective: What is a duality?\\ 
\hspace*{7pt} {2.2.}      The traditional approach to quantum dualities\\ \\
\hspace*{-2pt} {3.}       Bond-algebraic approach to quantum dualities\\ 
\hspace*{7pt} {3.1.}      Bond algebras and the concept of locality\\
\hspace*{7pt} {3.2.}      Some mathematical aspects of bond algebras\\ 
\hspace*{7pt} {3.3.}      Dualities as isomorphisms of bond algebras\\
\hspace*{7pt} {3.4.}      Connection to the traditional approach:
                                         Determination of dual variables\\
\hspace*{7pt} {3.5.}      Abelian versus non-Abelian dualities: the Heisenberg model\\ 
\hspace*{7pt} {3.6.}      Exact dualities for finite systems\\ 
\hspace*{7pt} {3.7.}      Dualities as unitary transformations\\ 
\hspace*{7pt} {3.8.}      Dualities and quantum symmetries\\ 
\hspace*{7pt} {3.9.}      Order and disorder variables for self-dual models\\
\hspace*{7pt} {3.10.}     Emergent dualities\\
\hspace*{7pt} {3.11.}     Elimination of gauge symmetries by bond-algebraic dualities\\
\hspace*{7pt} {3.12.}     Unifying classical and quantum dualities\\ \\
\hspace*{-2pt} {4.}        Quantum self-dualities by example: Lattice models\\ 
\hspace*{7pt} {4.1.}      Self-dualities in the Potts, vector Potts, and 
                                 \(\mathbb{Z}_p\) clock models\\
\hspace*{7pt} {4.2.}      Dualities in some limits and related approximations\\
\hspace*{7pt} {4.3.}      The Xu-Moore model\\ \\
\hspace*{-2pt} {5.}        Quantum dualities by example: Lattice models\\ 
\hspace*{7pt} {5.1.}      XY/solid-on-solid models\\
\hspace*{7pt} {5.2.}      Xu-Moore/planar orbital compass models \\
\hspace*{7pt} {5.3.}      Two-dimensional  \(\mathbb{Z}_p\) gauge/vector Potts models\\
\hspace*{7pt} {5.4.}      Two-dimensional compact QED and XY models\\
\hspace*{7pt} {5.5.}      Toric code/\(\mathbb{Z}_2\) Higgs models \\ \\
\hspace*{-2pt} {6.}       Bond-algebraic dualities in quantum field theory\\ 
\hspace*{7pt} {6.1.}      One-dimensional free and externally coupled bosonic field, and the Kibble model\\
\hspace*{7pt} {6.2.}      The Luttinger model \\
\hspace*{7pt} {6.3.}      QED without sources, compact QED,  and \(\Z_p\) gauge theories\\
\hspace*{7pt} {6.4.}      QED without vector potential\\
\hspace*{7pt} {6.5.}      Abelian and $\mathbb{Z}_p$ Higgs, and St\"uckelberg models\\
\hspace*{7pt} {6.6.}      The self-dual St\"uckelberg model\\ 
\hspace*{7pt} {6.7.}      Field theory and dimensional reduction\\ \\
\hspace*{-2pt} {7.}       Bond-algebraic approach to classical dualities \\
\hspace*{7pt} {7.1.}      The classical  Ising model in the Utiyama lattice\\
\hspace*{7pt} {7.2.}      The classical vector Potts model \\
\hspace*{7pt} {7.3.}      The classical eight-vertex model\\
\hspace*{7pt} {7.4.}      Classical dualities in $D=3$ and $D=4$ dimensions\\ 
\hspace*{7pt} {7.5.}      Dualities for continuum models of classical statistical mechanics\\  
\hspace*{7pt} {7.5}       Classical disorder variables from quantum ones \\ \\ 
\hspace*{-2pt} {8.}       Applications of dualities \\ 
\hspace*{7pt} {8.1.}      Self-dualities and phase transitions \\
\hspace*{7pt} {8.2.}      Correlation functions\\
\hspace*{7pt} {8.3.}      Fermionization as a duality \\ 
\hspace*{7pt} {8.4.}      Self-dualities and quantum integrability\\
\hspace*{7pt} {8.5.}      Duality, topological quantum order, and dimensional reduction\\ \\
\hspace*{-2pt} {9.}       Appendices \\ 
\hspace*{7pt} {Appendix A:}    Duality by Fourier transformation \\ 
\hspace*{7pt} {Appendix B:}    Conditions for spectral equivalence \\ 
\hspace*{7pt} {Appendix C:}    Lattice quantum field theory\\ 
\hspace*{7pt} {Appendix D:}    Bond-algebraic dualities in finite-size systems \\ 
\hspace*{7pt} {Appendix E:}    Classical Poisson dualities \\
\hspace*{7pt} {Appendix F:}    Exponential of operators in the Weyl group algebra 

     }} 
\end{abstract}

\newpage
\section{Introduction: The power of dualities}
\label{sec1}

The term {\it duality} is pervasive in physics, mathematics and 
philosophy. In general, a duality connects and contrasts two aspects or 
realizations of a given entity. In this article we will be concerned
only with dualities in physics. These are  specific mathematical
transformations that we will uncover as  we proceed connecting seemingly
unrelated {\it physical phenomena}.

Over time, dualities have appeared in various guises in nearly all
disciplines of  physics \cite{witten}. The electromagnetic (EM) duality
of Maxwell's equations in the absence of sources, noticed by Heaviside
in 1884, is probably the oldest well-known duality in modern physics.
Later, the wave-particle duality of quantum mechanics \cite{dirac}
became a fundamental tenet of the modern physical description of
reality.  This Fourier transform-based duality has since  appeared in
numerous arenas,  including in recent years various branches of quantum
statistical  mechanics and  field theory  (see, for example, 
\cite{shastry}), and it is likely to  continue to play an ever
increasing role. Kramers and Wannier introduced dualities in statistical
mechanics  in their foundational 1941 paper  \cite{KW}. These authors
discovered an  elegant relation between the  two-dimensional classical
Ising model on a square lattice {\it at high temperature}, and the same
model {\it at low temperature}  (hence the origin of the name
``self-duality"  in this case), and used it to determine that model's
{\it exact critical temperature}  some years before Onsager
\cite{onsager} published its exact solution. This first {\it
quantitative} success was followed by other similar ones, and so
dualities became a standard tool in statistical mechanics since they
could also provide {\it qualitative} insight. The  spectacular
cross-fertilization between statistical mechanics and quantum field
theory   (QFT) of the 1970's brought dualities to the attention of high
energy theorists, and soon it became apparent that dualities in QFT
combined features of the EM and statistical mechanics dualities while
retaining their distinct  capability to produce weakly-coupled
representations of strongly-coupled problems.
 
Dualities can provide reliable qualitative  or even exact quantitative
information about  systems that need not  be exactly solvable, partly
because they can put constraints on the  phase boundaries  and the exact
location of some critical or multi-critical points. Thus
(self-)dualities have been essential for investigating the phase
diagrams of numerous models in statistical mechanics and field theory
\cite{savit,malyshev}. This aspect encompasses some of the most
spectacular applications  of dualities, and  constitutes the legacy of 
Kramers and Wannier. However, dualities can become even more potent when
fused with other tools, such as perturbation theory.  A case in point is
the AdS-CFT correspondence, a topic that  remains the focus of intense
research efforts.  Since its original formulation in high energy physics
\cite{maldacena, GKP, AdS-Witten, AdS-review},  this {\it conjecture of
a duality} between a weakly-coupled five-dimensional gravity theory (on
an {\it Anti de Sitter} (AdS) background)  and a strongly-coupled
four-dimensional conformal field theory (CFT) has been generalized and
exploited in other branches of physics. At present, the range of
applications of the AdS-CFT correspondence includes problems as diverse
as electronic transport properties \cite{electron}, quantum critical
dynamics \cite{understanding_qpt}, and the physics of strongly coupled
quark-gluon plasmas \cite{hydro}. The efforts to apply the AdS-CFT
correspondence to other strongly-coupled models continue, but the
problem of pushing it beyond a conjecture into a rigorous mathematical
statement  remains. We think that turning the AdS-CFT  correspondence
into a mathematically  rigorous  duality is essential to understanding
its  potential generalizations.

These examples (old and new) attest to the power of dualities and
justifies the efforts of numerous researchers to exploit them to address
hard problems by simple, elegant means. This article is a self-contained
exposition of extensive original  developments,  including many new
dualities and  self-dual models,  in the general bond-algebraic theory
of dualities first introduced in Reference \cite{con} (bond algebras
have also been employed in the  analysis and spectral resolution of
exactly solvable models  \cite{bondDec08,OQC}, but this application is
not discussed in this paper).  In the context of some specific models,
quantum dualities have been well understood for many years.  Reference
\cite{con} introduced a coherent framework supporting a systematic study
of both quantum {\it and} classical  dualities on an equal footing, and
providing a systematic way to compute dual (disorder) variables. The
new, unified, theory  that  emerges is rigorous, easy to use,
algorithmic, and of practical  significance to theoretical studies and
numerical simulations. It has  the potential to provide invaluable
insight into a myriad of pressing problems that are beyond perturbation
theory.  Our theory of dualities rests on a single key observation: The
{\it bonds  or interactions} of a Hamiltonian or transfer matrix are
more relevant to a  duality transformation than the elementary degrees
of freedom. Those bonds, or interaction terms, are  organized into a
{\it bond algebra}.  In contrast, symmetries or properties of the
elementary  degrees of freedom  are largely irrelevant  from a duality
mapping perspective \cite{bondDec08,con}.  Our bond-algebraic approach
to  dualities may shed light on problems like the characterization and
classification of  collective (topological) excitations of lattice
models, and the AdS-CFT correspondence \cite{noc}, when complemented
with recent  developments in the theory of dimensional reduction
\cite{batista_nussinov, tqo}. 

 
Hamiltonians that meet basic physical requirements like causality are
usually realized by adding together (or integrating with a given
measure  in QFT) simple local or  quasi-local operators, like few-body
interactions and kinetic energy single-body operators. These operators
(or some simple function of them) constitute the {\it bond operators},
or bonds for short. Together they generate, in a sense to be specified
in Section \ref{sec3.2}, an operator  von Neumann algebra that we call a
{bond algebra}.  Bond algebras are model-specific algebras of
interactions that can  be {\it radically different} from the algebras
that embody elementary degrees of freedom like bosons, fermions, or
spins. Our new theory states that dualities are structure-preserving 
mappings (homomorphisms) of bond algebras, {\it that are typically local
in the bonds}. This {\it purely algebraic} characterization is
physically meaningful because homomorphism of bond algebras  are {\it
always} equivalent to a unitary transformation  (or, if gauge symmetries
are involved, to a partial isometry) connecting the Hilbert spaces on
which the two models connected by a duality are defined.  Thus quantum
dualities are revealed as unitary transformations, and dual models must
share identical spectra and level degeneracies; that is, provided gauge
symmetries are not involved. The precise connection between gauge
symmetries and dualities will be one of the central themes of this
paper. 

Remarkably, these ideas can be extended to  include classical dualities 
\cite{con}, thus unifying the theory of  classical and quantum
dualities  in a way that had been overlooked up to now.  The key is to
notice that many problems in classical statistical physics can be
formulated in terms of a transfer matrix or operator. In this context,
physical requirements like locality become manifest in the {\it
multiplicative} structure of the transfer matrix or operator, that will
be in general a {\it product} of  local or quasi-local bonds. Once the
bonds are identified, the theory of dualities proceeds as before. Thus,
a duality mapping for the bond algebra of the transfer matrix realizes a
dual transfer matrix that determines a dual representation \(\pf^D\) of
the partition function \(\pf\). But, since dualities are  unitary
transformations, it follows that \(\pf \propto \pf^D\). This is, in
short,  the bond-algebraic approach to classical dualities. In this way
one can show, for instance, that the self-duality of the two-dimensional
classical Ising model \cite{KW, bookEPTCP},  and the self-duality of the
quantum Ising chain \cite{fradkin_susskind} are two manifestations of a
symmetry possessed by one and the same bond algebra. But more 
importantly, we will illustrate in this paper that the bond-algebraic
approach  has the potential to extend classical dualities beyond its
present boundaries, as set  for example in the reviews 
\cite{savit,malyshev} (see Appendix \ref{appA}).  We think that this
strongly indicates that our technique can be relevant  to solving the
open problem of constructing dual representations of non-Abelian 
lattice models, although so far we have only some  examples to  support
this claim. 

Bond algebras are most efficient in revealing dualities because they are
model (Hamiltonian/transfer matrix)-specific, and thus capture, in the
form of a local mapping, the precise information  that makes two
operators dual.  This is different from techniques based on essentially
fixed mappings, like generalized  Jordan-Wigner transformations
\cite{GJW}. These transformations are isomorphisms (that we call  {\it
dictionaries}) connecting operator algebras (i.e., {\it languages}),
typically representing elementary degrees  of freedom.  Since these
dictionaries are insensitive to the specific  structure of physical
models, they can easily spoil important physical  characteristics, as
illustrated by standard fermionization in more than one  space
dimension.   However, it is important to underscore that a bond algebra
is   {\it not} uniquely determined by the Hamiltonian/transfer matrix
that motivates its definition, but rather by the choice of bond
decomposition of those operators.  Since that choice is not unique, (i)
any single Hamiltonian may produce many different bond algebras, and
(ii) any one bond algebra may be related to many different Hamiltonians.
This flexibility is essential for classification purposes and  to the
success of our technique. It enables us to explore a variety of bond
algebras for one and the same Hamiltonian/transfer matrix to capture its
full range of dual  representations. Similarly, it allow us to apply one
and  the same duality mapping to different problems.   The self-duality
for the spin \(1/2\) XY model in a transverse field of Section
\ref{sec3.10.2} provides an  example of  the first observation (i)
above.
In this case, the self-duality {\it emerges} on a special line in
coupling space, and it takes a special choice of bonds to generate the
bond algebra that makes this self-duality apparent. The second
observation (ii)  is illustrated by the new dualities of Section 
\ref{sec3.6} for the quantum Heisenberg model in any number of space 
dimensions \(d\). These dualities for a {\it non-Abelian} model are
based  on dualities for the (Abelian!) $d$-dimensional quantum Ising
model  and highlight some difficulties with the standard notion of
non-Abelian dualities \cite{savit}. They point to the fact that the
concepts  of symmetry and duality are quite often inaccurately related. 
While symmetries of a theory represent isomorphisms leaving  the
Hamiltonian invariant, (self-)dualities do not  preserve the form of
the  theory, but rather preserve its spectrum and level degeneracies.

The relevance of bonds over elementary degrees of freedom  is further
emphasized by the fact that dualities are {\it local} transformations in
terms of bonds, while they are  {\it non-local} when described in terms
of elementary degrees of freedom.  This follows because one can invert
the relations between bonds and elementary degrees of freedom to obtain
the latter as non-local functions of the bonds. Once this is done,  one
can compute the action of the duality mapping on the elementary degrees
of freedom, to obtain their dual image.  Since dualities are
structure-preserving mappings, these {\it dual}  elementary degrees of
freedom are guaranteed to be equivalent to the original ones. This {\it
systematic derivation} of  (non-local in general) dual variables
(elementary degrees of freedom) from more basic, local objects is
extremely  important, not only because it establishes a bridge between
the bond-algebraic and the traditional approach  to quantum dualities,
but also because dual variables can have a fundamental physical meaning
as (generalized)  {\it disorder  variables}
\cite{Kadanoff_Ceva,fradkin_susskind}.

Starting with any specific bond algebra, one can proceed to look
systematically for its alternative realizations that feature local
representations of the bond generators. {\it Any of these realizations
defines a duality} (realizations that feature non-local representations
of the bond  generators can appear, but would typically be discarded in
practice). This simple premise can lead to surprising results, as we
will see often in this paper. Exact dimensional reduction can appear as
a duality, and, as illustrated by our new derivation of the
Jordan-Wigner  transformation, {\it statistical transmutation} can
appear as a duality too. {\it Symmetry transmutation} is also not
unusual as illustrated for example  by our new derivation of  the
duality between the  XY and solid-on-solid (SoS) models, obtained  here
for the first time  {\it without invoking the Villain approximation} 
({symmetry transmutation} here refers to the fact that the  \(U(1)\)
symmetry group of the XY model is not isomorphic,  but rather the
Pontryagin dual of the \(\mathbb{Z}\) symmetry group of the SoS model).
But perhaps  what is most surprising about bond-algebraic dualities is
their  capability to {\it dispose of gauge symmetries}, as we explain
next.

Gauge symmetries are constraints on the state Hilbert space,  encoded in
a large set of local operators that commute with the Hamiltonian, and
with any other (measurable) observable. Since bonds are in general
measurable, local observables, they generate a bond algebra of
gauge-invariant operators. This has the remarkable consequence that bond
algebras of gauge models can have dual representations on state spaces
of lower dimensionality  (a fact that should not be confused with an
{\it emergent} duality as described in Section \ref{sec3.11}). We call
these dualities gauge-reducing; they map the original gauge model to a
model with less, or simply without gauge symmetries, and are represented
by {\it homomorphisms} of bond algebras, rather than isomorphisms. 
Unlike ordinary dualities that are unitarily implementable, 
gauge-reducing dualities are implemented by partial isometries   (one
can think of a partial isometry as a rectangular unitary matrix, that
either maps a state to zero or to an isometric (equal norm) state).  But
in spite of these mathematical twists,  dualities are just as easy to
detect in gauge theories as in any other model. That is because, while
gauge symmetries do affect the  algebra of elementary degrees of
freedom, {\it they do not affect the algebraic relations between bonds}.
Thus bond-algebraic dualities may provide in some instances a practical
solution to the problem of eliminating  gauge constraints. This  paper
describes in detail the mathematics and multiple applications of these
ideas to models with {\it Abelian} gauge symmetries. {\it Non-Abelian}
gauge symmetries abide by the  same principles but are technically more
involved, so we defer their complete treatment to future publications. 

This article contains many new results and is organized as follows. 
Section \ref{sec2.1} starts with a discussion of the different physical
contexts in  which dualities have been introduced, followed in Section
\ref{sec2.2} by a more detailed discussion of what was known about
quantum dualities prior to the publication of Reference \cite{con}, the
so-called standard approach. These discussions should help the reader
put the subject of dualities, and some of the problems they address,  in
perspective. Section \ref{sec3} contains all the important formal
developments related  to the bond-algebraic approach to dualities.
Overall it is devoted  to quantum dualities and their intimate
connection to classical dualities, mainly  discussing every major new
idea and mathematical technique that is  of relevance. We start by
studying bond algebras in Sections \ref{sec3.2} and \ref{sec3.3}. Our
new definition of quantum dualities as mappings of bond algebras is
presented in Section \ref{sec3.4}, and the next section explains how the
standard approach  described in Section \ref{sec2.2} follows immediately
from the  bond-algebraic formalism (determination of dual variables). 
Section \ref{sec3.6} discusses critically the notion of non-Abelian
duality, in the light of new dualities for the quantum Heisenberg model,
and  Section \ref{sec3.7} develops new boundary conditions that preserve
duality properties {\it in finite size systems}.
Section \ref{sec3.8} expands, in more concrete 
terms, on the connection between bond algebra mappings and 
unitary transformations. The remaining  sections are devoted to deriving
and explaining  the precise relations between  self-dualities, standard
quantum symmetries, Section \ref{sec3.9},  and disorder variables,
Section \ref{sec3.10}, and the novel concept of emergent  duality,
Section \ref{sec3.11}. Disorder variables can be systematically
determined  from  bond-algebraic dual mappings and they share an
interesting relation  with topological excitations. Section
\ref{sec3.12} explains how bond-algebraic dualities afford a practical
way to eliminate gauge symmetries completely, and finally  Section
\ref{classical&quantum} explains how many classical dualities follow from
bond-algebraic techniques. 

The rest of the article is dedicated to unveil old and discover new
dualities  in a broad spectrum of problems of physical relevance. There
are so many  examples that in the following we only indicate a few of
them.  Sections \ref{sec4} and \ref{sec5} exemplify the many ideas ideas
and techniques developed in Section \ref{sec3} with self-dualities and
dualities in quantum lattice models, of arbitrary spatial dimensions,
mainly of interest in condensed matter  physics. For example, Section
\ref{sec5.5} describes a new duality for the extended Kitaev toric code
model in arbitrary space dimensions to a  well known model of
Hamiltonian lattice field theory,  the \(\mathbb{Z}_2\) Abelian Higgs
model \cite{fradkin_shenker}.   The application of bond-algebraic
techniques to QFT is developed in Section \ref{sec6}.
Sometimes we show results directly in the continuum but we want to 
stress the fact that the lattice Hamiltonian formalism is the most
convenient approach for interacting field theories. We study in full
detail  compact and non-compact versions of quantum electrodynamics in
various  spatial dimensions and make an attempt to introduce a version
of  quantum electrodynamics without vector potential.  In Section
\ref{sec6.4} we introduce a new family of self-dual models related to
the Abelian Higgs model, we prove the quantum St\"uckelberg model to  be
self-dual in two-dimensions in Section \ref{sec_stuckel},  and Section
\ref{6.8} discusses self-dual field  theories that display the
phenomenon of {\it dimensional reduction} of prime interest in the
theory of  topological quantum order  \cite{tqo}.

Section \ref{sec8} presents several problems of classical  statistical
mechanics whose duality properties are uncovered by our  bond-algebraic
approach. Section \ref{sec8.2} describes duality  properties of the
Ising model in the Utiyama lattice, while the  vector Potts model in two
dimensions is not only studied in great detail in  Section \ref{sec8.3}
but also it is shown how to modify it to make it  self-dual for
arbitrary couplings and states. Building upon these ideas we introduce a
new self-dual \(p\)-state approximation to four-dimensional lattice 
electrodynamics in Section \ref{sec8.4}. 
Section \ref{eightvertexM} establishes a link between the eight-vertex
model (in its Ashkin-Teller representation) and the quantum anisotropic
Heisenberg model, based on duality mappings rather than integrability.  
The general connection between
quantum and classical disorder variables is discussed in Section 
\ref{appB}. Finally, Section \ref{sec9} presents  several important
applications and consequences of dualities. Those include  general
properties of the spectrum of self-dual theories and the way to  extract
exact relevant  information in presence of phase transitions,  and most 
importantly a new way to look at {\it fermionization in arbitrary
spatial dimensions}  as a general duality mapping. In particular,  we
demonstrate how the Jordan-Wigner transformation is a consequence of bond
algebraic duality mappings  when these are applied to nearest-neighbor
spinless fermion and  spin $S=1/2$ systems on a chain. We further show,
how {\em no such duality can map} all (real-space) nearest-neighbor
spinless Fermi hopping terms to all spin $S=1/2$ exchange terms on
general lattices (and viceversa) in spatial dimensions $d>1$. That is,
in general, no extension of the $d=1$ Jordan-Wigner transformation that
connects all such individual {\it local terms} can appear in  higher
dimensions. We show, however, that notwithstanding it is possible to
fermionize spin systems in $d>1$ dimensions  (such as the $d=2$ quantum
Ising model) via a gauge reducing duality. Thus we further extend and
formalize, via the systematics of bond algebras, the scope  of known
systems that can be fermionized. In Section \ref{sec_tqo_dim} we examine
the connection between  the phenomenon of dimensional reduction and
duality in the context  of topological quantum order systems \cite{tqo}.
We also indicate the most natural way to classify these systems in terms
of so-called $d$-dimensional  gauge-like symmetries as opposed to using
a particular measure of  entanglement since no single measure can
uniquely characterize a quantum state.

The appendices, a total of five,  contain not only technical details
omitted in the  text for pedagogical reasons  but also some relevant
mathematical expressions not found in the literature, so we advice the
reader to consider consulting them if he or she  wants to get a deeper
understanding of the subject.

\section{Traditional approaches to dualities}
\label{sec2}

This short section discusses dualities in physics from a historical
perspective. Mainstream, traditional ideas about dualities in  
field theory, classical statistical mechanics and quantum mechanics
are summarized and illustrated with key examples.

\subsection{Dualities in perspective: What is a duality?}
\label{sec2.1}

The first appearance of the term {\it duality} 
can be traced back to the early days of 
electromagnetism  as a field theory.
In 1884, Heaviside recast Maxwell's equations in vector form
\cite{nahin}, in the absence of sources, 
\begin{eqnarray}
\nabla\cdot \vec{E}&=&0,\ \ \ \ \ \ \ \ \ \ \ \ \ \ \ \ \ \ \ \,
\nabla\cdot\vec{B}=0,\label{maxwell's}\\
\nabla\times\vec{E}&=&-\partial_t\vec{B},\ \ \ \ \ \ \ \ \ \ \ \
\nabla\times\vec{B}=\partial_t\vec{E}\nonumber
\end{eqnarray}
(in rationalized Heaviside-Lorentz units, and the speed of 
light $c=1$), and pointed out that the {\it duality} 
\begin{equation}\label{em_duality}
\vec{E}\  \rightarrow\  \hat{\vec{E}}=\vec{B},\ \ \ \ \ \ \ \ \ \ \ \
\vec{B}\  \rightarrow\  \hat{\vec{B}}=-\vec{E}
\end{equation}
maps solutions of these equations  to other solutions.
This notion of duality stresses {\it symmetries of 
the equations of motion} showing that different 
physical quantities are interchangeable.  
 
Two of Maxwell's equations can be solved by introducing a four-vector potential 
\(A_\mu\) (\(\mu,\nu=0,1,2,3\)), that is related to $\vec{E}=(E_1,E_2,E_3)$
and $\vec{B}=(B_1,B_2,B_3)$ through the antisymmetric field-strength tensor 
\begin{equation}
F_{\mu \nu}=\partial_\mu A_\nu-\partial_\nu A_\mu,
\ \ \ \ \ \ 
F_{\mu \nu}=
\begin{pmatrix}
0& E_1& E_2& E_3\\
-E_1& 0& -B_3& B_2\\
-E_2& B_3& 0& -B_1\\
-E_3& -B_2& B_1& 0
\end{pmatrix} .
\end{equation}
Then the remaining Maxwell's equations read (\(\eta={\sf diag}(1,-1,-1,-1)\))
\begin{equation}\label{meqa}
\eta^{\mu\nu}\partial_\mu F_{\nu \gamma}=(
\eta^{\mu\nu}\partial_\mu\partial_\nu)A_\gamma-\partial_\mu(\eta^{\nu\gamma}
\partial_\nu A_\gamma)=0\ .
\end{equation}
This formalism is essential for well know reasons,
but the (self-)duality of Maxwell's equations 
is not readily reflected by Equation \eqref{meqa}, partly
because the gauge transformation, with arbitrary scalar function $\chi(\x)$, 
\begin{equation}
A_\mu(\x) \mapsto {A'}_\mu(\x)= A_\mu(\x)+\partial_\mu \chi(\x), 
\end{equation}
leaves the EM field \(F_{\mu \nu}\) unchanged but renders \(A_\mu\)
unobservable. This motivates one of this paper's central theme: the 
interplay between dualities and gauge symmetries.

A  second, seemingly very different, concept of duality was developed 
in the early 1940s in the context
of {\it classical} statistical mechanics. It originated in
the work of Kramers and Wannier \cite{KW} on the  
two-dimensional ($D=2$) Ising model on a square lattice,
with partition function
\begin{equation}
\mathcal{Z}_{\sf I}(K)=\sum_{\{\sigma_\r\}}\ \exp \Bigl[K
\sum_\r\sum_{\nu=1,2}\ \sigma_{\r+\bm{e_\nu}}\sigma_\r\ \Bigr] .
\end{equation}
Here $\sigma_\r=\pm 1$ stands for a classical spin degree of freedom 
located on the vertex (site) $\r=(r^1,r^2)=r^1 \bm{e_1} +r^2  \bm{e_2}$, $r^1,r^2\in
\mathbb{Z}$,  of a
square lattice with unit vectors $\bm{e_1}, \bm{e_2}$, see Figure 
\ref{notation_links}. The exchange constant is $J=K k_BT$, with $T$ 
temperature and $k_B$ Boltzmann's constant. 
Kramers and Wannier  noticed that \(\pf_{\sf I}\) satisfies a relation
\begin{equation}\label{first_time_statistics}
\pf_{\sf I}(K)=A(K,K^*)\pf_{\sf I}(K^*),
\end{equation}
that, since
\begin{equation}\label{critical_K_ising}
K^*=-\frac{1}{2}\ln \tanh K,
\end{equation}
connects the high-temperature (weak \(K\)) behavior of the Ising
model to its low temperature (strong $K$) behavior 
(\(A(K,K^*)\) is a known, analytic (non-singular) proportionality factor). 
It follows that every singularity 
of  \(\mathcal{Z}_{\sf I}\) at coupling  \(K\) must be matched
by another singularity at a corresponding dual coupling \(K^*\), so that if 
\(\pf_{\sf I}\) has only one phase transition (one singularity) at
a critical coupling \(K_c\), it must be located at the self-dual point 
\(K_c=K_c^*\). Then from Equation \eqref{critical_K_ising}
\begin{equation}
K_c=\frac{1}{2}\ln(1+\sqrt{2}),
\end{equation}
which determines the {\it exact} critical temperature of the Ising model
$T_c=2/\ln(1+\sqrt{2})$ (in units of $J/k_B$).

This spectacular result drew considerable attention to duality
transformations,
and soon after the publication of Reference \cite{KW}, Wannier 
\cite{wannier} showed how to obtain duality relations of the form
\begin{equation}\label{genidclass}
\pf_{\Lambda}(K)=A(K,K^*)\pf_{\Lambda^*}(K^*),
\end{equation}
for Ising models defined on nearly arbitrary planar lattices \(\Lambda\). 
These transformations where dubbed dualities \cite{wannier}
because \(\Lambda^*\) stands for the lattice dual to \(\Lambda\) 
({\it dual} here is used in the sense of old algebraic topology, see for 
instance, Reference \cite{bredon}, Chapter IV, Section 6.)
Dual lattices are defined in Appendix \ref{appA}.
We see from Equation \eqref{genidclass} that the Ising model on a square 
lattice is self-dual partly because so is the square lattice, 
while the Ising model on  a triangular lattice is dual to that same model 
on a hexagonal lattice (and viceversa). This last duality 
is part of a simple approach to determining these models' critical 
temperature \cite{baxter}. 

Potts extended the duality transformation of Wannier to some of the models now known 
under his name in his 1952 paper \cite{potts}.
It was Wegner, however, who uncovered the general structure underlying
dualities, by showing in his 1973 paper 
\cite{wegner} that duality transformations 
could be obtained for a wide class of models
(including models in more than two dimensions) by considering the (group-theoretical)
Fourier transform of individual Boltzmann weights. 
The 1975 paper  \cite{wu_wang} popularized Wegner's approach to
the point that it became the standard, traditional approach to classical dualities. 
This traditional approach is described thoroughly in the
review articles by Savit \cite{savit} and Malyshev and Petrova 
\cite{malyshev}, but it is also summarized in Appendix \ref{appA} 
(see Reference \cite{bookEPTCP} for a pedagogical introduction).

The key fact that the notion of a Fourier transform exists for {\it any}
group allows for application of the traditional approach to produce dual representations
of any {\it Abelian} model displaying appropriate interactions, but it is impossible
to take the same traditional approach to produce dual representations
for lattice {\it non-Abelian} models (that is, models like
Wilson's non-Abelian lattice gauge theories). To construct dual representations
of such models remains one of the most difficult open problems in the 
theory of dualities \cite{savit} (let us point out that {\it Abelian} and
{\it non-Abelian} here does not refer to the symmetry group of the model of
interest, but rather to the nature of that model's degrees of freedom, 
see Section \ref{sec3.6}).
 
Yet a third concept of duality was developed for 
{\it quantum} many-body problems in the late 1970s, mainly in a series
of papers  \cite{fradkin_susskind,horn_yankielowicz,horn,kogut,shankar}
concerned  with Hamiltonian lattice
quantum field theory (LQFT). Starting with Reference \cite{fradkin_susskind}, 
these papers settled a notion of 
quantum duality that we interpret and formalize as follows:
two Hamiltonians  
\begin{equation}
\label{h1h2ok}
H_1[\lambda_1,\lambda_2,\cdots](o_\Gamma),\ \ \ \ \ \ \ \ \
H_2[\lambda_1,\lambda_2,\cdots](o_\Gamma),
\end{equation}
that feature coupling constants $\lambda_\nu$, $\nu=1,2,\cdots$, 
and elementary degrees of freedom $\{{o}_\Gamma\}$ labelled by some labels \(\Gamma\)
are {\it dual} if there
exists an {\it  alternative representation} $\{\hat{o}_\Gamma\}$ of the
algebra of the $\{{o}_\Gamma\}$ such that
\begin{equation}\label{d_condition}
H_1[\lambda_1,\lambda_2,\cdots](o_\Gamma)=
H_2[\lambda_1^*,\lambda_2^*,\cdots](\hat{o}_\Gamma).
\end{equation}
That is, \(H_1\) and \(H_2\) are dual if they are equal up to  an {\it
operator change of variables} \(o_\Gamma \rightarrow \hat{o}_\Gamma\), together
with a readjustment  of couplings \(\lambda_\nu\rightarrow\lambda_\nu^*\). 

In summary, we have described notions of duality that arose 
in three different fields 
of physics. It is not clear a {\it a priori} that they are related
beyond general, conceptual features, but we will show in this paper
that there is a common mathematical background underlying all three of them.

\subsection{The traditional approach to quantum dualities}  
\label{sec2.2}

As described near the end of the last section, quantum dualities 
have the following far reaching consequence: If Equation
\eqref{d_condition} holds, it must be that the energy spectra of the
two dual Hamiltonians satisfy
\begin{equation}\label{same_levels}
E_1(\lambda_1,\lambda_2,\cdots)=E_2(\lambda_1^*,\lambda_2^*,\cdots).
\end{equation}
This suggests that a quantum duality may be a unitary equivalence of the form
\begin{equation}\label{d_as_u}
\mathcal{U}_{\sf d}\, H_1[\lambda_1,\lambda_2,\cdots](o_\Gamma)\,
\mathcal{U}^{\dagger}_{\sf d}=H_2[\lambda_1^*,\lambda_2^*,\cdots](\hat{o}_\Gamma),
\end{equation}
that would imply Equation \eqref{same_levels} right away. Surprisingly, 
the definition of quantum dualities just introduced 
seems to be incompatible with Equation \eqref{d_as_u}, as we
will explain below.

Let us consider a specific, non-trivial example in detail, the quantum,
one-dimensional ($d=1$)  Ising model in a transverse field 
(or ``quantum Ising chain'' for short), specified by the Hamiltonian
\begin{equation}
H_{\sf I}[h,J](\sigma)=\ \sum_i\ (h\sigma^x_i+\ J\sigma^z_i\sigma^z_{i+1})
\label{infinite_ising_transverse}.
\end{equation}
\(H_{\sf I}\) features  \(S=1/2\) spins located at each
site $i\in\mathbb{Z}$ of a chain, represented
by Pauli matrices \(\sigma^x_i, \sigma^z_i\). They constitute
the elementary degrees of freedom that were denoted by \(o_\Gamma\)
near the end of last section. The goal is to show, along the lines of 
Equation \eqref{d_condition} that \(H_{\sf I}\) is 
self-dual (that is, dual to itself). So
we must find a new representation $\mu^x_i,\ \mu^z_i$ of the Pauli matrices
and new values \(J^*,\ h^*\) of the couplings so that \(H_{\sf I}[h,J](\sigma)=
H_{\sf I}[h^*,J^*](\mu)\). Since we know from the exact solution of \(H_{\sf I}\)
\cite{pfeuty} that its  energy levels are symmetric in \(J,\ h\) (so that
\(E_{\sf I}(J,h)=E_{\sf I}(h,J)\)), we set \(J^*=h\) and \(h^*=J\).
We are left then with the problem of finding an appropriate dual representation 
of the Pauli matrices.

The equality
\begin{equation}\label{sdisingold}
\sum_{i}\ (h\sigma^x_i+J\sigma^z_i\sigma^z_{i+1})\ =\ \sum_{i}\ (J\mu^x_i
+h\mu^z_i\mu^z_{i+1}),
\end{equation}
suggests setting up the relations,
\begin{eqnarray}
\label{ldis}
\mu^z_i\mu^z_{i+1}&=&\sigma^x_{i+m},\ \ \ \ \ \ \ \  \ \ \ \ \ \  \
m=?\nonumber\\ \mu^x_i&=&\sigma^z_{i+m'}\sigma^z_{i+1+m'}, \ \ \ \ m'=?.
\end{eqnarray}
As underscored by the question marks in Equation \eqref{ldis}, we have
to decide what $m$ and $m'$  should be. The obvious choice $m=m'=0$
leads to
\begin{equation}\label{baddualising}
\mu^z_i\mu^z_{i+1}=\sigma^x_{i},\ \
\ \ \ \ \ \ 
\mu^x_i=\sigma^z_{i}\sigma^z_{i+1}.
\end{equation}
On the other hand, if the new spin variables $\mu$ exist at all,
they must satisfy ${(\mu^z_i)}^2=1$.  Thus,
\begin{equation}\label{mustring}
\mu^z_i=\mu^z_i\mu^z_{i+1}\times
\mu^z_{i+1}\mu^z_{i+2}\times\cdots=\prod_{m=i}^\infty\sigma^x_m.
\end{equation}
But then we see from Equations \eqref{baddualising} and \eqref{mustring} that
\(\mu^x_i\) {\it commutes} with \(\mu^z_i\).  
Let us set then \(m=0\) and \(m'=-1\) in Equation \eqref{ldis}, so that
\begin{equation}
\mu^x_i=\sigma^z_{i-1}\sigma^z_i\ \ \ \ \ \ \mu^z_i\mu^z_{i+1}=\sigma^x_i.
\end{equation}
The solution to {\it this} set of equations is
\begin{equation}
\label{muxmuz}
\mu^x_i=\sigma^z_{i-1}\sigma^z_i,\ \ \ \ \ \ \mu^z_i=
\prod_{m=i}^\infty\sigma^x_m, 
\end{equation}
and now \(\mu^x_i\), \(\mu^z_i\) do satisfy the correct spin-1/2 algebra. 
This completes the proof in the traditional approach to quantum dualities
that $H_{\sf I}$ is self-dual \cite{fradkin_susskind}.

Admitting that this example is a fair representation of an
``average" quantum duality, we can infer that
\begin{enumerate}
\item{quantum dualities need not be strong-coupling/weak coupling
relations;}
\item{quantum dualities are {\it ``fundamentally" non-local},}
\item{quantum dualities are {\it not} unitarily implementable.}
\end{enumerate}
The last statement follows from this simple observation. Suppose
we could recast the self-duality of the Ising chain as a unitary
equivalence \(\mathcal{U}_{\sf d}H_{\sf I}[h,J]\mathcal{U}_{\sf d}^\dagger=
H_{\sf I}[J,h]\). Then we would have that
\begin{equation}
\label{paradox}
\mathcal{U}_{\sf d} H_{\sf I}[0,J]\mathcal{U}_{\sf d}^\dagger=
\mathcal{U}_{\sf d}\left(\sum_i\ J\sigma^z_i\sigma^z_{i+1}\right)
\mathcal{U}_{\sf d}^\dagger=H_{\sf I}[J,0]=\sum_i\ J\sigma^x_i,
\end{equation}
but this cannot possibly be right, because \(H_{\sf I}[0,J]\)
and \(H_{\sf I}[J,0]\) {\it have different level degeneracies}.
This is not to say, however, that the traditional approach, based
on operator changes of variables, goes successfully
beyond unitary equivalence. Equation \eqref{sdisingold}
implies that 
\begin{equation}
\sum_i\ J\sigma^z_i\sigma^z_{i+1}=\sum_i\ J\mu^x_i.
\end{equation}
If the $\mu^x_i,\mu^z_i$ are truly an alternative representation of the
Pauli matrices, this equation cannot possibly be right either, for the
same reasons as before. 


\section{Bond-algebraic approach to quantum dualities}
\label{sec3}

This section is devoted to explaining our theory of quantum and classical 
dualities based on bond algebras, and it
discusses every major new idea and mathematical technique that 
we introduce to the subject (some of them  advanced in Reference \cite{con}).
We will argue that
\begin{enumerate}
\item{quantum and (a very large class of) classical dualities are unitary 
equivalences (or projective unitary
equivalences if the duality eliminates gauge symmetries), and that}
\item{the easiest way to search for dualities 
is to look for structure-preserving mappings between Hamiltonian-dependent 
{\it bond algebras}, Section \ref{sec3.8}.}
\end{enumerate}
Conceptually, (1) is more important than (2), yet mathematically it
follows from (2), as basically does everything else in this paper. 




\subsection{Bond algebras and the concept of locality}
\label{sec3.2}

It is a basic fact of physics that the Hamiltonian of a system determines
its dynamics and thermodynamics (some important consequences of this
statement are reviewed in Appendix \ref{sec3.1}). 
Bond algebras \cite{bondDec08} were devised to exploit 
a simplifying feature common to most Hamiltonians, and rooted in
fundamental physical principles: {\it Hamiltonians are  sums
of (possibly a huge number of) simple, local terms} 
({\it local} is used in a broad sense in this paper, either to indicate 
that interactions are local in space/space-time in the sense of field
theory, or that they involve only a few degrees of freedom). 
Take for example the Hamiltonian for $N$ electrons of 
charge $e$ and mass $m$ in an external potential,
\begin{equation}
\label{general_cond-mat}
H_e=\sum_{i=1}^{N}\left(\frac{1}{2m}{\bm p}_i^2+V(\x_i)\right)
+\sum_{i\neq j}\frac{e^2}{\vert \x_i-\x_j\vert}.
\end{equation}
The elementary degrees of freedom \(x_i^\mu, p_i^\nu\), satisfy
the Heisenberg relations
\begin{equation}\label{heisenberg}
[x_i^\mu,p_i^\nu]=i\hbar\delta_{\mu,\nu},\ \ \ \ \ \ \mu,\nu=1,2,3,
\end{equation}
or commute otherwise (from now on, we set \(\hbar=1\)).
This is in no way specific to the problem at hand (understanding
\(H_e\)), but is just a general fact. On the other
hand, \(H_e\) is the sum of \(2N+ N(N-1)/2\)  individual operators
\begin{equation}
{\bm p}_i^2,\ \ \ \ \ \ V(\x_i),\ \ \ \ \ \ \frac{1}{\vert \x_i-\x_j\vert},
\end{equation}
that we would like to consider altogether on an equal footing, and so we 
call them generically {\it bonds}. Taken individually bonds
are still elementary to understand, but, in contrast to elementary degrees of 
freedom, they satisfy algebraic relations that are {\it specific to the 
problem at hand}. Also, there is
{\it an algebraic notion of connectivity} for bonds that 
reflects  locality, in the sense that if two bonds commute, they can
only influence each other {\it indirectly}.  
Thus the algebraic relations between bonds strike a balance
between general physical principles and model specificity 
suggesting that algebras of bonds  (as opposed to algebras of elementary 
degrees of freedom) could potentially become important mathematical 
tools. 
This idea was pioneered in  Reference 
\cite{bondDec08}, where algebras of bonds where exploited to solve exactly 
several quantum lattice models of interest in the context of topological quantum order.
Let us introduce next the formal definition of a {\it  bond algebra}.

Consider a Hamiltonian operator $H$, written as
a sum of bond operators  $h_\Gamma$
\begin{equation}
H=\sum_\Gamma\ \lambda_\Gamma\ h_\Gamma.
\label{bae}
\end{equation}
with c-number coupling constants \(\lambda_\Gamma\).
The index ``\(\Gamma\)'' is completely general. It could stand for
a particle index, or for 
a site, a link, or some other subregions of a lattice $\Lambda$, or may
denote a point \(\x\) in space or a Fourier mode,
or may stand for any other suitable label one can think of.  
\begin{definition}\label{defba}
A bond algebra for the Hamiltonian \(H=\sum_\Gamma\ \lambda_\Gamma\ h_\Gamma\)
with bond decomposition \(\{h_\Gamma\}_\Gamma\),
is the von Neumann algebra ${\cal A}\{h_\Gamma\}$ generated
by the bonds.
\end{definition}
The basic mathematical aspects of this definition (including the
definition of a von Neumann algebra) are discussed in the next section.
The rest of this section is devoted to an informal discussion of bond algebras.
 
Intuitively speaking, \({\cal A}\{h_\Gamma\}\) is an algebra of 
operators generated by taking all possible {\it finite}, complex, linear 
combinations of 
powers and products of bonds, their Hermitian conjugates, and the 
identity operator \(\mathbb{1}\),
\begin{eqnarray}\label{oo}
\{\mathbb{1}, h_\Gamma,  h_\Gamma^\dagger, 
h_\Gamma h_{\Gamma'},  h_\Gamma^\dagger h_{\Gamma'},
h_{\Gamma'}^\dagger h_\Gamma, h_{\Gamma'}^\dagger h_\Gamma^\dagger, 
h_\Gamma h_{\Gamma'} h_{\Gamma''}, \cdots \ \}.
\end{eqnarray}
So by construction, if an operator   $\mathcal{O}
\in  {\cal A}\{h_\Gamma\}$ then $\mathcal{O}^\dagger
\in  {\cal A}\{h_\Gamma\}$ as well. This intuitive picture of bond
algebras suffices to understand most of the rest of the paper.

It is important to understand that a bond algebra 
is {\it not} determined by a Hamiltonian \(H\), but rather
by its bond decomposition \(\{h_\Gamma\}_\Gamma\). Any single 
Hamiltonian may produce many
different bond algebras, since different decompositions  
\begin{equation}
H=\sum_\Gamma\ \lambda_\Gamma\ h_\Gamma=\sum_\Sigma\ {\lambda'}_\Sigma 
\ {h'}_\Sigma
\end{equation}
define different, equally valid sets of  bonds that can potentially 
generate very different bond algebras. Conversely,
any one bond algebra may be related to many different Hamiltonians.
Consider for illustration the single site spin Hamiltonian
\begin{equation}
H_{\sf I}=h_x\sigma^x+h_y\sigma^y.
\end{equation} 
One can take \(\sigma^{x}\) and \(\sigma^{y}\)
as generating bonds, or the single bond
$(h_x\sigma^x+h_y\sigma^y)$. Then we get 
two bond algebras that are clearly different, 
\begin{equation}\label{hxhyex}
{\cal A}\{\sigma^x,\sigma^y\}\ \neq\ {\cal A}\{h_x\sigma^x+h_y\sigma^y\}, 
\end{equation}
since \({\cal A}\{h_x\sigma^x+h_y\sigma^y\}\) is commutative while 
\({\cal A}\{\sigma^x,\sigma^y\}\) is not.
This flexibility of the concept of bond algebra turns out to be an 
essential advantage. 
{\it Applications dictate what bond decomposition is best for
any given problem}.

The complexity of a bond algebra can vary, and a practical measure
of that complexity is simply afforded by considering
the Hilbert space \(\mathcal{H}\) on which the bond algebra acts on.  
Then one can recognize three increasingly difficult (in the number of 
resources) scenarios:
\begin{enumerate}
\item{\(\mathcal{H}=\mathcal{H}_1\otimes\cdots\otimes\mathcal{H}_N\), where each factor 
in the tensor product is finite dimensional;}
\item{\(\mathcal{H}=\mathcal{H}_1\otimes\cdots\otimes\mathcal{H}_N\), where some or all
factors are infinite dimensional;}
\item{ \(\mathcal{H}=\bigotimes_{\alpha\in I} \mathcal{H}_\alpha\), 
where the index set \(I\) is infinite, so that \(\mathcal{H}\) is an {\it
infinite tensor product} \cite{vonneumannII}. For example, 
\(\mathcal{H}=\bigotimes_{i\in\mathbb{Z}} \mathbb{C}^2_i\)}.
\end{enumerate}
The first scenario (1) is elementary (bond algebras are then just matrix 
algebras), and (2) is moderately simple, 
but (3) is directly connected to the thermodynamic limit
and/or  the continuum limit of QFTs \cite{sewell,barton},
and it is the source of endless fascination and complications. 
In practice, problems in (3) must be regularized (turned into problems
in (1) or (2)) before any progress can be made (see 
Appendix \ref{appE} on LQFT). But, as long
as bonds are chosen to be {\it local} (specifically 
in the sense that they act non trivially only
on a {\it finite} number of factors), bond algebras are perfectly well defined 
\cite{vonneumannII} even if the state space were as complicated an object
as (3) above.

Let us notice next that the generators listed in 
Equation \eqref{oo} need not be in general linearly independent. 
Then one can find a (potentially much) smaller basis 
\(\{\mathcal{O}_{\alpha}\}\) for the bond algebra \({\cal A} \{h_\Gamma\}\), 
and  decompose products of bonds as
\begin{equation}
h_{\Gamma_1}\cdots h_{\Gamma_N}\cdots=
\sum_\alpha\ c_\alpha{\cal{O}}_{\alpha}.
\end{equation} 
It is interesting to recognize that bond algebras must have a basis, because
this shows that a kind of ``reducibility hypothesis" \cite{kadanoff,Kadanoff_Ceva}, or
``operator product expansion"  formula 
\begin{eqnarray}\label{Aijk}
{\cal{O}}_{\alpha} {\cal{O}}_{\beta} = \sum_{\gamma}
 A_{\alpha \beta}^\gamma\ {\cal{O}}_{\gamma}
\end{eqnarray}
holds. The  c-numbers \(A_{\alpha \beta}^\gamma\) are {\it structure constants} 
for the bond algebra. 

We can get a feeling for the physical meaning 
of the structure constants by taking the expectation value of Equation \eqref{Aijk} 
(vacuum expectation value, or thermal
average, etc.),
\begin{equation}\label{algebra_averages}
\langle{\cal{O}}_{\alpha} {\cal{O}}_{\beta}\rangle = \sum_{\gamma}
 A_{\alpha \beta}^\gamma\ \langle{\cal{O}}_{\gamma}\rangle.
\end{equation}
This shows that bond algebras afford a partial realization of the idea
of algebras of fluctuating variables (see \cite{kadanoff, Kadanoff_Ceva}, 
and references therein).
{}From a different perspective, the structure constants 
\(A_{\alpha \beta}^\gamma\) can be seen 
as generalized constants of motion.
In the Heisenberg picture, the basis of the bond algebra evolves as
\begin{equation}
\mathcal{O}_{\alpha}(t)=\mathcal{U}(t)^\dagger\, \mathcal{O}_{\alpha}\,
\mathcal{U}(t),
\end{equation}
where \(\ \mathcal{U}(t)=\widehat{T}\ e^{-i\int_0^t H dt'}\), and $\widehat{T}$
is the  time-ordering symbol. Then,
\begin{equation}
{\cal{O}}_{\alpha}(t) {\cal{O}}_{\beta}(t)\ =\ \sum_{\gamma}
A_{\alpha \beta}^\gamma\ {\cal{O}}_{\gamma}(t)
\end{equation}
{\it with the same structure constants} as in Equation \eqref{Aijk} at
time $t=0$.

\subsection{Some mathematical aspects of bond algebras}
\label{sec3.3}

By definition, bond algebras are von Neumann algebras of operators. 
In this section we spell out the meaning and far reaching 
consequences of this requirement. We start by recalling the definition of a 
von Neumann algebra \cite{vonneumannI,jones}.
Let \(\mathcal{H}\) be a Hilbert space (the space of quantum states), 
and let \(B(\mathcal{H})\) denote
the algebra of bounded operators on \(\mathcal{H}\) (an operator \(\mathcal{O}\)
is bounded if there is some number \(0\leq C<\infty\) such that 
\(||\mathcal{O} v||\leq C||v||\), for every vector \(v\in \mathcal{H}\)). 
If \(\mathcal{S}\subset B(\mathcal{H})\) is an arbitrary subset, its commutant
\(\mathcal{S}'\in B(\mathcal{H})\) is the subalgebra defined by
\begin{equation}
\mathcal{S}'=\{\mathcal{O}\in B(\mathcal{H})\ |\ 
\forall\mathcal{R}\in\mathcal{S},\ \ \ 
\mathcal{O}\mathcal{R}=\mathcal{R}\mathcal{O}\}.
\end{equation}
\begin{definition}
A subalgebra \({\cal A}\subset B(\mathcal{H})\) is a {\it von Neumann algebra}
if it satisfies three algebraic conditions \cite{jones}:
\begin{itemize}
\item{It contains the identity operator, \(\mathbb{1}\in{\cal A}\).}
\item{ It is closed under Hermitian conjugation, if \(\mathcal{O}\in{\cal A}\),
then \(\mathcal{O}^\dagger \in  {\cal A}\) as well.}
\item{It is equal to its bycommutant, \({\cal A}={\cal A}''\).}
\end{itemize}
\end{definition}
Since von Neumann algebras are algebras of {\it bounded} operators,
the sense in which a bond decomposition \(\{h_\Gamma\}_\Gamma\)
generates a (von Neumann) bond algebra \(\mathcal{A}\{h_\Gamma\}\)
varies according to whether the bonds are bounded operators or not.
If the bonds are all bounded operators, the bond algebra they generate
is simply the smallest von Neumann algebra \({\cal A}\{h_\Gamma\}\subset
B(\mathcal{H})\) that contains every bond.
If the bonds are {\it not} all bounded operators, this notion needs to be refined.
An operator \(\mathcal{O}\) (not necessarily bounded) is {\it affiliated}
to a von Neumann algebra \(\mathcal{A}\) if it commutes \(\mathcal{O}U=
U\mathcal{O}\) with every unitary operator \(U\in\mathcal{A}'\). 
Every operator that is affiliated and bounded belongs to $\mathcal{A}$.
This notion is useful for the following reason.
Suppose \(\mathcal{O}\) is unbounded and affiliated to \(\mathcal{A}\), and
suppose also that \(\mathcal{O}\) admits an spectral decomposition, so that
we can construct operators \(f(\mathcal{O})\) that are functions of \(\mathcal{O}\)
in the usual way. Then one can show \cite{vonneumannI} that every {\it bounded} 
\(f(\mathcal{O})\) {\it is} an operator  in \(\mathcal{A}\), 
\(f(\mathcal{O})\in\mathcal{A}\), 
even though \(\mathcal{O}\) itself is not. So we define: 
If the set of bonds generators \(\{h_\Gamma\}_\Gamma\)
includes unbounded operators, then the bond algebra they generate
is the {\it smallest} von Neumann algebra \({\cal A}\{h_\Gamma\}\subset
B(\mathcal{H})\) such that every bond \(h_\Gamma\) is affiliated to 
\({\cal A}\{h_\Gamma\}\) (such an algebra always exists \cite{sunder}).  
In summary, whatever the nature of the bonds may
be, their bond algebra is a convenient (since it contains only bounded operators)
yet faithful representative of the structure of the interactions that are
embodied in the bonds. 

A mapping of von Neumann algebras \(\Phi:\mathcal{A}_1\rightarrow\mathcal{A}_2\)
is an {\it homomorphism} if 
\begin{eqnarray}
\Phi(\mathbb{1})&=& \mathbb{1},\ \ \ \ \ \ \ \ \ \ \ \ \ \ \ \ \ 
\ \ \ \ \ \ \ \ \ \ \ \ \, \Phi(\mathcal{O}^\dagger)=
\Phi(\mathcal{O})^\dagger,\label{homomorphism}\\
\Phi({\cal O}_1{\cal O}_2)&=&\Phi({\cal O}_1)\Phi({\cal O}_2),
\ \ \ \ \ \ \Phi({\cal O}_1+\lambda {\cal O}_2)=
\Phi({\cal O}_1)+\lambda\Phi({\cal O}_2).\nonumber
\end{eqnarray}
If \(\Phi\) is also one-to-one and onto, then it is called
an {\it isomorphism}. As mentioned before, one of the main goals of this paper is 
to establish a connection between dualities and unitary transformations. The following
theorem \cite{jones} then explains to a great extent our insistence in embedding bonds
in a von Neumann algebra.
\begin{theorem}\label{isoarespatial}
Let \(\mathcal{A}_i\) be von Neumann algebras of operators on the Hilbert spaces
\(\mathcal{H}_i\), for \(i=1,2\). If \(\Phi:\mathcal{A}_1\rightarrow\mathcal{A}_2\)
is an isomorphism, then there exists
\begin{itemize}
\item{a Hilbert space \(\mathcal{M}\), and}
\item{a unitary transformation \(\mathcal{U}:\mathcal{H}_1\otimes\mathcal{M}
\rightarrow\mathcal{H}_2\otimes\mathcal{M}\) such that}
\end{itemize}
\begin{equation}\label{spatial}
\Phi({\cal O})\otimes\mathbb{1}=\mathcal{U}({\cal O}\otimes\mathbb{1})\mathcal{U}^\dagger ,
\end{equation}
\end{theorem}
\noindent
where \(\mathbb{1}\) 
stands for the identity operator on \(\mathcal{M}\). In this paper we will often
be able to take \(\mathcal{M}=
\mathbb{C}\), so that Equation \eqref{spatial} simplifies to 
\(\Phi({\cal O})=\mathcal{U}{\cal O}\mathcal{U}^\dagger\). Then we say that \(\Phi\)
is unitarily implementable.

\subsection{Dualities as isomorphisms of bond algebras}
\label{sec3.4}

In this section we introduce and illustrate our definition of 
quantum duality based on bond algebras. It will be refined in Section 
\ref{sec3.12} to include models with gauge symmetries. Classical
dualities will be defined similarly in
Section \ref{classical&quantum}, after we discuss
how to associate bond algebras to classical models
of statistical mechanics. 

Our new approach to dualities is based on the recognition that, if we exclude
models with gauge symmetries for the moment, 
{\it quantum dualities are isomorphisms of bonds algebras}. More precisely \cite{con}, 
\begin{definition}\label{algebraically_dual}
Two Hamiltonians $H_{1}$ and $H_2$ are dual if there is a bond algebra
\(\mathcal{A}_{H_1}\) for \(H_1\) isomorphic to some bond algebra
\(\mathcal{A}_{H_2}\) for \(H_2\), and if the isomorphism 
$\Phi_\d:\mathcal{A}_{H_1}\rightarrow\mathcal{A}_{H_2}$ maps 
\(H_1\) to \(H_2\). 
\end{definition}
Since \(H_1\) and \(H_2\) are self-adjoint, Equation \eqref{spatial}
implies that these Hamiltonians share identical spectra and level degeneracies, 
and so
\begin{equation}\label{isoduu}
H_2=\mathcal{U}_\d H_1\mathcal{U}^\dagger_\d.
\end{equation}
A Hamiltonian \(H[\lambda]\) which depends on some set
of coupling parameters $\lambda=(\lambda_1,\lambda_2,\cdots)$ is
self-dual if it is dual to itself, up to a change in the coupling
$\lambda\rightarrow\lambda^*$, with \(\lambda^*\) the {\it dual couplings}. 
Notice that by Equation \eqref{homomorphism}, a bond algebra
homomorphism $\Phi_\d$ preserves the equations of motion of an arbitrary
observable ${\cal O}$
\begin{equation}\label{heisenberg_covariant}
\frac{d {\cal O}}{dt}-i[H_1,{\cal O}]=0\ \ \ \ \stackrel{\Phi_\d}{\longrightarrow}\ \ \ \
\frac{d\Phi_\d({\cal O})}{dt}-i[H_2,\Phi_\d({\cal O})]=0.
\end{equation}

Now that we have a precise definition of duality, we need to:
\begin{itemize}
\item{show that it includes the known dualities, and}
\item{show that it is useful.}
\end{itemize}
To start with, let us show that the 
quantum Ising chain is self-dual in the sense of definition \ref{algebraically_dual}.
Take the basic bonds in \(H_I\) of Equation \eqref{infinite_ising_transverse}
to be $ \{\sigma^z_i\sigma^z_{i+1}\},\{ \sigma^x_i \}$.
They generate a bond algebra $\mathcal{A}_{\sf I}$
that we can characterize in terms of relations:
\begin{enumerate} \label{relations}
\item{\((\sigma^z_i\sigma^z_{i+1})^2=\mathbb{1}=(\sigma^{x}_i)^2\);}
\item{Any bond $\sigma^x_i$ {\it anti-commutes} with two other 
bonds, $\sigma^z_{i-1}\sigma^z_i$ and $\sigma^z_{i}\sigma^z_{i+1}$, and 
{\it commutes} with all other bonds;}
\item{Any bond $\sigma^z_{i}\sigma^z_{i+1}$ {\it anti-commutes} with two
other bonds, $\sigma^x_i$  and $\sigma^x_{i+1}$, and {\it commutes} with
all other bonds.}
\end{enumerate}
We will assume that these relations characterize the bond algebra.
While this may seem plausible it is far from obvious, since it is not 
hard to argue that $\mathcal{A}_{\sf I}$ is reducible. There
are, however, consistency checks that we can run on the results
that we will obtain from this assumption. Also let us point out that the bond algebra
$\mathcal{A}_{\sf I}$ is a well defined algebra of bounded operators, in spite
of the fact that it is generated by an infinite number of bonds. This follows 
because the bonds act locally on the infinite tensor product 
\(\bigotimes_{i\in \mathbb{Z}}\mathbb{C}^2_i\) \cite{vonneumannII}.

\begin{figure}[h]
\begin{center}
\includegraphics[width=0.75\columnwidth]{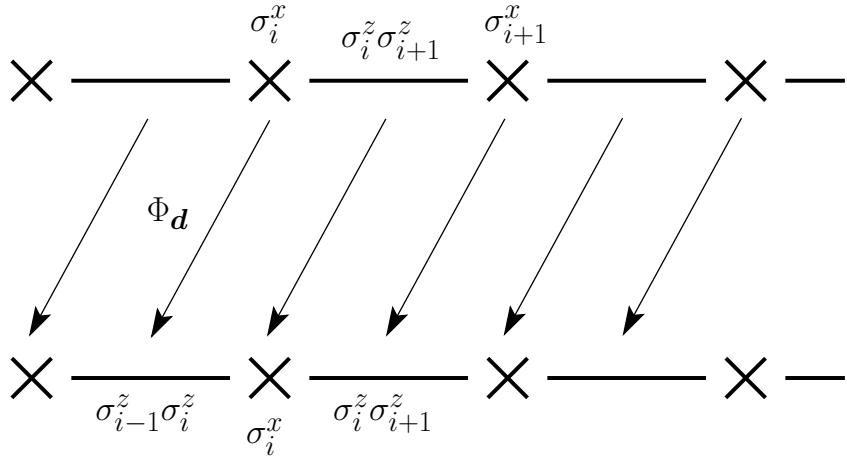}
\end{center}
\caption{A graphic representation of two quantum Ising chains, 
connected by the self-duality isomorphism \(\Phi_\d\) of
Equation \eqref{aut_ising1}. The crosses
$\times$ represent the bonds $\sigma^x_i$, and the thick lines
between crosses represent the bonds $\sigma^z_i\sigma^z_{i+1}$.
\(\Phi_\d\) exchanges the two while preserving all algebraic relations.}
\label{sdinfiniteising}
\end{figure}

Coming back to the set of relations above, we see that 
\(\sigma^x_i\) and \(\sigma^z_{i}\sigma^z_{i+1}\) play
perfectly symmetrical roles, and so we can set up the relation-preserving
mapping
\begin{equation}\label{aut_ising1}
\Phi_{\sf d}(\sigma^z_{i}\sigma^z_{i+1} )=\sigma^x_i,\ \ \ \ \ \ \ \
\Phi_{\sf d}(\sigma^x_{i})=\sigma^z_{i-1}\sigma^z_i.
\end{equation}
These equations define the action of  \(\Phi_{\sf d}\) on bonds alone
but, since it preserves all the algebraic relations among them, it extends
to a unique isomorphism of the {\it full} bond algebra $\mathcal{A}_{\sf I}$. 
\(\Phi_{\sf d}\) is illustrated in Figure \ref{sdinfiniteising}.
Furthermore, \(\Phi_\d\) maps \(H_{\sf I}[h,J]\) to \(H_{\sf I}[J,h]\). It follows
that the quantum Ising chain is {\it self-dual} in the sense of 
definition \ref{algebraically_dual}.
It is not hard to see that \(\Phi_\d\) is unitarily implementable (recall the definition
after Theorem \ref{isoarespatial}), so that there is a \(\mathcal{U}_\d\) such that
\begin{equation}
\mathcal{U}_\d\ \sigma^z_{i}\sigma^z_{i+1}\ \mathcal{U}_\d^\dagger =\sigma^x_i,
\ \ \ \ \ \ 
\mathcal{U}_\d\ \sigma^x_{i}\ \mathcal{U}_\d=\sigma^z_{i-1}\sigma^z_i, \ \ \ \ \ \
\forall i\in \mathbb{Z} .
\end{equation}

The homomorphism  \(\Phi_{\d}\) reveals that the Ising chain is
self-dual due to a {\it local mapping} that reflects a symmetry of its {\it local} 
interactions. In contrast, the traditional approach
seems to imply that dualities are of necessity non-local, because
it focuses on non-local transformations of
elementary degrees of freedom. Notice also that the self-duality mapping
\(\Phi_{\d}\) determines a large family of perturbations of \(H_{\sf I}\) 
that preserve its self-dual character. For example,
\begin{equation}
H=H_{\sf I}+\lambda \sum_i\ ( \sigma^y_i\sigma^z_{i+1}
+\sigma^z_i\sigma^y_{i+1}) 
\end{equation}
is self-dual because the perturbation (the term proportional to
\(\lambda\)) is invariant under \(\Phi_{\d}\) (the action of 
\(\Phi_{\d}\) on \(\sigma^y_i\sigma^z_{i+1}\) for instance can
be determined by factoring \(\sigma^y_i\sigma^z_{i+1}=
-i \sigma^z_i\sigma^z_{i+1} \sigma^x_i\)). Also
we can apply \(\Phi_{\d}\) to Hamiltonians other than \(H_{\sf I}\),
as long as they are affiliated to $\mathcal{A}_{\sf I}$. 
Consider the \(d=1\) dimensional, spin \(S=1/2\) XY-model,
\begin{equation}
H_{\sf XY}=\sum_i\ (J_x\sigma^x_i\sigma^x_{i+1}+
J_z\sigma^z_i\sigma^z_{i+1}).
\end{equation}
The bonds \(\sigma^z_i\sigma^z_{i+1}\)  of
$H_{\sf XY}$ are already bonds of \(H_{\sf I}\).  The
\(\sigma^x_i\sigma^x_{i+1}\) in $H_{\sf XY}$ are the products  of two
bonds of \(H_{\sf I}\). Thus it is  possible to use the isomorphism of
the quantum Ising model to compute a dual form of the XY-model. As
\(\sigma^x_i\sigma^x_{i+1}\ \stackrel{\Phi_{\d}}{\longrightarrow}\ 
\sigma^z_{i-1}\sigma^z_{i} \sigma^z_{i} \sigma^z_{i+1}\), we find that
\begin{equation}
H_{\sf XY}\ \stackrel{\Phi_{\d}}{\longrightarrow}\
H_{\sf Inn}=\sum_i\ (J_x\sigma^z_{i-1}\sigma^z_{i+1}+J_z\sigma^x_i) .
\end{equation}
The fact that,  in \(d=1\), \(H_{\sf XY}\) and \(H_{\sf Inn}\) share the 
same energy spectra was first noticed in Reference \cite{pfeuty},
but explained only later in Reference \cite{grady} 
(\(H_{\sf Inn}\) is trivially dual to two decoupled Ising chains).  

Next we would like to establish the 
precise connection between the bond-algebraic and traditional
approach to quantum dualities of Section \ref{sec2.2}. 
 
\subsection{Connection to the traditional approach: Determination of
dual variables} 
\label{sec3.5}

The traditional approach to dualities (Section \ref{sec2.2})
focuses on dual variables, that is, on
operator change of variables that are non-local in general. In
contrast, the bond-algebraic approach to dualities
of the previous section focuses on local mappings of bonds.
How can the two be related? As it turns out, the isomorphism
of bond algebras determines uniquely the dual variables of the problem.
This is the bridge between the bond-algebraic and the traditional approach.
 
Let us illustrate this point with the quantum
Ising chain. To start with, consider the relation (see Equation
\eqref{aut_ising1})
\begin{equation}
H_{\sf I}[h,J]=\sum_i 
(h\Phi_{\sf d}(\sigma^z_i\sigma^z_{i+1})+J\Phi_{\sf d}(\sigma^x_i)).
\end{equation}
If the individuals spins \(\sigma^z_i\) happen to belong to the
bond algebra \(\mathcal{A}_{\sf I}\), then we can further write
\begin{equation}\label{first_step_dual_variables}
H_{\sf I}[h,J]=\sum_i (h\Phi_{\sf d}(\sigma^z_i)\Phi_{\sf
d}(\sigma^z_{i+1})+J \Phi_{\sf d}(\sigma^x_i)).
\end{equation}
If we now compare this last relation to Equation \eqref{sdisingold},
we see that the dual variables could be connected to the self-duality isomorphism
as
\begin{eqnarray}
\Phi_{\sf d}(\sigma^z_i)=\mu^z_i,\ \ \ \ \ \ 
\Phi_{\sf d}(\sigma^x_i)=\mu^x_i.
\end{eqnarray}
But \(\Phi_{\sf d}\) is defined by \eqref{aut_ising1}.  This makes
sense of $\mu^x_i$ as \(
\mu^x_i\equiv \Phi_{\sf d}(\sigma^x_i)= \sigma^z_{i-1}\sigma^z_i
\), but it is not clear what the action of \(\Phi_{\sf d}\) on \(\sigma^z_i\)
should be.  
Now, at least formally,
\begin{eqnarray}\label{infprodsz}
\sigma^z_i=\prod_{m=i}^\infty\sigma^z_m\sigma^z_{m+1}.
\end{eqnarray}
Unfortunately, this does not quite show that \(\sigma^z_i\in\mathcal{A}_{\sf I}\),
because the left-hand side of Equation \eqref{infprodsz} features an {\it infinite} 
product of bonds. An infinite combination (sum and/or product)
of bonds will only be an element in the
bond algebra if it converges to some bounded operator in the strong or weak
operator topology \cite{vonneumannI}. {\it Suppose though for now} that 
\(\sigma^z_i\in\mathcal{A}_{\sf I}\). Then we can compute
\begin{eqnarray}
\Phi_{\sf d}(\sigma^z_i)=\Phi_{\sf
d}\left(\prod_{m=i}^\infty\sigma^z_m\sigma^z_{m+1}
\right)=\prod_{m=i}^\infty\Phi_{\sf d}(\sigma^z_m\sigma^z_{m+1})=
\prod_{m=i}^\infty\sigma^x_m=\mu^z_i.
\end{eqnarray}
{\it Thus the expressions we obtain for the dual variables
\(\mu^x_i, \mu^z_i\) are identical to
the ones derived by traditional arguments in {\rm Section \ref{sec2.2}}}
[see Equation  \eqref{mustring} in particular]. Because $\Phi_{\sf d}$
is an algebra isomorphism, the  dual variables are guaranteed to
satisfy the same algebra as the original variables \(\sigma^x_i,\sigma^z_i\). 
But we can view this from a different perspective. The fact that the
dual variables satisfy the correct algebra affords an independent check
supporting that \(\Phi_\d\) is indeed an isomorphism, and thus the relations
that were assumed to characterize the bond algebra are complete. 
 
In summary, the structure of the bond
algebra  determines the self-duality homomorphism, and the
self-duality homomorphism enables us to compute the dual variables.
Thus, we have both a test for self-duality and an algorithm to construct
dual variables. 

Now that we have the intuitive picture, let us point out for the sake
of mathematical rigor that \(\sigma^z_i\notin\mathcal{A}_{\sf I}\) and
\(\Phi_\d(\sigma^z_i)\) is not defined. The reason is that formally we can 
also write \(\sigma^z_i=\prod_{m=-\infty}^{i-1}\sigma^z_m\sigma^z_{m+1}\). 
Then it would follow from computing the action of \(\Phi_{\sf d}\)
of both representations of \(\sigma^z_i\) that 
\begin{equation}
\prod_{m=i}^\infty\sigma^x_m\ \stackrel{?}{=}\ 
\prod^{i-1}_{m=-\infty}\sigma^x_m.
\end{equation}
But this  cannot possibly hold true. It is important to notice that 
this is {\it not} a limitation of the bond-algebraic approach to dualities
(that managed to establish in the previous section the self-duality of the Ising
model purely by well-defined manipulations involving bonds), but rather of the concept of 
dual variables in infinite systems. In practice, however, infinite systems are
studied as limits of finite ones, for which dual variables are
well defined and can be computed as above. 
We will come back to this issue in Section \ref{sec3.7}. 

\subsection{Abelian versus non-Abelian dualities: the Heisenberg model}
\label{sec3.6}

Lattice Non-Abelian dualities constitute one of the greatest challenges in the 
theory of dualities, the classical aspects of which are discussed in 
Appendix \ref{appA}. In this section, we present preliminary
contributions of bond algebras to understanding this difficult problem.

There is a broad, well established consensus among physicists that
a duality is {\it non-Abelian} if the dual models have  non-Abelian 
symmetries, and is Abelian otherwise \cite{savit}. Bond
algebras can realize duality mappings of an Abelian origin in models  with non-Abelian 
symmetries,
suggesting that this classification is not appropriate. 
The discussion of this 
section, based on a new duality for the Heisenberg model
in any space dimension $d$, may help sharpen the notion of non-Abelian 
duality beyond the somewhat inaccurate standard lore. 
 
The Heisenberg model,
\begin{equation}\label{anyd_heisenberg_model}
H_{\sf H}=J\sum_\r\sum_{\nu=1}^d\ (\sigma^x_\r\sigma^x_{\r+\bm{e_\nu}}+
\sigma^y_\r\sigma^y_{\r+\bm{e_\nu}}+\sigma^z_\r\sigma^z_{\r+\bm{e_\nu}}),
\end{equation}
is one the fundamental models of magnetism (its application
to cuprates  is reviewed in Reference \cite{manousakis}).
To our knowledge, exact dualities for the Heisenberg model have not been reported
before, and this may seem reasonable, since it has a non-Abelian
group of global symmetries (it is invariant under global $SU(2)$ rotations 
in spin space). Thus it is surprising to find out that,
with the help of bond algebras, we can write a duality for \(H_{\sf H}\)
right away.   
 
The starting point is the observation that the bond algebra
\begin{equation}
\mathcal{A}_{\sf H}\equiv \mathcal{A}\{\sigma^x_\r\sigma^x_{\r+\bm{e_\nu}},
\sigma^y_\r\sigma^y_{\r+\bm{e_\nu}},\sigma^z_\r\sigma^z_{\r+\bm{e_\nu}}\}
\end{equation}
is a sub-algebra of the bond algebra of the quantum Ising model in \(d\) dimensions,
\begin{equation}\label{anyd_ising}
H_{\sf I}=\sum_\r\ \left(h\sigma^x_\r +\sum_{\nu=1}^d\ 
J\sigma^z_\r\sigma^z_{\r+\bm{e_\nu}}\right),
\end{equation}
simply because the bonds of the Heisenberg model can be written 
as products of bonds of the Ising model. Then {\it any (self-)duality
for the Ising model can be translated into a duality for the Heisenberg 
model}. For example, we can use the self-duality mapping of the quantum 
Ising chain,  Equation  \eqref{aut_ising1}
to find a dual form for the $d=1$ Heisenberg model. Since
\begin{equation}
\sigma^y_i\sigma^y_{i+1}=\sigma^x_i\sigma^z_i\sigma^z_{i+1}
\sigma^x_{i+1}\ \stackrel{\Phi_{\d}}{\longrightarrow}\ 
-\sigma^z_{i-1}\sigma^x_i\sigma^z_{i+1},
\end{equation}
we find that
\begin{equation}\label{dual_heisenberg}
H_{\sf H}\ \stackrel{\Phi_{\d}}{\longrightarrow}\ 
H_{\sf H}^D=\frac{J}{4}\sum_i\ (\sigma^z_{i-1}\sigma^z_{i+1}
-\sigma^z_{i-1}\sigma^x_i\sigma^z_{i+1}+
\sigma^x_i).
\end{equation}
Appendix \ref{appG} describes
a version of this duality for finite systems that can be checked numerically. 
The Hamiltonian \(H_{\sf H}^D\) has an interesting connection to the eight-vertex
model that seems to have gone unnoticed to the best of our knowledge. 
Section \ref{eightvertexM} discusses the relation between the eight-vertex 
model and the anisotropic quantum Heisenberg model. In 
particular,  its is shown that \(H_{\sf H}^D\) is directly related to the 
quantum Ashkin-Teller model.


In contrast to what standard practice would suggest, 
we do not think that it is appropriate 
to call Equation \eqref{dual_heisenberg} a non-Abelian duality. 
The duality of 
Equation \eqref{dual_heisenberg}
is strictly based on the self-duality
of the quantum Ising chain, that is Abelian on at least two accounts. 
First, the Ising model has only Abelian symmetries. Second,
we will show (Section \ref{sec8})
that the self-duality of the quantum Ising chain is strictly equivalent to the
self-duality of the classical $D=2$ Ising model.  
Classical dualities are strictly based on very special properties of Abelian groups (see
Appendix \ref{appA}). It seems fair to say that  the duality of 
Equation \eqref{dual_heisenberg}
avoids the non-Abelian structure of the model, and thus
it is inappropriate to call it a non-Abelian duality.
In other words, this duality for the Heisenberg model
suggests  that {\it the group of symmetries of a model} is not the most
important factor in determining the character of a  duality. 

\begin{figure}[h]
\begin{center}
\includegraphics[width=0.9\columnwidth]{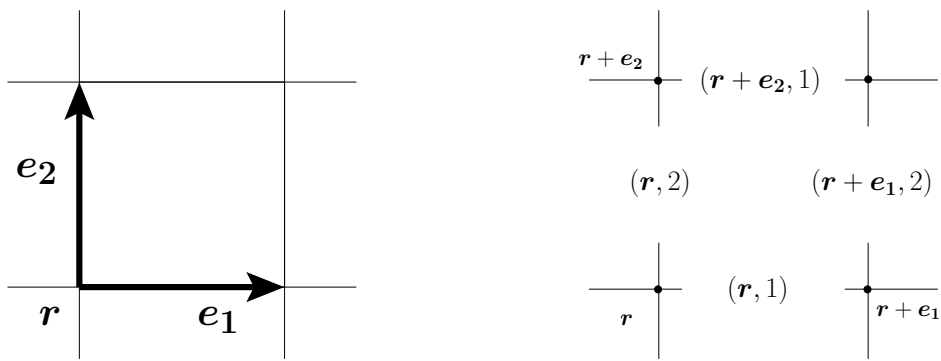}
\end{center}
\caption{(Left panel) Convention to denote vertices $\r=(r^1,r^2)=r^1
\bm{e_1}+ r^2\bm{e_2}$ in a two-dimensional square lattice with  unit vectors 
$\bm{e_1},  \bm{e_2}$, and (right panel) links, attached to a vertex 
$\r$, $(\r,\nu)$  with $\nu=1,2$.}
\label{notation_links}
\end{figure}

The results and ideas just discussed are not peculiar to one dimension,
but before discussing the higher-
dimensional analogues of Equation \eqref{dual_heisenberg},
we need to introduce a bit of notation to describe 
degrees of freedom on the {\it links} of a lattice. 
In general, to specify a
link \(\bm{l}\) of a {\it hyper-cubic lattice} of dimension \(d\), we
determine first the lattice site \(\r\) and direction \(\nu\) such that 
\(\bm{l}\) connects the two sites \(\r\) and \(\r+\bm{e_\nu}\). Then, we
denote \(\bm{l}\) by the pair \((\r,\nu)\), as shown in  Figure
\ref{notation_links}.
We can now place a spin \(S=1/2\) degree of freedom at each link 
\((\r,\nu)\), represented by  sets of Pauli matrices 
\(\sigma^\mu_{(\r,\nu)}\), \(\ \mu=x,y,z\). Let us introduce one more
piece of notation. Both the plaquette operator
\begin{equation}\label{general_plaquette}
B_{(\r,\mu\nu)}=\sigma^z_{(\r,\mu)}\sigma^z_{(\r+\bm{e_\mu},\nu)}
\sigma^z_{(\r+\bm{e_\nu},\mu)}\sigma^z_{(\r,\nu)},\ \ \ \ \ \ 
\mu\neq\nu=1,\cdots,d,
\end{equation}
that resides on the plaquette with vertex $\r$ and spanned by 
$(\bm{e_\mu},\bm{e_\nu})$, and the vertex operator
\begin{equation}
A_\r=\prod_{\nu=1}^{d}\
\sigma^x_{(\r,\nu)}\sigma^x_{(\r-\bm{e_\nu},\nu)}\ ,
\end{equation}
that resides on the lattice site $\r$,  will show up repeatedly in this article.
Also, in dimensions \(d=2\) or \(3\), 
we prefer a more compact notation for the plaquette operator,
\begin{equation}\label{gauge_plaquette}
B_{(\r,1)}\equiv B_{(\r,23)},\ \ \ \ B_{(\r,3)}\equiv B_{(\r,12)},
\ \ \ \ B_{(\r,2)}\equiv B_{(\r,31)}.
\end{equation}

With these conventions in place, we can introduce a model dual
to the Ising model in any dimension:
\begin{equation}\label{anyd_dual_ising}
H_{\sf I}^D=\sum_\r\ \left(h\ A_\r +\sum_{\nu=1}^d\ J\sigma^z_{(\r,\nu)}\right).
\end{equation}
The duality follows from these observations: 
The vertex operators \(A_\r\) anti-commute with exactly \(2d\) spins \(\sigma^z_{(\r,\nu)}\),
just as the spins \(\sigma^x_\r\) of the Ising model of Equation \eqref{anyd_ising}
anti-commute with exactly \(2d\) bonds \(\sigma^z_\r\sigma^z_{\r+\bm{e_\nu}}\).
Similarly, the spins  \(\sigma^z_{(\r,\nu)}\) anti-commute with just two
vertex operators \(A_\r\) and \(A_{\r+\bm{e_\nu}}\), just as  
\(\sigma^z_\r\sigma^z_{\r+\bm{e_\nu}}\) anti-commutes with \(\sigma^x_\r\)
and \(\sigma^x_{\r+\bm{e_\nu}}\) only (a classical analogue of this duality
was introduced by Wegner in Reference \cite{wegner_ising}).
On the other hand, the relation
\begin{eqnarray}
\sigma^y_\r\sigma^y_{\r+\bm{e_\nu}}=-\sigma^z_\r\sigma^z_{\r+\bm{e_\nu}}\ 
\sigma^x_\r\sigma^x_{\r+\bm{e_\nu}},
\end{eqnarray}  
shows that the bond algebra of the Heisenberg model of Equation 
\eqref{anyd_heisenberg_model} is a sub-algebra of the bond algebra of the 
Ising model. Hence we can transfer the duality of Equation \eqref{anyd_dual_ising} 
to the Heisenberg model:
\begin{equation}\label{dual_to_heisenberg}
H_{\sf H}^D=J\sum_\r\sum_{\nu=1}^d\  (A_\r A_{\r+\bm{e_\nu}}-
A_\r\sigma^z_{(\r,\nu)}A_{\r+\bm{e_\nu}}+\sigma^z_{(\r,\nu)}).
\end{equation} 
The duality mapping reads,
\begin{equation}
A_\r A_{\r+\bm{e_\nu}}\dual\sigma^x_\r\sigma^x_{\r+\bm{e_\nu}} 
,\ \ \ \ \ \ \ \ \ \ \ \ \
\sigma^z_{(\r,\nu)}\dual \sigma^z_\r\sigma^z_{\r+\bm{e_\nu}}.
\end{equation}

Notice  that \(H_{\sf H}\) has one
spin degree of freedom per lattice site, while \(H_{\sf H}^D\) has 
\(d\) (one per link). Thus \(H_{\sf H}\) and \(H_{\sf H}^D\) do not
act on state spaces of the same dimensionality when $d>1$, and cannot be dual in the
strict sense of definition \eqref{algebraically_dual}. In order to
resolve this dilemma, one must appreciate that \(H_{\sf H}^D\) has a large group of 
{\it gauge (local) symmetries} that is not shared by the Heisenberg 
model (that has only global symmetries). All of the plaquette operators
\(B_{(\r, \mu \nu)}\) defined in Equation \eqref{general_plaquette}
commute with \(H_{\sf H}^D\), 
\begin{equation}
[B_{(\r,\mu\nu)},\ H_{\sf H}^D]=0,
\end{equation}
and rigorously speaking, \(H_{\sf H}\) and \(H_{\sf H}^D\) are 
dual {\it up to the complete elimination of these gauge symmetries}
(Appendix \ref{appG} presents a version of this statement that can
be checked numerically. Note also that this discussion also applies to the 
duality of Equation \eqref{anyd_dual_ising} between $H_{\sf I}$ and $H_{\sf I}^D$
for $d>1$).
This crucial refinement of the concept of duality will be discussed 
at length in Section \ref{sec3.12} and will justify the need for homomorphisms, 
as opposed to isomorphisms, 
in the more general case. 

Let us notice in closing this line of arguments, that there are also 
several examples of {\it self-dual} models with a {\it non-Abelian symmetry group},
most notably the \(d=1\) vector Potts ($p$-clock) model of Section \ref{sec4.1},
and \(d=3\) \(\mathbb{Z}_p\) gauge theories of Section \ref{sec6.3}. 
The fact that these models
have non-Abelian symmetries seems to have gone unnoticed in the literature. 
We consider next a connection between self-dualities and non-Abelian 
groups of a completely different character. Self-duality isomorphisms 
connect one and the same Hamiltonian at different regions in coupling space, and
taken together they close a self-duality group, because the composition of 
two self-dualities is another self-duality. This group acts linearly on the
bond algebra of the Hamiltonian, and can be either Abelian or non-Abelian.

\begin{figure}[h]
\begin{center}
\includegraphics[width=5.2in]{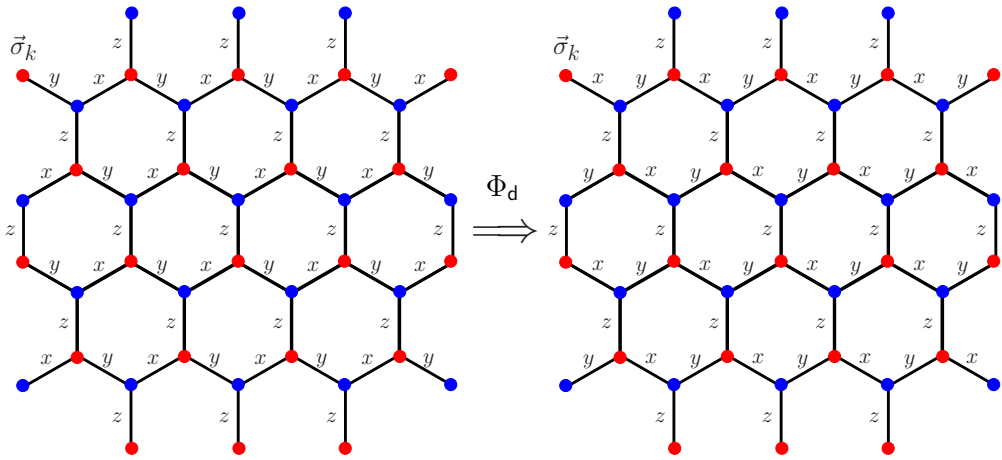}
\end{center}
\caption{Kitaev's honeycomb model features  $S=1/2$ spins
represented by a Pauli matrices $\vec{\sigma}_{k}$.  
The model has  three types of bonds, indicated by
the letter $\mu=x,y,z$, that represent the bond  operators
$\sigma^\mu_i \sigma^\mu_j$. $\Phi_{\sf d}$ stands
for the duality mapping that realizes the exchange \(
\sigma^x_i \sigma^x_j\leftrightarrow\sigma^y_i \sigma^y_j\), and 
that will be denoted in what follows as \(P_{yxz}\).}
\label{Kitaev_honeycomb}
\end{figure}

To illustrate the premise, consider Kitaev's ``honeycomb
model''  \cite{Kitaev2006, nussinov_chen, bondDec08, pachosannals}, 
defined by the $S=1/2$ Hamiltonian,
\begin{eqnarray}
H_{\sf Kh}&=&-J_x\sum_{x{\sf-bonds}}\ \sigma^x_{i}\sigma^x_{j} -J_y
\sum_{y{\sf -bonds}}\sigma^y_{i}\sigma^y_{j} -J_z\sum_{z{\sf -bonds}}
\sigma^z_{i}\sigma^z_{j} \nonumber \\ 
&=& - \sum_{\langle ij \rangle} J_{\mu} \sigma_{i}^{\mu} \sigma_{j}^{\mu},
\ \ \ \ \ \ {\bm e}_{\mu} || (\vec{j} - \vec{i}),
\label{HK}
\end{eqnarray}
on a honeycomb lattice.
Here the spins are located at the sites $i,j$ (we use 
the notation that is standard in the literature
to avoid any confusion). The three
nearest neighbor directions on the honeycomb lattice (at 120 degrees
relative to one another) are denoted by the
indices $\mu = x,y,z$ in Equation (\ref{HK}), see Figure
\ref{Kitaev_honeycomb}. 

The Hamiltonian \(H_{\sf Kh}\) admits several simple self-dualities
that exchange any two of its couplings \(J_x,\ J_y\), and \(J_z\), and more 
general permutations as well. Let \(\tau\in{\cal S}_3\), the group 
of permutations of three elements, be the 
permutation \(x,y,z\mapsto \tau(x), \tau(y),\tau(z)\). Then we denote
the corresponding self-duality mapping by $P_{\tau(x)\tau(y)\tau(z)}$,
that realizes the exchange \(J_x,J_y,J_z\mapsto J_{\tau(x)}, 
J_{\tau(y)},J_{\tau(z)}\)
(for example, the self-duality shown in Figure \ref{Kitaev_honeycomb} 
will be denoted by $P_{yxz}$). We see that this family of
self-dualities affords a representation of the non-Abelian group of permutations
in the space of bonds of the model, but this group does not commute
with \(H_{\sf Kh}\) unless the Hamiltonian is fine-tuned to be at the self-dual 
line \(J_x=J_y=J_z\). 

We can write down representations for the pairwise
permutations. For instance, a global rotation about the $\sigma^z$ axis
by 90 degrees will exchange $\sigma^x \sigma^x$ with $\sigma^y \sigma^y$
and viceversa. It follows that
\begin{equation}
P_{yxz} = \exp\left [i \frac{\pi}{4} \sum_{j=1}^{N} \sigma_{j}^{z}\right ].
\label{pair_permute}
\end{equation}
The three pairwise permutations  $\{P_{yxz}, P_{xzy}, P_{zyx}\}$ can be
represented as (non-commuting) rotations by 90 degrees. Any permutation 
of the three bond types (or any bonds more generally in other systems)
can, of course, be written as a product of pairwise permutation
operators of the form of Equation (\ref{pair_permute}). For instance, 
\begin{equation}
P_{zxy} = P_{zyx} P_{xzy}.
\end{equation}

Similar to
the group ${\cal S}_{3}$, we might embed other finite  groups as acting on
a finite number of bond types. 

\subsection{Exact dualities for finite systems}
\label{sec3.7}

Up to now we have only considered bond algebras of infinite systems.
This has advantages and disadvantages. On one hand, the
bond algebras are mathematically well defined, and the bond algebra
mappings of interest are typically simple. On the other hand, 
it is of great interest to study the action of these mappings on operators
that are infinite combinations of bonds (e.g., the Hamiltonian). 
This may be a concern because those operators need to be defined in the
infinite tensor product space where the bond algebra acts.  We saw in Section
\ref{sec3.5} some of the problems that can arise from trying to extend the 
action of bond algebra mappings to infinite combinations of bonds. Let us 
take a look at these problems from a slightly different perspective that 
will be useful later in this section.

It is standard practice to argue that the quantum Ising chain 
of Equation \eqref{infinite_ising_transverse} has a \(\Z_2\) symmetry generated by 
\begin{equation}
Q=\prod_{i=-\infty}^{\infty}\ \sigma^x_i,\ \ \ \ \ \ [H_{\sf I}, Q]=0.
\end{equation}
Now, since formally we can write \(\mathbb{1}=\prod_{i=-\infty}^{\infty}
\ \sigma^z_i\sigma^z_{i+1}\), it would seem that the mapping of 
Equation \eqref{aut_ising1} satisfies
\begin{equation}\label{phi_paradox}
\Phi_\d(\mathbb{1})=\prod_{i=-\infty}^{\infty}\ 
\Phi_\d(\sigma^z_i\sigma^z_{i+1})=Q.
\end{equation}
Since  \(\Phi_\d(\mathbb{1})=\mathbb{1}\) must 
hold true as well, it would seem that \(\Phi_\d\) is a multivalued mapping. 
This problem was already 
pointed out in Section \ref{sec2.2} from a different but equivalent perspective.
In general, duality mappings established in the
limit of infinite size, or in the continuum as in QFT (see
Section \ref{sec6}), are well defined on finite combinations of bonds, 
but have ill-defined actions on for example global symmetries that involve infinite 
combinations of bonds.

The practical solution to these problems is to work with bond algebras 
of finite-size systems, and eventually take the thermodynamic limit if 
one is interested in the infinite-size system.
But typically, standard boundary
conditions (BCs) (open, periodic, anti-periodic, etc.) 
may spoil duality properties that are apparent in the infinite- size limit. 
For example, both open and periodic BCs,  
\begin{eqnarray}
H_{\sf I}^o&=&\sum_{i=1}^{N}\ h\sigma^{x}_i + \sum_{i=1}^{N-1}\
J\sigma^z_{i}\sigma^z_{i+1}, \ \ \ \ \ \ (\mbox{open BCs}),\ 
\ \ \ \ \mbox{}\label{io}\\
H_{\sf I}^c&=&H_{\sf I}^o+J\sigma^z_N\sigma^z_1, \ \ \ \ \ \
(\mbox{periodic (toroidal) BCs}), \label{ic}
\end{eqnarray}
spoil the self-duality \(J\leftrightarrow h\) of the quantum Ising chain,
and the same happens with many other (self-)dualities.

On the other hand, BCs can help to restore
in finite-size systems properties of the thermodynamic limit like
translation invariance. Similarly, 
bond algebras can be exploited to find in a systematic way
{\it model specific} BCs that restore duality properties \cite{con}.
This is an impressive advantage of the bond-algebraic
over the traditional approach, because it puts models that are {\it exactly}
dual or self-dual at reach of computer simulations (the role
of BCs in connection to dualities was noticed from 
time to time in the literature in the context of specific models, 
see for example \cite{horn,hinrichsen}). Let us see how this works
with the models of Equations \eqref{io} and \eqref{ic}.
More complicated examples will be discussed in later sections.

\begin{figure}[h]
\begin{center}
\includegraphics[width=0.65\columnwidth]{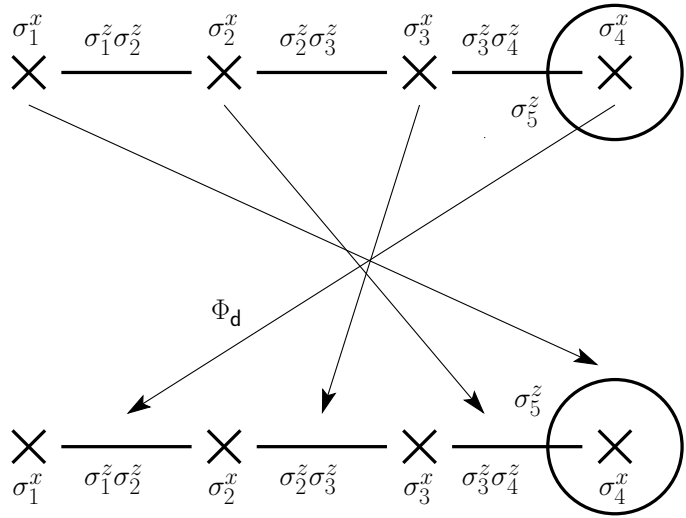}
\end{center}
\caption{Two finite-size ($N=4$ sites) quantum Ising chains with
self-dual BCs that break \(Z_2\) invariance, connected
by the self-duality isomorphism \(\Phi_\d\) of Equation \eqref{automorphismfi}.  
The big circle at the rightmost end of the chains represents the boundary correction
$\sigma^z_4$.}
\label{automorphismfising}
\end{figure}

Intuitively speaking, the Hamiltonian \(H_{\sf I}^o\) of Equation 
\eqref{io} should not be self dual, because
it has $N$ bonds $\sigma^x_i$, but only $N-1$ bonds $\sigma^z_i\sigma^z_{i+1}$.
This suggests  adding a bond
\begin{equation}\label{finiteisingsd}
H_{\sf I}^o\ \rightarrow\ 
\tilde{H}_{\sf I}^o=H_{\sf I}^o+\ J\sigma^z_N
\end{equation}
(see Figure \ref{automorphismfising}) that becomes {\it irrelevant in the
thermodynamic limit}, from the standpoint of bulk properties. 

The next step is to check that \(\tilde{H}_{\sf I}^o\)
is self-dual. To see this, we notice that if the model admits a
self-duality mapping \(\Phi_{\sf d}\), it must be that  
\begin{eqnarray}
\Phi_{\sf d}(\sigma^x_1)=\sigma^z_N,
\end{eqnarray}
due to the structure of relations among bonds.
Next, to compute $\Phi_{\sf d}(\sigma^z_1\sigma^z_2)$, notice that 
it must be one of the $\sigma^x$s, and that it must 
anti-commute with $\Phi_{\sf d}(\sigma^x_1)=\sigma^z_N$. Thus it must be 
that $\Phi_{\sf d}(\sigma^z_1\sigma^z_2)=\sigma^x_N$. Reasoning in this way,
we can reconstruct the full self-duality isomorphism
\begin{eqnarray}\label{automorphismfi}
\sigma^x_1&\ \stackrel{\Phi_{\sf d}}{\longrightarrow}\ & \sigma^z_N
\nonumber\\
\sigma^x_i&\ \stackrel{\Phi_{\sf d}}{\longrightarrow}\ &
\sigma^z_{r(i)}\sigma^z_{r(i)+1},\ \ \ \ \ \ i=2,3,\cdots,N\nonumber\\
\sigma^z_N&\ \stackrel{\Phi_{\sf d}}{\longrightarrow}\ &
\sigma^x_1\nonumber\\
\sigma^z_i \sigma^z_{i+1} & \ \stackrel{\Phi_{\sf d}}{\longrightarrow}\ &
 \sigma^x_{r(i)},\ \ \ \ \ \ \ \ \ \ \ \ \ \ \ i=1,2,\cdots,N-1,
\end{eqnarray}
$r(i)$ represents the {\it inversion} map
\begin{eqnarray}
r(i)=N+1-i.
\label{inversion_map_eq}
\end{eqnarray}
Notice that $\Phi_{\d}^2=\mathbb{1}$, the identity map. 
In general, as discussed in Section \ref{sec3.9}, 
$\Phi_{\d}^2$ is related to a symmetry of the model under consideration.

The boundary term $J \sigma^z_N$ also makes it possible  to compute
finite dual variables. The extra bond $\sigma^z_N$ guarantees that 
the individual spins $\sigma^z_i$, \(i=1,\cdots,N\) are elements
in the bond algebra, since we can write
\begin{eqnarray}
\sigma^z_i=\sigma^z_N\times \sigma^z_{N}\sigma^z_{N-1}\times
\cdots\times \sigma^z_{i+1}\sigma^z_i.
\end{eqnarray}
Then the dual variables \(\mu^{x,y,z}_i=\Phi_{\d}(\sigma^{x,y,z}_i)\)
are
\begin{eqnarray}
\mu_1^x&=&\sigma^z_N\nonumber\\
\mu^x_i&=&\sigma^z_{r(i)} \sigma^z_{r(i)+1},\ \ \ \ \ \ 
i=2,3,\cdots,N,\nonumber \\
\mu^z_i&=&\prod_{m=i}^N\sigma^x_{r(m)}=\prod_{m=1}^{r(i)}\sigma^x_m.
\label{ssstring}
\end{eqnarray} 

The mapping of Equation \eqref{automorphismfi} proves
that \(\tilde{H}_{\sf I}^o\) is indeed self-dual, and
is free of the mathematical inconsistencies embodied in 
Equation \eqref{phi_paradox}. In particular, the self-dual boundary
term breaks the \(\Z_2\) symmetry of the model, so the problem
inherent to Equation \eqref{phi_paradox} is no longer an issue. On the
other hand, one can find self-dual BCs that preserve the
\(\Z_2\) symmetry \(Q=\prod_{i=1}^N\ \sigma^x_i\), namely
\begin{equation}
H_{\sf I}^o\ \rightarrow\ {\tilde{H'}}^{o}_{\sf I}=H_{\sf I}^o-h\sigma^x_N.
\end{equation}
The self-duality mapping for \({\tilde{H'}}_{\sf I}^{o}\) can be constructed just
as before, starting with \(\Phi_\d(\sigma^x_1)=\sigma^z_{N-1}\sigma^z_N\), but
since \(\sigma^x_N\) is no longer in \({\tilde{H'}}_{\sf I}^{o}\) , this will not determine the action
of \(\Phi_\d\) on \(\sigma^x_N\). This is important because we would like 
to compute \(\Phi_\d(Q)\) and check that no inconsistency arises, and it is easy
to solve. The trick is to add \(\sigma^x_N\) to the list of bond generators, i.e., 
bond algebra,  
{\it but not to} \({\tilde{H'}}_{\sf I}^{o}\), and extend the action of \(\Phi_\d\)
consistently. In this case, the result is that \(\Phi_\d(\sigma^x_N)=\sigma^z_1\),
and so
\begin{equation}
Q^D\equiv \Phi_\d(Q)= \sigma^z_N,\ \ \ \ \mbox{so that}\ \ 
[Q^D,\ {\tilde{H'}}_{\sf I}^{o}]=0.
\end{equation}
The self-duality exchanges the two non-trivial symmetries of the model.

The discussion of previous paragraphs illustrates very general
features of the problem of constructing (self-)dual boundary terms, features
that we will find also in more complex models in higher dimensions.
In general, {\it (self-)dual BCs  are not unique}, and different
choices break and/or preserve different symmetries. This is intimately 
connected to the topic of Section \ref{sec3.9}, and it is important in 
practice to remember that the action of 
a duality on a non-local symmetry cannot be understood with any precision in the 
formal limit of infinite size or in the continuum (where, on the other hand, 
dualities are most easily spotted!). For that, one has to choose the
(self-)dual BC  that is best suited to the problem at hand,
to consider afterwards the action of the duality on the symmetries. Let us 
illustrate next self-dual BCs that preserve translation invariance.

\begin{figure}[h]
\begin{center}
\includegraphics[width=0.45\columnwidth]{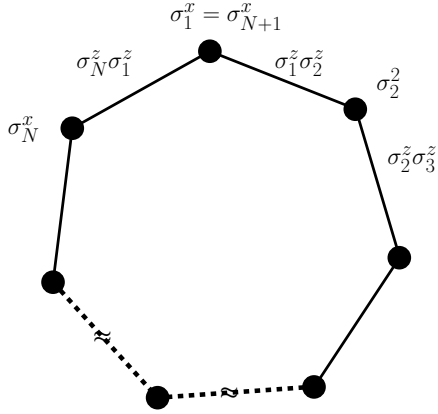}
\end{center}
\caption{Quantum Ising chain featuring \(N\) sites,
 with periodic BCs.}
\label{isingperiodic}
\end{figure}
In contrast to the open chain, the closed Ising chain 
\(H_{\sf I}^c\) of Equation \eqref{ic} does not suffer from any bond 
counting mismatch, so one could think that 
\begin{equation}\label{circle_half_tr}
\sigma^x_i\ \mapsto\ \sigma^z_i\sigma^z_{i+1},\ \ \ \ \ \ 
\sigma^z_i\sigma^z_{i+1}\ \mapsto\ \sigma^x_{i+1}, \ \ \ \ \ \ 
(\mbox{with } N+1\equiv1),
\end{equation} 
defines a self-duality. Unfortunately,
this is incorrect. If this mapping were an isomorphism, it would map
\begin{equation}
\mathbb{1}=\sigma^z_1\sigma^z_2\times\cdots\times\sigma^z_N
\sigma^z_1\ \mapsto\ \sigma^x_1\cdots\sigma^x_N,
\end{equation}
but isomorphisms can only map the identity \(\mathbb{1}\) to itself. 
Fortunately, this analysis suggests the solution to the problem. 
Let us change the boundary term as 
\begin{equation}
H_{\sf I}^c\ \rightarrow\ \tilde{H}_{\sf I}^c=H_{\sf I}^o
+J\sigma^z_N Q\sigma^z_1,
\end{equation}
with 
\begin{equation}
Q=\prod_{m=1}^N\sigma^x_m.
\end{equation}
Then, {\it with the understanding that \(\sigma^z_N\sigma^z_1\) 
should be replaced by \(\sigma^z_N Q \sigma^z_1\)},
the mapping of Equation \eqref{circle_half_tr} does define a self-duality
isomorphism for \(\tilde{H}_{\sf I}^c\). 

The boundary correction \(\sigma^z_N\sigma^z_1\rightarrow\sigma^z_NQ\sigma^z_1\)
seems to break the translational invariance of \(H^c_{\sf I}\). To see that 
this is not the case, let us fix a uniform notation \(z_i\equiv\sigma^z_i\sigma^z_{i+1},\ 
i=1,\cdots,N-1\), \(z_N\equiv\sigma^z_NQ\sigma^z_1\), and compute the action 
of \(\Phi_\d^2\) from Equation \eqref{circle_half_tr},
\begin{equation}
\Phi_\d^2(\sigma^x_i)=\sigma^x_{i+1}, \ \ \ \ \ \ 
\Phi_\d^2(z_i)=z_{i+1}, \ \ \ \ \ \ \mbox{with}\ \ i+N \equiv i.
\end{equation}
We see that \(\Phi_\d^2\) is in fact the generator of translations. When we
explain in the next section that \(\Phi_\d\) is unitarily implementable, this will
afford the proof that \(\tilde{H}_{\sf I}^c\) has translation invariance.

One easy way to check the correctness of the self-duality mapping
of Equation \eqref{circle_half_tr} (with corrected boundary term)
is to compute the dual variables, and 
check that they satisfy the correct Pauli algebra. The problem with this 
plan is that all the bonds in \(\tilde{H}_{\sf I}^c\) commute with the
symmetry \(Q\), and so the individual spins
\(\sigma^z_i\), \(i=1,\cdots, N\) cannot possibly be in the bond algebra.
The solution is to add one spin, say \(\sigma^z_N\) to
the set of bond generators, {\it but not to \(\tilde{H}_{\sf I}^c\)}.
In this way, we extend the bond algebra to include all the \(\sigma^z_i\)
without spoiling the fact that \(\tilde{H}_{\sf I}^c\) is self-dual,
provided we can extend the action of the self-duality isomorphism
to \(\sigma^z_N\) in a consistent fashion. 

For the sake of concreteness, let us see how this works when there
are only \(N=3\) spins in the chain. Then the self-duality isomorphism
\begin{eqnarray}
&&\sigma^x_1\dual \sigma^z_1\sigma^z_2,\ \ \ \ \ \ \ \ \ \ \ 
\sigma^z_1\sigma^z_2\dual\sigma^x_2,\nonumber\\
&&\sigma^x_2\dual \sigma^z_2\sigma^z_3,\ \ \ \ \ \ \ \ \ \ \ 
\sigma^z_2\sigma^z_3\dual\sigma^x_3,\\
&&\sigma^x_3\dual \sigma^z_3Q\sigma^z_1,\ \ \ \ \ \ 
\sigma^z_3Q\sigma^z_1\dual\sigma^x_1,\nonumber
\end{eqnarray}
can be extended to \(\sigma^z_3\) as 
\begin{equation}\label{pising_extension}
\sigma^z_3\dual \sigma^z_1,
\end{equation}
preserving all the algebraic relations.
The dual variables that follow read
\begin{eqnarray}
&&\mu^z_1\equiv\Phi_\d(\sigma^z_1)=\sigma^z_1\sigma^x_2\sigma^x_3,
\ \ \ \ \ \ 
\mu^x_1\equiv\Phi_\d(\sigma^x_1)=\sigma^z_1\sigma^z_2,\nonumber\\
&&\mu^z_2\equiv\Phi_\d(\sigma^z_2)=\sigma^z_1\sigma^x_3,
\ \ \ \ \ \ \ \ \ 
\mu^x_2\equiv\Phi_\d(\sigma^x_2)=\sigma^z_2\sigma^z_3,\label{duali_circle}\\
&&\mu^z_3\equiv\Phi_\d(\sigma^z_3)=\sigma^z_1,
\ \ \ \ \ \ \ \ \ \ \ \  
\mu^x_3\equiv\Phi_\d(\sigma^x_3)=\sigma^z_3Q\sigma^z_1.\nonumber
\end{eqnarray}
It is straightforward to extend this construction to \(N\) spins.

\subsection{Dualities as unitary transformations}
\label{sec3.8}

Last section's results strengthen our argument that the 
bond-algebraic approach to dualities is truly practical.
One can always take a duality between infinite models, recast it as 
a duality between finite renditions of those models, even check the 
homomorphism 
numerically, and be 
free of all potential inconsistencies. Also, 
the general definition of bond algebra and duality are meant 
to settle the connection between dualities and unitary transformations.
It is not clear how to use them to construct explicitly 
the unitaries that implement dualities (from now on, we use 
the word {\it unitary} as short for unitary transformation).
Let us show  for concreteness 
how to build the self-duality unitary of the simplest self-dual
quantum Ising chain with only two sites.

Consider first the
Hamiltonian of Equation \eqref{finiteisingsd} with just
two sites, \(N=2\). Then (from Equation \eqref{automorphismfi}),
the self-duality isomorphism reads
\begin{eqnarray}\label{2sites}
\sigma^x_1\ \longleftrightarrow \sigma^z_2,\ \ \ \ \ \ 
\sigma^x_2\ \longleftrightarrow\  \sigma^z_1\sigma^z_2.
\end{eqnarray}
Let
\begin{equation}
|--\rangle,\ \ \ |-+\rangle,\ \ \  |+-\rangle,\ \ \
|++\rangle,
\end{equation}
be the simultaneous eigenstates of \(\sigma^x_1\) and \(\sigma^x_2\),
and let 
\begin{eqnarray}
|-1\ -1\rangle&=&|\uparrow\downarrow\rangle,\ \ \ \ \ 
|-1\ 1\rangle=|\downarrow\uparrow\rangle,\\
|1,-1\rangle&=&|\downarrow\downarrow\rangle,
\ \ \ \ \ \ \ \ \ |1\ 1\rangle=|\uparrow\uparrow\rangle,\nonumber
\end{eqnarray}
be the simultaneous eigenstates of \(\sigma^z_1\sigma^z_2\) and 
\(\sigma^z_2\). These latter bonds are dual to 
\(\sigma^x_2\) and \(\sigma^x_1\) respectively,
thus we must have 
\begin{eqnarray}
\mathcal{U}_\d&&= |-1\ -1\rangle\langle --|+|-1\ 1\rangle
\langle+-|+\\
&&\ \ \ \ \ \ \ |1\ -1\rangle\langle-+|+|1\ 1\rangle
\langle++|,\nonumber
\end{eqnarray}
or, in matrix form,
\begin{equation}
\mathcal{U}_\d=\frac{1}{2}
\begin{bmatrix}
1 &1 &1 &1\\
1 &-1 &-1 &1\\
1 &-1 &1 &-1\\
1 &1 &-1 &-1
\end{bmatrix}.
\end{equation}
The isomorphism \(\Phi_\d\) of Equation \eqref{2sites}
is its own inverse, and correspondingly,
\(\mathcal{U}_\d^2=\mathbb{1}\).

\subsection{Dualities and quantum symmetries}
\label{sec3.9}

A symmetry transformation is a modification of the observer's point  of
view that does not change the outcome of an experiment performed on  the
same system. Mathematically, in quantum mechanics,  it is a mapping that
takes the Hilbert space of states $\cal H$ into an equivalent Hilbert
space. Wigner's theorem asserts that any transformation $\hat{T}$ which
preserves the transition probability between rays in the Hilbert space
$\cal H$,  $| \Psi_1 \rangle, | \Psi_2 \rangle$,
\begin{equation}
|\langle \hat{T}^\dagger \Psi_1 | \hat{T} \Psi_2 \rangle|^2= |\langle
\Psi_1 | \Psi_2 \rangle|^2
\end{equation}
can be represented by a linear unitary or anti-linear anti-unitary map $U$
($U^\dagger=U^{-1}$) on $\cal H$. The discrete operation of
time-reversal is one of the few relevant examples in physics which
involves an anti-unitary operator. A  symmetry transformation leaves the
Hamiltonian invariant. That is, $U \, H[\lambda] \, U^\dagger=
H[\lambda]$ or, equivalently, \([H,U]=0\).

Self-dualities have sometimes been alluded to as being symmetries 
\cite{witten}.  We think this is a misnomer for the following reasons:
As argued, self-dualities are usually unitarily implementable
transformations. However, while quantum symmetries are trivial isomorphisms
that leave  the Hamiltonian $H$ invariant, self-duality transformations
do not  preserve the form of $H$, but rather preserve its spectrum and 
level degeneracies. In a sense,  self-dualities capture  non-trivial 
isomorphisms aside from the
more trivial case of symmetries that leave $H$ itself invariant. 

One basic connection between symmetries and dualities is that
symmetries control the variety of ways in which dualities can
manifest themselves. A closer look at any duality reveals that it could 
be embodied in a wide variety of isomorphisms, but there is 
a common denominator: these are all related by symmetry. This is 
easy to understand on general grounds, 
now that we know that dualities are unitary
equivalences. For suppose that you have two different dualities
\(\mathcal{U}_\d\) and \({\mathcal{U}'}_\d\) that connect Hamiltonians
\(H_1\) and \(H_2\),
\begin{equation}\label{H12H2}
\mathcal{U}_\d H_1\mathcal{U}_\d^\dagger=H_2,\ \ \ \ \  \mbox{and}
\ \ \ \ \ {\mathcal{U}'}_\d H_1{\mathcal{U}'}_\d^\dagger=H_2.
\end{equation} 
Then
\begin{equation}
{\mathcal{U}'}_\d^\dagger\mathcal{U}_\d H_1\mathcal{U}_\d^\dagger{\mathcal{U}'}_\d
=H_1,\ \ \ \ \ \mbox{and}\ \ \ \ \  
{\mathcal{U}'}_\d\mathcal{U}_\d^\dagger H_2\mathcal{U}_\d{\mathcal{U}'}_\d^\dagger=H_2,
\end{equation}
so that \({\mathcal{U}'}_\d^\dagger\mathcal{U}_\d\) is a symmetry of \(H_1\),
and \({\mathcal{U}'}_\d\mathcal{U}_\d^\dagger\) is a symmetry of \(H_2\).
Conversely if \(U_1\) (\(U_2\)) is a symmetry of \(H_1\) (\(H_2\)), and 
\(\mathcal{U}_\d\) is a duality as above, then 
\begin{equation}
U_2\mathcal{U}_\d U_1
\end{equation} 
is a duality as well. These connections could be used to unveil
hidden symmetries through dualities.

The connection between dualities and symmetries is even stronger for
self-dual models \cite{con}. Suppose, for simplicity, that we have a Hamiltonian
$H[\lambda_1,\lambda_2,\cdots]$,  dependent upon a set of couplings
$\lambda_\nu$,  that is self-dual under the exchange
$\lambda_1\leftrightarrow\lambda_2$, that is,
\begin{eqnarray}\label{usym}
\mathcal{U}_{\sf
d}\, H[\lambda_1,\lambda_2,\cdots]\, \mathcal{U}^\dagger_\d=
H[\lambda_2,\lambda_1,\cdots],
\end{eqnarray}
with $\mathcal{U}_\d$  a unitary independent of the couplings.  
$\mathcal{U}_\d$ relates to symmetries of $H$ in two ways. 
First, it is  clear that
\begin{eqnarray}
[H[\lambda_1,\lambda_2,\cdots],\ \mathcal{U}^{2n}_\d]=0,
\end{eqnarray}
i.e., $\mathcal{U}^{2n}_\d$ are symmetries of $H$  for any
$n=1,2\cdots$ up to the power that  gives unity back. 
We see that loosely speaking, a self-duality can
be seen as the  square root of a symmetry.
Second, Equation \eqref{usym} shows that at the {\it self-dual point} 
\begin{equation}
\lambda_1=\lambda_2\ \ \ \ \ \ \ \ (\mbox{self-dual point}) ,
\end{equation}
$\mathcal{U}_\d$ itself commutes with $H$.  In other words, 
$\mathcal{U}_\d$ {\it emerges as a new symmetry at the 
self-dual point}. In fact, the full sequence of powers $\mathcal{U}_\d,
\mathcal{U}^2_\d, \mathcal{U}^3_\d, \cdots$
is a sequence of quantum symmetries at the self-dual point.

The results just described suggest that self-dualities may
increase the symmetry of a model drastically at the self-dual point, 
maybe even by becoming a continuous group of symmetries. 
This is an especially attractive
possibility for models that exhibit a phase transition at the self-dual
point, but in fact, it can be excluded on general principles. 
For suppose that one could find a self-duality transformation
$\mathcal{U}_\d(\theta)$ that depends on some set of continuous 
coordinates $\theta$, so that 
\begin{eqnarray}
\mathcal{U}_\d(\theta)\, H[\lambda_1,\lambda_2,\cdots]\, 
\mathcal{U}^\dagger_\d(\theta)=H[\lambda_2,\lambda_1,\cdots]
\end{eqnarray}
for any value of $\theta\neq 0$, {\it independently of the values of
the  couplings in $H$}. Such a group of self-duality unitaries would
become an extra continuous symmetry of the model at the self-dual
point. But this is impossible, because
$\mathcal{U}^2_\d(\theta)$ must be a symmetry always. Then, taking
$\theta=\epsilon$ infinitesimal so that  $\mathcal{U}^2_{\sf
d}(\epsilon) \approx \mathbb{1}+2i\epsilon\cdot \breve{T}$, we see that the
generators $\breve{T}$ must  always  commute with $H$. But then
$\mathcal{U}_{\sf d}(\epsilon)\approx \mathbb{1}+i\epsilon\cdot \breve{T}$ must
commute with $H$ as well, rather than represent a self-duality. 

The discussion above does not exclude the possibility that
$\mathcal{U}_\d$ may depend on the couplings in the Hamiltonian,
\begin{eqnarray}
\mathcal{U}_\d(\lambda_1,\lambda_2,\cdots) \, 
H[\lambda_1,\lambda_2,\cdots] \, \mathcal{U}^\dagger_{\sf
d}(\lambda_1,\lambda_2,\cdots)= H[\lambda_2,\lambda_1,\cdots],
\end{eqnarray}
and this is in fact the case for the spin \(S=1/2\) XY model
discussed in Section \ref{sec3.11}.
But this would not turn the self-duality at the self-dual point 
into continuous symmetry either. Rather, one would have  a
discrete set of symmetries, one discrete set for each value of the
self-dual coupling $\lambda\equiv \lambda_1=\lambda_2$. 

In closing, let us mention briefly two examples.
Consider first the finite, open, self-dual quantum Ising chain 
$\tilde{H}^o_{\sf I}$ introduced
in Section \ref{sec3.7}, Equation \eqref{finiteisingsd}. 
It is easy to verify that $\mathcal{U}^2_\d=\mathbb{1}$  (there is no
need to compute $\mathcal{U}_\d$ explicitly, just to note  that
$\mathcal{U}_\d$ implements the mapping defined in Equation
\eqref{automorphismfi}). At the self-dual point $J=h$, $\mathcal{U}_\d$
becomes a non-trivial discrete symmetry of the model, the generator of 
a $\mathbb{Z}_2$ symmetry group for $\tilde{H}^o_{\sf I}$.
This is especially interesting, since the standard \(\mathbb{Z}_2\) symmetry
of the Ising model is broken by the self-dual boundary term \(J\sigma^z_N\). 
For the {\it infinite} quantum Ising chain, we have from Equation
\eqref{aut_ising1} that
\begin{equation}
\Phi_{\sf d}^2(\sigma^x_i)=\sigma^x_{i-1},\ \ \ \ \ \ \ \ \ \ \ \
\Phi_{\sf d}^2(\sigma^z_i\sigma^z_{i+1})
=\sigma^z_{i-1}\sigma^z_{i}, 
\end{equation} 
Thus $\Phi_{\sf d}^{2}$ generates lattice translations to the left.

\subsection{Order and disorder variables for self-dual models}
\label{sec3.10}

Recognizing that self-dualities are unitary equivalences has consequences
that go beyond symmetry, and are intimately tied to the 
behavior of the quantum 
fluctuations that compete at a quantum phase transition. 
For self-dual models, there is a natural way to associate a disorder
parameter to any order parameter (and viceversa), through the self-duality
unitary, and moreover,  the eigenstates of the self-duality unitary are states at which
the expectation value of a pair of ``duality-conjugate" observables {\it are 
equal}. While these states are not specially meaningful at general couplings, 
at the self-dual point they can be chosen to be simultaneous eigenstates 
of the Hamiltonian 
(because the self-duality becomes a symmetry at the self-dual point).
 
The general setting we are going to consider in this section is that of 
a self-dual Hamiltonian \(H[\lambda]\)
depending on any number of parameters 
\(\lambda=(\lambda_1,\lambda_2,\cdots)\), with dual parameters \(\lambda^*\)
defined by
\begin{eqnarray}\label{dual_here}
H[\lambda^*]=\mathcal{U}_\d\,H[\lambda]\,\mathcal{U}_\d^\dagger.
\end{eqnarray}
The observable \({\cal O}_\d\) dual-conjugate to \({\cal O}\) 
is defined by the equation
\begin{equation}
{\cal O}_\d=\mathcal{U}_\d {\cal O} \mathcal{U}_{\d}^\dagger
\end{equation}
For example, \(H[\lambda^*]\) is the dual-conjugate of \(H[\lambda]\),
and for the Ising models studied in Section
\ref{sec3.7}, the dual-conjugates of the spin operators
\(\sigma^x_i,\ \sigma^z_i\) are the dual variables
\(\mu^x_i,\ \mu^z_i\) of 
Equations \eqref{ssstring} and \eqref{duali_circle}. 

The first interesting consequence of this definition is that,
{\it relative to self-duality eigenstates \(|\phi_j\rangle\)},
\begin{equation}
{\mathcal{U}}_\d|\phi_j\rangle=e^{i\phi_j} |{\phi_j}\rangle,\ \ \ \ \ \ 
j=1,\cdots,{\sf dim}\mathcal{H},
\end{equation}
pairs of observables that are dual-conjugate have identical
expectation values,
\begin{equation}
\langle\phi_j|{\cal O}_\d |\phi_j\rangle=
\langle\phi_j|\mathcal{U}_\d {\cal O} \mathcal{U}_{\d}^\dagger
|\phi_j\rangle=\langle\phi_j|{\cal O} |\phi_j\rangle.
\end{equation}
This is especially interesting at the self-dual point \(\lambda_{\sf sd}=
\lambda_{\sf sd}^*\), where the states \(|\phi_j\rangle\) can be chosen
to be simultaneous eigenstates of \(\mathcal{U}_\d\) and 
\(H[\lambda_{\sf sd}]\) (since
\(\mathcal{U}_\d\) is a symmetry of \(H\) at the self-dual point).

Next we would like to compare expectation values of dual-conjugate
pairs relative to arbitrary states \(|\psi\rangle\). For this it is convenient
to specialize the discussion to self-dualities that are their own inverses,
so that
\begin{equation}\label{very_sd}
\mathcal{U}_{\d}^\dagger=\mathcal{U}_\d .
\end{equation}
It is often possible to arrange for this to be the case, thanks to the
freedom in choosing \(\mathcal{U}_\d\) discussed in the previous section.
Then, we have on one hand that
\begin{equation}
\langle\psi|\mathcal{O}_\d|\psi\rangle=
\langle\psi_\d|\mathcal{O}|\psi_\d\rangle,
\end{equation}
where 
\begin{equation}
|\psi_\d\rangle\equiv \mathcal{U}_\d|\psi\rangle.
\end{equation}
But thanks to Equation \eqref{very_sd}, a completely analogous
relation holds for \(\mathcal{O}\):
\begin{equation}
\langle\psi|\mathcal{O}|\psi\rangle=
\langle\psi|{\mathcal{U}}_\d^2\mathcal{O}{\mathcal{U}}_\d^2|\psi\rangle=
\langle\psi_\d|\mathcal{O}_\d|\psi_\d\rangle,
\end{equation}
It is in this specific sense that \(\mathcal{O}\) and \(\mathcal{O}_\d\)
show perfectly complementary behavior.

Let us apply these general results to quantum phase transitions.
Let \(\vert \Omega; \lambda\rangle\) denote  a ground state for \(H[\lambda]\). 
Then \(\mathcal{U}_\d\vert \Omega; \lambda\rangle\) is a ground state
for \(H[\lambda^*]\), that we denote \(\vert \Omega;\lambda^*\rangle\). 
It follows that 
\begin{eqnarray}
\langle \Omega;\lambda\vert {\cal O} \vert \Omega;\lambda\rangle &=&\langle \Omega
;\lambda^*\vert {\cal O}_\d\vert \Omega;\lambda^*\rangle,\ \ \ \ \ \ \mbox{and}\\
\langle \Omega;\lambda\vert {\cal O}_\d \vert \Omega;\lambda\rangle &=&\langle \Omega
;\lambda^*\vert {\cal O}\vert \Omega;\lambda^*\rangle.\nonumber
\end{eqnarray}
Hence, if the mean value of ${\cal O}$ happens to be related to the {\it
order parameter} associated with a phase transition that takes place as 
the couplings \(\lambda\) are changed, it follows
immediately from the two relations above that \({\cal O}_\d\) represents
an operator  related to the {\it disorder parameter}. 

Consider for illustration the quantum Ising chain, and set
$\lambda \equiv J/h$, so that the dual $\lambda^*$, 
resulting from the self-duality transformation $h \leftrightarrow J$,  
is $\lambda^* = \lambda^{-1}$. Then we have that
\begin{equation}\label{sdsstring}
\langle 0;\lambda\vert \sigma^z_i\sigma^z_j\vert 0;\lambda\rangle=
\langle 0;\lambda^{-1}\vert \mu^z_i\mu^z_j\vert 0;\lambda^{-1}\rangle,\ \
\ \langle 0;\lambda\vert \mu^z_i\mu^z_j\vert 0;\lambda\rangle=
\langle 0;\lambda^{-1}\vert \sigma^z_i\sigma^z_j\vert 0;\lambda^{-1}\rangle.
\end{equation}
Equation \eqref{sdsstring} demonstrates that the string operator of
Equation \eqref{ssstring} is the disorder variable conjugate to the
order variable \(\sigma^z_i\) \cite{fradkin_susskind}. The relation
between our bond-algebraic  approach to {\it quantum disorder variables}
and the work of Kadanoff and Ceva \cite{Kadanoff_Ceva} on (commutative)
algebras in the {\it classical} $D=2$ Ising model is elaborated on in
Section \ref{appB}. Our bond-algebraic approach generalizes the 
work of \cite{Kadanoff_Ceva}.  Here we would like to point out that in Reference
\cite{Kadanoff_Ceva}, it was argued (only in the context of the $D=2$
Ising model) that the product of an order and a neighboring disorder
variable should behave as  a fermion. This is largely satisfied by our
operator order and disorder quantum variables for the quantum Ising
chain. If we define
\begin{eqnarray}
\gamma_i \equiv \sigma^z_i\mu^z_{i+1},
\end{eqnarray}
then it will be easy to check that the operator $\gamma_{i}$ represents
a Majorana fermion, that is, a Dirac fermion
that is self-conjugate (and consequently, its own anti-particle).  For the
current purposes, it suffices to mention that Majorana fermions can be
expressed in terms of (spinless) fermion creation/annihilation  
operators and satisfy the following anti-commutation relations
\begin{eqnarray}
\label{majorana_f}
\{\gamma_i, \gamma_j\}=2\delta_{i,j}.
\end{eqnarray}
This enables the standard Jordan-Wigner transformation \cite{GJW} that
maps $S=1/2$ spin degrees of freedom into spinless fermions (and
viceversa). 

It seems to be a general feature of $d=1$ quantum models
that the product of the order and disorder variables satisfies simple and
interesting algebraic relations. 
Unfortunately, this pattern seems to break down in higher dimensions.

\subsection{Emergent dualities}
\label{sec3.11}

Two standard ways to simplify models in condensed matter physics
are to restrict couplings to take very special values, and/or project
out some states of the full state space. A typical example is the
$t$-$J$ model \cite{auerbach}, that is obtained as a projection from the
Hubbard model in the strong-coupling limit. In this section, 
we explain how emergent (self-)dual properties
can appear in the effective models that come out of such manipulations \cite{con}, 
even when the starting models are not (self-)dual to
start with.
We discuss the two scenarios (special couplings versus reduced state space)
separately for simplicity, but examples more complex than the ones
we are going to consider can well present a blend of both. 

\subsubsection{Projective emergent dualities}

The projection of a Hamiltonian, and a corresponding bond algebra, 
into a sector (subspace) \(\mathcal{W}\) of the full state space 
\(\mathcal{H}\), produces a {\it new} bond algebra that may,
or may not, have new algebraic and duality properties. 
For instance, bonds that do not  commute in
general may commute when projected onto certain sectors, or vanish 
(thus reducing the number of relations that characterize the algebra).
Hence it may well be that the projected bond algebra enjoys (self-)dualities 
that are not available for the full model.   

More precisely, a
projection will always change the structure of the 
bond  algebra, unless the projector
\(P_\mathcal{W}=P_\mathcal{W}^2\) commutes with all the bonds. To see this,
notice that the following relation always holds
\begin{equation}
P_\mathcal{W}(h_\Gamma+\lambda h_{\Gamma'})P_\mathcal{W}\ =\ 
P_\mathcal{W}h_\Gamma P_\mathcal{W}+\lambda P_\mathcal{W} h_{\Gamma'}
P_\mathcal{W}.
\end{equation}
So, projection always preserve the linear structure. Problems
can develop with respect to the multiplicative structure, since
\begin{equation}
P_\mathcal{W}(h_\Gamma h_{\Gamma'})P_\mathcal{W} = P_\mathcal{W}h_\Gamma
P_\mathcal{W}\ P_\mathcal{W} h_{\Gamma'}P_\mathcal{W}
\end{equation}
will generally hold only if for all $\Gamma$, 
\begin{equation}
\label{hpw}
[h_\Gamma,P_\mathcal{W}]=0.\ \ \ \ \ \ \ \ 
\end{equation} 
If this is the case, then
\begin{eqnarray}
P_\mathcal{W}(h_\Gamma h_{\Gamma'})P_\mathcal{W}\ =\ 
P_\mathcal{W}^2(h_\Gamma h_{\Gamma'})P_\mathcal{W}^2\ =\
P_\mathcal{W}h_\Gamma P_\mathcal{W}\ P_\mathcal{W}
h_{\Gamma'}P_\mathcal{W},
\label{ppp}
\end{eqnarray}
and the projection process preserves (to some extent) 
the structure of the bond algebra (in other words,
the mapping \(\Phi(h_\Gamma)=P_\mathcal{W}h_\Gamma P_\mathcal{W}=
h_\Gamma P_\mathcal{W}\) is an algebra homomorphism).

However, it is unlikely that the projections of physical
interest will preserve the bond algebra
as in Equation \eqref{ppp}, and so we can expect that the 
effective, projected model will have a different bond algebra
(Gauge models represent 
the most important exception to this rule, see the next section).
We call {\it emergent dualities} those dualities that are brought
about by this change in the bond algebra due to a projection
(or a special fixing of the couplings, like in the next subsection),  
to stress that these dualities {\it emerge} in some sector of the
theory  but need not be exact relations for the full system. 

In many
instances,  the sector of interest is that of low energies. At low
temperatures,  the system becomes more and more confined to this Hilbert
space sector (especially so when  spectral gaps are present between the
low energy sector of $H$ and all other excited states).  It is important
to appreciate that emergent dualities must be  unitarily implementable,
just as ordinary dualities, with the  extra freedom that the unitary
transformations need only be defined on certain subspaces such as that
spanned by the low energy states. 

To make this lucid, we now consider two examples. The first 
example is afforded by the elementary  Hamiltonian
\begin{equation}
\label{hsl}
H_{\sf L}=L_{z}+\frac{1}{2},  
\end{equation}
with the angular momentum operator \(L_{z}=-i \partial/\partial\theta\).
If one takes its domain to be the full Hilbert space of wave-functions on the
circle \(\langle \theta|\psi\rangle=\psi(\theta)\in{\cal L}^2(U(1))\), 
then \(H_{\sf L}\) is  not bounded below. 
However, $H_{\sf L}$ has an {\it emergent} duality to the standard harmonic
oscillator
\begin{equation}
H_{\sf HO}=a^\dagger a+\frac{1}{2}, \ \ \ \ \ \ \ \ \ \ 
[a,a^\dagger]=1,
\end{equation}
on the sector of states of non-negative angular momentum. 

To see how this works, consider the algebra,
\begin{equation}
[L_{z},\ A]=-A,\ \ \ \ \ \ \ \ [L_{z},\ A^\dagger]=A^\dagger,
\ \ \ \ \ \ \ \ [A,A^\dagger]=0
\end{equation}
where \(A\) and \(A^\dagger\) are the ladder operators associated with
\(L_{z}\), and act on wave-functions by multiplication, 
\(\langle\theta\vert A\vert\psi\rangle=e^{-i\theta}\psi(\theta)\).
Then, if we let \( \{\vert m\rangle \}\) denote angular  momentum eigenstates, 
\begin{eqnarray}
L_{z}\vert m\rangle=\ m\vert m\rangle,\ \ \ \ \ \ \ \ \ \ 
\langle\theta\vert m\rangle=\frac{1}{\sqrt{2\pi}}e^{im\theta},
\end{eqnarray}
we have that \(\langle\theta\vert A|m\rangle=\langle\theta\vert m-1\rangle\) and
\(\langle\theta\vert A^\dagger|m\rangle=\langle\theta\vert m+1\rangle\).

We would like to study this algebra on the subspace spanned by the
states \(\vert m\rangle\) of non-negative angular momentum only, \(m=0,1,2,
\cdots\ \). If \(P\) denotes the orthogonal projector onto this
subspace, then  it is not difficult to check that 
\begin{equation}
\label{LPAP}
[L_P,\ A_P]=-A_P,\ \ \ \ \ \ \ \ [L_P,\ A_P^\dagger]=A_P^\dagger,
\ \ \ \ \ \ \ \ [A_P,\ A_P^\dagger]=\ P_0,\ \ \ 
\end{equation}
where \(P_0=\vert 0 \rangle\langle 0\vert\) denotes the projector onto
the eigenspace of \(0\) angular momentum (the ground state sector
for $PH_{\sf L}P$).  In Equation \eqref{LPAP}, we employ the
shorthand \(M_P=PMP\) for general projected operators (and we further
abbreviate, $L_{P} = P L_{z} P$).  The algebra of Equation \eqref{LPAP}
is isomorphic to that of the harmonic
oscillator \cite{remark_HS}, as the mapping
\begin{eqnarray}
&&L_P \ \stackrel{\Phi_\d}{\longrightarrow}\ a^\dagger a,\nonumber\\
&& A_P\ \stackrel{\Phi_\d}{\longrightarrow}\ (a^\dagger a+1)^{-1/2}a,
\ \ \ \ \ \ \ \ \ \ 
A_P^\dagger\ \stackrel{\Phi_\d}{\longrightarrow}\ a^\dagger (a^\dagger
a+1)^{-1/2}.
\label{Losc}
\end{eqnarray} 
shows. Thus \(\Phi_\d(PH_{\sf L}P)=H_{\sf HO}\) embodies an
elementary  {\it emergent duality}.

\begin{figure}[h]
\begin{center}
\includegraphics[width=.43\columnwidth]{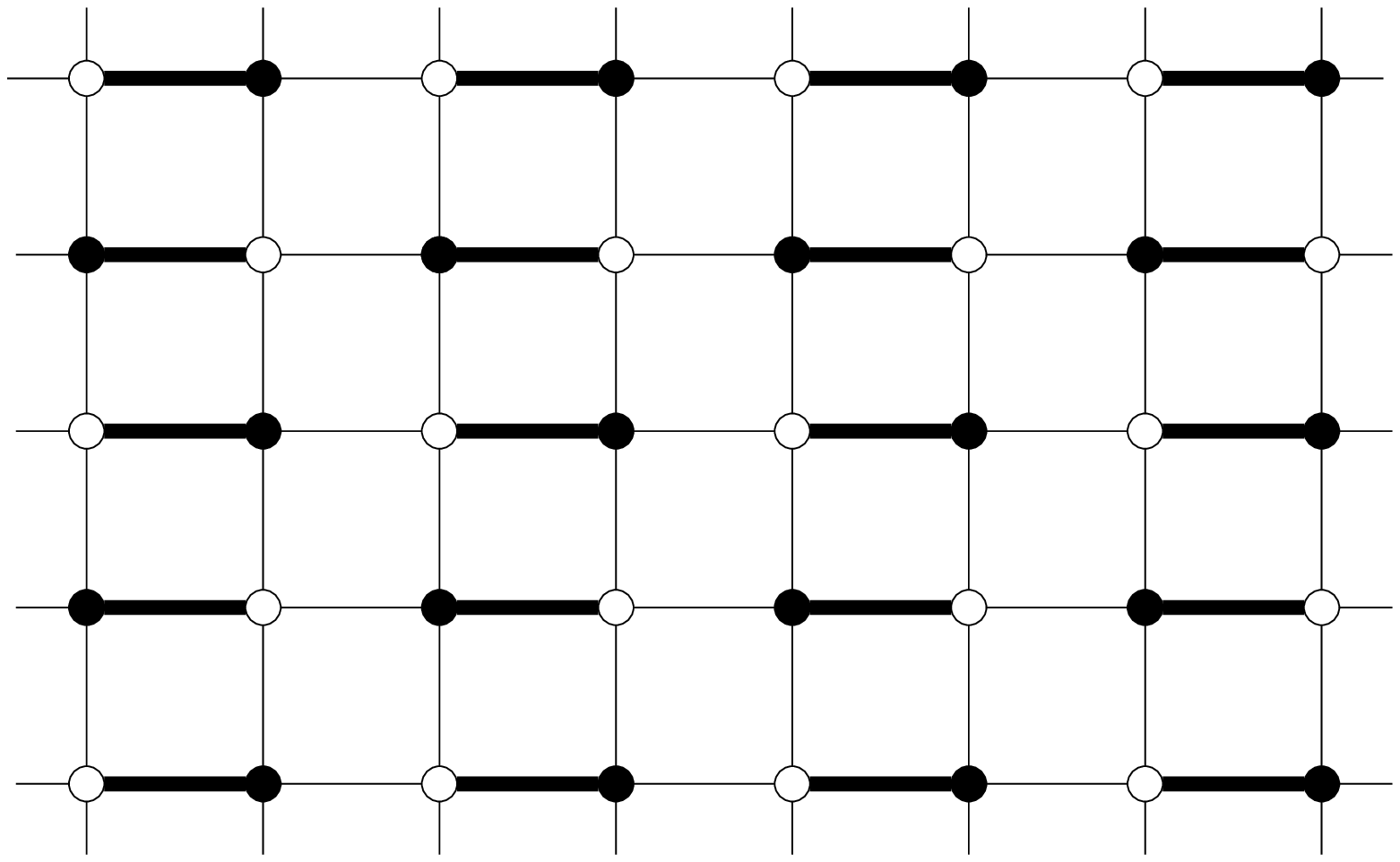} \hspace*{1cm}
\includegraphics[width=.43\columnwidth]{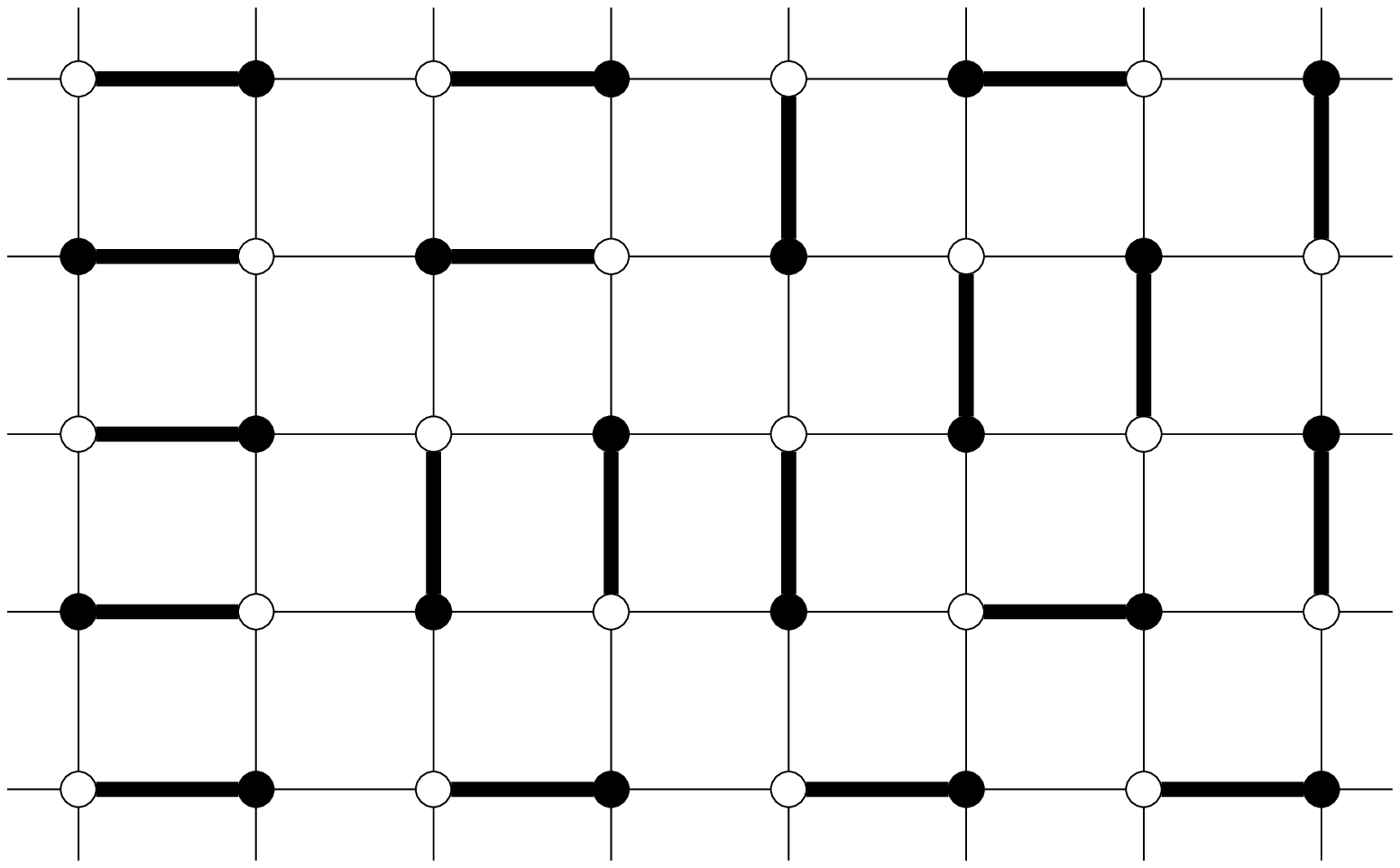}
\end{center}
\caption{Two dimer coverings of the square lattice. Dimer coverings label
an orthonormal basis of states for the state space of the quantum dimer model
defined in the text.}
\label{qdm1}
\end{figure}

A more interesting example of an {\it emergent self-duality}  is
afforded  by  the Quantum Dimer Model \cite{con}. This model's 
Hamiltonian \cite{RK}
\begin{eqnarray}
H_{\sf QDM}&=&\sum_{\Box} \left[
-t\left(\left|\vv\right\rangle\left\langle\hh\right | +
\left|\hh\right\rangle\left\langle\vv \right | \right)\right . _{\Box} 
\nonumber\\
&+&\left.v\left(\left|\vv\right\rangle\left\langle\vv\right|+
\left|\hh\right\rangle\left\langle\hh \right|\right)_{\Box} \right],
\label{QDM}
\end{eqnarray}
acts on a state space spanned by  orthonormal states labelled by dense dimer
coverings of a lattice, see Figure \ref{qdm1}  (the sum 
\(\sum_{\Box}\) must include all the elementary plaquettes $\Box$).  It
contains both a kinetic $t$ term that flips one dimer tiling of any
plaquette to another (a horizontal covering to a vertical one and
viceversa),  and a potential $v$ term. On every plaquette, the potential
term operator is equal to the square of the kinetic term \cite{nn}. 

At the so-called Rokhsar-Kivelson (RK) point $t=v$ \cite{RK}, the ground
states are equal amplitude superpositions of dimer coverings. If $P_{g}$
is the projection operator onto the ground state sector, then on any
plaquette, 
\begin{eqnarray} \!\!\!\!\!\!\! \!\!\!\!\!
P_{g} [\left(\left|\vv\right\rangle\left\langle\hh\right| +
\left|\hh\right\rangle\left\langle\vv \right|\right)]_{\Box} P_{g}
= P_{g} [\left(\left|\vv\right\rangle\left\langle\vv\right|+
\left|\hh\right\rangle\left\langle\hh \right|\right)]_{\Box} P_{g} =
x_{\Box} P_{g} ,
\end{eqnarray}
with $x_{\Box} = 0$ or $1$ on the particular plaquette $\Box$.  At the 
RK point, the projected Hamiltonian becomes $P_{g} H_{\sf QDM} P_{g}
=0$. Since both the kinetic ($t$) and potential ($v$) terms are given by
$x_{\Box} P_{g}$ within the ground state sector, they can be
interchanged without affecting the bond algebra. This self-duality
emerges exclusively in the ground state sector of the model at the RK
point.

\subsubsection{Coupling-dependent emergent dualities}
\label{sec3.10.2}

Next we discuss dualities that emerge in some specific region 
of the space of parameters of some models. To some extent, this 
notion is already included in the general concept of bond-algebraic 
duality, 
but the examples we have studied do not emphasize it sufficiently. 
The point to notice
is that we can choose the bond generators to 
include the coupling and external fields in a non-trivial fashion. 
Then the structure of the corresponding {\it bond algebra depends on 
those couplings} as well, and varies with them. Thus it is possible that
some specific values of the couplings will afford (self-)dual 
properties that may be absent in general. 

Let us present an example to clarify this idea. In what follows, we 
show that the spin \(S=1/2\) XY model 
\begin{equation}\label{hxyn}
H_{\sf XY}= \sum_i\ (J_x\sigma^x_i\sigma^x_{i+1}+\
J_y\sigma^y_i\sigma^y_{i+1} +\ \bar{h} \sigma^z_i)
\end{equation}
presents an {\it emergent self-duality on the surface \(J_x=J=1/J_y\)} 
in coupling space \((J_x,J_y,\bar{h}\)) \cite{hinrichsen}. 

The reason is that on this surface we can re-write
\begin{equation}
H_{\sf XY}[J,h]=\sum_i\ (A_i(J,h)+B_i(J^{-1},h^{-1}))
\end{equation} 
as a sum of bonds 
\begin{equation}
A_i(J,h)\equiv J\sigma^x_{i}\sigma^x_{i+1}+\ h\sigma^z_i,\ \ \ \ \ \ 
B_i(J^{-1},h^{-1})=J^{-1}\sigma^y_i\sigma^y_{i+1}+\ h^{-1}\sigma^z_{i+1},
\end{equation}
provided we split \(\bar{h}=h+h^{-1}\),
thus generating a bond algebra symmetrical under the exchange $J\leftrightarrow h$
(so that after the duality transformation the dual magnetic field reads \(\bar{h}^*=J+J^{-1}\)).
In what follows, we write \(A_i, B_i\) for \(A_i(J,h), B_i(J^{-1},h^{-1})\),
and \(\hat{A}_i, \hat{B}_i\) for \(A_i(h,J), B_i(h^{-1},J^{-1})\).

The self-duality \(J\leftrightarrow h\) is already intuitively clear from the set 
of relations that characterize the bond algebra generated by the \(A_i, B_i\):
\begin{eqnarray}
&&\ \ \ \ \ \ A_i^2=J^2+h^2,\ \ \ \ B_i^2=J^{-2}+h^{-2},\nonumber\\
&&\ \ \ \ \ \ \{B_i,A_i\}=0, \ \ \ \  \{B_i,A_{i+1}\}=2,\nonumber\\
&& A_i\ A_{i+1}\ A_i\ A_{i+1}\ +\ A_{i+1}\ A_i\ A_{i+1}\ A_i =\ 
2(J^4+h^4)\label{xy_relation}\\
&&B_i\ B_{i+1}\ B_i\ B_{i+1}\ +\ B_{i+1}\ B_i\ B_{i+1}\ B_i=\
2(J^{-4}+h^{-4})\nonumber\\
&&A_i\ B_{i+1}\ A_i\ B_{i+1}\ +\  B_{i+1}\ A_i\  B_{i+1}\ A_i=\
2(J^2h^{-2}+h^2J^{-2})\nonumber
\end{eqnarray}
(all other pairs of generators commute). Rigorously, the self-duality
follows from the observation that the mapping
\begin{equation}\label{iso_xy}
A_i\dual\hat{A}_i,\ \ \ \ \ \ B_i\dual\hat{B}_i,
\end{equation}
is an isomorphism, simply because the dual bonds
\(\hat{A}_i, \hat{B}_i\)
satisfy the same relations of Equation \eqref{xy_relation},
so that \(H_{\sf XY}[h,J]=\sum_i\ (\hat{A}_i+\hat{B}_i)\) is unitarily
equivalent to  \(H_{\sf XY}[J,h]\).

The self-duality isomorphism of Equation \eqref{iso_xy} reduces to the 
identity map at the self-dual point \(J=h\), because the bonds  
\(\hat{A}_i, \hat{B}_i\) become identical to the \(A_i, B_i\) 
there. This has interesting consequences in the dual variables 
to be computed below. Also, the product
\begin{eqnarray}
iA_jB_j=(Jh^{-1}+hJ^{-1})\sigma^x_j\sigma^y_{j+1}=
i\hat{A}_j\hat{B}_j,
\end{eqnarray}
is invariant under duality and can be added to the Hamiltonian
with arbitrary couplings without spoiling its self-dual structure.

The dual variables for this problem are particularly interesting 
because they depend on the coupling parameters, and have an additive
and multiplicative structure (while every other set of dual variables
considered in this paper are purely multiplicative). We will indicate
how to construct them in general, and write them explicitly for the
simplest case of just two sites (\(N=2\)). 

The starting point is to notice
that the isomorphism of Equation \eqref{iso_xy}
works just as well when restricted to a finite bond
algebra generated by the \(2N\) bonds \(A_i, B_i\) with \(i=1,\cdots,N\).
This shows that the {\it finite} rendition \(H_{\sf XY}=\sum_{i=1}^{N-1}
\ (A_i+B_i)\) is self-dual (because of self-dual BCs, this
Hamiltonian has \(\sigma^z_1\) coupled to \(h\) only, and \(\sigma^z_{N}\)
coupled only to \(h^{-1}\)). The next step is to figure out whether
the single spins \(\sigma^x_i, \sigma^y_i\), \(i=1,\cdots,N\)
are elements in the bond algebra generated by \(A_i, B_i\), \(i=1,\cdots,N-1\).
Clearly they are not, since every one of these bonds commutes
with \(\prod_{i=1}^{N}\ \sigma^z_i\). As in Section \ref{sec3.7},
the solution to this problem is to enlarge  the bond algebra by
adding generators that do not spoil the symmetry under exchange of  $J$
and $h$.

A simple analysis shows that two such operators are 
$\sigma^y_1$ and $\sigma^x_N$, since they
commute with almost every other bond, except
for $A_{1}$ and $B_{N-1}$,
\begin{equation}
\{A_1,\sigma^y_1\}=0,\ \ \ \ \ \ \{B_{N-1}, \sigma^x_N\}=0,
\end{equation}
and these relations (being independent of the couplings) 
preserve the symmetry in $J$ and $h$.
In other words, the extended Hamiltonian 
\begin{equation}\label{sdXY}
H_{\sf XY}[J,h,\tilde{h}^y,\tilde{h}^x]=\tilde{h}^y\sigma^y_1+
\tilde{h}^x\sigma^x_N+ \sum_{i=1}^{N-1}\ 
(J\sigma^x_i\sigma^x_{i+1}+h\sigma^z_i+ J^{-1}\sigma^y_i\sigma^y_{i+1}+
h^{-1}\sigma^z_{i+1})
\end{equation}
is self-dual under the exchange $J\leftrightarrow h$, provided that the
real constants $\tilde{h}^y$ and  $\tilde{h}^x$ are kept fixed, as follows
from the isomorphism of Equation \eqref{iso_xy} extended as
\begin{equation}\label{automorphism_sdXY}
\sigma^y_1\dual \sigma^y_1,\ \ \ \ \ \ \sigma^x_N\dual\sigma^x_N.
\end{equation}
This is
a {\it hidden self-duality} that will not be apparent if we consider
only bond algebras independent  of the couplings, and that will not hold
true unless \(J_x=J=1/J_y\). 

In the special case with only two sites, \(N=2\), the inclusion
of \(\sigma^y_1\) in the list of generators suffices to compute
dual variables. {}From Equations \eqref{iso_xy} and \eqref{automorphism_sdXY},
we get
\begin{eqnarray}
&&\mu^y_1=\Phi_\d(\sigma^y_1)=\sigma^y_1 ,\nonumber\\
&&\mu^x_1=\Phi_\d(\sigma^x_1)=
(2Jh\sigma^x_1+(J^2-h^2)\sigma^z_1\sigma^x_2)/(J^2+h^2),\nonumber\\
&&\  \\
&&\mu^y_2=\Phi_\d(\sigma^y_2)=
(2(Jh)^{-1}\sigma^y_2+(J^{-2}-h^{-2})\sigma^y_1\sigma^z_2)/(J^{-2}+h^{-2}),\nonumber\\
&&\mu^x_2=\Phi_\d(\sigma^x_2)=\sigma^x_2.
\nonumber
\end{eqnarray}
These dual variables become identical to the original ones at the self-dual point, 
as expected, since the self-duality map of Equation \eqref{iso_xy}
reduces to the identity map.

It is interesting to point out, in the light of this self-duality, an argument that
has been put forward
to show that non-Abelian self-dualities cannot possibly exist. In Reference 
\cite{orland}, it is argued that self-dualities are unitary transformations
{\it that exchange the kinetic with the potential energy term} in a Hamiltonian,
and since these two terms must have different spectra for a non-Abelian theory, 
a non-Abelian self-duality cannot exist. The emergent self-duality
of the XY model does not explicitly contradict this reasoning, but 
suggests a way to escape its conclusion:
an emergent non-Abelian self-duality may appear as a property
of bonds that are {\it combinations of kinetic and 
potential energy terms}, since such combinations can have matching spectra.

\subsection{Elimination of gauge symmetries by bond-algebraic dualities}
\label{sec3.12}

In this section we explain an extension of the notion of duality
established in Section \ref{sec3.8} that can accommodate changes in
the dimension of the state space, and show its use  
to eliminate gauge symmetry constraints. In practice, however, we can potentially 
eliminate any {\it local} (or in the language
of Reference \cite{tqo}, \(d=0\) gauge-like) symmetry 
in this way, so it is important to keep in mind
that a local symmetry need not always be a gauge constraint that
can be disposed of. We term dualities that 
eliminates gauge symmetries gauge-reducing dualities.
We start with a brief reminder of the distinction between ordinary (Wigner) and gauge
quantum symmetries, before discussing gauge-reducing dualities in detail. 
In principle, the ideas that follow apply equally well to Abelian and non-Abelian 
gauge theories,
but non-Abelian models present technical complications that put them at the frontier
of bond-algebraic studies, and thus beyond the scope
of this paper.

\subsubsection{Ordinary versus gauge symmetries}

Quantum symmetries are always embodied in one and the same mathematical
statement: they are unitary or anti-unitary mappings that commute with the Hamiltonian
(see the discussion at the beginning of Section \ref{sec3.9}).
But this is not to say that all symmetries have the same physical meaning,
nor the same mathematical consequences. There is a distinction 
between ordinary symmetries, like rotations in space, and gauge symmetries. 
Ordinary symmetries  have direct physical impact, since they 
can influence the 
level degeneracy of a Hamiltonian (and with it, its thermal physics, see Appendix
\ref{sec3.1}) and constrain transition amplitudes to satisfy stringent selection 
rules. On the other hand, gauge symmetries are {\it constraints} pointing to
a fundamental redundancy. The state space of a model
with gauge symmetries is {\it larger than physical}, meaning that it contains
states that cannot be prepared or observed by experimental means (or may even
contain states of negative norm). The
{\it sector of physical states} is precisely that sector that 
is invariant under the action of all the gauge symmetries. 
Similarly, Hermitian operators that do not
commute with the gauge symmetries are not observables, in the sense
that its eigenvalues do not represent measurable quantities (think, for 
example, of the vector potential in QED). In a gauge
theory, an observable must be Hermitian, and commute with all the gauge
symmetries. 

In perspective, gauge symmetries are better thought of as constraints,
and it may seem desirable to keep them conceptually far apart from 
ordinary symmetries.
But it is unavoidable on first principles that quantum constraints
may look just like a symmetry. For suppose \(C\) is an operator representing
a quantum constraint. Then it must be that 
\begin{equation}
\frac{dC}{dt}=0=i[H,C],\ \ \ \ (\mbox{consistent constrained dynamics}),
\end{equation}
to ensure that the dynamics generated by \(H\) is consistent with 
the constraint. Then if \(C\) is (Hermitian)
unitary, it will look just like (the generator of) a symmetry.

In practice, physical input is required to set apart
constraints from symmetries. Take, for example,
the specification of the quantum statistics of identical particles.
Until  Pauli proposed his exclusion principle, it would have
been natural to argue that the many-body Schr\"odinger equation
for indistinguishable particles had the group of permutations among its {\it symmetries}. 
It took the introduction of a {\it new physical principle} 
to show that these symmetries were in fact {\it constraints}, 
or superselection rules, 
that select the fermionic or bosonic sector of Fock space as the
sector of physical states.  

If a quantum gauge theory arises from the quantization of a {\it classical} gauge theory, 
then there is no risk of confusing gauge and ordinary symmetries.
On the other hand, recent developments in condensed matter physics are 
fostering the development of quantum models that do not show an obvious classical
limit, and that posses local symmetries that look much like gauge symmetries (see, for 
example, Kitaev's honeycomb model, discussed at the end of Section \ref{sec3.6}, 
and Section \ref{subsec3.12.3}). Then, one is forced to face the problem of deciding 
whether these should be treated as ordinary symmetries, or 
as gauge symmetries (constraints), in part because the (self)-dualities available
will depend drastically on which one it is. 
The problem could easily show up for effective theories
of strongly correlated systems, where emergent symmetries \cite{GJW} could
well be local. 

\subsubsection{Gauge-reducing dualities}

In the light of the previous discussion, it would seem natural to assume that
a duality between a model with and a model without gauge symmetries 
(a gauge-reducing duality) should
be emergent, in the sense of Section \ref{sec3.11}. 
That is, it would seem that one should 
project the bond algebra into the subspace of gauge invariant states first,
\(h_\Gamma \rightarrow P_{\sf GI}h_\Gamma P_{\sf GI}\), in order to figure out
the algebraic relations between physical (gauge-invariant) bonds, and then look 
for a duality. 

But as it turns out, the bond-algebraic approach does not require the
elimination of gauge symmetries to work, and that is why bond-algebraic 
dualities are practical tools for removing gauge symmetries. If one chooses
the bond  algebra of the gauge model wisely, one can find mappings that preserve all the 
algebraic relations to models that do not
have any gauge symmetries.  

To make these ideas more precise, let \(H_{\sf G}\) be the Hamiltonian for the gauge model,
with gauge symmetries \(G_\Gamma\), \([H_{\sf G},G_\Gamma]=0\), and let \(H_{\sf GR}\)
be the dual, completely gauge-reduced model. Then the gauge-reducing duality 
maps
\begin{equation}\label{projective_unitary_I}
\Phi_\d(H_{\sf G})=H_{\sf GR},\ \ \ \ \ \ \mbox{and}\ \ \ \ 
\Phi_\d(G_\Gamma)=\mathbb{1},\ \  \forall \Gamma,
\end{equation}
thus rendering all the gauge symmetries trivial. Notice that \(\Phi_\d\) is not
an isomorphisms as for ordinary dualities, but rather an
{\it homomorphism} (homomorphisms preserve all the algebraic relations, 
but need not be one-to-one). To be quantum-mechanically 
meaningful, \(\Phi_\d\) must be implementable as an operator \(U_\d\)
(called a projective unitary) 
that preserves the norm of gauge-invariant states, and projects other states out.
In formulas,
\begin{equation}
\Phi_\d(\mathcal{O})=U_\d \mathcal{O}U_\d^\dagger,
\end{equation}
 with
\begin{equation}\label{projective_unitary_II}
U_\d U_\d^\dagger=\mathbb{1},\ \ \ \ \ \  U_\d^\dagger U_\d=P_{\sf GI},
\end{equation}
where \(P_{\sf GI}=P_{\sf GI}^2=P_{\sf GI}^\dagger\) is the orthogonal
projector onto the subspace of gauge invariant states \(|\psi\rangle\) 
that satisfy
\begin{equation}
G_\Gamma|\psi\rangle=|\psi\rangle,\ \  \forall \Gamma.
\end{equation}
This completes the definition of a gauge-reducing duality. 

Appendix \ref{appG}
describes a
concrete example of projective unitaries \(U_\d\). 
Unlike ordinary unitaries, projective unitaries are represented by {\it rectangular
matrices}. In the particular case described in Equation 
\eqref{projective_unitary_I}, \(U_\d\) has dimension \({\sf dim}\mathcal{H}_{\sf GR}
\times{\sf dim}\mathcal{H}_{\sf G}\), where \(\mathcal{H}_{\sf G}\) 
and \(\mathcal{H}_{\sf GR}\) are the state spaces of the
gauge and gauge-reduced models, respectively.

Our bond-algebraic approach constitutes an excellent technique for detecting 
gauge-reducing dualities, in part because 
gauge symmetries are {\it local} in general.
This makes possible  to choose a set of bond generators such that
each bond {\it individually} commutes with the gauge symmetries.
In this way we gain direct access to the physical (gauge-invariant)
algebra of interactions, and we can look for bond algebraic dualities to other
representations that show {\it no gauge symmetries}.  We saw already
a duality along these lines in Section \ref{sec3.6}, when we studied 
a new duality for the Abelian quantum Ising and non-Abelian Heisenberg models
in arbitrary dimension $d$. Let us study next
a simpler example, well-known in the literature \cite{fradkin_susskind,kogut},
from our new perspective.

\begin{figure}[h]
\begin{center}
\includegraphics[width=0.5\columnwidth]{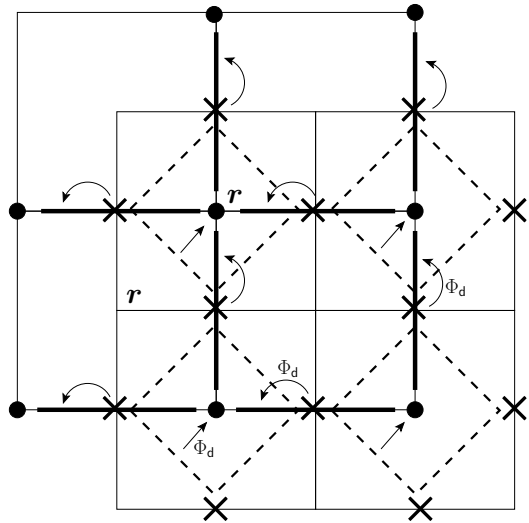}
\end{center}
\caption{A joint (superimposed) representation of the \(d=2\) dimensional Ising and
$\mathbb{Z}_2$ (Ising) gauge models.
The bonds of the  Ising model are indicated by heavy
bullets and rods; and the bonds of the $\mathbb{Z}_2$ gauge model by 
crosses and dashed diamonds. The square lattices for the quantum Ising  and
$\mathbb{Z}_2$ (Ising) gauge models are shown displaced relative to each
other (they are lattices dual to each other), to clarify the geometric aspects 
of the duality homomorphism \(\Phi_\d\) of Equation \eqref{gauge_to_ising}.
}
\label{duality_gauge_ising}
\end{figure}

The \(\mathbb{Z}_2\), \(d=2\) dimensional gauge model \cite{fradkin_susskind},
\begin{equation}\label{ising_gauge}
H_{\sf G}=\sum_\r\ (\sigma^x_{(\r,1)}+\sigma^x_{(\r,2)}+
\lambda\ B_{(\r,3)}),
\end{equation}
features spin $S=1/2$ degrees of freedom residing on links of a square
lattice, and plaquette operators $B_{(\r,3)}$ defined in Equation 
\eqref{gauge_plaquette}. 
Its group of gauge symmetries is generated by the unitaries
\begin{equation}
G_\r= \sigma^x_{(\r,1)}\sigma^x_{(\r,2)} \sigma^x_{(\r-\i,1)}
\sigma^x_{(\r-\j,2)},
\end{equation}
that not only commute with \(H_{\sf G}\), \([H_{\sf G},G_\r]=0\), but 
commute with each one of the bonds  
\begin{equation}
\sigma^x_{(\r,1)},\ \ \ \ \ \ \sigma^x_{(\r,2)}, \ \ \ \ \ \ B_{(\r,3)},
\end{equation}
individually. In other words, the bond algebra they generate is 
gauge-invariant, and it is further characterized by simple relations:
(i) all the bonds square to the identity, 
(ii) each spin \(\sigma^x\) anti-commutes with two adjacent plaquettes
\(B_{(.,3)}\), and 
(iii) each plaquette \(B_{(.,3)}\) anti-commutes
with four spins \(\sigma^x\). 
This set of relations is identical to the one found in the \(d=2\)
dimensional  quantum Ising model of Equation \eqref{anyd_ising}, and 
the mapping 
\begin{eqnarray}
\sigma^x_{(\r,1)}\dual \sigma^z_{\r-\j}\sigma^z_\r,\ \ \ \ \ \
\sigma^x_{(\r,2)}\dual \sigma^z_{\r-\i}\sigma^z_\r,\ \ \ \ \ \
B_{(\r,3)}\dual\sigma^x_\r,
\label{gauge_to_ising}
\end{eqnarray}
illustrated in  Figure \ref{duality_gauge_ising}, shows
that the two are homomorphic. Thus \(\Phi_\d\) 
maps $H_{\sf G}$ to $H_{\sf I}$, provided we identify the constants 
$\lambda\leftrightarrow h$ and  $1 \leftrightarrow J$.

On the other hand, the mapping  \eqref{gauge_to_ising} 
is intriguing, because the Ising model does not have any gauge
symmetries. What happened to all the gauge symmetries? To answer this
question, we compute
\begin{eqnarray}
&&\Phi_\d(G_\r)=\Phi_\d(\sigma^x_{(\r,1)}\sigma^x_{(\r,2)}
\sigma^x_{(\r-\i,1)} \sigma^x_{(\r-\j,2)})=\nonumber\\
&&\sigma^z_{\r-\j}\sigma^z_\r\times\sigma^z_{\r-\i}\sigma^z_\r\times
\sigma^z_{\r-\j-\i}\sigma^z_{\r-\i}\times\sigma^z_{\r-\i-\j}\sigma^z_{\r-\j}=
\mathbb{1}.
\end{eqnarray}
Thus we see that \(\Phi_\d\) is a gauge-reducing duality homomorphism, and that
\(H_{\sf I}\) represents all the physics contained in \(H_{\sf G}\), but 
without all the gauge redundancies. 
 
Since a gauge-invariant bond algebra
captures the purely physical properties of the interactions, the
equivalence (duality) between the two theories is self-evident in our
formalism. The bond algebraic approach naturally includes gauge
invariant Wilson loops, Aharonov-Bohm phases, and the local fields 
\(F^{\mu \nu}\)  as gauge invariant quantities constructed out of the
individual bonds of the gauge theory. 

Let us explain briefly how to construct
the projective unitary that implements the mapping of Equation \eqref{gauge_to_ising}.
Notice that the complete commuting set of observables
\(\sigma^x_{(\r,\nu)}\), \(\nu=1,2\), is mapped to the set
\(\sigma^z_\r\sigma^z_{\r+\bm{e_\nu}}\). Thus we can write
the projective unitary \(U_{\sf d}\)  
in terms of the basis \(|x\rangle\)  
of simultaneous eigenstates of \(\sigma^x_{(\r,\nu)}\), and the spanning set
\(|z\rangle\) of simultaneous eigenstates of \(\sigma^z_\r\sigma^z_{\r+\bm{e_\nu}}\),
\begin{eqnarray}
\sigma^x_{(\r,\nu)}|x\rangle&=&x_{(\r,\nu)}|x\rangle,
\ \ \ \ \ \ x_{(\r,\nu)}=\pm 1,\\
\sigma^z_\r\sigma^z_{\r+\bm{e_\nu}}|z\rangle&=&z_{(\r,\nu)}|z\rangle,
\ \ \ \ \ \ z_{(\r,\nu)}=\pm 1 . \nonumber
\end{eqnarray}
The set \(\{|z\rangle\}\) is over-complete, because the same vector shows
up $m_z$ times with different labels, and it is convenient to renormalize
the  \(|z\rangle\) so that 
\begin{equation}
\langle z|z'\rangle=\frac{1}{m_z}\tilde{\delta}(z,z'),
\end{equation}
\(\tilde{\delta}(z,z')\) equals \(1\) if \(z\) and \(z'\) label the same 
vector, and zero otherwise. 

With these conventions in place, we can describe \(U_{\sf d}\) explicitly:
\begin{equation}
U_{\sf d}^\dagger=\sum_{x_{(\r,\bar{\nu})}=
z_{(\r-\bm{e_{{\nu}}},{\nu})}} |x\rangle
\langle z|,
\end{equation}
where \(\bar{\nu}=2\) if \(\nu=1\), and \(\bar{\nu}=1\) if \(\nu=2\).
The condition \({x_{(\r,\bar{\nu})}=z_{(\r-\bm{e_{{\nu}}},{\nu})}}\) 
follows from the duality homomorphism that maps
\begin{equation}
\sigma^x_{(\r,\bar{\nu})}\dual \sigma^z_{\r-\bm{e_\nu}}\sigma^z_\r.
\end{equation}
In Appendix \ref{appG}, we describe this same construction for 
finite renditions of the model that can be checked numerically.

\subsubsection{Ordinary versus gauge symmetries II: An example}
\label{subsec3.12.3}

The bond-algebraic elimination of gauge symmetries can proceed
just as easily for {\it any local symmetry}, but if the symmetry that gets
discarded is not gauge, the dual model is not a faithful representation
of the physics of the original model. This fact forces us
to reconsider critically the concept of a gauge symmetry. What sets an
ordinary local symmetry (that should not be eliminated)
apart from a gauge symmetry?  Is there any {\it intrinsic} property of
\(H_{\sf G}\) and/or its symmetries that distinguish some
of them as gauge symmetries?  Unfortunately, we do not know the answer to this
question, and, as we understand them know, 
it seems that dualities cannot set apart gauge from ordinary
local symmetries. This is suggested by the example that we discuss next.

\begin{figure}[h]
\begin{center}
\includegraphics[width=0.6\columnwidth]{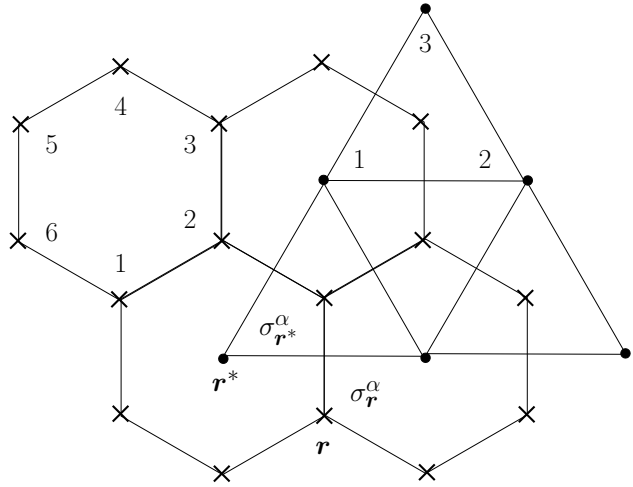}
\end{center}
\caption{Honeycomb and triangular dual lattices with vertices $\r$ and
$\r^*$, respectively.  There are  spins $S=1/2$ degrees of freedom placed on
the vertices of both lattices, that are  labelled either by the site labels
\(\r\) or $\r^*$, or by an integer \(i\) that indicates
their relative position within elementary hexagonal or triangular plaquettes. 
Notice also that \(\r\) ($\r^*$) can be used to label these plaquettes.}
\label{hex_tr}
\end{figure}

Consider a honeycomb lattice with spins \(S=1/2\) residing on its
vertices \(\r\),  and a dual triangular lattice with sites \(\r^*\). As
shown in Figure \ref{hex_tr}, we can use the sites \(\r^*\) to label the
elementary hexagons of the honeycomb lattice. 
With this convention and notation in place, we can introduce the
Hamiltonian 
\begin{equation}
H_{\sf honeycomb}[h,J]= \sum_\r\ h\ \sigma^x_\r+\ \sum_{\r^*}\  J\
(\sigma^z_1\sigma^z_2\sigma^z_3 \sigma^z_4\sigma^z_5\sigma^z_6)_{\r^*}
\end{equation}
that feature  plaquette interactions among the spins laying on every
single elementary hexagon. 
It is ideal to illustrate the dilemma brought
about just now, because $H_{\sf honeycomb}[h,J]$ does commutes with a large
set of local unitaries
\begin{equation}\label{sym_honey}
S_{\r^*}=(\sigma^x_1\sigma^x_2\sigma^x_3\sigma^x_4\sigma^x_5\sigma^x_6)_{\r^*},
\ \ \ \ \ \ \ \ \ \ \ [S_{\r^*},\ H_{\sf honeycomb}]=0,
\end{equation}
but does not have an obvious classical limit. This face us with the 
problem of deciding whether we should treat the \(S_{\r^*}\) as 
gauge symmetries and try to eliminate them, or as ordinary symmetries.  

In any case, there is a duality that maps all of the
operators \(S_{\r^*}\) to the identity. The dual model is most easily
described by  placing spins \(S=1/2\) on the sites \({\r^*}\) of the
dual triangular lattice,  as in Figure \ref{hex_tr}. Its Hamiltonian
then reads
\begin{eqnarray}
H_{\sf tr}[J,h]= \sum_{\r^*}\ J\ \sigma^x_{\r^*}+\ \sum_{\r}\  h\ 
(\sigma^z_1\sigma^z_2\sigma^z_3)_{\r},
\end{eqnarray}
and features  plaquette interactions among the spins laying on every
single elementary triangle. It is straightforward to check that the
mapping 
\begin{equation}
\sigma^x_\r \dual(\sigma^z_1\sigma^z_2\sigma^z_3)_{\r}\ , 
\ \ \ \ \ \ \ \ 
(\sigma^z_1\sigma^z_2\sigma^z_3 \sigma^z_4\sigma^z_5\sigma^z_6)_{\r^*}
\dual \sigma^x_{\r^*}
\end{equation}
is a homomorphism of bond algebras such that \(\Phi_\d(H_{\sf
honeycomb}[h,J])=H_{\sf tr}[J,h]\), {\it but \(H_{\sf tr}\) has no local
symmetries}. In particular,
\begin{equation}
\Phi_\d(S_{\r^*})=\prod_{i=1}^6\ \Phi_\d(\sigma^x_i)=\prod_{i=1}^6
\ (\sigma^z_1\sigma^z_2\sigma^z_3)_i=\mathbb{1},
\end{equation}
since \(\Phi_\d\) maps the six spins \(\sigma^x\) on the vertices of an
hexagon to the six plaquette terms \(\sigma^z_1\sigma^z_2\sigma^z_3\)
that share the  center point of that hexagon, a vertex of the dual
lattice. 

Whether this duality is physically meaningful or not rests on deciding
whether the symmetries of Equation \eqref{sym_honey} should be 
discarded or not.

\subsubsection{Systematic determination of gauge-reduced dual models}
\label{sec3.12.4}

There is a systematic way to construct completely gauge-reduced duals
of gauge models that has the unpleasant feature of 
requiring the introduction of non-local bonds in the dual model.
While there are intuitive 
physical arguments to justify this (namely, Gauss' law permits to
measure the charge of a particle 
by measuring its electric field on the surface of 
a sphere {\it arbitrarily far away from it}), the fact is that sometimes
completely gauge-reducing duals with local bonds 
are {\it readily available} (like in the example of the previous section),
and there is no need to exploit the systematic construction 
(see Sections \ref{sec5.3}, \ref{sec5.4}, \ref{sec5.5}, and \ref{sec6.4}). But sometimes
the systematic construction seems to be the best we can do (see Section 
\ref{sec6.3}).

In this section, we describe the systematic construction for the \(\mathbb{Z}_2\) 
gauge model of Equation \eqref{ising_gauge}, to illustrate the ideas on which
it rests (these ideas are a reinterpretation and generalization
in terms of bond algebras of an scheme introduced in 
Reference \cite{fradkin_susskind} to study the
\(d=3\), \(\mathbb{Z}_2\) gauge theory). The generalization to higher dimensions,
other (Abelian) gauge groups, and matter-coupled gauge theories is straightforward,
and will be illustrated in other parts of this paper (see for example
Section \ref{sec6.4}).  

Given $H_{\sf G}$, we are looking for a model \(H_{\sf GR}\),
and a mapping \(\Phi_\d\), \(\Phi_\d(H_{\sf G})=H_{\sf GR}\), that satisfy some 
stringent conditions. The bonds of the dual model \(H_{\sf GR}\) are going to be 
the operators
\begin{equation}
\Phi_\d(\sigma^x_{(\r,1)}),\ \ \ \ \ \Phi_\d(\sigma^x_{(\r,2)}),
\ \ \ \ \ \Phi_\d(B_{(\r,3)}),
\end{equation}
that satisfy
\begin{equation}\label{byebye_gauge}
\Phi_\d(G_\r)=\Phi_\d(\sigma^x_{(\r,1)})\Phi_\d(\sigma^x_{(\r,2)})
\Phi_\d(\sigma^x_{(\r-\i,1)})\Phi_\d(\sigma^x_{(\r-\j,2)})=\mathbb{1},
\end{equation}
and our job is to determine them (the splitting \( \Phi_\d(\sigma^x_{(\r,1)}\sigma^x_{(\r,2)}
\sigma^x_{(\r-\i,1)}\sigma^x_{(\r-\j,2)})=\Phi_\d(\sigma^x_{(\r,1)})\Phi_\d(\sigma^x_{(\r,2)})
\Phi_\d(\sigma^x_{(\r-\i,1)})\Phi_\d(\sigma^x_{(\r-\j,2)})\) is 
correct because the \(\sigma^x_{(\r,\nu)}\) are gauge-invariant operator, see 
next section, Section \ref{no_dual_vars}). 

\begin{figure}[h]
\begin{center}
\includegraphics[width=0.65\columnwidth]{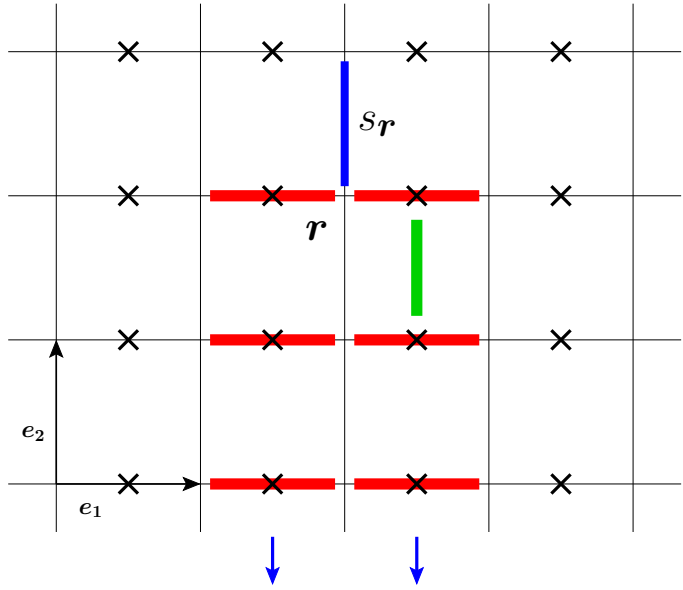}
\end{center}
\caption{The gauge-reduced dual of the \(\Z_2\) gauge theory constructed
in this section features spins \(S=1/2\) on the {\it horizontal} links of 
a square lattice (denoted by crosses on those links),
but {\it not} on the vertical links. Also it features a non-local bond,
the string operator \(s_\r\) of Equation \eqref{bye_gauge_string}, constructed
as an infinite product of all the spins \(\sigma^x_{(\r',1)}\) that lie
below the link \((\r,2)\), and to the immediate left and right of the straight line
containing it (some of them are shown in red). 
In spite of being non-local, \(s_\r\) commutes with most other bonds. For example, 
it commutes with \(\sigma^z_{(\r,1)}\sigma^z_{(\r-\j,1)}\), shown in green 
in the figure.}
\label{string2dop}
\end{figure}

Now, since \(\Phi_\d(\sigma^{x\ \ 2}_{(\r,\nu)})=\Phi_\d(\sigma^{x}_{(\r,\nu)})^2=\mathbb{1}\),
Equation \eqref{byebye_gauge} shows that \(\Phi_\d(\sigma^x_{(\r,2)})\) satisfies
a recurrence relation,
\begin{equation}
\Phi_\d(\sigma^x_{(\r,2)})=\left(\Phi_\d(\sigma^x_{(\r-\i,1)}) 
\Phi_\d(\sigma^x_{(\r,1)})\right)\Phi_\d(\sigma^x_{(\r-\j,2)}),
\end{equation}
so that 
\begin{equation}\label{bye_gauge_string}
\Phi_\d(\sigma^x_{(\r,2)})=\prod_{m=0}^\infty\ \left(\Phi_\d(\sigma^x_{(\r-m\j-\i,1)}) 
\Phi_\d(\sigma^x_{(\r-m\j,1)})\right)\equiv s_{(\r,2)}.
\end{equation}
The form of the string operator \(s_{(\r,2)}\)
 suggests that we should look for a representation of 
\(\Phi_\d(\sigma^x_{(\r,1)})\) and \(\Phi_\d(B_{(\r,3)})\) in terms 
of \(\sigma^x_{(\r,1)}\) and \(\sigma^z_{(\r,1)}\) {\it alone},
because then the dual model will feature only {\it half as many}
degrees of freedom (or equivalently, will act on a state space
exponentially smaller). It follows that 
\begin{equation}\label{bye_gaugeII}
\Phi_\d(\sigma^x_{(\r,1)})=\sigma^x_{(\r,1)},\ \ \ \ \ \ 
\Phi_\d(B_{(\r,3)})=\sigma^z_{(\r,1)}\sigma^z_{(\r+\j,1)},
\end{equation}
and the dual, completely gauge-reduced model reads
\begin{equation}\label{systematic_gr}
H_{\sf GR}=\sum_\r\ (\sigma^x_{(\r,1)}+ s_{(\r,2)}+
\lambda\ \sigma^z_{(\r,1)}\sigma^z_{(\r+\j,1)})=H_0+\sum_\r\ s_{(\r,2)}.
\end{equation}
It is interesting to notice that \(H_0\) describes a bundle of 
horizontal quantum Ising chains that do not interact with each other. 
The interactions between chains are carried exclusively
by the term \(\sum_\r\ s_{(\r,2)}\) that contains all the 
non-local bonds.

The traditional approach to the duality between the \(\mathbb{Z}_2\) 
gauge model and the Ising model relies more or less implicitly on the 
construction we have just described \cite{kogut}, because the 
Hamiltonian of Equation \eqref{systematic_gr} can be mapped to 
\(H_{\sf I}\) in terms of dual variables, while \(H_{\sf G}\) cannot
(this will be explained in detail in the next section). Let us show how 
this works. The duality (which is now an ordinary, unitarily implementable duality)
between \( H_{\sf GR}\) and \(H_{\sf I}\) is established by the Isomorphism
\begin{equation}
\label{em_ising_isoIII}
\sigma^x_{(\r,1)}\dual\sigma^z_{\r-\j}\sigma^z_{\r},\ \ \ \ \ \
\sigma^z_{(\r,1)}\sigma^z_{(\r+\j,1)}\dual \sigma^x_\r
\end{equation}
Since \(s_{(\r,2)}\) is a function of \(\sigma^x_{(\r,1)}\), the relation 
\begin{equation}
s_{(\r,2)}\dual\sigma^z_{\r-\i}\sigma^z_{\r}
\end{equation}
follows from the ones just listed. The dual variables that follow
from the isomorphism of Equation \eqref{em_ising_isoIII} are
\begin{equation}
\mu^x_{(\r,1)}=\sigma^z_{\r-\j}\sigma^z_{\r},\ \ \ \ \ \
\mu^z_{(\r,1)}=\prod_{m=0}^\infty\ \sigma^x_{\r+m\j},
\end{equation}
that clearly satisfy the correct Pauli algebra, and are essentially the
same as the dual variable used in Reference \cite{kogut}.  

\subsubsection{Dual variables in gauge-reducing dualities}\label{no_dual_vars}

The bond-algebraic approach to dualities
of gauge models is remarkably simpler than the traditional approach 
\cite{fradkin_susskind,kogut},
and it is also unrelated, because the homomorphism \(\Phi_\d\) cannot
be extended to define dual variables. To see this, suppose to the 
contrary that we can extend the action
of the gauge-reducing  homomorphism 
\(\Phi_\d\) defined in Equation \eqref{gauge_to_ising} to include
the spins \(\sigma^z_{(\r,\nu)}\),  \(\nu=1,2\), so that 
\begin{equation}
\sigma^z_{(\r,\nu)}\dual \Phi_\d(\sigma^z_{(\r,\nu)})
\end{equation}
becomes a meaningful operation. Then, on one hand, this would 
imply that
\begin{equation}
\Phi_\d([\sigma^z_{(\r,\nu)},G_\r])=[\Phi_\d(\sigma^z_{(\r,\nu)}),\mathbb{1}]=0.
\end{equation}
But, on the other hand,
\begin{equation}
[\sigma^z_{(\r,\nu)},G_\r]=-2\sigma^z_{(\r,\nu),\r}G_\r
\end{equation}
which, together with the previous equation, implies that
\begin{equation}
\Phi_\d(\sigma^z_{(\r,\nu)})=0.
\end{equation}
Since this is in contradiction with the fact that \(\Phi_\d(B_{(\r,3)})\neq 0\), 
we see that we cannot extend the action of \(\Phi_\d\) 
to the spins \(\sigma^z_{(\r,\nu)}\) in a consistent way. In other words,
\(\Phi_\d\) cannot be used to define dual spins \(\mu^z\).

This conclusion
seems paradoxical because  \(\Phi_\d\) must be projectively unitarily implementable. Then, 
it would seem natural to define
\begin{equation}\label{dual_zz}
\Phi_\d(\sigma^z_{(\r,\nu)})\equiv U_\d\sigma^z_{(\r,\nu)}U_\d^\dagger
\stackrel{?}{=}\mu^z_{(\r,\nu)}.
\end{equation}
Notice, however,  that this extension of \(\Phi_\d\) to operators
that are not gauge-invariant is {\it not multiplicative}.
In other words, the relation
\begin{equation}\label{multiplicative_on_gi}
U_\d\mathcal{O}\mathcal{O}'U_\d^\dagger =
U_\d\mathcal{O}U_\d^\dagger\  U_\d\mathcal{O}'U_\d^\dagger\ \ \ \ \ \ 
\ \ \ (\mathcal{O},\ \mathcal{O}'\ \mbox{gauge invariant}),
\end{equation}
only holds true if \(\mathcal{O}\) and \(\mathcal{O}'\) commute
with all the gauge symmetries. This means that
we should not expect the operator of Equation \eqref{dual_zz} to satisfy 
the correct anti-commutation relations with \(\Phi_\d(\sigma^x_{(\r,\nu)})\),
and also explains the paradoxical conclusion of the last paragraph
(\(\Phi_\d(\sigma^z_{(\r,\nu)})=0\)),
that was derived precisely on the assumption that Equation 
\eqref{multiplicative_on_gi} holds true {\it in general}. 

The fact that Equation \eqref{multiplicative_on_gi} only works for 
gauge-invariant operators follows from the general 
relations of Equations \eqref{projective_unitary_I} and 
\eqref{projective_unitary_II}. First notice that 
\begin{eqnarray}
P_{\sf GI}(\mathbb{1}-P_{\sf GI})&=&U_\d^\dagger U_\d(\mathbb{1}-P_{\sf GI})=0\
\ \ \rightarrow \ \ \ U_\d(\mathbb{1}-P_{\sf GI})=0,\\
(\mathbb{1}-P_{\sf GI})P_{\sf GI}&=&(\mathbb{1}-P_{\sf GI})U_\d^\dagger U_\d=0
\ \ \ \rightarrow \ \ \ (\mathbb{1}-P_{\sf GI})U_\d^\dagger=0
\end{eqnarray}
(since \(U_\d U_\d^\dagger=\mathbb{1}\)). Then, it follows from 
the decomposition
\begin{equation}
\mathcal{O}=P_{\sf GI}\mathcal{O}P_{\sf GI}+P_{\sf GI}\mathcal{O}
(\mathbb{1}-P_{\sf GI})+
(\mathbb{1}-P_{\sf GI})\mathcal{O}P_{\sf GI}+(\mathbb{1}-P_{\sf GI})
\mathcal{O}(\mathbb{1}-P_{\sf GI}) ,
\end{equation}
that
\begin{equation}
U_\d\mathcal{O}\mathcal{O}'U_\d^\dagger=
U_\d P_{\sf GI}\mathcal{O}\mathcal{O}'P_{\sf GI}U_\d^\dagger.
\end{equation}
But, if and only if, \(\mathcal{O}\) and \(\mathcal{O}'\) commute
with all the gauge symmetries, then they also commute with \(P_{\sf GI}\)
as well, so that we can further write
\begin{equation}
U_\d P_{\sf GI}\mathcal{O}\mathcal{O}'P_{\sf GI}U_\d^\dagger=
U_\d\mathcal{O}P_{\sf GI}\mathcal{O}'U_\d^\dagger=
U_\d\mathcal{O}U_\d^\dagger U_\d\mathcal{O}'U_\d^\dagger.
\end{equation}

In summary, \(\Phi_\d(\mathcal{O})=U_\d\mathcal{O}U_\d^\dagger\) is defined
on any operator, {\it but it acts as an algebra 
homomorphism  only on gauge-invariant operators}.

\subsection{Unifying classical and quantum dualities}
\label{classical&quantum}

This section explains one of the most important new results 
in this paper: a bond-algebraic approach to dualities in classical 
statistical mechanics \cite{con} (classical dualities described in
Section \ref{sec2.1} and Appendix \ref{appA}). 
To our knowledge, this section's results
have been completely overlooked in the literature,
possibly because the key fact that dualities are unitary or projective unitary 
mappings \cite{con}, has been unclear or overlooked up to now as well. 
Some advantages of the bond-algebraic over
the traditional approach of Appendix \ref{appA} to classical dualities 
will be discussed in detail
in Section \ref{sec8}, but we present 
here for illustration a toy example, a {\it new self-duality}
for the \(D=1=d+1\) classical Ising chain in an external magnetic field.
Elementary as it is, this duality is noteworthy because the traditional 
approach to classical dualities described in Appendix \ref{appA} 
{\it fails in the presence of a minimally coupled external field}, and could not
have been used to derive it. 
The bond-algebraic
approach to classical dualities is only rigorously applicable to models that admit a transfer
matrix formulation. While this covers a very wide range of interesting models, 
of finite of infinite size extent, 
there are dualities for models outside this category, most famously, the
duality of the solid-on-solid (SoS) model to a coulomb gas \cite{bookEPTCP}. 
We explain these dualities in Appendix \ref{poisson}.

We associate bond algebras to partition functions of classical models through
the transfer matrix. The transfer matrix formalism
\cite{onsager,baxter} permits to recast partition functions $\cal Z$ as traces of 
linear operators, and can come in several different flavors. For example, 
row-to-row transfer matrices permit to write
\begin{equation}\label{rowtorow}
\pf=\tr (T_1\cdots T_s)^N ,
\end{equation}
where \(N\) is an integer related to the number of sites 
in one of the lattice directions, that 
one can think of as the {\it Euclidean time direction}, and
\(T_1, \cdots, T_s\) contain information about the directions transverse to the
time direction, and how constant-time sections of the lattice
are connected from one time to the next.
Written as in Equation \eqref{rowtorow}, \(\pf\) must satisfy periodic boundary
conditions in the time direction. In contrast, corner transfer matrices \cite{baxter}
permit to write
\begin{equation}\label{cmt}
\pf=\tr C_1\cdots C_s ,
\end{equation}
where \(s\) is fixed, the size of the lattice is completely encoded in 
the \(C_i\), and the BCs are not fixed by the structure 
of Equation \eqref{cmt}.

This paper will only consider row-to-row transfer matrices,
so from now on  we focus on Equation \eqref{rowtorow},
keeping in mind though that it is possible to extend our formalism
to other types of transfer matrices. The general arguments of Section
\ref{sec3.2} concerning the additive bond structure of physical Hamiltonians
can be repeated {\it verbatim} for transfer matrices, with the only difference
that transfer matrices display a {\it multiplicative} rather than
additive bond structure:
\begin{equation}
T_i=\prod_\Gamma\ T_{i\Gamma}
\end{equation}
(\(\Gamma\) stands for a general index). The 
\(\{T_{i\Gamma}\}_{\Gamma, i=1,\cdots,s}\) are now the bonds of interest,
and the definition of bond algebra proceeds as before, see Section \ref{sec3.2}.

Suppose next that you have an isomorphic representation of the bond algebra
\(\mathcal{A}\{T_{i\Gamma}\}_{\Gamma, i=1,\cdots,s}\), generated by a set 
of dual bonds \(\{T_{i\Gamma}^D\}_{\Gamma, i=1,\cdots,s}\). The dual 
transfer matrices \(T_i^D={\cal U}_{\sf d } T_i {\cal U}^\dagger_{\sf d }=
\prod_\Gamma\ T^D_{i\Gamma}\) will define,
through Equation \eqref{rowtorow}, a partition function \(\pf^D\) that may
look very different from \(\pf\). However,
\begin{equation}
\pf^D=\tr (T_1^D\cdots T_s^D)^N=
\tr (\mathcal{U}_\d T_1\cdots T_s\mathcal{U}^\dagger_\d)^N=\pf.
\end{equation} 
We call this relation between partition functions, obtained in this way, 
a classical bond-algebraic duality. Much as with quantum bond-algebraic dualities,
we would like to show that:
\begin{itemize}
\item{classical bond-algebraic dualities include traditional classical dualities
(at least for models that admit a transfer matrix) \cite{con}, and that}
\item{classical bond-algebraic dualities are useful.}
\end{itemize}
Both points will be discussed at length in Section \ref{sec8} where
we derive classical dualities, old and new, by bond-algebraic methods.
We would like, however, to advance here some support for the second point,
by illustrating our ideas with a new self-duality {\it that is elementary, and  
yet cannot be derived by traditional (Fourier-based) means} (Appendix \ref{appA}). 

The partition function of the periodic  Ising chain of length $N$ 
($\sigma_{i+N}\equiv\sigma_i$) is
\begin{equation}\label{classical_ichain}
\mathcal{Z}_{\sf I}(K,\tilde{h})=\sum_{\{\sigma_i\}}\ \exp \left [ \sum_{i=1}^{N}
(K\sigma_{i}\sigma_{i+1}+\tilde{h}\sigma_i)\right ]= 
\tr (T_1T_2)^N,
\end{equation}
where 
\begin{equation}
\label{tsfl} 
T_1=e^K+e^{-K}\ \sigma^x,\ \ \ \ \ \
T_2=e^{\tilde{h}\ \sigma^z}=\cosh(\tilde{h})+\sinh(\tilde{h})\ \sigma^z,
\end{equation}
and we assume that \(K,\tilde{h}\geq0\). In this $D=1$ case, the
row-to-row transfer matrix connects rows that feature one Ising degree of freedom only.
The bond algebra of \(T_1\) and \(T_2\) is symmetric under the exchange
\(\sigma^x\leftrightarrow\sigma^z\) (this is just a rotation in spin space). 
It follows that
\begin{eqnarray}\label{dualt0dising}
T_1^D=e^K+e^{-K}\ \sigma^z=A\ e^{\tilde{h}^*\sigma^z},
\ \ \ \ \ \ 
T_2^D=e^{\tilde{h}\ \sigma^x}=B(e^{K^*}+ e^{-K^*}\ \sigma^x),
\end{eqnarray} 
provided the dual couplings \(\tilde{h}^*, K^*\) satisfy
\begin{equation}\label{dc_icf}
\sinh(2K)\sinh(2\tilde{h}^*)=1,\ \ \ \ \ \
\sinh(2K^*)\sinh(2\tilde{h})=1,
\end{equation}
and 
\begin{equation}
A^2=1/(2\sinh(2\tilde{h}^*)),\ \ \ \ \ \  B^2=2\sinh(2\tilde{h}).
\end{equation}
Then, Equations \eqref{tsfl} and \eqref{dualt0dising} put together 
define the bond-algebraic classical duality
\begin{equation}\label{ZKTH}
\frac{\mathcal{Z}_{\sf I}(K,\tilde{h})}{(2\sinh(2\tilde{h}))^{N/2}}
=\frac{\tr(T_1T_2)^N}{(2\sinh(2\tilde{h}))^{N/2}}=
\frac{\tr(T_2^DT_1^D)^N}{(2\sinh(2\tilde{h}))^{N/2}}=
\frac{\mathcal{Z}_{\sf I}(K^*,\tilde{h}^*)}{(2\sinh(2\tilde{h}^*))^{N/2}},
\end{equation}
where we have used the cyclic property of the trace. 
Notice how {\it linear} bond algebraic operations 
(the exchange \(\sigma^x\leftrightarrow\sigma^z\)) produce
highly {\it non-linear} relations between
classical couplings.
The classical self-dual line determined by \(\tilde{h}^*=\tilde{h}\) and \(K^*=K\)
is characterized by 
\begin{equation}
\sinh(2K)\sinh(2\tilde{h})=1.
\end{equation}
This self-dual line is only {\it critical} when $ \tilde{h}=0$ and 
$K \rightarrow \infty$, i.e., at zero temperature.

To our knowledge, the classical self-duality embodied in Equations \eqref{ZKTH} 
and \eqref{dc_icf}, valid for any $N$,   has not been written before in the literature 
despite its simplicity.
It has, however, one remarkable feature: it is a duality for a model {\it 
in a minimally-coupled external field}, and dualities for such models are
beyond the traditional approach as described in Appendix \ref{appA}, and 
References \cite{wegner,wu_wang,savit,malyshev}. 
The reason is that the standard approach
relies on the Fourier transform technique for establishing dualities, {\it and},
for these models with two-body interactions, 
on having individual Boltzmann weights that define {\it circulant matrices}
\cite{aldrovandi}. But the Boltzmann weights in \(\mathcal{Z}_{\sf I}(K,\tilde{h})\)
cannot be chosen to be circulant except if \(\tilde{h}=0\). Therefore, the self-duality 
of Equation \eqref{ZKTH} is not attainable by the Fourier transform method. 
We think that this indicates that our  bond-algebraic approach to classical
dualities may push its scope beyond the standard paradigm of 
Fourier transforms, perhaps even to include classical non-Abelian dualities, 
though this is a matter under study. 

There is a different albeit related way to connect bond-algebraic quantum
dualities to classical dualities. It exploits the well-known relation between 
partition functions of classical problems in $D=d+1$ dimensions and 
quantum Hamiltonian problems in $d$ dimensions.
Quantum mechanical problems in Euclidean time (or equivalently, at finite temperature) 
can be mapped to a classical partition function problem by use of 
Feynman's path integral 
for the case of quantum particles \cite{schulman} and fields \cite{rivers},
or by use of the closely related Suzuki-Trotter-Lie (STL) decomposition for 
quantum lattice models \cite{suzuki,bookEPTCP}.
This {\it quantum-classical} mapping takes the general form  \cite{bookEPTCP}
\begin{equation}\label{quantum_to_classical}
\pf(K)=\tr e^{-H[\lambda]} ,
\end{equation}  
where \(\pf(K)\) stands for the path integral/partition function, and the
classical, $K$, and quantum, $\lambda$, couplings  typically 
connected by non-linear functional relations. 
In general, this quantum-classical mapping 
takes quantum problems $H$ in \(d\) dimensions into classical problems 
$\pf$ in 
\(D=d+1\) dimensions where the extra dimension, because of the 
construction,  attains periodic BCs \cite{bookEPTCP}.

Now we can translate duality properties of \(H\) into properties of \(\pf(K)\).
Suppose that
the Hamiltonian \(H_1[\lambda]\) is dual to another Hamiltonian
\(H_2[\lambda^*]\) as  in Equations \eqref{H12H2}, 
$H_2[\lambda^*]=\mathcal{U}_\d H_1[\lambda]\mathcal{U}_\d^\dagger$. 
Then, with the aid of the identity  \(\tr(AB)=\tr(BA)\), 
\begin{equation}
\tr e^{-H_1[\lambda]}=\tr \!(\mathcal{U}_\d^\dagger e^{-H_2[\lambda^*]}\mathcal{U}_\d)
= \tr e^{-H_2[\lambda^*]}, 
\end{equation}
which implies that 
\begin{equation}\label{classd_from_q}
\pf_1(K)=A(K,K^*) \pf_2(K^*),
\end{equation}
with an analytic proportionality factor $A(K,K^*)$ that is model specific. 
Thus, {\it every quantum  duality translates into a classical duality} \cite{con}.  
Notice that it is crucial that $\mathcal{U}_\d$ is
a  unitary or projective unitary, a fact that was overlooked in the past \cite{con}.  

Due to the results of Section \ref{sec3.7}, the duality 
relation \eqref{classd_from_q} can be made exact for {\it any} finite
size system $N$, and {\it not} just for the thermodynamic limit 
$N \rightarrow \infty$. 
Rigorously speaking this is always the case, with quantum and classical 
dualities representing two sides of the same coin. On the other hand, 
there is a useful 
practical algorithm, that will be presented in Section \ref{sec8}, that makes 
use of the  STL decomposition and where the 
thermodynamic limit is, in principle, required \cite{bookEPTCP}.
Typically, the use of the STL decomposition implies that the derived
classical model presents infinitesimally weak and infinitely-strong couplings,
and so the classical dualities derived in this way are less clearly established.
It is also important to keep in mind that the classical model \(\pf\) of Equation 
\eqref{quantum_to_classical} depends on the quantum couplings \(\lambda\) not just
through \(K\), in the sense that different quantum couplings
can relate to qualitatively very different classical models (and not just
the same \(\pf\) at different values of \(K\)). For example, 
the STL decomposition maps  
a single quantum spin \(H=h_x\sigma^x+h_z\sigma^z\) to  the model
of Equation \eqref{classical_ichain} (in the limit \(N\rightarrow\infty\),
{\it provided \(h_x<0\)}), and not otherwise. Similar constraints apply 
elsewhere. As dualities relate different regions in the coupling spaces of 
quantum problems, it is essential to keep track of exactly which classical
problems may correspond to a given quantum  system for each set of
couplings.

\section{Quantum self-dualities by example: Lattice models}
\label{sec4}

\subsection{Self-dualities in the Potts, vector Potts, and
\(\mathbb{Z}_p\) clock models}
\label{sec4.1}

In this section, we will study two self-dual  generalizations of the
\(d=1\) dimensional quantum Ising chain: the quantum Potts (P) and {\it vector}
Potts (VP) models (also known as \(p\)-clock model \cite{bookEPTCP,potts}). 
These models are just two special examples of
a large class of \(\mathbb{Z}_p\) clock models \cite{cardy}, that we discuss
too, but very briefly. The fact that the family of \(\mathbb{Z}_p\) clock models
contains so many self-dual members is remarkable because \(\mathbb{Z}_p\) clock models
have {\it non-Abelian} symmetries {\it for \(p\geq3\)}. In other words, this
section discusses important examples of, inappropriately called,  {\it non-Abelian self-dualities}
(see Section \ref{sec3.6} for a critical discussion of the notion 
of non-Abelian duality). To the best of 
our knowledge, the fact that these models display non-Abelian symmetries has been 
overlooked up to now.

Potts models feature spins confined to a plain that can exist in any one of  
$p\geq 2$ different states, and if $p=2$, they reduce to the  Ising model.
We know already that in that case, the unitary implementing the self-duality squares 
to one. But if $p\geq 3$  the self-duality is the square root of a {\it non-trivial} 
discrete symmetry. This illustrates some ideas discussed 
in Section \ref{sec3.9}.
We will first discuss the VP model in detail, since then the 
self-dual properties of the P 
and general \(\mathbb{Z}_p\) clock models will easily follow.

\subsubsection{The vector Potts model} \label{sec4.1.1}

The VP model is a popular test ground  for {\it exotic} critical
behavior, such as  phase transitions without long-range order
\cite{taroni,ortiz}. The configurations of
the classical, $D=2$ VP model are specified by a set of discretized angles
$\theta_{\r}$
\begin{equation}
\theta_\r=\theta_{r^1,r^2}=\frac{2\pi s_\r}{p},\ \ \ \ \ \ \ \ \ \
s_\r=0,1,\cdots,p-1,
\end{equation} 
situated at the sites  ${\bm r}=(r^1,r^2)$ of a square
lattice, and its partition function reads 
\begin{equation}\label{classicalVP}
\pf_{\sf VP}=\sum_{\{\theta_\r\}}\ \exp\left[\sum_{\r}\sum_{\mu=1,2}\ 
K_\mu\cos(\theta_{\r+\bm{e_\mu}}-\theta_\r)\right].
\end{equation}
  
The statistical mechanics of this model has been the subject of
research for many years, but there are still open problems. For 
example, the VP model is known to exhibit a Kosterlitz-Thouless (KT) 
phase transition
\cite{kosterlitz_thouless}, {\it for sufficiently large $p$}
\cite{frohlich}, yet to determine the smallest $p$ required for a 
KT transition is a difficult  problem, currently under debate \cite{ortiz}. 
Reference \cite{ortiz} discusses the phase diagram, the nature of the 
transitions and topological excitations of the VP model, and unveils the 
U(1) symmetry, that emerges for $p\geq 5$,  associated 
with the appearance of discrete vortices and KT transitions. 

In preparation for
writing the \(d=1\) quantum model corresponding to the classical
VP model by the quantum-classical mapping \cite{bookEPTCP}, 
we introduce some basic facts about the Weyl group algebra  
\cite{schwinger}. 
Its generators $U$ and
$V$ are operators characterized by the relations
\begin{equation}\label{weyl_group_algebra}
VU=\omega UV,\ \ \ \ \ \ V^p=1=U^p, 
\end{equation}
where $ \omega=e^{2\pi i/p}$ is a $p$th root of unity. Equations
(\ref{weyl_group_algebra}) completely determine the irreducible, finite
dimensional, representations of $U$ and $V$. A $(p\times p)$ matrix
representation is given by
\begin{eqnarray} 
{V}=
\begin{pmatrix}
0& 1& 0& \cdots& 0\\
0& 0& 1& \cdots& 0\\
\vdots& \vdots& \vdots&      & \vdots \\
0& 0& 0& \cdots& 1\\
1& 0& 0& \cdots& 0\\
\end{pmatrix} \ , \mbox{ and } {U}={\sf
diag}(1,\omega,\omega^2,\cdots,\omega^{p-1}).
\label{VsandUs}
\end{eqnarray}
$V$ is sometimes called the fundamental circulant (unitary) matrix because
it generates the (commutative) algebra of circulant matrices
(meaning that any circulant matrix \(C\) is of the form \(C=\sum_{j=0}^{p-1}a_jV^j,\ 
a_i\in\mathbb{C}\) \cite{aldrovandi}). \(U\) and \(V\) together generate the full algebra of 
$(p\times p)$ complex matrices, that we continue to call the Weyl group algebra, 
to emphasize that we are working with a distinguished set of generators.

The Weyl group algebra admits a unitary automorphism $\Phi$  in the form
of a discrete Fourier transform \(F\) \cite{schwinger} that,
essentially, interchanges the two types of operators ($U,V$ and their
Hermitian conjugates) of the Weyl algebra. A  direct calculation reveals that
the unitary and symmetric Fourier matrix  \(F_{m
n}^\dagger=\omega^{mn}/\sqrt{p}\) with $m,n=0,1,\cdots,p-1$ maps
\begin{equation}\label{wga_aut}
\Phi(U)=V^{\dagger}=FUF^\dagger,\ \ \ \ \ \ \Phi(V)=U=FVF^\dagger.
\end{equation}

With these notations in place, we can proceed to introduce the 
{\it quantum VP model} \cite{elitzur}, 
\begin{equation}\label{infinite_vector_potts}
H_{\sf VP}[\lambda]= -\frac{1}{2}\sum_i\ 
(V_i+\ \lambda\ U_iU_{i+1}^{\dagger}+ {\sf h.c.}) 
\end{equation}
(the link to its classical counterpart is worked out Section \ref{sec8.3}),
and to study its bond algebra \(\mathcal{A}_{\sf VP}\) generated by     
\begin{eqnarray}
U_i U_{i+1}^{\dagger},\ \  U_i^{\dagger} U_{i+1},\ \ V_i,\ \
V_i^{\dagger}; \ \ \ \ \ i\in\mathbb{Z}.
\end{eqnarray}
It is readily
verified, either from Equation \eqref{weyl_group_algebra}  or Equation
\eqref{wga_aut}, that the mapping
\begin{eqnarray}\label{infinite_sd_vector_potts}
U_{i-1}^{\dagger}U_i\  \stackrel{\Phi_{\d}}{\longrightarrow}\
V_i^{\dagger}, \ \ \ \ \ \ \ \ 
V_i^{\dagger} \ &\stackrel{\Phi_{\d}}{\longrightarrow}& \  
U_i^{\dagger}U_{i+1},
\end{eqnarray}
(together with the corresponding relations for the Hermitian conjugate
bonds) defines a self-duality isomorphism of \(\mathcal{A}_{{\sf
VP}}\) that maps \(\Phi_{\d}(H_{\sf VP}[\lambda])=\lambda H_{\sf
VP}[1/\lambda]\).
  
The existence of the self-duality automorphism $\Phi_{\sf d}$ is
intimately connected to (i) the {\it
connectivity} of the system, as determined by the structure of 
interactions (specifically, in this case, by the non-commutativity
amongst bonds), and (ii) the local  degrees
of freedom posses a local symmetry which, in this case, is captured by
the automorphism $\Phi$ of Equation \eqref{wga_aut}. 
The fact that these two ingredients can be combined consistently
renders \(H_{\sf VP}\) self-dual. Although not all (self-) dualities 
follow from this pattern, many do. An
interesting exception was presented in Section \ref{sec3.11}.

As explained in Section \ref{sec3.5}, $\Phi_{\sf d}$ determines the dual
variables for the problem.  Since $U_i$ can we written in terms of bonds
as
\begin{equation}\label{UVUtVt}
U_i=\cdots (U_{i-3}^{\dagger}U_{i-2}) (U_{i-2}^{\dagger}U_{i-1})
(U_{i-1}^{\dagger}U_{i})=\prod_{j\leq i}
(U_{j-1}^{\dagger}U_j),
\end{equation}
the dual variables are
\begin{eqnarray}
\label{UVUtVt1}
\hat{V}_i&&\equiv \Phi_{\sf d}(V_i)=U_i^{\dagger}U_{i+1},\\
\hat{U}_i&&\equiv \Phi_{\sf d}(U_i)=\Phi_{\sf d}(\prod_{j\leq i}
U_{j-1}^{\dagger}U_j)=\prod_{j\leq i}\Phi_{\sf d}(
U_{j-1}^{\dagger}U_j)=\prod_{j\leq i} V_j^{\dagger}.\nonumber
\end{eqnarray}
The fact that \(\hat{U}_i\), \(\hat{V}_i\) satisfy the
same algebra as the \(U_i\), \(V_i\) affords a useful independent check
of the correctness of $\Phi_{\d}$ as a bond algebra isomorphism.

The product \( \Gamma_i =
U_i\tilde{U}_{i-1}^\dagger\) of a direct degree of freedom \(U_i\) and its
neighbor dual \(\tilde{U}_{i-1}^\dagger\) satisfy the non-local algebra
\begin{equation}\label{P_order_disorder}
\Gamma_i\Gamma_j=\omega\Gamma_j\Gamma_i,\ \ \ \ \ \ \mbox{if}\ \ i\neq
j, \ \ \ \mbox{and}\ \ \ \Gamma_i\Gamma_i^\dagger=\mathbb{1}.
\end{equation}
This suggests that if \(p>2\), the excitations of the model  are
governed by the parafermionic statistics that were described in Reference
\cite{rajabpour_cardy} for classical \(p\)-state models, see the discussion
at the end of Section \ref{sec3.10}. For \(p=2\), Equation \eqref{P_order_disorder} 
reduces to the fermionic algebra associated with the Ising model.

Next we show 
that the group of symmetries of the VP model is non-Abelian for
\(p\geq3\). The first step
is to introduce two new operators \(C_{0 i}, C_{1 i}\) that 
act on the basis states \(|s_i\rangle, s_i=0,\cdots, p-1\) of the
VP model at site \(i\) as follows:
\begin{equation}\label{defc0c1}
C_{0 i}|s_i\rangle=|-s_i\rangle, \ \ \ \ \ \ \ \ 
C_{1 i}|1-s_i\rangle=|1-s_i\rangle
\end{equation}
The arithmetic in these definitions is modular, modulo \(p\). 
For example, if \(p=7\), then \( C_{0 i}|0\rangle=|-0\rangle=|0\rangle\),
\( C_{0 i}|1\rangle=|-1\rangle=|6\rangle\), and so on.
If \(p=2\), \( C_{0 i}=\mathbb{1}\) and \(C_{1 i}=\sigma^x\), so we 
assume in what follows that \(p\geq3\).

Let us define \(\mathcal{C}_0=\prod_i C_{0 i}\) and 
\(\mathcal{C}_1=\prod_i C_{1 i}\). \(\mathcal{C}_0\) is known
in the literature as the charge conjugation operator \cite{Henkel}, 
but, to the best of our knowledge, \(\mathcal{C}_1\) has not been 
discussed before. Both  \(\mathcal{C}_0\) and 
\(\mathcal{C}_1\) are unitary {\it and} Hermitian, see Equation 
\eqref{polyhedral1} below. Since the operators \(U_i\) 
and \(V_i\) act basis states \(|s_i\rangle\) as
\begin{eqnarray}
U_i|s_i\rangle=\omega^{s_i}|s_i\rangle,\ \ \ \ \ \ \ \ \ \ 
V_i|s_i\rangle=|s_i-1\rangle,
\end{eqnarray}
it follows from Equation \eqref{defc0c1} that 
\begin{eqnarray}\label{c0c1}
\mathcal{C}_0V_i\mathcal{C}_0&=&V_i^\dagger,\ \ \ \ \ \ \ \ \ \  
\mathcal{C}_1V_i\mathcal{C}_1= V_i^\dagger, \\
\mathcal{C}_0U_i\mathcal{C}_0&=&U_i^\dagger,\ \ \ \ \ \ \ \ \ \ 
\mathcal{C}_1U_i\mathcal{C}_1=\omega U_i^\dagger.
\end{eqnarray}

Toghether with their Hermitian conjugates, 
the relations of Equation \eqref{c0c1} can be used to show that 
\(\mathcal{C}_0=\prod_i C_{0 i}\) and 
\(\mathcal{C}_1=\prod_i C_{1 i}\) commute with \(H_{\sf VP}\).
Moreover,
\begin{equation}\label{polyhedral1}
\mathcal{C}_0^2=\mathcal{C}_1^2=(\mathcal{C}_0\mathcal{C}_1)^p=\mathbb{1}.
\end{equation}
This means \cite{rotman} that the group of symmetries of the VP model
generated by \(\mathcal{C}_0\), \(\mathcal{C}_1\) 
provides a representation of the {\it non-Abelian} polyhedral group 
\(P(2,2,p)\). The unitary \(\mathcal{C}_0\mathcal{C}_1\) generates
the well-known \(\mathbb{Z}_p\) subgroup of symmetries, since
\(\mathcal{C}_0\mathcal{C}_1=\prod_iV_i\).
As it turns out, many \(\mathbb{Z}_p\) models,
including the \(\mathbb{Z}_p\) gauge theories discussed in Sections 
\ref{sec5.3} and \ref{sec6.3}, have this group among its 
symmetries.  

Our discussion of the VP model presented above
has the disadvantage that the dual 
variables of Equation \eqref{UVUtVt1} are not strictly speaking
well defined operators (see Section \ref{sec3.5}). We remedy this by 
considering a finite-size chain with self-dual BCs 
\begin{equation}\label{finitesdvp}
H_{\sf VP}^N[\lambda]= -\frac{1}{2}\sum_{i=1}^N\ 
V_i-\frac{1}{2}\sum_{i=1}^{N-1}\ \lambda\ U_iU_{i+1}^{\dagger}- 
\frac{\lambda}{2}U_N 
+{\sf h.c.}\ \ .
\end{equation} 

The first step in constructing the finite self-duality isomorphism
$\Phi_{\sf d}$ is to match the bonds at the boundaries of the chain
\begin{equation}\label{aut_finite_VP1}
U_N \ \stackrel{\Phi_{\sf d}}{\longrightarrow}\ \Phi(U_1)= V^{\dagger}_1,
\ \ \ \ \ \ \ V^{\dagger}_1 \ \stackrel{\Phi_{\sf d}}{\longrightarrow}\  
\Phi(V^{\dagger}_N)=U_N^{\dagger} ,
\end{equation}
where \(\Phi\) was defined in Equation \eqref{wga_aut}. Next, we
extend the mapping to the remaining bonds, 
\begin{eqnarray}
U_iU_{i+1}^{\dagger}\ &\stackrel{\Phi_{\sf d}}{\longrightarrow}& \ 
\Phi(U_{r(i)})=V^{\dagger}_{r(i)},\ \ \ \ \ \ \ \ \ \ \ \ \ \ \ \ \ \ \
\ \ \ \ \  i=1,\cdots,N-1 \label{aut_finite_VP2}\nonumber \\
V^{\dagger}_{i}\ &\stackrel{\Phi_{\sf d}}{\longrightarrow}& \ 
\Phi(V^{\dagger}_{r(i)})U_{r(i)+1}=U^{\dagger}_{r(i)}U_{r(i)+1},
\ \ \ \ \ \ i=2,\cdots,N,
\label{uiui}
\end{eqnarray}
guided by the  fact that we
must preserve the connectivity of the interactions and exploit the local
symmetry of the Weyl group algebra.
In Equation \eqref{uiui}, the reflection map $r$ of site $i$ is defined
as in Equation \eqref{inversion_map_eq}  (i.e., $r(i)=N+1-i$). 
Notice that the unitary \(\mathcal{U}_{\d}\) that implements
the mapping \(\Phi_\d\) is not
just the  discrete Fourier transform $F^\dagger=\prod_{i=1}^N
F_i^\dagger$. The latter maps  $H_{\sf VP}^N$ into $\tilde{H}_{\sf
VP}=F^\dagger {H}_{\sf VP}^N F$, 
\begin{equation}
\tilde{H}_{\sf VP}^N[\lambda]= -\frac{1}{2}\sum_{i=1}^N\ 
U_i-\frac{1}{2}\sum_{i=1}^{N-1}\ \lambda\ V_iV_{i+1}^{\dagger}+ {\sf
h.c.}\ \ \ .
\end{equation}
Finally, we use \(\Phi_{\d}\) to compute well defined dual variables,
\begin{eqnarray}
\hat{V}^{\dagger}_{1} &\equiv& \Phi_\d(V^{\dagger}_1)=U^{\dagger}_N\nonumber\\
\hat{V}^{\dagger}_i &\equiv& \Phi_\d(V_i^{\dagger})=
U^{\dagger}_{r(i)}U_{r(i)+1}, \ \  \ \ \ \ i=2,\cdots,N,\\
\hat{U}_N &\equiv&  \Phi_\d(U_N)=V_1^{\dagger},\nonumber\\
\hat{U}_i &\equiv&  \Phi_{\sf d}(U_i)=V^{\dagger}_{r(i)}V^{\dagger}_{r(i)-1}\cdots 
V^{\dagger}_{2}V^{\dagger}_{1},\ \ \ \ \ \ 
i=1,\cdots,N-1.\nonumber
\end{eqnarray}

According to the general ideas of Section \ref{sec3.9}, the square of the
self-duality unitary \(\mathcal{U}_{\d}\) commutes with the 
Hamiltonian. One of the advantages of the bond algebraic approach is
that we do not need to compute \(\mathcal{U}_{\d}\) explicitly to figure
out the action of $\mathcal{U}^2_\d$ as an operator.  
Since, by construction,
conjugation by $\mathcal{U}^2_\d$ amounts to applying $\Phi_{\sf d}$
twice, it follows from Equations \eqref{aut_finite_VP1} and
\eqref{aut_finite_VP2} that
\begin{equation}
\mathcal{U}^2_\d\ V_i\ \mathcal{U}^{2\ \dagger}_{\sf
d}=V^{\dagger}_i, \ \ \ \ \ \  \mathcal{U}^2_\d\ U_i\
\mathcal{U}^{2\ \dagger}_\d= U^{\dagger}_i,\ \ \ \ 
\ \ \ \ i=1,\cdots, N.
\label{UVC}
\end{equation}
Thus comparing with Equation \eqref{c0c1}, we see that
\(\mathcal{U}^2_\d =\mathcal{C}_0\), where  \(\mathcal{C}_0\) is the
charge conjugation symmetry of the VP model as discussed above. 
Notice also
that the self-dual BCs of Equation \eqref{finitesdvp} spoil the 
\(\mathcal{C}_1\) symmetry of the (infinite) VP model. We could have used 
self-dual BCs that preserve both symmetries, in agreement with the techniques
of Section \ref{sec3.7} and Appendix \ref{appG}.


\subsubsection{The Potts model and general \(\mathbb{Z}_p\) clock models} 
\label{potts}

The Potts (P) model  
\begin{eqnarray}\label{Pclassical}
\pf_{\sf P}=\sum_{\{\theta_\r\}}\ \exp\left[\sum_{\r}\sum_{\mu=1,2}\ 
K_\mu \delta(s_{\r}, s_{\r+\bm{e_\mu}})\right],
 \ \ \ \ \ \ s_\r=0,1,\cdots,p-1,
\end{eqnarray}
in \(D=2\) dimensions is yet another 
\(p\)-state generalization of the Ising model that, unlike 
the VP model of the previous section, has a very well understood
statistical behavior \cite{martin} (\(\delta(s,s')=1\) if \(s=s'\),
and \(\delta(s,s')=0\) otherwise). The P model has a group of 
global {\it non-Abelian} symmetries (if \(p>2\)), 
the group \(\mathfrak{S}_p\) of  permutations of \(p\) elements. 

The \(d=1\) quantum rendition of the P
model \cite{solyom} reads 
\begin{equation}
H_{\sf P}[\lambda]=\sum_i\sum_{m=0}^{[p/2]} \left[(V_i^{m}+V_i^{\dagger m})
+\lambda\ (U_i^m U_{i+1}^{\dagger m}+U_i^{\dagger m} U_{i+1}^{m})\right],
\end{equation}
where \([p/2]\) denotes the integer part of \(p/2\), that is, 
the largest integer \(\leq p/2\). It follows that the P and VP model
coincide for \(p=2,3\). But even for arbitrary \(p\), since the bonds
in \(H_{\sf P}\) are powers of the bonds in \(H_{\sf VP}\), it is not
hard to see that the self-duality of the VP model, Equation
\eqref{infinite_sd_vector_potts}, implies the self-duality of 
the P model.

The form of the Hamiltonians \(H_{\sf P}\) and \(H_{\sf VP}\) suggests
introducing a family of models that generalize them and interpolates between
them, the \(\mathbb{Z}_p\) clock models. They are defined by the Hamiltonians 
\begin{equation}
H_{\sf GP}[\{\alpha_m\},\{\lambda_m\}]=
\sum_i\sum_{m=0}^{[p/2]} \left[\alpha_m(V_i^{m}+V_i^{\dagger m})
+\lambda_m\ (U_i^m U_{i+1}^{\dagger m}+U_i^{\dagger m} U_{i+1}^{m})\right],
\end{equation}
with \(\alpha_m,\lambda_m\in\mathbb{R}\), \(m=0,\cdots, [p/2]\), arbitrary
real coupling constants. The duality properties of some of these models
where studied in References \cite{rittenberg}, and the duality properties
of their classical counterparts where studied in Reference \cite{cardy}.

The \(\mathbb{Z}_p\) clock models do not have in general as many symmetries as their
special case the P model, {\it but they are all non-Abelian},
as the operators \(\mathcal{C}_0,\mathcal{C}_1\) of Equation \eqref{c0c1}
commute with \(H_{\sf GP}\) for any value of its couplings. 
Next we can use our experience with the P and VP models
to identify immediately the \(\mathbb{Z}_p\) clock models that are
self-dual: The Hamiltonian \(H_{\sf GP}\) is self-dual 
in those regions in coupling space 
where \(\alpha_m\) vanishes if and only if \(\lambda_m\) vanishes as well.  
Clearly, the P and VP model belong to two such regions.



\subsection{Dualities in some limits and related approximations}
\label{sec4.3}

It is standard and useful practice to think of some models as related through 
a limit of some parameter. The VP model of Equation \eqref{classicalVP} is a good case
in point. It is intuitively reasonable to expect that its behavior
must approach that of the XY model as the discrete angle
\(\theta_\r\) becomes very dense in the limit \(p\rightarrow\infty\), and
in fact this is know to be the case in many respects \cite{frohlich}.
Dualities, however, can show some counter-intuitive behavior with respect
to limits. This is the topic of this section, with the VP model
as main illustration.  

Weyl's group algebra describes states on a finite, equidistant set of points on
a unit circle. If we think of the roots of unity
\(\omega^0,\omega^1,\cdots, \omega^{p-1}\) as these points, then $U$ and
$U^\dagger$ will play the role of position operators, while  \(V^\dagger\)
(\(V\)) acts as a clockwise (counter-clockwise) rotation from
any of the \(p\) roots to one of its neighbors.  Following Schwinger
\cite{schwinger}, one can define position $\hat{\bf q}$, and momentum
$\hat{\bf p}$ Hermitian operators, such that
\begin{eqnarray}
{U}=e^{i \epsilon \hat{\bf q}} \ , \ {V}=e^{i \epsilon \hat{\bf p}} ,
\end{eqnarray}
with eigenvalues ${q}_r({p}_r)=0,\epsilon, 2\epsilon,
\cdots,(p-1)\epsilon$, and  $\epsilon^2=2\pi/p$. In matrix form,
$\hat{\bf q}=\epsilon \, {\sf diag}(0,1,2,\cdots,p-1)$, and 
\begin{eqnarray}
\hat{\bf p}=\epsilon \begin{pmatrix}
\frac{p-1}{2}& {\cal P}^*_1& {\cal P}^*_2& \cdots& {\cal P}^*_{p-1}\\
{\cal P}_1& \frac{p-1}{2}& {\cal P}^*_1& \cdots& {\cal P}^*_{p-2}\\
\vdots& \vdots& \vdots&      & \vdots \\
{\cal P}_{p-2}& {\cal P}_{p-3}& {\cal P}_{p-4}& \cdots& {\cal P}^*_1\\
{\cal P}_{p-1}& {\cal P}_{p-2}& {\cal P}_{p-3}& \cdots& \frac{p-1}{2}\\
\end{pmatrix} = {F}^\dagger \hat{\bf q} {F},
\end{eqnarray}
with ${\cal P}_m=\frac{1}{p}\sum_{n=1}^{p-1} n \, \omega^{m n}$. In the
$p \rightarrow \infty$  limit,  Weyl's algebra relates to the {\it
continuous} circle and  it can be shown that $\epsilon \hat{\bf q}
\rightarrow \theta \in [0,2\pi)$, and  $ \hat{\bf p} \rightarrow -i
\epsilon \partial/\partial \theta$.  In this subtle limit,   \(H_{\sf
VP}\) of Equation \eqref{infinite_vector_potts}  converges to the 
quantum version of the classical $D=2$ XY model \cite{bookEPTCP}  (up to
irrelevant constants)
\begin{equation}\label{KT_H} 
H_{\sf KT}=\sum_i\  (\frac{1}{2}L_i^2- \lambda \cos(\theta_{i+1}
-\theta_i) ),
\end{equation}
where \(L_j=-i \partial/\partial\theta_j\). Albeit being the $p \to
\infty$ limit of the self-dual VP model limit, $H_{\sf KT}$ {\it is not
self-dual}, but rather dual to the quantum SoS model (see 
Section \ref{sec5.1}). It follows that {\it the limit of a sequence of 
self-dual models  needs not be self-dual}.

There is a profound difference between compact theories (such as those
of variables defined on a circle or the compact $U(1)$ fields of
electromagnetism) and  non-compact theories (e.g., a theory whose fields
are defined on a line). An example of the latter which shares some
relation to  $H_{\sf KT}$ is a linear chain of coupled harmonic
oscillators representing  acoustic phonons \cite{kittel}
\begin{eqnarray}\label{phonons_H}
H_{\sf Ph}=\sum_i\  (\frac{p_i^{2}}{2m} +\ \frac{1}{2}m\omega^2
({x}_{i+1}-{x}_i)^2)\ , \ \ \ \ \ \ \ \ [x_i,p_j]=\ i\delta_{i,j} .
\end{eqnarray}  
However, unlike $H_{\sf KT}$ the  Hamiltonian \(H_{\sf Ph}\)
{\it is} self-dual under the exchange \(m\omega^2\leftrightarrow1/m\), 
as the isomorphism
\begin{eqnarray}\label{phonons_aut}
{x}_{i+1}-{x}_{i}\ \stackrel{\Phi_{\d}}{\longrightarrow} \ {p}_{i+1},
\ \ \ \ \ \ \ \ \ 
{p}_i\ \stackrel{\Phi_{\d}}{\longrightarrow} \ {x}_{i+1}-{x}_{i} 
\end{eqnarray}
shows. This self-duality arises due an interplay between the connectivity of
the model and the automorphism \(\Phi\)  of the Heisenberg algebra
\begin{equation}
x\ \stackrel{\Phi}{\longrightarrow}\ p,\ \ \ \ \ \ \ \ \
p\ \stackrel{\Phi}{\longrightarrow}\ -x ,
\end{equation}
and generates the set of dual variables 
\begin{eqnarray}\label{phonons_dual_var}
\hat{p}_i={x}_{i+1}-{x}_{i},\ \ \ \ \ \ \ \ \ \ \ \ \ \ 
\hat{x}_i=\sum_{m=-\infty}^i\ p_m.
\end{eqnarray}
The ground state of \(H_{\sf Ph}\) does not show a  phase transition 
at the self-dual point \(m\omega=1\), or anywhere else for that matter. 
Thus we see that 
while self-dualities can constrain phase transitions greatly, they
cannot guarantee by themselves that a phase transition will occur.

\subsection{The Xu-Moore model}
\label{sec4.4}

The \(d=2\) Xu-Moore (XM) Hamiltonian, 
\begin{equation}\label{XM}
H_{\sf XM}[J,h]= -\sum_{\r}\ (J\square\sigma^z_\r + h\sigma^x_\r),
\end{equation}
with 
\begin{equation}\label{plaquetteXM}
\square\sigma^z_\r =\sigma^z_\r  \sigma^z_{\bm{r+e_2}}
\sigma^z_{\bm{r-e_1+e_2}} \sigma^z_{\bm{r-e_1}},
\end{equation}
was introduced in Reference \cite{XM}
as an effective model to study ordering in 
arrays of Josephson-coupled \(p\pm ip\)
superconducting grains. \(H_{\sf XM}\) looks similar 
to the \(\mathbb{Z}_2\) gauge theory studied Section \ref{sec3.12}, 
but the fact that the spins \(S=1/2\)  are
now located at the {\it vertices} of a square lattice (see Figure
\ref{notation_links}) rather than at its links, makes the symmetries
and properties of the two models very different. In fact, 
unlike the  \(\mathbb{Z}_2\) gauge theory, the XM model
is self-dual \cite{XM}, and displays \(d=1\) dimensional
gauge-like symmetries \cite{tqo},
\begin{equation}
G_{r^1}=\prod_m\ \sigma^x_{r^1,m},\ \ \ \ \ \ G_{r^2}=\prod_m\ \sigma^x_{m,r^2},
\end{equation}
\([G_{r^i},\ H_{\sf XM}]=0\), that make the model a toy example
of topological quantum order and dimensional reduction \cite{tqo} 
(see also Section \ref{6.8}).
Its self-duality is the subject of this section.

\begin{figure}[h]
\begin{center}
\includegraphics[width=0.5\columnwidth]{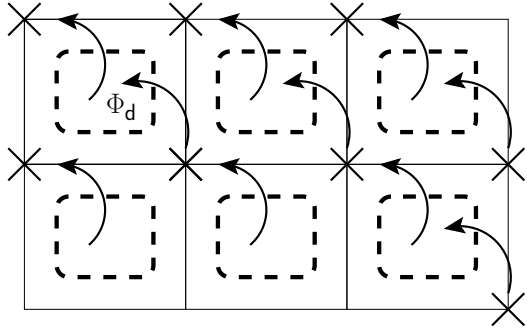}
\end{center}
\caption{ In this representation of the XM model, 
the heavy  crosses stand for the bonds \(\sigma^x_\r\) and
the dashed squares for the bonds \(\square\sigma^z_\r\). The self-duality
isomorphism \(\Phi_\d\) of Equation \eqref{sdXM} exchanges the 
two types of bonds. Notice that the net result of applying the mapping
\(\Phi_\d\) {\it twice} amounts to a translation by \(\bm{d}=-\i+\j\), 
a symmetry of the model. }
\label{sd_inf_XM}
\end{figure}

The bond algebra generated by the set of bonds  \(\{\square\sigma^z_\r,\sigma^x_\r\}_\r\) 
is characterized by three relations: 
Every bond (i) squares to one, (ii) anti-commutes with
four other neighboring bonds,  and (iii) commutes with every other bond.
It follows that there is an isomorphism
\begin{eqnarray}\label{sdXM}
\square\sigma^z_\r\ \xrightarrow{\Phi_{\sf d}}\ \sigma^x_{\bm{r-
e_1+e_2}}, 
\ \ \ \ \ \ \ \ 
\sigma^x_\r\ \xrightarrow{\Phi_{\sf d}}\ \square\sigma^z_\r,
\end{eqnarray}
illustrated in Figure \ref{sd_inf_XM}, that establishes the 
self-duality of the XM model.

As it should, $\Phi_\d^2$ is a symmetry of the model, 
\begin{equation}
\square\sigma^z_\r\ \xrightarrow{\Phi_{\sf d}^2}\
\square\sigma^z_{\bm{r-e_1+e_2}}, \ \ \ \ \ \ \ \ \ \  \sigma^x_\r\
\xrightarrow{\Phi_{\sf d}^2}\   \sigma^x_{\bm{r-e_1+e_2}},
\end{equation}
but it shows also unsettling features that are typical
of working directly in the limit of infinite size. 
Since formally \(\mathbb{1}=\prod_m\ \square\sigma^z_{r^1,m}=
\prod_m\ \square\sigma^z_{m,r^2}\), it seems that one could argue that
\(\Phi_\d\) is in fact a multivalued mapping,
\begin{equation}
\Phi_\d(\mathbb{1})=\mathbb{1},\ \ \mbox{or}\ G_{r^1},\ \ \mbox{or}\ G_{r^2}.
\end{equation}
This is not a problem, however, in the light of the general discussion
of Section \ref{sec3.7}.

To make contact with the traditional approach described in Section
\ref{sec2.1} and exploited in Reference \cite{XM}, 
we need to employ the ideas of Section \ref{sec3.5}
to compute dual variables. Since
\begin{eqnarray}
\sigma^z_\r=\ \prod_{m^1 \leq r^1,\ r^2\leq m^2}
\ \square\sigma^z_{\bm{m}},
\end{eqnarray}
where $\bm{m}=m^1\bm{ e_1}+m^2\bm{e_2}$, it follows that
\begin{eqnarray}\label{dualXM}
\mu^x_\r \equiv \Phi_{\sf d}(\sigma^x_\r)=\square\sigma^z_\r
,\ \ \ \ \ \ \ \mu^z_\r \equiv \Phi_{\sf d}(\sigma^z_\r)=\ 
\prod_{m^1\leq r^1-1,\ r^2+1\leq m^2 }\ \sigma^x_{\bm{m}}. 
\end{eqnarray}
This completes the calculation.
 
Let us discuss next self-dual, open BCs for the
XM Hamiltonian. Consider an square portion of the infinite model,
featuring \(N^2\) spins, 
\begin{equation}\label {XM_H_finite}
H_{\sf XM}^o=-J\sum_{r^1=1}^{N-1}\sum_{r^2=0}^{N-2} \,
\square\sigma^z_{r^1,r^2}  -h\sum_{r^1,r^2=0}^{N-1}\ \sigma^x_{r^1,r^2}
\end{equation} 
($r^1,r^2 = 0,1,\cdots, N-1$). 
The goal is to determine boundary corrections  to make \(H_{\sf
XM}^o\) self-dual, and as usual, this could  be accomplished by a systematic
study of \(H_{\sf XM}^o\)'s bond algebra. But this task grows increasingly
harder with dimension, so it is important to realize that 
there is a simple recipe to construct self-dual boundary terms that
works well in general (but not always), and that we illustrate
next with the XM model. The starting point is the 
self-duality mapping for the infinitely large model.  
\begin{figure}[h]
\begin{center}
\includegraphics[width=0.55\columnwidth]{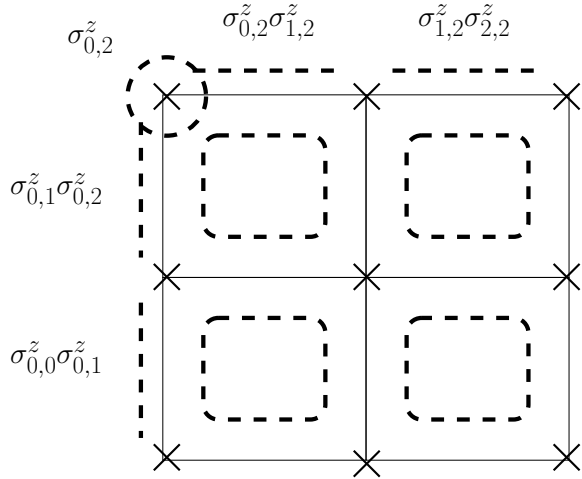}
\end{center}
\caption{One possible self-dual BC for the 
the finite-size XM model on a \(3\times3\) portion of the 
lattice ($N=3$). The bonds shown on the upper and left edge 
(indicated with broken lines), and on the left upper corner (indicated
with a broken circle) render the the finite-size XM Hamiltonian self-dual,
but also break the \(d=1\)-dimensional gauge-like symmetries of the model.
There are other self-dual BCs that preserve some or all of
the gauge-like symmetries, but are not as convenient for computing dual
variables.}
\label{f_XM}
\end{figure}

So imagine that we try to apply the self-duality isomorphism of  
Equation \eqref{sdXM} to the bonds \(\sigma^x_\r\) in \(H_{\sf XM}^o\). 
The problem is that, 
as defined in Equation \eqref {XM_H_finite}, \(H_{\sf XM}^o\)  does 
not have any plaquettes to which we can map the spins \(\sigma^x_{0,r^2}\) 
at the left edge and the spins \(\sigma^x_{r^1,N-1}\) at the top edge (see
Figure \ref{f_XM}). This suggests that to restore self-duality, we 
should re-introduce the missing plaquettes as boundary corrections,
or maybe incomplete versions of these plaquettes, in order to 
preserve the number of spins. 

As it turns out, this idea works perfectly, so it is convenient to 
introduce some notation to describe complete and incomplete plaquettes
in a unified manner.  We
write \(\square^I\sigma^z_{r^1,r^2}\) 
for the plaquette, or {\it incomplete} plaquette, that results from simplifying the
standard plaquette of Equation \eqref{plaquetteXM} by discarding the
spins {\it outside} the square region encompassed by \(H^o_{\sf XM}\). So, for example,
\begin{equation}
\square^I\sigma^z_{0,0}=\sigma^z_{0,0}\sigma^z_{0,1},\ \ \ \
\square^I\sigma^z_{1,0}=\square\sigma^z_{1,0},\ \  
\mbox{and}\ \ \ \square^I\sigma^z_{0,N-1}=\sigma^z_{0,N-1}.
\end{equation}
Our proposal is that
\begin{equation}\label{fselfdXM}
\tilde{H}_{\sf XM}^o=-\sum_{r^1,r^2=0}^{N-1}\ (J\square^I\sigma^z_{r^1,r^2}+
h\sigma^x_{r^1,r^2})
\end{equation} 
is a finite self-dual rendition of the XM model (\(\tilde{H}_{\sf XM}^o\)
is illustrated in Figure \ref{f_XM}, for \(N=3\)). To prove this, 
we must construct the self-duality isomorphism, starting by noticing that
\(\square^I\sigma^z_{0,N-1}=\sigma^z_{0,N-1}\) can only map to 
\(\sigma^x_{N-1,0}\). The full isomorphism follows right away,
\begin{equation}\label{iso_finite_XM}
\square^I\sigma^z_{r^1,r^2}\dual\sigma^x_{r^2,r^1},
\ \ \ \ \ \ \sigma^x_{r^1,r^2}\dual\square^I\sigma^z_{r^2,r^1}.
\end{equation}
Geometrically, it is a reflection along the main diagonal, and
$\Phi_{\sf d}^2=\mathbb{1}$.

The self-dual BC illustrated in Figure \ref{f_XM}
breaks the global \(\Z_2\), and all the gauge-like \cite{tqo} symmetries
of the XM model, and one could have chosen among several other ones
that preserve some or all of these symmetries (see Section \ref{sec3.7}). 
In particular, if we
just replace in Equations \eqref{fselfdXM} and \eqref{iso_finite_XM} the 
incomplete plaquettes \(\square^I\sigma^z\) with standard, complete ones
\(\square\sigma^z\), the resulting model is self-dual under the same 
mapping, and moreover has all the global and gauge-like symmetries
of the XM model. On the other hand, the self-dual BCs 
that we chose guarantee that all of the spins
$\sigma^z_{(r^1,r^2)}$ contained in the square region encompassed by 
$\tilde{H}_{\sf XM}^o$ are elements in its bond algebra. Thus we 
can use the finite isomorphism just defined in 
Equation \eqref{iso_finite_XM} to compute finite dual variables. 
It is convenient to do so, at least for small system size, to check
the algebraic consistency of Equation \eqref{iso_finite_XM}. For \(N=3\), 
\begin{eqnarray}
\mu^x_{0,2}&=&\square\sigma^z_{2,0},\ \ \ \ \ \ \ \ \ \ \ \  
\mu^z_{0,2}=\sigma^x_{2,0},\nonumber\\
\mu^x_{0,1}&=&\square\sigma^z_{1,0},\ \ \ \ \ \ \ \ \ \ \ \   
\mu^z_{0,1}=\sigma^x_{1,0} \sigma^x_{2,0},\nonumber\\
\mu^x_{0,0}&=&\sigma^z_{0,0}\sigma^z_{0,1},\ \ \ \ \ \ \ \ \ \
\mu^z_{0,0}=\sigma^x_{0,0}\sigma^x_{1,0} \sigma^x_{2,0},\nonumber\\
\mu^x_{1,2}&=&\square\sigma^z_{2,1},\ \ \ \ \ \ \ \ \ \ \ \ 
\mu^z_{1,2}=\sigma^x_{2,0}\sigma^x_{2,1},\\
\mu^x_{1,1}&=&\square\sigma^z_{1,1},\ \ \ \ \ \ \ \ \ \ \ \
\mu^z_{1,1}=\sigma^x_{1,0}\sigma^x_{1,1}\sigma^x_{2,1}\sigma^x_{2,0},
\nonumber\\
\mu^x_{1,0}&=&\sigma^z_{0,1}\sigma^z_{0,2},\ \ \ \ \ \ \ \ \ \
\mu^z_{1,0}=\sigma^x_{0,0}\sigma^x_{1,0}\sigma^x_{2,0}
\sigma^x_{0,1}\sigma^x_{1,1} \sigma^x_{2,1},\nonumber\\
\mu^x_{2,2}&=&\sigma^z_{1,2} \sigma^z_{2,2},\ \ \ \ \ \ \ \ \ \
\mu^z_{2,2}=\sigma^x_{2,0}\sigma^x_{2,1}\sigma^x_{2,2},\nonumber\\
\mu^x_{2,1}&=&\sigma^z_{0,2} \sigma^z_{1,2},\ \ \ \ \ \ \ \ \ \ 
\mu^z_{2,1}=\sigma^x_{1,0}\sigma^x_{1,1}\sigma^x_{1,2}
\sigma^x_{2,0}\sigma^x_{2,1}\sigma^x_{2,2},\nonumber\\
\mu^x_{2,0}&=&\sigma^z_{0,2},\ \ \ \ \ \ \ \ \ \ \ \ \ \ \
\mu^z_{2,0}=\sigma^x_{0,0}\sigma^x_{0,1} \sigma^x_{0,2}
\sigma^x_{1,0}\sigma^x_{1,1}\sigma^x_{1,2}
\sigma^x_{2,0}\sigma^x_{2,1}\sigma^x_{2,2}.\nonumber
\end{eqnarray}

\section{Quantum dualities by example: Lattice models}
\label{sec5}

\subsection{{\rm \bf XY}/solid-on-solid models}
\label{sec5.2}

The best developed approach to the Kosterlitz-Thouless  phase transition
of the classical $D=2$ XY model exploits two duality transformations
(see for example Reference \cite{bookEPTCP}). First one maps the XY model,
via a Villain approximation, to the SoS model, to proceed
afterwards to map the SoS model to a classical $D=2$ Coulomb gas. 
The two steps combined leads to the mapping between the XY model and a
$D=2$ Coulomb gas for which deconfinement of charges can be shown to occur at
sufficiently high temperatures. 

In this section,  we will study the first of these dualities to the
SoS model from a quantum, bond-algebraic perspective. The $D=2$
SoS model is specified by the partition function 
\begin{equation}
\mathcal{Z}_{\sf SS}=\sum_{\{m_\r\}}\
\exp\left[\sum_\r\sum_{\nu=1,2}\ \ 
K_\nu(m_{\r+\bm{e_\nu}}-m_\r)^2\right],\ \ \ \ \ \ \ \ m_\r \in
\mathbb{Z},
\label{SOSpart}
\end{equation}
where the classical degrees of freedom $m_\r$ reside on the vertices  of
a square lattice (see Figure \ref{notation_links}).  

The $d=1$ quantum version of the SoS model must have  states
labelled by integers \(\{\vert m\rangle\}\). Three operators are then
required to describe its quantum dynamics:  a position operator \(X\), and
left/right shift operators \(R/R^\dagger\),
\begin{equation}
X\vert m\rangle=\ m\, \vert m\rangle,\ \ \ \ \ \ \ \ 
R\vert m\rangle=\ \vert m-1\rangle,\ \ \ \ \ \ \ \
R^\dagger \vert m\rangle=\ \vert m+1\rangle,
\end{equation}
that satisfy
\begin{equation}\label{Z}
[X,R^\dagger]=R^\dagger,\ \ \ \ \ \ \ \ \ \ [X,R]=-R.
\end{equation}
\(X,\ R,\) and \(R^\dagger\) generate an algebra isomorphic to 
the one satisfied by the 
elementary degree  of freedom  associated to Equation \eqref{KT_H}, 
$[L,e^{\pm  i\theta}]=\pm  e^{\pm i\theta}$. 
In particular, \(X\) and \(L\) have  identical
spectra. 

The duality mapping between the quantum versions of the  XY  and the
SoS models can be derived by  comparing the bond
algebra generated by 
\begin{equation}
L_i,\ \ \ \ \ \ \ \ e^{i(\theta_{i+1}-\theta_{i})},
\ \ \ \ \ \ \ \  e^{-i(\theta_{i+1}-\theta_{i})}, 
\end{equation}
which is characterized by the relations
\begin{equation}
[L_i,\ e^{\pm i(\theta_{j+1}-\theta_{j})}]=\ (\pm \delta_{i, j+1} \mp 
\delta_{i, j})\  e^{\pm i(\theta_{j+1}-\theta_{j})},
\end{equation}
and the one generated by the SoS operators
\begin{eqnarray}
[(X_{i+1}-X_i),
R_{j+1}]&=&(\delta_{i,j+1}-\delta_{i,j})R_{j+1},\nonumber\\
{[}(X_{i+1}-X_i),
{R^\dagger_{j+1}}]&=&(-\delta_{i,j+1}+\delta_{i,j})R^\dagger_{j+1}.
\end{eqnarray}
Then, the isomorphism
\begin{equation} \label{iso_XY_ss}
L_i\ \stackrel{\Phi_\d}{\longrightarrow}\ (X_{i+1}-X_i),\ \ \ \ \ \ \ \ 
e^{i(\theta_{i+1}-\theta_{i})}\ \stackrel{\Phi_\d}{\longrightarrow}\ 
R_{i+1}
\end{equation}
establishes a duality between the two models (a related mapping in terms of 
less well-defined operators was described in Reference \cite{mattis}).
The dual form of \(H_{\sf KT}\) reads
\begin{eqnarray}
H_{\sf SS}= \frac{1}{2}\sum_i\ \left (- \lambda(R_i +
R^\dagger_i)+(X_{i+1}-X_i)^2\right )\ .
\end{eqnarray}
which to our knowledge has not been reported in the literature before.
One can verify, using the  standard quantum-classical mapping
\cite{bookEPTCP}, that the classical counterpart  of \(H_{\sf SS}\) is
indeed the classical $D=2$ SoS model of Equation
\eqref{SOSpart}.

This duality is similar, in spirit, to the self-duality of the quantum
Ising chain as seen through the dual variables
\begin{eqnarray}
L_i\dual\widehat{L}_i=X_{i+1}-X_i, \ \
\ \ \ \ \  \ \ \  e^{i\theta_i}\ \stackrel{\Phi_\d}{\longrightarrow}\
\widehat{e^{i\theta_i}} =\prod_{j<i}\ R_j\ .
\end{eqnarray}
The inverse \(\Phi_\d^{-1}\) of the isomorphism of Equation \eqref{iso_XY_ss}
defines a reciprocal set of dual variables.

\subsection{Xu-Moore/planar orbital compass models}
\label{sec5.1}

The planar orbital compass (POC) model of orbital ordering 
\cite{orbital_compass}, 
\begin{equation}
H_{\sf POC}[J_x,J_z]=\ \sum_\r\  (J_x\sigma^x_\r\sigma^x_{\r+\bm{e_1}}+ 
J_z\sigma^z_\r\sigma^z_{\r+\bm{e_2}}),
\label{3dOC}
\end{equation}
features spins $S=1/2$ residing on the vertices of a  square
lattice (see Figure \ref{notation_links}), and interacting
in such a way that strongly correlates directions in real and spin space.
The POC model is dual to the XM model studied in Section \ref{sec4.3}
\cite{NF} (for another duality see \cite{vidal_Thomale_dusuel_schmidt}), 
as follows from the duality isomorphism
\begin{equation}
\label{xme}
\square\sigma^z_\r\ \stackrel{\Phi_{\sf d}}{\longrightarrow} \
\sigma^x_{\bm{r-e_1}}\sigma^x_\r,\ \ \ \ \ \ \
\  \sigma^x_\r \ \stackrel{\Phi_{\sf d}}{\longrightarrow} \
\sigma^z_{\bm{r-e_2}}\sigma^z_\r\ , 
\end{equation}
so that \( \Phi_{\sf d}(H_{\sf XM}[J,h])=H_{\sf POC}[J,h] \). 
If follows that {\it the POC model is self-dual}.

The inverse duality transformation reads
\begin{equation}
\label{xme-}
\sigma^x_\r \sigma^x_{\bm{r+e_1}}\ 
\stackrel{\Phi_{\sf d}^{-1}}{\longrightarrow} \
\square\sigma^z_{\bm{r+e_1}}, \ \ \ \ \ \ \ \ 
\sigma^z_\r \sigma^z_{\bm{r+e_2}} \
\stackrel{\Phi_{\sf d}^{-1}}{\longrightarrow} \
\sigma^x_{\bm{r+e_2}},
\end{equation}
and both $\Phi_{\d}$ and $\Phi_{\d}^{-1}$ define their own set
of variables. {}From Equation \eqref{xme},
\begin{eqnarray}
\mu^x_\r =\sigma^z_{\bm{r-e_2}}\sigma^z_\r,
\ \ \ \ \ \ 
\mu^z_\r=\Phi_\d \left(\prod_{m^1\leq
r^1,r^2\leq m^2}\square\sigma^z_{\bm{m}}\right)= 
\prod_{n=0}^{\infty}\ \sigma^x_{\bm{r}+n\bm{e_2}},
\nonumber
\end{eqnarray}
so that $H_{\sf XM}(\mu)= H_{\sf POC}(\sigma)$ ($H(\cdot)$ means the 
Hamiltonian $H$ written in term of the variables inside the brackets). Similarly, from
Equation \eqref{xme-},
\begin{eqnarray}
\tau^z_\r&=& \Phi_{\sf d}^{-1}(\sigma^z_\r)= \Phi_\d
^{-1}\left(\prod_{n=0}^{\infty}\sigma^z_{\bm{r}-(n+1)\bm{e_2}}
\sigma^z_{\bm{r}-n\bm{e_2}}\right)=\prod_{n=0}^\infty\ 
\sigma^x_{\bm{r}-n\bm{e_2}},\\ 
\tau^x_\r&=& \Phi_{\sf d}^{-1}(\sigma^x_\r)= \Phi_\d
^{-1}\left(\prod_{n=0}^{\infty}\sigma^x_{\bm{r}-(n+1)\bm{e_1}}
\sigma^x_{\bm{r}-n\bm{e_1}}\right)=\prod_{n=0}^{\infty}\ 
\square\sigma^z_{\bm{r}-n\bm{e_1}}=\sigma^z_\r
\sigma^z_{\bm{r+e_2}},\nonumber
\end{eqnarray} 
so that $H_{\sf POC}(\tau)=H_{\sf XM}(\sigma)$. 

\begin{figure}[t]
\begin{center}
\includegraphics[width=0.85\columnwidth]{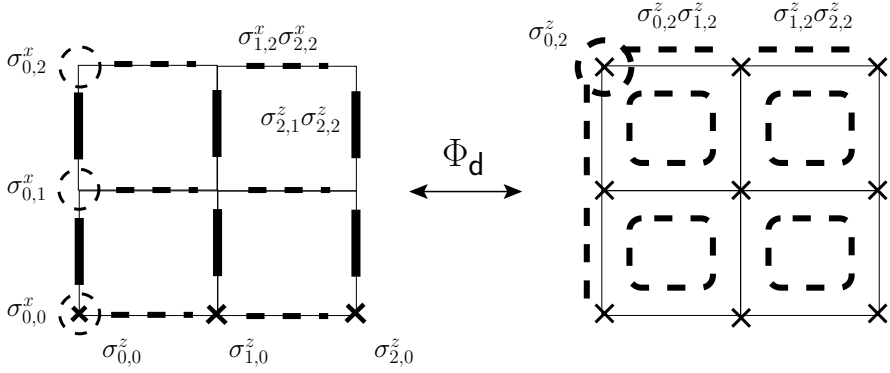}
\end{center}
\caption{The POC model (left panel) restricted to a finite square
section of its lattice is dual to the XM model (right panel)
restricted to the  same square section, provided both models are endowed
with suitable BCs. On the left panel (POC model), the 
broken circles represent spins \(\sigma^x_{0,i}, \ i=0,1,2\), and the
crosses on the lower edge represent spins  \(\sigma^z_{i,0}, \ i=0,1,2\), while
the broken (solid) links represent bonds \(\sigma^x_{i,j} \sigma^x_{i+1,j}\)
(\(\sigma^z_{i,j} \sigma^z_{i,j+1}\)). The right panel was explained in Figure
\ref{f_XM}.}
\label{XM_2dOC}
\end{figure}
Boundary corrections are
required to preserve the duality  for finite lattices. One possible set
of boundary terms is  shown in Figure \ref{XM_2dOC}, for a $3\times 3$
lattice.  Both models, including the boundary corrections,  have now
nine $\sigma^z$-like and nine  $\sigma^x$-like bonds. The duality
isomorphism is given below,
\begin{alignat}{3}
\square\sigma^z_{1,0}&\ \leftrightarrow\ \sigma^x_{0,0}\sigma^x_{1,0},&\qquad
\square\sigma^z_{2,0}\ & \leftrightarrow\ \sigma^x_{1,0}\sigma^x_{2,0}, &\qquad
\square\sigma^z_{1,1}\ & \leftrightarrow\ \sigma^x_{0,1}\sigma^x_{1,1},\nonumber\\
\square\sigma^z_{1,2}&\ \leftrightarrow \ \sigma^x_{1,1}\sigma^x_{1,2},&\qquad
\sigma^z_{1,2}\sigma^z_{2,2}\ &\leftrightarrow\ \sigma^x_{1,2}\sigma^x_{2,2},&\qquad
\sigma^z_{0,2}\sigma^z_{1,2}\ & \leftrightarrow\ \sigma^x_{0,2}\sigma^x_{1,2}, \nonumber\\
\sigma^z_{0,2}&\ \leftrightarrow \ \sigma^x_{0,2}, &\qquad
\sigma^z_{0,1}\sigma^z_{0,2} \ &\leftrightarrow\ \sigma^x_{0,1},&\qquad
\sigma^z_{0,0}\sigma^z_{0,1}\ &\leftrightarrow\ \sigma^x_{0,0},\nonumber \\
\sigma^x_{0,0}&\ \leftrightarrow \ \sigma^z_{0,0},&\qquad
\sigma^x_{1,0}\ &\leftrightarrow\ \sigma^z_{1,0},&\qquad
\sigma^x_{2,0}\ & \leftrightarrow\ \sigma^z_{2,0},\nonumber\\
\sigma^x_{0,1}&\ \leftrightarrow \ \sigma^z_{0,0}\sigma^z_{0,1},&\qquad
\sigma^x_{1,1}\ & \leftrightarrow\ \sigma^z_{1,0}\sigma^z_{1,1},&\qquad
\sigma^x_{2,1}\ & \leftrightarrow\ \sigma^z_{2,0}\sigma^z_{2,1},\nonumber\\
\sigma^x_{0,2}&\ \leftrightarrow \ \sigma^z_{0,1}\sigma^z_{0,2},&\qquad
\sigma^x_{1,2}\ & \leftrightarrow\ \sigma^z_{1,1}\sigma^z_{1,2},&\qquad
\sigma^x_{2,2}\ & \leftrightarrow\ \sigma^z_{2,1}\sigma^z_{2,2}.
\label{longtra}
\end{alignat}
In Equation  \eqref{longtra}, the bonds to the left of the double arrow
are those of the XM model and those to the right  denote bonds of
the orbital compass model. Albeit being tedious, it is straightforward
to extend Equation \eqref{longtra} to  an $N \times N$ square lattice. 
This explicit transformation enables the computation of all dual
variables.

\subsection{Two-dimensional \(\Z_p\) gauge/vector Potts models}
\label{sec5.3}

In this section we study a gauge-reducing duality, along the
lines of Section \ref{sec3.12}. 

The $d=2$  dimensional \(\mathbb{Z}_p\)  gauge theory \cite{horn},
\begin{equation}
H_{\sf G}=\frac{1}{2}\sum_\r\ \left(V_{(\r,1)}+V_{(\r,2)}+\
\lambda\  B_{(\r,3)}\ +\ {\sf h.c.}\right),
\end{equation}
with
\begin{equation}\label{B_plaquette}
B_{(\r,3)}\equiv\ U_{(\r,1)}U_{(\r+\bm{e_1},2)}U^\dagger_{(\r+\bm{e_2},1)}
U^\dagger_{(\r,2)},
\end{equation}
is a generalization of the Ising gauge model studied in Section \ref{sec3.12},
Equation \eqref{ising_gauge}.
The operators \(U_{(\r,\nu)}, V_{(\r,\nu)}\) (\(\nu=1,2\)), located at
the links (see Figure \ref{notation_links}) of a square lattice, commute
 on different links, and satisfy the algebra described
in Section \ref{sec4.1.1} otherwise. As the name of the model suggests,
\(H_{\sf G}\) displays a gauge \(\mathbb{Z}_p\) symmetry
as realized by the local symmetry operators
\begin{equation}
G_\r=V_{(\r,1)}V_{(\r,2)}V^\dagger_{(\r-\i,1)}V^\dagger_{(\r-\j,2)}.
\end{equation}
However, the {\it global} symmetries of the model are {\it non-Abelian},
as can be seen by folowing the discussion in Section \ref{sec4.1.1}
of the symmetries of the VP model.

We want to find a transformation to a dual Hamiltonian which is free of 
gauge symmetries.  The $d=2$ dimensional quantum VP model  (a
higher-dimensional version of Equation \eqref{infinite_vector_potts})
\begin{equation}
H_{\sf VP}= \frac{1}{2}\sum_\r\  \left(\lambda V_\r+\  U_\r
U_{\r+\i}^\dagger+  U_\r U_{\r+\j}^\dagger\ +\ {\sf h.c.}\right),
\end{equation}
is a natural candidate, because it is a
generalization of the $d=2$ Ising model in a transverse field in 
terms of Weyl group algebra operators (defined on the vertices of the  
lattice, see Figure \ref{notation_links}). It is not hard to 
find a bond algebra homomorphism that confirms the duality,
\begin{eqnarray}\label{I_G_aut}
B_{(\r,3)}\dual V_\r,\ \ \ \ V_{(\r,1)}\dual U_{\r-\j}U^\dagger_\r,\ \ \ \
V_{(\r,2)}\dual U_{\r-\i}^\dagger U_\r ,
\end{eqnarray}
{\it up to the complete elimination of gauge symmetries}, 
\begin{eqnarray}\label{G_I_trivial}
\Phi_\d(G_\r)= U_{\r-\j}U_\r^\dagger\times U_{\r-\i}^\dagger U_\r
\times U_{\r-\j-\i}^\dagger U_{\r-\i}\times
U_{\r-\i-\j}U_{\r-\j}^\dagger =\mathbb{1}. 
\end{eqnarray}

The homomorphism of Equation \eqref{I_G_aut} affords a simple and
conceptually clarifying proof that \(H_{\sf VP}\) encodes the observable, gauge-invariant,
physics of \(H_{\sf G}\). On the other hand, it
cannot be used to define dual variables, see Section \ref{no_dual_vars}. 

\subsection{Two-dimensional compact QED and {\rm \bf XY} models}
\label{sec5.4}

The elimination of gauge symmetries has a slightly different flavor
for models that feature continuous (or merely infinite) degrees of freedom, 
essentially because it becomes more convenient to work
with the Hermitian generators of the gauge symmetries, rather than with the
unitary symmetries themselves. This section presents two examples of this kind,
a gauge-reducing duality for \(d=2\) compact QED to the \(d=2\)
SoS model, and a duality from a gauge SoS model to
the \(d=2\) XY model. These {\it quantum} dualities are new to the best of our 
knowledge, but a classical (\(D=3\)) relative (performed by way of the Villain 
approximation) of the duality for the \(d=2\) XY model  was first
used in Reference \cite{peskin}.

The Hamiltonian for \(d=2\) compact QED that follows from 
Wilson's lattice QED \cite{wilson} can be worked out along the
lines of Reference \cite{kogut_susskind},
\begin{equation}\label{LEMh}
H_{\sf LEM}=\sum_\r\ (\frac{1}{2}\ L_{(\r,1)}^2+\ 
\frac{1}{2}\ L_{(\r,2)}^2-\lambda\ \cos \Theta_{(\r,3)}),
\end{equation}
It features continuous angular variables $\theta_{(\r,\nu)} \in
[0,2\pi)$ defined on the  links of a square lattice, with
\begin{equation}
L_{(\r,\nu)}=-i\frac{\partial}{\partial\theta_{(\r,\nu)}}\ ,
\ \ \ \ \ \ \ \ \Theta_{(\r,3)}\ =\ \theta_{(\r,1)}+\theta_{(\r+\i,2)}-
\theta_{(\r+\j,1)}-\theta_{(\r,2)},
\end{equation}
and where the elementary degrees of freedom satisfy the operator algebra
\begin{eqnarray}
[L_{(\r,\mu)},e^{\pm i \theta_{(\r',\nu)}}]=\pm \delta_{\r,\r'} \delta_{\mu,\nu} 
e^{\pm i \theta_{(\r',\nu)}}.
\label{CQEDcomr}
\end{eqnarray}
The gauge symmetries of Wilson's action translates in the Hamiltonian
language to the fact that
\(H_{\sf LEM}\) commutes with all the star operators
\begin{equation}
g_\r=L_{(\r,1)}+L_{(\r,2)}-L_{(\r-\i,1)}-L_{(\r-\j,2)}.
\end{equation}
These are the infinitesimal generators of gauge symmetries.

The gauge-reducing duality that we describe next is a hybrid between the dualities
presented in Sections \ref{sec5.2}  and \ref{sec5.3}. It allow 
us to recast the gauge invariant information contained in \(H_{\sf LEM}\) in 
a form that is free of gauge redundancies. The dual, completely gauge
reduced model is the SoS model
\begin{equation}
H_{\sf SS}=\frac{1}{2}\sum_\r\ (-\lambda\ (R_\r+R_\r^\dagger)+
(X_{\r+\bm{e_1}}-X_\r)^2+\
(X_{\r+\bm{e_2}}-X_\r)^2),
\end{equation}
that features elementary degrees of freedom placed on the sites \(\r\) of
a  square lattice, and can be connected to the classical 
\(D=3\) SoS model through a STL decomposition (see 
Section \ref{classical&quantum}). The operators \(R_\r, R_\r^\dagger, X_\r\) 
at site \(\r\) satisfy 
the relations of Equation \eqref{Z}, and commute with the operators on
other sites. Notice that \(H_{\sf SS}\) possesses global symmetries only.

The connection between \(H_{\sf LEM}\) and 
\(H_{\sf SS}\) is established through the homomorphism of bond algebras
\begin{eqnarray}
e^{i\Theta_{(\r,3)}}\ & \stackrel{\Phi_\d}{\longrightarrow}&\ R_\r,
\ \ \ \ \ \ \ \ \ \ \ \ \ \ \ \ \ \ \ \ \ \ \ \ 
e^{-i\Theta_{(\r,3)}}\ \stackrel{\Phi_\d}{\longrightarrow} \ R_\r^\dagger,
\\
L_{(\r,2)} \
&\stackrel{\Phi_\d}{\longrightarrow}& \ (X_\r-X_{\r-\i}),\ \ \ \ \ \ \ \ \ \ \ \ \, 
L_{(\r,1)}\ \stackrel{\Phi_\d}{\longrightarrow} \ -(X_\r-X_{\r-\j}).
\nonumber
\end{eqnarray}
That the gauge symmetries of  \(H_{\sf LEM}\) are
trivialized (eliminated) by the duality follows from the computation  
\begin{eqnarray}
\Phi_\d(g_\r)&=&-(X_\r-X_{\r-\j})+(X_\r-X_{\r-\i})+\nonumber\\
&&\ \ \ \ (X_{\r-\i}-X_{\r-\j-\i})-(X_{\r-\j}-X_{\r-\i-\j})=0.
\end{eqnarray} 



Next we consider a duality that is in some sense complementary to the 
previous one. The $d=2$ XY model
\begin{equation}
H_{\sf KT}=\sum_\r\ (\frac{1}{2}L_\r^2\ -\lambda
\cos(\theta_{\r+\i}-\theta_\r) - \lambda \cos(\theta_{\r+\j}-\theta_\r)
)
\end{equation} 
is the completely gauge-reduced version of a gauge SoS
model,
\begin{equation}
H_{\sf GSS}=\frac{1}{2}\sum_\r\ (-\lambda
(R_{(\r,1)}+R_{(\r,1)}^\dagger)-  \lambda
(R_{(\r,2)}+R_{(\r,2)}^\dagger)+ b_{(\r,3)}^2),
\end{equation}
where \(b_{(\r,3)}=X_{(\r,1)} +X_{(\r+\i,2)}-X_{(\r+\j,1)}-X_{(\r,2)}\). 
The generators of the group of gauge symmetries of \(H_{\sf GSS}\) are
\begin{equation}
G_\r=R_{(\r,1)}R_{(\r,2)}R^\dagger_{(\r-\i,1)}R^\dagger_{(\r-\j,2)},
\ \ \ \ \ \ \ \ \ \ \ \ \mbox{ and } \ G_\r^\dagger,
\end{equation}
but these are not infinitesimal. The gauge-reducing duality to the
XY model is established by the homomorphism of bond algebras
\begin{eqnarray}
&&X_{(\r,1)}+X_{(\r+\i,2)}-X_{(\r+\j,1)}-X_{(\r,2)} \dual L_\r,\\
&&R_{(\r+\j,1)}\dual e^{-i(\theta_{\r+\j}-\theta_\r)},  \ \ \ \ 
R^\dagger_{(\r+\i,2)}\dual e^{-i(\theta_{\r+\i}-\theta_\r)},\nonumber
\end{eqnarray}
that satisfies  
\begin{equation}
\Phi_\d(G_\r)=e^{-i(\theta_\r-\theta_{\r-\j})}e^{i(\theta_\r-\theta_{\r-\i})}
e^{i(\theta_{\r-\i}-\theta_{\r-\j-\i})}e^{-i(\theta_{\r-\j}-\theta_{\r-\i-\j})}=
\mathbb{1}.
\end{equation}

\subsection{Toric code/\(\mathbb{Z}_2\) Higgs models}
\label{sec5.5}

Over the last fifteen years quantum computation has become a 
well  developed theoretical discipline, fostering a
paradigm-breaking new understanding of computational complexity and 
quantum mechanics \cite{mermin}. In contrast, the technological
problem of building a quantum computer remains essentially unsolved,
and one of the biggest challenges is the realization of quantum memories.

Kitaev's toric code (TC) model \cite{toric_code} (see Equation 
\eqref{etch}) is an excellent example of the virtues and pitfalls of one
of the most popular approaches to the problem of storing quantum
information: the use of topological quantum order \cite{tqo}.
While the TC model is a good example of topological quantum
order, it fails as a quantum memory at any finite temperature
\cite{tqo}. This surprising result,  known as {\it thermal fragility}
\cite{tqo,3detc},  could be proved and probed in detail thanks to the
realization \cite{tqo} that the TC model is
(exactly) dual to two decoupled Ising chains in zero  magnetic field,
a duality established by arguments that are direct precursors of the
bond-algebraic machinery. 

Whether or not topological quantum order turns out to be  key to the
implementation of quantum memories or the  quantum computer, it is clear
by now that, as an order of matter that goes beyond the  Landau
symmetry-breakdown paradigm, it is worth studying in its  own right.
In this section we show that two popular models to study 
topological quantum order, the {\it extended} toric code (ETC) model
in two \cite{extended_tc} and three space dimensions, are
dual to the \(\mathbb{Z}_2\) Higgs model \cite{fradkin_shenker}. While
the duality in $d=2$ dimensions is already exploited in 
Reference \cite{extended_tc} to help the numerical simulation
of the ETC model, 
the duality in $d=3$ dimensions is one of the most interesting
new dualities reported in this paper. Both dualities are in fact special
cases of a general gauge-reducing duality for the \(\mathbb{Z}_2\) Higgs model
{\it that works in any number \(d\) of space dimensions}, and that 
is also special in that it does not necessitates the introduction of 
non-local string operators (recall the discussion of Section \ref{sec3.12.4}). 
Other aspects of this duality will be discussed in Section \ref{sec6.3}.
Section \ref{sec_tqo_dim} presents a broader discussion of the role of dualities 
in the study of topological quantum order.

The \(\mathbb{Z}_2\) Higgs model in \(d\) spatial dimensions
\begin{eqnarray}\label{z2_higgs}
H_{\sf dH}=-\sum_\r ( J_x\sigma^x_\r +
\sum_{\nu=1}^d (h_z\sigma^z_{\r}
\sigma^z_{(\r,\nu)}\sigma^z_{\r+\bm{e_\nu}}+h_x\sigma^x_{(\r,\nu)} ) 
+  J_z\sum_{\nu<\mu}\ B_{(\r,\nu \mu)}), 
\end{eqnarray}
features spin \(1/2\) degrees of freedom on the sites and links
of a hyper-cubic lattice. It can be thought of as a lattice, 
two-state approximation to a complex Higgs field \(\phi\) of fixed modulus
\(\phi\phi^*=1\)  (or in its broken symmetry phase \cite{savit}),  in interaction 
with electromagnetism, and it represents one of the best understood models of 
confinement in dimensions \(d=2\) and \(d=3\).
Its gauge symmetries are
\begin{equation}
G_\r=\sigma^x_\r\ \prod_{\nu=1}^d\
\sigma^x_{(\r,\nu)}\sigma^x_{(\r-\bm{e_\nu},\nu)} \equiv \sigma^x_\r
A_\r,\ \ \ \ \ \ \ \ \ \ [H_{\sf dH},G_\r]=0.
\end{equation}
The goal of this section is to find a completely gauge-reducing duality to take care
of this gauge redundancy. 

The structure of \(H_{\sf dH}\)s bond algebra suggests the model
\begin{equation}
\label{KTCMd3}
H_{\sf dGRH}=-\sum_\r (J_x A_\r+
\sum_{\nu=1}^d\  ( h_z\sigma^z_{(\r,\nu)}  + h_x\sigma^x_{(\r,\nu)})
+J_z\sum_{\nu<\mu}B_{(\r,\nu \mu)} ),
\end{equation}
for a gauge-reduced dual. \(H_{\sf dGRH}\) shows no local symmetries, 
due to the couplings to external magnetic fields, and 
the bond algebra homomorphism that connects it to \(H_{\sf dH}\)
follows naturally,
\begin{eqnarray}
\sigma^x_\r &\dual& A_\r,\ \ \ \ \ \ \ \ \ \ 
\sigma^z_{\r}\sigma^z_{(\r,\nu)}\sigma^z_{\r+\bm{e_\nu}}\dual 
\sigma^z_{(\r,\nu)}\nonumber\\
\ \nonumber\\
B_{(\r,\nu\mu)}&\dual& B_{(\r,\nu\mu)},\ \ \ \ \ \ \ \ \ \  \ \ \ \ \ \, 
\sigma^x_{(\r,\nu)}\dual \sigma^x_{(\r,\nu)}.
\end{eqnarray} 
To check that \(\Phi_\d\) is indeed gauge reducing, we compute
\begin{equation}
\Phi_\d(G_\r)=\Phi_\d(\sigma^x_\r) \Phi_\d(A_\r)=A_\r A_\r=\mathbb{1}.
\end{equation}
Notice that the completely gauge-reduced model \(H_{\sf dGRH}\) features only 
degrees of  freedom on the links of the lattice, and local bonds.

The duality just described works in any space dimension $d$. In
$d=1,2,3$, the models \(H_{\sf dGRH}\) have well known physical
meanings.  The  $d=1$ \(\mathbb{Z}_2\) Higgs model 
\begin{equation}
H_{\sf 1H}=-\sum_i (J_x\sigma^x_i+
h_z\sigma^z_i\sigma^z_{(i,1)}\sigma^z_{i+1}+h_x\sigma^x_{(i,1)})
\end{equation}
is dual to 
\begin{equation}
H_{\sf 1GRH}=-\sum_i (J_x \sigma^x_{(i-1,1)}\sigma^x_{(i,1)}
+h_z\sigma^z_{(i,1)}+h_x\sigma^x_{(i,1)} ),
\end{equation}
which is just an Ising chain in the presence of transverse and
longitudinal fields \cite{fradkin_shenker}.  This means that \(H_{\sf
1H}\) has no phase transition when \(h_x\neq0\).  We see that the gauge
field has opened a mass gap in the model. 

In $d=2$ dimensions, the gauge reduced form of the \(\mathbb{Z}_2\)
Higgs model reads
\begin{equation}\label{etch}
H_{\sf 2GRH}=H_{\sf ETC}=-\sum_\r (J_x A_\r+
\sum_{\nu=1}^2 ( h_x\sigma^x_{(\r,\nu)}+h_z\sigma^z_{(\r,\nu)} )
+J_zB_{(\r,3)})
\end{equation}
(\(B_{(\r,3)}\) was defined in Equation \eqref{gauge_plaquette}).
This is exactly the ETC model of Reference \cite{extended_tc}
(if we further set  \(h_x=h_z=0\),  we recover Kitaev's TC model).
The duality maps the Coulomb phase \cite{fradkin_shenker} of the \(\mathbb{Z}_2\) Higgs
model to the  topological quantum ordered state of the ETC model.

In $d=2$ dimensions, the \(\mathbb{Z}_2\) Higgs model is self-dual \cite{horn_yankielowicz},
which implies that the ETC model is self-dual as well \cite{extended_tc,con}. Let us check
this in the latter model. The self-duality isomorphism \cite{con} reads
\begin{equation}\label{ETC_aut}
\sigma^{x,z}_{(\r,1)}\ \stackrel{\Phi_\d}{\longrightarrow}
\ \sigma^{z,x}_{(\r+\bm{e_1},2)},\ \ \ \ \ \ \ \ 
\sigma^{x,z}_{(\r,2)}\ \stackrel{\Phi_\d}{\longrightarrow}
\ \sigma^{z,x}_{(\r+\bm{e_2},1)}
\end{equation}
that exchanges $J_x$ with $J_z$, and simultaneously $h_x$ with $h_z$ in 
\(H_{\sf ETC}\). This is one of the rare instances where a self-duality
mapping is {\it local in the spins}. Other interesting dualities for 
the ETC model are reported in \cite{con} (and its supplemental material).

The gauge-reduced version of \(H_{\sf 3H}\) is again intimately
connected to a $d=3$ dimensional generalization of the ETC model 
\begin{eqnarray}
H_{\sf 3GRH}&=&-\sum_\r
\sum_{\nu=1}^3\ (h_x\sigma^x_{(\r,\nu)}+h_z\sigma^z_{(\r,\nu)} )
-\nonumber\\
&&\ \  \,
\sum_\r\ (J_x A_\r+ J_z (B_{(\r,1)}+\ B_{(\r,2)}+\ B_{(\r,3)})),
\end{eqnarray}
studied in Reference \cite{3detc} in the case of vanishing
magnetic fields. The  phase diagram of the full model,
that we  call the $d=3$ ETC model, can be obtained from the
literature on the Higgs model (see, for example, Reference
\cite{fradkin_shenker}, Figure 2).

\section{Bond-algebraic dualities in quantum field theory}
\label{sec6}

Over the years, some of the most interesting and ambitious
dualities have been conjectured in the context of quantum field 
theory (QFT)
\cite{montonen_olive, AdS-review},
and any progress in the theory of non-Abelian dualities
should be tested against QCD (quantum chromodynamics).  
The functional approach to QFT \cite{rivers} puts QFTs in a 
language that resembles closely that of classical statistical mechanics.
Therefore there have been some attempts \cite{thooftI} at dualities 
for path integrals of QFTs that resemble that of Kramers and Wannier 
introduced in Section \ref{sec2.1}. However, 
the progress in this direction has been limited (see though
References \cite{alvarez,non_Abelian,lozano}).
The situation improves considerably if the (Euclidean) path integral for the QFT of 
interest is regularized by replacing the continuum for a lattice \cite{wilson}
(see Appendix \ref{appE}). 
Then, for Abelian theories, one can use the machinery of Appendix \ref{appA} to
construct systematically regularized dual (Euclidean) field theories \cite{savit}.
But, as discussed in Section \ref{classical&quantum}, and illustrated
in Section \ref{sec8}, {\it this path-integral based, lattice approach to
dualities in QFT is covered, and in fact simplified} by the bond-algebraic techniques
of this paper \cite{con}. In this light, many of the dualities we have seen already
can be interpreted as bond-algebraic dualities for QFTs.

In contrast, this section aims to explore the extension of bond algebraic techniques
to operator-based quantizations of field theories. This is perfectly feasible
for some QFTs, but in general we do not know yet how to construct
a complete operator quantization of an {\it interacting} field theory, and
so in many cases the {\it bond algebra of an interacting QFT is not 
well defined}. This means for instance that we could have trouble
deciding whether two operators in a QFT should commute or not, as
exemplified by the Schwinger term in QED \cite{schwinger_term}: 
The canonical quantization of electromagnetism in interaction with the Dirac electron field
dictates the charge density operator should commute with the current, but in
fact this is inconsistent with the requirement that the theory should have
a ground state \cite{schwinger_term}. It follows that this 
commutation relation must be changed relative to its canonical form. 

There is an approximate approach to the operator quantization 
of field theories that is specially compatible with bond-algebraic
techniques, and was popularized in the literature on confinement
under the name of {\it lattice Hamiltonian formalism} 
\cite{kogut_susskind, kogut}.
The idea is to discretize the {\it classical} field theory first {\it only
in space} (that is, to
approximate it with a classical mechanical model featuring degrees of freedom
on a spatial lattice), and then proceed to quantization, that can now 
be carried through by standard means (see Appendix \ref{appE}). 
The resulting many-body quantum theory typically features non-relativistic bosons,
fermions, or rigid rotators in interaction, and contains a new parameter, the lattice
spacing \(a\). Of course, it lacks the symmetries characteristic
of the continuum, {\it but has in exchange a well defined operator content}, and 
typically internal and gauge symmetries are well represented.
In what follows, we study the duality properties of 
several QFTs  in this approximation. But when possible,
we work also directly in the continuum, and show that the two approaches give 
compatible results when a naive continuum limit \(a\rightarrow 0\) is taken.  

\subsection{One-dimensional free and externally coupled bosonic field, and 
the Kibble model}
\label{sec6.1}

The free, massless bosonic field in $d=1$ (i.e., 1+1 space-time dimensions)
affords the simplest example of a self-dual QFT \cite{witten,con}.
In dimensions \(d\geq 2\), a complete operator quantization is always available
for {\it free} fields \cite{takahashi}. This is not the case in dimension \(d=1\).
In particular, the Green's function for the massless boson field is too singular to
be interpreted in the sense of distributions. Still, its bond-algebra based on 
canonical quantization 
reflects its self-dual properties, and we can check them in the lattice Hamiltonian
approximation. In this approximation, the bosonic field reduces to a self-dual model of one
dimensional phonons, Section \ref{sec4.3}, and the phononic lattice dual variables
converge to the bosonic dual variables in the continuum.
Next we consider two simple extensions,  the $d=1$ Kibble model in
Section \ref{1dKibblemodel},  and a multiplet of bosonic fields in interaction
with an external driving forces. 
 
\subsubsection{Free, Massless bosonic field}
\label{triv_boson_sub}
   
The massless, free bosonic field in \(1+1\) dimensions, ($\mu,\nu=0,1$)
is described by the action
\begin{equation}
\label{1+1sb}
S_{\sf FB}=\int d^2x\ \frac{1}{2}\eta^{\mu \nu}
\partial_\mu\phi\partial_\nu\phi ,
\end{equation}  
where \(\eta={\sf diag}[1,-1]\), 
$x^{0} = t$ stands for the time coordinate, and
$x^1= x$ for the spatial coordinate.
Its canonical quantization proceeds by defining
the Hamiltonian
\begin{equation}\label{boson_H}
H_{\sf FB}=\int dx\ \left
(\frac{1}{2}\pi^2+\frac{1}{2}\left(\partial_1  {\phi}\right)^2 \right ),
\end{equation}
together with the equal-time commutation relations 
\begin{equation}
[\phi(x),\pi(y)] =\ i\delta(x-y).
\end{equation}
One may think of $H_{\sf FB}$ as  the quantum theory of a continuous
elastic line (see Reference \cite{kittel}, Chapter 2). 

{}From the perspective of bond algebras, the new feature is
that we have an uncountable infinity of bonds, two bonds 
\(\pi^2(x),\ \left(\partial_1  {\phi}\right)^2(x)\) per space point.
It is easier to characterize the bond algebra
in terms of \(\pi(x)\) and \(\partial_1\phi(x)\),
\begin{equation}
[\partial_1\phi(x),\pi(y)]=\ i\delta'(x-y),
\label{can_com}
\end{equation}
where $\delta'(x-y)=\partial_1 \delta(x-y)=\partial_x \delta(x-y)$  is
the spatial derivative of the Dirac delta function. 
It is apparent from this relation that 
\begin{equation}\label{boson_aut}
\partial_1\phi(x)\ \stackrel{\Phi_{\d}}{\longrightarrow}\
\pi(x),\ \ \ \ \ \ \ \ 
\pi(x) \ \stackrel{\Phi_{\d}}{\longrightarrow}\ \partial_1\phi(x),
\end{equation}
is a self-duality isomorphism, since
\begin{equation}
[\Phi_{\d}(\partial_1\phi(x)),\Phi_{\d}(\pi(y))]=
[\pi(x),\partial_1\phi(y)]= -i\delta'(y-x)=
i\delta'(x-y), 
\end{equation}
and \(\Phi_\d(H_{\sf FB})= H_{\sf FB}\). 

Next we use Equation \eqref{boson_aut} to compute dual variables.
Since 
\begin{equation}
\label{phi_int}
\phi(x)=\int_{-\infty}^x dy\ \partial_y\phi(y),
\end{equation}
the dual field variables are 
\begin{eqnarray}\label{boson_dual_var}
\pi(x)&&\stackrel{\Phi_\d}{\longrightarrow}\ 
\hat{\pi}(x)=\ \partial_1\phi(x),\\
\phi(x)&&\stackrel{\Phi_\d}{\longrightarrow}\ 
\hat{\phi}(x)=\int_{-\infty}^x dy\ \widehat{\partial_1\phi}(y)=
\int_{-\infty}^x dy\ \pi(y). \nonumber
\end{eqnarray}
We may regard these dual fields as toy examples of Mandelstam variables
\cite{mandelstam_variables}, the variables that appear in the bosonization 
of $d=1$ theories \cite{bosonization}.

Let us compare next the calculations in the continuum to the
predictions of the lattice Hamiltonian approach. If we discretize
in space the action of Equation \eqref{1+1sb}, and quantize afterwards,
we get a quantum model specified by 
\begin{equation}
H_{\sf LFB}=\frac{1}{2a}\sum_i\ (\pi_i^2+(\phi_{i+1}-\phi_i)^2),\ \ \ \ \ \
\ \ [\pi_m,\phi_n]=i\delta_{m,n},
\end{equation}
with \(a\) the lattice spacing.
This model is essentially identical to the self-dual model of phonons studied
in Section \ref{sec4.3}, and thus self-dual as well (notice that while
\(\pi_i\) is dimensionless, \(\pi(x)\) has dimensions of \(1/a\)). 
Its self-duality mapping  can be read from Equation \eqref{phonons_aut},
\begin{equation}
\frac{\pi_i}{a}\dual\frac{\phi_{i+1}-\phi_i}{a},\ \ \ \ 
\frac{\phi_{i+1}-\phi_i}{a}\dual\frac{\pi_{i+1}}{a},
\end{equation} 
and it clearly converges to the corresponding mapping in the continuum,
Equation \eqref{boson_aut}, in the naive continuum limit \(a\rightarrow 0\).

\subsubsection{The one-dimensional Kibble model}
\label{1dKibblemodel}

The self-duality mapping investigated in the previous section 
is readily applicable to the Kibble model \cite{aitchison},
\begin{eqnarray}
H_{\sf K}=\frac{1}{2}\int dx\ \left(\pi^2+(\partial_1\phi)^2\right)
+ \frac{1}{2}\int dx\ dy\ \ \pi(x)V(x-y)\pi(y),
\end{eqnarray}
in \(d=1\) dimension. The $d=3$ version of this
model with Coulomb potential $V(\bm{x})  =e^2/|\bm{x}|$ has been
studied as an example of a  model that violates Goldstone  theorem
\cite{bookEPTCP}, due to the long-range nature of the Coulomb interaction.

The Kibble model has the unusual feature that the momenta $\pi(x)$
participate in the interaction term. In \(d=1\) this is remedied
by the duality of the previous section. The mapping of Equation 
\eqref{boson_aut} shows that  \(H_{\sf K}\) admits a dual representation
\begin{align}
H_{\sf K}^D= \frac{1}{2}  \int dx\ \left(\pi^2+(\partial_1\phi)^2\right)
+\frac{1}{2}\int dx\ dy\ \ \partial_1\phi(x)\ V(x-y)\ \partial_1\phi(y),
\end{align}
so that now the interaction term involves only spatial gradients.

\subsubsection{Massless bosonic field coupled to classical external sources}

The self-duality of the free boson field survives even after we
have coupled it to classical, external sources \(A(x^0,x^1)\) and \(J(x^0,x^1)\),
\begin{equation}\label{boson_H}
H_{\sf B}=\int dx\ \left
(\frac{1}{2}(\pi-A)^2+\frac{1}{2}\left(\partial_1  {\phi}\right)^2+J\phi \right),
\end{equation}
that vanish outside some finite interval.
While the coupling \(J\phi\) is standard, the coupling \(\pi-A\) is not,
because \(A\) is not the vector potential for an EM field (for one thing,
\(\phi=\phi^\dagger\) is not charged). We introduce it anyways, because the 
self-duality we are going to describe exchanges \(J\) and \(A\).

The next step is to apply to \(H_{\sf B}\) the self-duality mapping for
the free boson field, Equation \eqref{boson_dual_var}. After some 
rearrangements (that include discarding a boundary term, since 
$A$ has compact support), the resulting Hamiltonian reads
\begin{equation}
H_{\sf B}^D=\int dx\ \left
(\frac{1}{2}(\pi-A^D)^2+\frac{1}{2}\left(\partial_1  
{\phi}\right)^2+J^D\phi \right)+ \frac{1}{2}\int dx\ \left(A^2-A^{D 2}
\right),
\end{equation}
with
\begin{equation}
A^{D}(x^0,x^1)=-\int_{x^1}^{\infty}dy\ J(x^0,y),\ \ \ \ \ \
J^D(x^0,x^1)=\frac{\partial A}{\partial x^1}(x^0,x^1).
\end{equation}
Since \(H_{\sf B}^D\) has the same structure as \(H_{\sf B}\) (up to
an additive c-number), we see that \(H_{\sf B}\) is still self-dual as
in the free case. Notice that \(\Phi_\d^2=\mathbb{1}\), since 
\(A^{DD}=A\) and \(J^{DD}=J\).

\subsection{The Luttinger model}
\label{sec6.7}

Next we describe a duality for fermions in one dimension.
The Luttinger model describes a $d=1$ dimensional interacting
many-electron system  in a box of size $\ell$. Its  {\it Fermi surface}
consists of only two points, $k= \pm k_{F}$, corresponding to two types
of electrons moving to the right/left. 
Henceforth, we denote the right
and left moving electrons by  the fields $\psi_{1}$ and $\psi_{2}$
(which anti-commute) and construct a two component field $\psi^{\dagger}
= (\psi_{1}^{\dagger}, \psi_{2}^{\dagger})$. In the vicinity of the two
Fermi points $\pm k_F$ (i.e., for small $|q|$ and small $|k-k_{F}|$ for
$k>0$ or small $|k+k_{F}|$ for $k<0$), the free electron dispersion may
be linearised to read $\epsilon_{k+q} - \epsilon_{k} = \pm v_{F} q$ with
$v_{F}$ the Fermi velocity.  The (spinless) Luttinger model is defined
by the Hamiltonian
\begin{equation}
\label{lutt_model}
H_L=\int_0^\ell dx \, \psi^\dagger\sigma^z\left(-i\frac{\partial}
{\partial x}\right)\psi + \int_0^\ell dx dy\ \psi^\dagger_1(x)
\psi_1(x)V(x-y)\psi^\dagger_2(y)\psi_2(y) .
\end{equation}
In Equation \eqref{lutt_model}, the first term describes the free
electron dispersion (with $v_{F}$ set to unity). This is augmented, in
the second term of Equation \eqref{lutt_model},  by density-density
interactions that couple left ($k<0$) and right ($k>0$) movers.  By
explicitly constructing unitary transformations, Luttinger showed that
$H_L$ is unitarily equivalent to a non-interacting model, and thus it is
exactly solvable  (see, for example, \cite{mattis_lieb} and references
therein). It has, however,  the non-physical characteristic that in the
thermodynamic limit it displays an  infinite reservoir of negative
energy states.

We will next re-derive this unitary equivalence within the bond algebraic
framework. The Hamiltonian \(H_L\) commutes with  particle number
operators for each fermion species (left and right movers), and can be
written in a first-quantized form as
\begin{equation}\label{luttinger_first_quantized}
H_L=\sum_{m=1}^M\ p_m -\sum_{n=1}^N\ P_n + \sum_{m=1}^M\sum_{n=1}^N 
V(x_m-y_n),
\end{equation}
where 
\((p_m=-i\partial/\partial x_m, x_m)\) and \((P_n=-i\partial/\partial
y_n, y_n)\) are the momenta and  positions associated with the right (a
total of $M$) and left (a total of $N$)  movers.  The wave functions on
which $H_L$ operates must be  totally antisymmetric because of Fermi
statistics. Luttinger's  result amounts to the statement that \(H_L\) is
dual to 
\begin{equation}
H^D_L=\sum_{m=1}^M\ p_m -\sum_{n=1}^N\ P_n. 
\end{equation}
To prove this result from a bond-algebraic perspective,  we introduce
the bonds
\begin{equation}
A_m= p_m+\frac{1}{2}\sum_{n=1}^N\ V(x_m-y_n),
\ \ \ \ \ \ B_n=P_n-\frac{1}{2} \sum_{m=1}^M V(x_m-y_n),
\end{equation}
so that 
\begin{eqnarray}
H_L=\sum_{m=1}^M\ A_m - \sum_{n=1}^N B_n.
\end{eqnarray}
It is immediate that
\begin{equation}
[A_m, A_{m'}]=0,\ \ \ \ \ \ \ [B_n, B_{n'}]=0, \ \ \ \ \ \ \ \mbox{and }
\ \  [A_m, B_n]=0,
\end{equation}
since
\begin{equation}
[A_{m}, B_{n}]= - \frac{1}{2}\left[p_{m}, \sum_{m'=1}^M\ 
V(x_{m'}-y_{n})\right]
+\frac{1}{2}\left[\sum_{n'=1}^N\ V(x_{m}-y_{n'}), P_{n}\right]=0.
\end{equation}
Thus, putting all the pieces together, we establish the duality
isomorphism
\begin{equation}
A_m\ \stackrel{\Phi_{\sf d}}{\longrightarrow}\ p_m,\ \ \ \ \ \
B_n\ \stackrel{\Phi_{\sf d}}{\longrightarrow} P_n.
\end{equation}
The above demonstration illustrates that Luttinger's assertion holds
for  arbitrary interactions  \(V(x-y)\) in Equation \eqref{lutt_model}.

\subsection{QED without sources, compact QED, and \(\Z_p\) gauge theories}
\label{sec6.2}

\subsubsection{QED without sources}\label{noncompactQED}
The quantization of the EM field suffers from well known 
complications due to gauge invariance, and very different from the
complications that arise in the quantization of the \(d=1\) free boson.
They are much less of a problem from a bond algebraic perspective, because
in the end, at least in the absence of sources, we do know how to construct
a full operator quantization of the EM field. The resulting vacuum QED is the
starting point to quantum optics \cite{glauber}, 
and is self-dual under the exchange of the quantum
electric and magnetic field operators.

We start by setting up the version of QED that we are going to work with.
The starting point is the gauge-invariant action for the vector potential,
\begin{equation}
S_{\sf EM}=-\frac{1}{4}\int dx^0d^3x\ (\partial_\mu A_\alpha-\partial_\alpha A_\mu)
(\partial_\nu A_\beta-\partial_\beta A_\nu)\eta^{\mu\nu}\eta^{\alpha\beta}
\end{equation}
(the connection of the vector potential to the electric \(\vec{E}\)
and magnetic \(\vec{B}\) fields was described in Section \ref{sec2.1}).
To proceed with canonical quantization, we need to partially fix the gauge.
If we choose the condition $A_{0}=0$, called the axial gauge, we can 
complete the canonical quantization prescription easily. The resulting $d=3$ QFT 
reads
\begin{equation}\label{QED_H}
H_{\sf EM}=\int d^3x\ \left ( \frac{1}{2}\ \vec{\Pi}^{2}(\x)\ +\ 
\frac{1}{2}(\nabla\times \vec{A}(\x))^2 \right ),
\end{equation}
together with
\begin{equation}\label{comm_EM}
[A_\mu(\x),\ \Pi_\nu(\x')]=\ i\ \delta_{\mu,\nu}\ \delta(\x-\x'),\ \ \ \ \
\ \mu,\nu=1,2,3,
\end{equation}
and the \(A_\mu(\x),\ \Pi_\nu(\x')\) can be realized as well defined
operators (i.e., operator-valued distributions) action on a Hilbert 
state space. Notice that in the axial gauge we are using,  
\(\vec{E}=-\partial_t\vec{A}=-\vec{\Pi}\). There is, however, and issue
left from the remaining gauge symmetry of the theory defined by 
Equations \eqref{QED_H} and \eqref{comm_EM}. The state space is 
larger-than-physical, and only the states \(\vert \Psi \rangle\) 
that satisfy the Gauss constraint
\begin{equation}\label{gauss_constraint}
\nabla\cdot\vec{\Pi}\ \ \vert \Psi \rangle=0,
\end{equation}
can be prepared and observed by experimental means. The reason is 
that \(\nabla\cdot\vec{\Pi}\) is the generator of the residual 
gauge symmetries that where not fixed by the axial condition 
$A_{0}=0$, and so Equation \eqref{gauss_constraint} amounts to
the statement that only gauge invariant states are physical.

Let us consider next the bond algebra characterized by 
\begin{equation}\label{bonda_em}
[(\nabla\times \vec{A})_\mu(\x),\ \Pi_\nu(\x')]=
i\delta_{\mu,\nu}(\nabla\times \delta(\x-\x'))_\mu.
\end{equation}
It is easy to check that the mapping 
\begin{equation}\label{sd_emcont}
\Pi_\mu(\x)\dual(\nabla\times \vec{A})_\mu(\x),\ \ \ \ \ \ 
(\nabla\times \vec{A})_\mu(\x)\dual -\Pi_\mu(\x),
\end{equation}
preserves the relations of Equation \eqref{bonda_em}, but there is
a subtlety of interpretation. \(\Phi_\d\) maps \(\nabla\times \vec{A}\),
that is automatically divergenceless, to \(\vec{\Pi}\), that is not.
However, \(\vec{\Pi}\) is divergenceless in the gauge invariant 
subspace of physical states, and so \(\Phi_\d\) is a true isomorphism
of bond algebras in that subspace. In the language of Section 
\ref{sec3.11}, and for this particular quantization of the EM field,
{\it its self-duality is an emergent one}. In other words, the 
EM duality is truly a
duality between the observable physical electric and magnetic fields, and
not a more general property of the vector potential. 

We can use the mapping \(\Phi_\d\) of Equation \eqref{sd_emcont} to
compute dual variables on the subspace of physical states,
\begin{eqnarray}
&&\vec{\Pi}(\x) \ \stackrel{\Phi_\d}{\longrightarrow}\ 
\widehat{\vec{\Pi}}(\x)=\nabla\times\vec{A}(\x),\nonumber\\
&&\vec{A}(\x)\ \stackrel{\Phi_\d}{\longrightarrow}\ 
\widehat{\vec{A}}(\x)=-\frac{1}{4\pi}\nabla\times\int d^3x'\
\frac{\vec{\Pi} (\x')}{\vert \x-\x'\vert}.
\end{eqnarray}
This completes our
discussion of the self-duality of QED in the continuum. The next step
is to check it against the lattice Hamiltonian formalism. This analysis 
will  facilitate later our discussions of duality in other
Hamiltonian lattice gauge field theories that are closely connected to
QED and to the dynamics of center vortices in QCD \cite{fradkin_susskind, horn}.

\begin{figure}[t]
\begin{center}
\includegraphics[width=0.55\columnwidth]{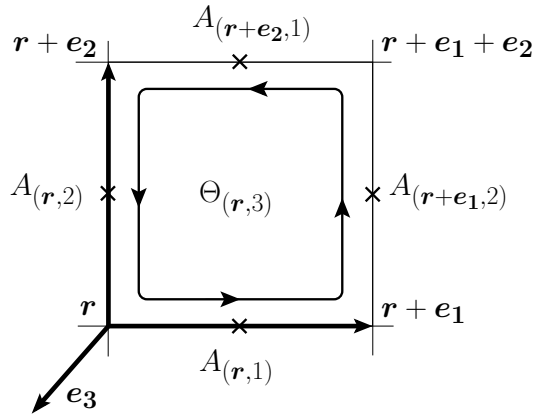}
\end{center}
\caption{Gauge field degrees of freedom $A_{(\r,\nu)}$ ($\nu=1,2,3$) for
$d=3$ non-compact lattice QED are placed on the links of a cubic 
lattice (of lattice spacing \(a\)), as indicated by the crosses.  
The plaquette variable $\Theta_{(\r,3)}=A_{(\r,1)}+A_{(\r+\i,2)}-
A_{(\r+\j,1)}-A_{(\r,2)}$
resides in the plaquette  indicated in the $(\i,\j)$ plane. Other
symbols' meaning are defined in Figure \ref{notation_links}.}
\label{plaquette}
\end{figure}
If we discretize the EM field as we discretized the free boson field near the 
end of Section \ref{triv_boson_sub}, we get
\begin{equation}
H_{\sf LEM}=\sum_\r\ \left (  \frac{1}{2a^3}\ \vec{\Pi}_\r^2\ +\ 
\frac{1}{2}\ a\vec{\Theta}_\r^2 \right ),
\end{equation}
where  $\vec{\Pi}_\r=(\Pi_{(\r,1)}, \Pi_{(\r,2)},\Pi_{(\r,3)})$, and 
$\vec{\Theta}_\r$ stands for the discretized curl,
\begin{eqnarray}
\label{theta_fields}
\Theta_{(\r,1)}&=&A_{(\r,2)}+A_{(\r+\j,3)}-A_{(\r+\k,2)}-A_{(\r,3)},
\nonumber\\
\Theta_{(\r,2)}&=&A_{(\r,3)}+A_{(\r+\k,1)}-A_{(\r+\i,3)}-A_{(\r,1)},
\label{discrete_curl}\\
\Theta_{(\r,3)}&=&A_{(\r,1)}+A_{(\r+\i,2)}-A_{(\r+\j,1)}-A_{(\r,2)}
\nonumber
\end{eqnarray}
(see Figure \ref{plaquette}). The fact that the theory in the continuum
features a vector field \(\vec{A}\) is reflected in that the lattice
degrees of freedom reside {\it on links}, with commutation relations 
\begin{eqnarray}
[A_{(\r,\mu)}\ ,\ \Pi_{(\r',\nu)}]=\ i\ \delta_{\mu,\nu}\
\delta_{\r,{\r'}}.
\label{AP_comm}
\end{eqnarray}
Discretizing the theory spoils its space-time symmetries, but gauge
symmetries remain almost unchanged. The generators of gauge symmetries are
\begin{equation}
g_\r=\sum_{\nu=1}^{3}\ (\Pi_{(\r,\nu)}-\Pi_{(\r-\bm{e_\nu},\nu)}),\ \ \
\ \ \ \ \ \ \ [g_\r, H_{\sf LEM}]=0,
\end{equation}
and physical states are characterized as before as \(g_\r|\Psi\rangle=0, \
\forall \r\).

\begin{figure}[b]
\begin{center}
\includegraphics[width=0.55\columnwidth]{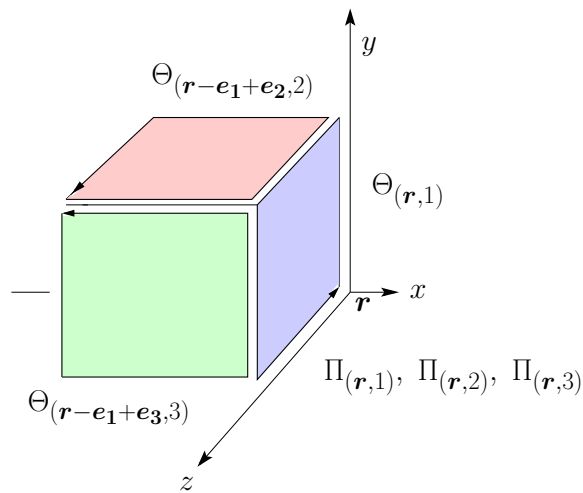}
\end{center}
\caption{The effect of the exchange duality $\Phi_{\sf d}$ of  Equation
\eqref{EM_duality_auto} on the three   $\Pi$ fields at site $\r$. The
directions $x, y, z$ are associated to the unit vectors  $\i,\j,\k$,
respectively. Notice that the $\Pi$ fields, although associated to  the
vertex $\r$,  reside on the links of the lattice.}
\label{pi_to_delta}
\end{figure}
\begin{figure}[h]
\begin{center}
\includegraphics[width=.6\columnwidth]{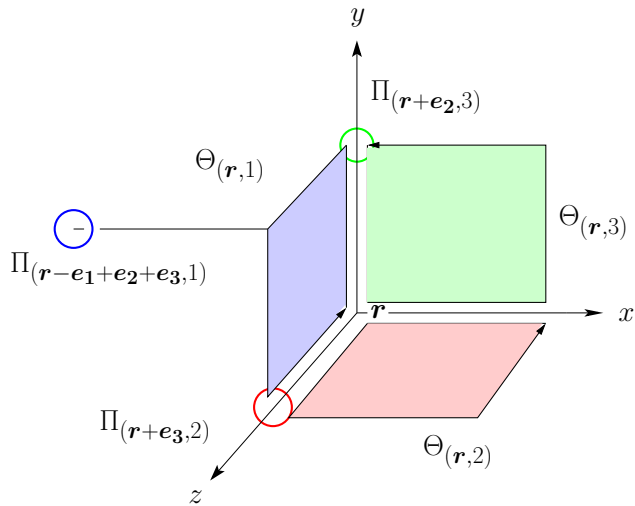}
\end{center}
\caption{The effect of the exchange duality $\Phi_{\sf d}$ of  Equation
\eqref{EM_duality_auto} on the three   $\Theta$ fields at site $\r$. The
directions $x, y, z$ are associated to the unit vectors  $\i,\j,\k$,
respectively. Notice that the $\Pi$ fields associated to  the indicated
vertices  reside on the links of the lattice, while the $\Theta$ fields 
reside on the corresponding plaquettes. Each colored plaquette map to
the $-\Pi$ at the correspondingly colored site.}
\label{delta_to_pi}
\end{figure}


\(H_{\sf LEM}\) is self-dual, just as its counterpart in the continuum.
The mapping 
\begin{eqnarray}
\label{EM_duality_auto}
\Pi_{(\r,1)}&\stackrel{\Phi_\d}{\longrightarrow}&\  \Theta_{(\r,1)},\ \
\ \ \ \ \ \ \ \ \ \ \ \ \ \ \ \  \ \Theta_{(\r,1)}
\stackrel{\Phi_\d}{\longrightarrow}\ -\Pi_{(\r-\i+\j+\k,1)},\nonumber\\
\Pi_{(\r,2)}&\stackrel{\Phi_\d}{\longrightarrow}& \
\Theta_{(\r-\i+\j,2)}, \ \ \ \ \ \ \ \ \ \ \, \Theta_{(\r,2)}
\stackrel{\Phi_\d}{\longrightarrow}\ -\Pi_{(\r+\k,2)},\nonumber\\
\Pi_{(\r,3)}& \stackrel{\Phi_\d}{\longrightarrow}& \
\Theta_{(\r-\i+\k,3)},\ \ \ \ \ \ \ \ \ \ \, \Theta_{(\r,3)}
\stackrel{\Phi_\d}{\longrightarrow}\ -\Pi_{(\r+\j,3)}.
\end{eqnarray}
defines a self-duality isomorphism {\it in the subspace of physical states},
that exchanges the lattice electric \(\vec{\Pi}_\r\) and magnetic \(\vec{\Theta}_\r\)
operators, see Figures \ref{pi_to_delta} and \ref{delta_to_pi}. 
\(\Phi_\d\) is not well defined outside the subspace of physical states,
because as it stands in Equation \eqref{EM_duality_auto}, it can be shown to be a 
many-valued mapping, \(\mathbb{1}\dual g_\r,\ \forall \r\). It is easy to check
that the lattice self-duality converges exactly to its counterpart in the continuum
in the naive limit \(a\rightarrow 0\).

It is remarkable that, in the end, the
EM duality has the same origin as the  self-duality of the
Ising model: a symmetry of a bond algebra.

\subsubsection{Compact QED and \(\Z_p\) gauge theories}

The self-dual lattice rendition of QED that we studied in the previous section
is non-standard. In
contrast, the standard Hamiltonian lattice field theory for QED
\cite{kogut,kogut_susskind} (see, for example, Section 6 of
\cite{kogut}),  
\begin{equation}
\label{lem}
H_{\sf CEM}=\sum_\r\sum_{\nu=1}^3\ \left (
\frac{1}{2}L_{(\r,\nu)}^2-\lambda \cos \Theta_{(\r,\nu)} \right ),
\end{equation} 
arises from the quantization \cite{kogut} of {\it compact} (lattice) QED as
defined by Wilson \cite{wilson}.  Here, the plaquette term 
$\Theta_{(\r,\nu)}$ is formally defined as in Equation \eqref{theta_fields} 
up to the replacement $A_{(\r,\nu)} \rightarrow \theta_{(\r,\nu)}$,
and the latter satisfy the commutation relations of Equation  
\eqref{CQEDcomr}.  

The Hamiltonians \(H_{\sf LEM}\) and \(H_{\sf CEM}\) exhibit radically
different phase diagrams, simply because \(H_{\sf LEM}\) describes a system of 
harmonic oscillators, while \(H_{\sf CEM}\) features plane rotors in interaction. 
In particular, \(H_{\sf CEM}\) {\it is not self-dual}. On the other hand, 
one can use the techniques of Sections \ref{sec4.3} and \ref{sec5.4} to set up
a dual gauge model in terms of integer valued degrees of freedom,
\begin{equation}
H_{\sf CEM}^D=\sum_\r\sum_{\nu=1}^3\ \left (
\frac{1}{2}b_{(\r,\nu)}^2-\frac{\lambda}{2}
(R_{(\r,\nu)}+R_{(\r,\nu)}^\dagger)\right )
\end{equation}
(the plaquette term 
$b_{(\r,\nu)}$ is formally defined as in Equation \eqref{theta_fields} 
up to the replacement $A_{(\r,\nu)} \rightarrow X_{(\r,\nu)}$, the operators 
\(X,\ R,\ R^\dagger\) where introduced in Section \ref{sec4.3}). A close classical 
relative of this duality to integer valued fields
was found of great use in the latest comprehensive 
Monte-Carlo simulation of \(H_{\sf CEM}\) \cite{panero}.

We can also use the  mathematics introduced in Sections \ref{sec4.1} and
\ref{sec4.3} to  write down a self-dual \(p\)-state approximation 
to \(H_{\sf CEM}\) 
\begin{equation}\label{N_state_gauge_H}
H_{\sf G}= \frac{1}{2}\sum_\r \sum_{\nu=1}^{3}\  (
V_{(\r,\nu)}\ +\  \lambda B_{(\r,\nu)}+\ {\sf h.c.}),
\end{equation}
that features generators \(V_{(\r,\nu)}\) and \(U_{(\r,\nu)}\) 
of Weyl's group at each link, and  
\begin{eqnarray}
 B_{(\r,1)}&=&U_{(\r,2)}U_{(\r+\j,3)}U^\dagger_{(\r+\k,2)}
U_{(\r,3)}^\dagger , \nonumber\\
 B_{(\r,2)}&=&U_{(\r,3)}U_{(\r+\k,1)}U_{(\r+\i,3)}^\dagger
U_{(\r,1)}^\dagger , \label{3d_Uplaquettes}\\
 B_{(\r,3)}&=&U_{(\r,1)}U_{(\r+\i,2)}U_{(\r+\j,1)}^\dagger
U_{(\r,2)}^\dagger. \nonumber
\end{eqnarray}
The self-duality mapping for this model was discussed in 
Reference \cite{con}, and is given by  
\begin{eqnarray}
V_{(\r,1)}&\dual&B_{(\r,1)},\ \ \ \ \ \ \ \ \
\ \ \ \ \ \ \ \ \  B_{(\r,1)}\dual
V^\dagger_{(\r-\i+\j+\k,1)},\nonumber\\
V_{(\r,2)}&\dual& \ B_{(\r-\i+\j,2)}, \ \
\ \ \ \ \ \ \ \, B_{(\r,2)}\dual V^\dagger_{(\r+\k,2)},\label{map_gauge_zp}\\
V_{(\r,3)}&\dual& \ B_{(\r-\i+\k,3)},\ \
\ \ \ \ \ \ \ \, B_{(\r,3)}\dual V^\dagger_{(\r+\j,3)}.\nonumber 
\end{eqnarray}
(and similarly for Hermitian conjugate bonds).  
Similar to the VP model
of Section \ref{sec4.1}, the squared duality isomorphism \(\Phi_\d^2\)
is a non-trivial, discrete charge conjugation symmetry \(\mathcal{C}\)
of the model, 
that maps \(V_{(\r,\nu)}\) and \(U_{(\r,\nu)}\) to
\(V_{(\r,\nu)}^\dagger\) and \(U_{(\r,\nu)}^\dagger\), up to a
translation by \(\bm{d}= -\i+\j+\k\). When  \(p\rightarrow\infty\), 
the phase structure of \(H_{\sf G}\) approaches that of \(H_{\sf LEM}\) 
\cite{frohlich_gauge}, even though \(H_{\sf LEM}\) is not self-dual 
(see Section \ref{sec4.3}). 

In this section, we introduced \(H_{\sf G}\) in connection to  QED.
Actually this model has been intensively studied  in the literature
\cite{yoneya,green,horn} in connection to confinement in QCD. This is so
as \(H_{\sf G}\) for \(p\) states affords a simple effective theory  to
study center vortices in QCD with \(p\) flavors. This was discussed, to
some extent, by 't Hooft \cite{thooft} where the relevance of center
vortices to confinement was first elucidated, but the specific Hamiltonian
\(H_{\sf G}\) was proposed in Reference \cite{horn}. {}From this point of view,
however, this lattice model suffers from its inability to incorporate
magnetic monopoles.  Other aspects of \(H_{\sf G}\) will be discussed in 
Section \ref{classical_zpgauge}.

\subsection{QED without a vector potential}
\label{sec6.3}

In classical physics, the EM vector potential  $A_\mu$ 
is a technical advantage, but otherwise expendable, essentially because
the interaction of classical charged particles with the EM field can be
described purely in terms of \(\vec{E}\) and 
\(\vec{B}\). In contrast, the vector potential is unavoidable
at the quantum  level, and 
the best illustrations of this fact is the Aharonov-Bohm effect
\cite{aharonov_bohm} that is non-local in \(\vec{E}\) and 
\(\vec{B}\). But even disregarding the interaction to
charges, it is difficult to set up the quantum mechanics of the EM field 
alone without introducing a vector potential, though Mandelstam
\cite{mandelstam} managed to put forward a consistent scheme.
The standard quantizations of the EM field, however, rely on 
the unobservable $A_\mu$, and suffer from well-known difficulties \cite{das},
that depend on the gauge fixing condition of choice. In the axial gauge
of the previous section, the state space that emerges is a Hilbert space, 
but it is redundant, due to the presence of gauge symmetries inherited
upon quantization. 

On the other hand, as explained in Section \ref{sec3.12}, we can
use bond-algebraic dualities to find a dual representation that features
no gauge redundancies.
In this section, we illustrate these ideas for compact QED, because it is
the model of greatest relevance in numerical studies of QED.

Ideally, we would like to find a gauge reducing duality for the
Hamiltonian \(H_{\sf LEM}\) of Equation \eqref{lem} to a model
that features local bonds. We saw in Section \ref{sec5.4} 
that this is possible in \(d=2\), but it does not seem to be 
possible in \(d=3\). Hence we are left with the systematic approach
of Section \ref{sec3.12.4}, that is known to introduce non-local bonds
in the dual model. The starting point is to recognize the generators
of gauge symmetries, in this case
\begin{equation}\label{gtbk}
g_\r=\sum_{\nu=1}^3\ (L_{(\r,\nu)}-L_{(\r-\bm{e_\nu},\nu)}).
\end{equation}
Now we can proceed with the general techniques described in 
Section \ref{sec3.12.4}. The gauge-reducing duality should satisfy
\begin{equation}
\Phi_\d(g_\r)=0=\sum_{\nu=1}^3\ (\Phi_\d(L_{(\r,\nu)})-
\Phi_\d(L_{(\r-\bm{e_\nu},\nu)})),
\end{equation}
so that in the completely gauge-reduced dual the bonds \(\Phi_\d(L_{(\r,3)})\)
{\it are not independent}, but we can write instead
\( 
\Phi_\d(L_{(\r,3)})=\ \sum_{n\geq1}\
(\Phi_\d(L_{(\r+n\k,1)})+\Phi_\d(L_{(\r+n\k,2)}) -\Phi_\d(L_{(\r+n\k-\i,1)})
-\Phi_\d(L_{(\r+n\k-\j,2)}))
\). {}From this point on, the reasoning follows through just as in Section 
\ref{sec3.12.4}. The gauge-reducing duality homomorphism  reads
\begin{eqnarray}
&&L_{(\r,1)}\dual L_{(\r,1)}, \ \ \ L_{(\r,2)}\dual L_{(\r,2)},\ \ \ 
L_{(\r,3)}\dual {\cal E}_{(\r,3)}, \nonumber\\
&&e^{i\Theta_{(\r,1)}}\dual e^{i(\theta_{(\r+\k,2)}-\theta_{(\r,2)})},\ \
\  e^{i\Theta_{(\r,2)}}\dual
 e^{i(\theta_{(\r+\k,1)}-\theta_{(\r,1)})},\\
&&e^{i\Theta_{(\r,3)}}\dual e^{i\Theta_{(\r,3)}}\nonumber
\end{eqnarray}
(the plaquette angle \(\Theta_{(\r,\nu)}\) was defined right 
after Equation \eqref{lem}). The electric string operator \({\cal E}_{(\r,3)}\) is defined as
\begin{equation}
{\cal E}_{(\r,3)} \equiv
\sum_{n\geq1} (L_{(\r+n\k,1)}+L_{(\r+n\k,2)} -
L_{(\r+n\k-\i,1)}-L_{(\r+n\k-\j,2)}),
\end{equation}
and carries the full weight of the non-locality that seems to be unavoidable
in \(d=3\), if gauge constraints are to be eliminated. The completely gauge-reduced
dual Hamiltonian reads
\begin{eqnarray}
&&H_{\sf GRCEM}=H^0+\sum_{\r}\ \left({\cal E}_{(\r,3)}-\lambda \cos\Theta_{(\r,3)}
\right),\\
&&H^0=\sum_\r \sum_{\nu=1,2}\ \left(\frac{1}{2}L_{(\r,\nu)}-\lambda
\cos(\theta_{(\r+\k,\nu)}-\theta_{(\r,\nu)})\right).\nonumber
\end{eqnarray}
It is remarkable that \(H^0\) describes a stalk of non-interacting (independent) 
\(d=1\), XY models. The idea that the physics of gauge fields in \(D=4\) is 
closely connected to the physics of spin models in \(D=2\) has been put
forward many times over the years (see \cite{fradkin_susskind}, and references 
therein). The duality just presented is a new indication/realization of this
connection. Moreover, it afford us  {\it a theory of QED
without  a vector potential}, and consequently, without gauge symmetries
or unwanted non-physical states. Stated differently, \(H_{\sf GREM}\) is
the quantum rendition  of Maxwell's dynamics purely in terms of the
electric and magnetic fields.  The cost to be paid is the introduction
of non-local bonds. 

{}From a practical point of view, it is important to notice that this 
gauge-reducing duality can  be
restricted to finite systems, and the advantage for numerical
simulations of the gauge reduced  rendition of the theory  could be
enormous. For one thing, since there are no degrees of freedom associated with the 
links along the \(\k\) direction, the Hilbert space of \(H_{\sf
GRCEM}\) reduces to   
\begin{equation}
\mathcal{H}_{\sf GRCEM}=\bigotimes_\r\ \left({\cal L}^2(S^1)_{(\r,1)}\otimes 
{\cal L}^2(S^1)_{(\r,2)}\right).
\end{equation}
that is much smaller than the one for \(H_{\sf LEM}\) (\({\cal L}^2(S^1)\) denotes
the Hilbert space of square integrable functions on the circle). Physically speaking,
\(\mathcal{H}_{\sf GRCEM}\) exhibits two ``states of polarization" per lattice site.

\subsection{Abelian and $\mathbb{Z}_p$ Higgs models}
\label{sec6.4}

In this section  we study the duality properties of the Abelian Higgs (AH)
QFT {\it in its broken symmetry state}. 
The AH model is complex 
enough that an operator treatment in the continuum is not well
defined, so we proceed directly to the lattice Hamiltonian formalism.
We uncover a new duality in \(d=3\) to a {\it local}, completely 
gauge-reduced model, introduce new \(p>2\)-state approximations to the lattice
AH model that we call \(\mathbb{Z}_p\) Higgs models, 
and discuss their (self-)dual properties in \(d=2\) and \(d=3\) 
(the \(p=2\) case \cite{fradkin_susskind} 
is of importance to the theory of topological quantum order and storage of quantum
information, see Section \ref{sec5.5}). 
We start with some general comments to put the AH model in perspective.

Both in condensed matter and high energy physics,  the success of
QFTs in describing interactions
hinges to a large extent on the principle of
gauge invariance and the Higgs  mechanism. The reason for the 
latter is that, if the gauge group is compact, gauge invariance requires
gauge fields to be  {\it massless}, restricting 
in principle their applicability to the
description of  {\it long-range} interactions. The Higgs mechanism affords
a way out of this restriction, since it is a
process  by which gauge fields acquire mass through the spontaneous
breakdown  of a continuous symmetry with no Goldstone bosons. In this way,
gauge fields become capable of describing {\it short-range} interactions as 
well, at the expense of introducing a {\it Higgs field}. 

The AH model,  
\begin{equation}\label{AH}
S_{\sf AH}= \int d^4x\ \left ( -\frac{1}{4}F_{\mu \nu}F^{\mu \nu}+
\frac{1}{2}(D_\mu\phi)(D^\mu\phi)^*+
\lambda(\phi\phi^*-v^2)^2 \right),
\end{equation} 
features a complex scalar Higgs field $\phi$ of charge \(q\), in
interaction with the EM field ($F_{\mu\nu}=\partial_\mu A_\nu - \partial_\nu A_\mu$ 
is the EM field, and  \(D_\mu\phi\equiv(\partial_\mu-iqA_\mu)\phi\)), and it 
is the simplest field theory that
combines both the principle of gauge invariance and  the Higgs
mechanism. When $v^2$ is positive,
the (classical) potential energy is minimized by setting  $\phi = v e^{i
\theta}$, and the ground state  breaks
(spontaneously) the \(U(1)\) symmetry. The reader familiar with 
superconductivity will recognize the resulting action
as the starting point for the phenomenological
Ginzburg-Landau theory of superconductivity (wherein \(\phi\) represents
the superconducting order parameter). In the light of QFT, however, the
spontaneous breakdown translates into
a particle spectrum containing one {\it massive photon}
and one {\it massive, real} scalar  (see, for instance, \cite{das}).
This is the  Higgs mechanism of mass generation: the would be Goldstone
boson associated to the spontaneous symmetry breakdown is reabsorbed as
an extra degree of freedom of the gauge field $A_\mu$. It is {\it the} mechanism
of mass generation in the standard model of particle physics, but it
is not the only possible one, as we will see in the next section when we
study the St\"uckelberg model. 

The lattice Hamiltonian for the AH model in its broken symmetry phase reads
\begin{eqnarray}
H_{\sf LAH}=&&\sum_\r\sum_{\nu=1}^3 (\ \frac{1}{2}L_{(\r,\nu)}^2
- \lambda\ \cos(\Theta_{(\r,\nu)}))  \nonumber  \\ 
+&& \sum_\r (\ \frac{1}{2}L_{\r}^2-\sum_{\nu=1}^3\ 
\kappa\ \cos(\theta_{\r+\bm{e_\nu}}-\theta_\r-q\theta_{(\r,\nu)})), 
\label{hlah}
\end{eqnarray}
with the notation of Section \ref{sec6.3} for $L_{(\r,\nu)}$ and
$\Theta_{(\r,\nu)}$. The Higgs, or matter, field is represented
by degrees of freedom $L_\r$ and
$\theta_{\r}$ on the sites of a cubic lattice, while the gauge
field is represented by degrees of freedom $\theta_{(\r, \nu)}$,
$L_{(\r,\nu)}$ on its links. \(H_{\sf LAH}\) follows from the classical
lattice action introduced in Reference \cite{fradkin_shenker}, treated
according to the quantization techniques of Reference \cite{kogut_susskind}.
Notice that \(q\) is now constrained to take integer values.
As discussed in Reference
\cite{fradkin_shenker}, the physics (phase diagram) of the
AH model depends strongly upon whether \(q=1\) or has some other value, and
in fact \(q\) is an explicit parameter in the duality mappings to be discussed.

\(H_{\sf LAH}\) defines a gauge theory, with gauge symmetries generated 
by 
\begin{equation}
g_\r = -qL_\r +\sum_{\nu=1}^3\ (L_{(\r,\nu)}-L_{(\r-\bm{e_\nu},\nu)}).
\end{equation} 
It is a remarkable feature that these gauge symmetries are completely
eliminated by a duality to a {\it local} model,  
\begin{eqnarray}
H_{\sf GRLAH}= &&\frac{1}{2} \sum_\r \sum_{\nu=1}^3
(X_{(\r,\nu)}^2-
 \lambda\ (B_{(\r,\nu)}+B_{(\r,\nu)}^\dagger)) \nonumber \\
+&&\frac{1}{2} \sum_\r\ (\frac{1}{q^2}A_\r^2-
\sum_{\nu=1}^3 \kappa\ q(R_{(\r,\nu)}+R_{(\r,\nu)}^\dagger)) ,
\end{eqnarray}
that features integer-valued degrees of freedom and local bonds. 
The plaquette operators \(B_{(\r,\nu)}\) are defined just as 
in \eqref{3d_Uplaquettes}, up to the replacement \(U_{(\r,\nu)}
\rightarrow R_{(\r,\nu)}\), and \(
A_\r=\sum_{\nu=1}^3\ (X_{(\r,\nu)}-X_{(\r-\bm{e_\nu},\nu)})\)
(the operators \(R\), \(R^\dagger\) and \(X\) were  introduced in
Section \ref{sec5.2}). It follows that \(H_{\sf GRLAH}\) does not 
feature any degrees of freedom on the {\it sites} of the lattice. 
The duality homomorphism reads
\begin{eqnarray}
L_\r\ &\stackrel{\Phi_\d}{\longrightarrow}&\ \frac{1}{q}A_{\r},
\ \ \ \ \ \ \ \ \ \ \ \ 
e^{i(\theta_{\r+\bm{e_\nu}}-\theta_\r-q\theta_{(\r,\nu)})}
\ \stackrel{\Phi_\d}{\longrightarrow}\ qR_{(\r,\nu)}, \nonumber\\
L_{(\r,\nu)}\ &\stackrel{\Phi_\d}{\longrightarrow}&\ X_{(\r,\nu)},
\ \ \ \ \ \ \ \ \ \ \ \ \ \ \ \ \ \ \ \,
e^{-i\Theta_{(\r,\nu)}}\ \stackrel{\Phi_\d}{\longrightarrow}\ 
B_{(\r,\nu)},
\end{eqnarray}
and it is straightforward to check the trivialization of the 
infinitesimal generators of gauge symmetries,
\begin{equation}
\Phi_\d(g_\r)=-A_\r+\sum_{\nu=1}^3\ (\Phi_\d(L_{(\r,\nu)})
-\Phi_\d(L_{(\r-\bm{e_\nu},\nu)})=0.
\end{equation}
\(H_{\sf GRLAH}\) has no  local symmetries. Let us point out
without elaborating the details that the
AH model admits a {\it local gauge-reducing duality} along these
lines {\it in any dimension \(d\)}.

We can take a different approach to the study of the AH model,
and write \(p\)-state approximations to the
Hamiltonian of Equation \eqref{hlah}, in terms of generators
of the Weyl group algebra introduced in Sections \ref{sec4.1} (this is 
similar in spirit to the approximation of the \(d=1\) XY model by a 
VP model). We call these $p$-state 
quantum models $\mathbb{Z}_p$  Higgs models, and to our knowledge they
have not been studied before, for \(p\geq3\) 
(the $p=2$ case  \cite{fradkin_shenker,horn_yankielowicz,lamont}
was discussed in Section \ref{sec5.5} from the perspective of 
topological quantum order).
The $\mathbb{Z}_p$  Higgs models admit completely gauge-reducing 
dualities, and the dual models that arise are natural generalizations
of the ETC model of Section  \ref{sec5.5}, 
to any number of states $p$ and 
any number of dimensions \(d\). Moreover, in \(d=2\), they define a 
{\it new class of self-dual models}. Let us focus on this case for
the rest of the section. 

The Hamiltonian for the \(d=2\)-dimensional $\mathbb{Z}_p$  Higgs model 
reads
\begin{eqnarray}\label{AH_N_H}
H_{\sf pAH}&=&\frac{1}{2} \sum_\r\  ( V_\r-\lambda B_{(\r,3)}\ +\ {\sf
h.c.}) \\
&+&\frac{1}{2} \sum_\r\sum_{\nu=1,2}\  (V_{(\r,\nu)}- \kappa\
U^\dagger_{\r}(U^\dagger_{(\r,\nu)})^qU_{\bm{r+e_\nu}}+\ {\sf
h.c.}),\nonumber
\end{eqnarray}
with gauge symmetries
\begin{equation}
G_\r=(V^\dagger_\r)^q V_{(\r,1)} V_{(\r,2)}V^\dagger_{(\r-\i,1)}
V^\dagger_{(\r-\j,2)},\ \ \ \ \ \ [G_\r,H_{\sf G}]=0,
\end{equation}
and \(G^\dagger_\r\).
As usual, $U_{\r}, V_{\r}$ denote site (vertex) operators, and
$U_{(\r, \nu)}, V_{(\r, \nu)}$ reside at the link $(\r,\nu)$.
Notice also that the charge \(q\) is a $\mathbb{Z}_p$ charge now,
and can only take one of the values \(q=0,\cdots,p-1\). The self-duality
mapping reads
\begin{eqnarray}
V_\r &\dual& B_{(\r,3)}, \ \ \ \ \ \ \ \ \ \ \ \ B_{(\r,3)}\dual V_{\r+\i+\j},
\label{zphiggs_sd}\\ 
U^\dagger_\r (U^\dagger_{(\r,1)})^q U_{\r+\i}&\dual& V_{(\r+\i,2)},\ \ \ \ \ \ \ \ \
V_{(\r,1)}\dual U_{\r+\i}U^q_{(\r+\i,2)}U^\dagger_{\r+\i+\j}
,\nonumber\\
U^\dagger_\r (U^\dagger_{(\r,2)})^q U_{\r+\j}&\dual&\ V^\dagger_{(\r+\j,1)},\ \ \ \ \ \ \ \
V_{(\r,2)}\dual U^\dagger_{\r+\j}(U^\dagger_{(\r+\j,1)})^qU_{\r+\i+\j}. \nonumber
\end{eqnarray}
Strictly speaking, \(\Phi_\d\) is an isomorphism only on the restriction
of the bond algebra to the subspace of gauge-invariant states, so this
is another example of an {\it emergent self-duality} (we encountered
a similar situation in \(d=3\) QED, Section \ref{sec6.2}). To see this,
notice that one can expand the identity operator \(\mathbb{1}\) as a product of
bonds in many different ways, and that all these different expansions are mapped to
products of gauge symmetries. It follows that \(\Phi_\d\) is a {\it multivalued} 
homomorphism, unless it is restricted to the subspace of gauge-invariant states.

Since \(H_{\sf pAH}\) features both {\it matter and gauge} 
fields in interaction, it is interesting to compare the  self-duality 
of Equation \eqref{zphiggs_sd} to the self-duality properties {\it expected}
of QED in the presence of suitable sources. 
In general, it is argued that by introducing
magnetic charges, the self-duality of QED
in the absence of sources, Section \ref{sec6.2},
could be extended to include sources as well. This putative
self-duality, however, would not mix matter fields with gauge fields. In
contrast, in the Higgs case,  the self-duality establishes an
equivalence between matter and gauge fields.

The \(p\)-state models $H_{\sf pAH}$ can be completely gauge-reduced 
by dualities that are very similar to that for \(p=2\), worked
out in Section \ref{sec5.5}. There is also a variation of the 
\(\Z_p\) Higgs models where the VP-like interactions are 
replaced by P-like interactions \cite{kogut_potts}. This Potts-Higgs model
has the advantage that its 
phase diagram can be studied analytically in a  \(1/p\) expansion.

\subsection{The self-dual St\"uckelberg model}\label{sec_stuckel}
In this section we discuss the  St\"uckelberg model of mass
generation, and show that it is self-dual in \(d=2\) dimensions.

The massless free boson of Section \ref{sec6.1}
\begin{equation}
S_{\sf FB}=\int d^Dx\ \frac{1}{2}\eta^{\mu \nu}
\partial_\mu\phi\partial_\nu\phi
\end{equation}  
(now in any dimension \(D=d+1\)) has a global, internal continuous symmetry 
of the form \(
\phi(\x) \mapsto \phi(\x)+\alpha\),
\(\alpha\in\mathbb{R}\), but this fact is rarely of interest because
the conservation law that follows is tantamount to the equation of motion.
Things become more interesting if apply the gauge principle
to this symmetry, and gauge it to make it local, at the expense of introducing
a vector potential \(A_\mu\),
\begin{equation}\label{s_action}
S_{\sf S}=\int d^Dx\ \frac{1}{2}\eta^{\mu \nu}
(\partial_\mu\phi-mA_\mu)(\partial_\nu\phi-mA_\nu)-\frac{1}{4}F^{\mu \nu}F_{\mu \nu}.
\end{equation}
This is the {\it St\"uckelberg model of mass generation} (see \cite{das}, and 
references therein), proposed by St\"uckelberg in 1958. 
The gauge symmetries of \(S_{\sf S}\) are
\begin{equation}
\phi(\x)\ \mapsto\ \phi(\x)+\alpha(\x),\ \ \ \ 
A_\mu(\x)\ \mapsto\ A_\mu(\x)+\frac{1}{m}\partial_\mu\alpha(\x),
\end{equation}
and show that {\it both} the \(\phi\) and \(A_\mu\) field are {\it not
observable}. Combined, however, they describe {\it in a gauge invariant fashion}
a {\it massive} vector field (a Proca
field) of bare mass \(m\) (to see this, just impose the 
gauge-fixing condition \(\phi={\sf constant}\) in Equation \eqref{s_action}). 
Since the Proca field 
can be completely quantized in operator form \cite{takahashi}, we expect
that the same holds for \(S_{\sf S}\), and  
we work directly in the continuum. Still, let us point out that analogous
calculations in the lattice Hamiltonian approach 
return results that are perfectly compatible with the continuum, 
in the naive continuum limit $a \rightarrow 0$. 

The St\"uckelberg
model is self-dual only in \(d=2\), as will become clear soon, 
so from now on we work out just that case.
The canonical quantization of \(S_{\sf S}\) has the usual complications 
coming from gauge-invariance. The simplest way to proceed is to partially
fix the gauge by imposing the axial constraint \(A_0=0\). This allows to put 
\(S_{\sf S}\) in canonical form, so that we can apply the standard quantization 
procedures to get
\begin{eqnarray}
&&H_{\sf S}=\frac{1}{2}\int d^2x\left ( \vec{\Pi}^2+\ 
\left(\partial_1 A_2-\partial_2 A_1\right)^2+\ \pi^2+\ \left(
\nabla\phi-m\vec{A}\right)^2 \right ),\\
&&[\phi(\x),\ \pi(\x')]=\ i\ \delta(\x-\x'),\ \ \ \ \ \ 
[A_\mu(\x),\ \Pi_\nu(\x')]=\ i\ \delta_{\mu,\nu}\ \delta(\x-\x'),\nonumber
\end{eqnarray}
with \(\mu,\nu=1,2\), and every other commutator set equal to zero.
The subspace of physical, gauge-invariant states is characterized by
\begin{equation}
(-\frac{1}{m}\nabla\cdot\vec{\Pi}+\pi)|\Psi\rangle=0.
\end{equation}
That \(H_{\sf S}\) is self-dual follows from the mapping
\begin{eqnarray}
\pi(\x)&\dual& \left(\partial_1 A_2-\partial_2 A_1\right)(\x),\ \  
\left(\partial_1 A_2-\partial_2 A_1\right)(\x)\dual \pi(\x), \nonumber
\\
\Pi_1(\x)&\dual& -(\partial_2\phi-mA_2)(\x),\ \ \ \ \  
(\partial_2\phi-mA_2)(\x)\dual -\Pi_1(\x),
\\
\Pi_2(\x)&\dual& (\partial_1\phi-mA_1)(\x),\ \ \ \ \ \ \ 
(\partial_1\phi-mA_1)(\x)\dual \Pi_2(\x).\nonumber
\end{eqnarray}
Since we can write \(0=-\partial_1(\partial_2\phi-mA_2)+\partial_2
(\partial_1\phi-mA_1)-m(\partial_1A_2-\partial_2A_1)\) (among other 
possibilities), we have that  
\begin{eqnarray}
\Phi_\d(0)=\nabla\cdot\vec{\Pi}-m\pi.
\end{eqnarray}
This means that just as in QED, Section \ref{noncompactQED}, 
\(\Phi_\d\) represents a self-duality isomorphism only
in the sector of gauge-invariant states, and is multi-valued
in the full, gauge-redundant state space.

This completes our discussion of the St\"uckelberg model, but let us 
point out in closing that all of the previous results follows just as easily
in the lattice Hamiltonian approach. 

\subsection{Field theory and dimensional reduction}
\label{6.8}

There is an intimate connection between $d$-dimensional systems
possessing   $\bar{d}$-dimensional gauge-like symmetries \cite{tqo},
and the phenomenon of {\it dimensional reduction}, where the physical 
system in $d$ dimensions behaves
in many ways as if it had effectively a smaller $\bar{d}<
d$ number of dimensions \cite{tqo,3detc}. Mathematically,  this
connection results  from establishing bounds for the correlation
functions of the $d$-dimensional theory in terms of another theory in
$\bar{d}$  dimensions.  A very broad and exciting field where
dimensional reductions, $\bar{d}$-dimensional gauge-like symmetries, 
and dualities come to the fore is that of topological quantum order 
\cite{tqo,3detc}. In topologically ordered systems,  the state of the
system cannot be characterized by local measurements but rather by
topological quantities. 

In this paper, we have considered the duality properties of several 
lattice models that display topological quantum order, including the XM and 
POC models of Sections \ref{sec4.4} and \ref{sec5.1} \cite{tqo},
and the paradigmatic ETC model of Section \ref{sec5.5}.
Here we develop continuum (field-theoretic)
versions of those lattice models, where dimensional reduction occurs
because of the existence of  $\bar{d}$-dimensional gauge-like
symmetries. Consider then the non-relativistic, $d=2$-dimensional QFT 
\begin{equation}\label{bosonicCXM}
H_{\sf P}=\frac{1}{2}\int d^2x\ ( \pi^2+ \lambda 
\left(\partial_1\partial_2\phi\right)^2),\ \ \ \ \ \ 
[\phi(\x),\ \pi(\x')]=i\delta(\x-\x').
\end{equation}
By construction, this model is invariant under the \(\bar{d}=1\) gauge-like symmetry
\(\phi(x)\rightarrow \phi(x)+\alpha(x^1)+\beta(x^2)\), where \(\alpha,\ \beta\) are
smooth, real functions of {\it one} variable.
Also, it is self-dual, as follows from the mapping
\begin{eqnarray}\label{dual_bosonicXM}
\!\!\!\!\!\!\!\!\!\!\!\!\!\!
\phi(\x)\dual -\int_{-\infty}^{x^1}\int_{x^2}^{\infty} d^2x'\ \ \pi(\x'), 
\ \ \ \ \ \ \
\pi(\x)\dual -\partial_1\partial_2\phi(\x)\ ,
\end{eqnarray}
that defines the dual variables of the problem. Notice that formally,
\(g_i(\alpha)\dual 0\). This is the standard manifestation (seen many times
in formally infinite lattice models, see the discussion in Section \ref{sec3.7})
of the  fact that to make \(\Phi_\d\) rigorous we need to specify 
self-dual BCs.

In what follows we are going to study these results 
in some detail on the lattice, to showcase the connection of 
the model of Equation \eqref{bosonicCXM} to other lattice 
models with gauge-like symmetries. The lattice Hamiltonian 
approach applied to Equation \eqref{bosonicCXM} returns
\begin{equation}\label{bosonicXM}
H_{\sf PL}=\frac{1}{2a^2}\sum_\r\ (\pi^{2}_\r +\ \lambda (\square
\phi_\r)^2 ), \ \ \ \ \ \ [\phi_\r,\ \pi_{\r'}]=i\delta_{\r,\r'},
\end{equation}
with $\square \phi_\r \equiv \phi_\r-\phi_{\bm{r+e_2}}
+\phi_{\bm{r-e_1+e_2}}-\phi_{\bm{r-e_1}}$, that is self-dual by virtue of 
\begin{equation}
\pi_\r\dual-\square\phi_{\bm{r-e_1+e_2}}, \ \ \ \ \ \ 
\square \x_\r\dual \pi_{\bm{r-e_1+e_2}}. 
\end{equation}
The structure of \(H_{\sf PL}\) suggests setting up a \(p\)-state model
of the form
\begin{equation}
H_{\sf pPL}=\frac{1}{2}\sum_\r\ \left (V_\r+\ \lambda\square U_\r
+\ {\sf h.c.} \right),
\end{equation}
with $\square U_\r=U_\r
U^\dagger_{\bm{r+e_2}}U_{\bm{r-e_1+e_2}}U^\dagger_{\bm{r-e_1}}$ 
(the \(U\) and \(V\) operators where introduced in Section \ref{sec4.4}).
For \(p\geq3\), \(H_{\sf pPL}\) defines a {\it class 
of self-dual models that has not been studied before} to the best of
our knowledge. For \(p=2\), \(H_{\sf pPL}\) becomes
identical to \(H_{\sf XM}\), the XM model of Section \ref{sec4.4}.
The connection between \(H_{\sf P}\), \(H_{\sf PL}\), and 
\(H_{\sf pPL}\) stands on their common self-dual structure  {\it and the
shared  presence of \(\bar{d}=1\) gauge-like symmetries} (see Section
\ref{sec4.4} for a discussion of the gauge-like symmetries of the XM
model). Thus these models
afford an excellent scenario to study
the role of dimensional reduction and topological quantum order in 
more general settings, where the structure of the elementary degrees
of freedom are varied in a controlled fashion.

Since \(H_{\sf PL}\) shares some formal similarities with the XM model,
we expect it to show a duality to a model analogous to the POC
model of Section \ref{sec5.2}. The dual model turns out to be
\begin{equation}\label{BPOC}
H_{\sf PL}^D=\frac{1}{2a^2}\sum_{\r}\ \left((\pi_{\r+\i}-\pi_\r)^2+\lambda
(\phi_{\r+\j}-\phi_\r)^2\right),
\end{equation} 
as follows from the mapping
\begin{equation}
\label{bpoc2bxm}
\pi_{\r+\i}-\pi_\r\dual -\square \phi_{\bm{r+e_1}}, 
\ \ \ \ \ \ \ \ \ \ 
\phi_{\r+\j}-\phi_\r\dual \pi_{\r+\j}.
\end{equation}
Interestingly, it is easy to take the naive continuum limit \(a\rightarrow 0\) 
of \(H_{\sf PL}^D\), 
\begin{equation}
H_{\sf P}^D= \frac{1}{2}\int d^2x\ ((\partial_1 \pi)^2+\lambda
(\partial_2 \phi)^2),\ \ \ \ \ \ [\phi(\x),\ \pi(\x')]=i\delta(\x-\x').
\end{equation}
That \(H_{\sf P}^D\) is indeed dual to the QFT defined in Equation
\eqref{bosonicCXM} follows from the isomorphism
\begin{equation}
\partial_1 \pi(\x)\dual\partial_1\partial_2\phi(\x),
\ \ \ \ \ \ \ \ \ \ \  
\partial_2\phi(\x)\dual\pi(\x).
\end{equation}

\section{Bond-algebraic approach to classical dualities}
\label{sec8}

In this section we establish dualities for models of 
classical statistical mechanics, exploiting the theory of 
Section \ref{classical&quantum}, and introduce a new 
classical gauge \(\mathbb{Z}_p\) model that is self-dual for any \(p\). 
The aim is to illustrate the unification of the theory of dualities
in the framework of bond algebras, and the advantages of approaching classical
dualities from this new perspective. On the other hand, it is important to
keep in sight the related fact that quantum dualities are fundamentally related to 
classical ones by  path integrals or the
the STL decomposition (see Section \ref{classical&quantum}). 
These widely used techniques approximates the exponential of a sum of 
(non-commuting) operators by products of exponentials
\begin{eqnarray}
e^{-(H_{1} + H_{2})/N} =  e^{-H_{1}/N} e^{-H_{2}/N} 
+ O(1/N^2),
\label{explain_suzuki}
\end{eqnarray}
with an error assumed to be bounded \cite{schulman}, and, combined
with bond-algebraic techniques, they can afford a simple fast way
to connect quantum dualities to classical dualities \cite{con,bookEPTCP}.
In the following, however, we will bypass the use of the STL 
decomposition technique, to obtain exact duality mappings 
between finite or infinite models of classical statistical physics.  

\subsection{The classical Ising model in the Utiyama lattice}
\label{sec8.2}

\begin{figure}[h]
\begin{center}
\includegraphics[width=0.40\columnwidth]{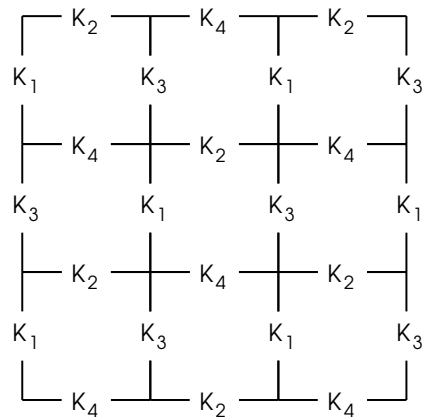}
\end{center}
\caption{Utiyama's version of the $D=2$-dimensional (classical)
Ising model features four different coupling constants \(K_1,\cdots,K_4\)
distributed in checkerboard fashion. Sometimes this arrangement
is referred to as the Ising model in the Utiyama lattice.}
\label{2Dutiyama}
\end{figure}

As mentioned in Section \ref{sec2.1}, one of the most celebrated 
self-dualities discovered long ago by Kramers and Wannier 
enabled a quantitative prediction \cite{KW} of the 
critical temperature of the classical Ising model in a square lattice.
{}From a bond-algebraic perspective this self duality is a consequence of 
the self-duality of the quantum Ising chain of Section \ref{sec3.7}. 
Since this calculation is already
available in textbook form  \cite{bookEPTCP}, we 
present in this section the most general case of a duality for the 
classical Ising model 
in the Utiyama lattice  \cite{utiyama}  defined in Figure \ref{2Dutiyama}. 

The partition function that describes Utiyama's anisotropic, bipartite
$D=2$ classical Ising model 
\begin{eqnarray}
&&\ \ \ \pf_{\sf U}(K_1,K_2,K_3,K_4)=\\
&&\sum_{\{\sigma_\r\}}\ \exp \left[\sum_{\r\in {\sf even}}\ (K_4\sigma_{\r+\i}\sigma_\r+
K_1\sigma_{\r+\j}\sigma_\r)+
 \sum_{\r\in {\sf odd}}\ (K_2\sigma_{\r+\i}\sigma_\r+
K_3\sigma_{\r+\j}\sigma_\r)\right] \ ,\nonumber
\end{eqnarray}
features four different nearest-neighbor couplings, $K_\mu$, arranged as
shown in Figure \ref{2Dutiyama} (a point \(\r\) is {\sf even} 
if  \(r^1+r^2={\sf even}\), and {\sf odd} otherwise).
The advantage in studying \(\mathcal{Z}_{\sf U}(K_1,K_2,K_3,K_4)\) is that
it describes in a unified fashion several important renditions of the \(D=2\)
Ising model. In particular,
\begin{itemize}
\item 
$K_1=K_3$ and $K_2=K_4$  corresponds to the  Ising model on a square lattice,
\begin{equation}\label{isquare}
\pf_{\sf I}(K_1,K_2)=\pf_{\sf U}(K_1,K_2,K_1,K_2).
\end{equation}
\item 
$K_4 \rightarrow \infty$ corresponds to the Ising model on a triangular lattice,
\begin{equation}\label{itr}
\pf_{\sf IT}(K_1,K_2,K_3)=\lim_{K_4 \rightarrow \infty}\pf_{\sf U}(K_1,K_2,K_3,K_4).
\end{equation}
\item 
$K_3 \rightarrow 0$ corresponds to the Ising model on an hexagonal lattice,
\begin{equation}\label{ihex}
\pf_{\sf IH}(K_1,K_2,K_4)=\lim_{K_3 \rightarrow 0}\pf_{\sf U}(K_1,K_2,K_3,K_4).
\end{equation}
\end{itemize}
The Ising model
on a triangular lattice is dual to the Ising model on an hexagonal lattice,
and the Ising model on a square lattice is self-dual. As will be shown next, both
of these results can be determined at once from a bond-algebraic analysis
of Utiyama's partition function. 

We start by recasting
\(\pf_{\sf U}(K_1,K_2,K_3,K_4)\) for a $2M\times 2N$ lattice in terms
of a transfer matrix, $T$, to later relate it to a quantum Hamiltonian
problem. 
The transfer matrix elements, most convenient to our purposes, read 
\begin{eqnarray}\label{transfer_uti}
&& \langle \sigma'|T(K_1,K_2,K_3,K_4)|\sigma\rangle=\exp\Big [K_2\sigma_{2M}^{\;}+
K_4\sigma_{2M-1}\sigma_{2M}\\
&+&\sum_{i=1}^{M-1} \left(K_4\sigma_{2i-1}\sigma_{2i}+K_2\sigma_{2i}\sigma_{2i+1} \right )
+\sum_{i=1}^{M} \left (K_1\sigma_{2i-1}'\sigma_{2i-1}+
K_3\sigma_{2i}'\sigma_{2i}\right) \Big ],\nonumber
\end{eqnarray}
where \(2M\) is the number of sites along the horizontal direction subject to
open BCs. The Ising spin variable $\sigma_j=\pm 1$ 
belongs to the horizontal row, while \(\sigma_{j}'\) 
denotes the spin immediately above \(\sigma_{j}\) (that
is, on the next horizontal line). Notice also that we have introduced a boundary 
term \(K_2\sigma_{2M}\), inconsequential in the thermodynamic limit, but
that will turn out to be essential to define classical self-dual BCs. 

If we impose periodic BCs in the
vertical  direction (so that any column contains \(2N\) sites), we can write
\begin{equation}\label{pf_uti}
\tilde{\pf}_{\sf U}(K_1,K_2,K_3,K_4)=\tr \left[
T(K_1,K_2,K_3,K_4)T(K_3,K_4,K_1,K_2)\right]^N
\end{equation}
for the Utiyama-Ising model with self-dual BCs. The 
next step is to  write \(T\) using techniques similar to those of Reference \cite{schultz}
\begin{eqnarray}
&& \frac{T(K_1,K_2,K_3,K_4)}{\left(4\sinh(2K_1) 
\sinh(2K_3)\right)^{M/2}}=e^{-H^1[h_1,h_3]}e^{-H^0[K_4,K_2]},
\end{eqnarray}
where
\begin{eqnarray}\label{hs_uti}
H^1[h_1,h_3]&=&-\sum_{i=1}^{M}\ (h_1\sigma^x_{2i-1}+h_3\sigma^x_{2i}),
\\
H^0[K_4,K_2]&=&-K_2\sigma^z_{2M}-K_4\sigma^z_{2M-1}\sigma^z_{2M}-\sum_{i=1}^{M-1}\ 
(K_4\sigma^z_{2i-1}
\sigma^z_{2i}+K_2\sigma^z_{2i}\sigma^z_{2i+1}) \nonumber ,
\end{eqnarray}
with \(h_\nu=-\frac{1}{2}\ln\tanh K_\nu\), \(\ \nu=1,3\).
At this point one can apply the bond-algebraic results of  
Section \ref{sec3.7} to show that \(H^1\)
is dual to \(H^0\),
\begin{equation}
H^1[h_1,h_3]=\mathcal{U}_\d^\dagger H^0[h_3,h_1]\mathcal{U}_\d,\ \ \ \
H^0[K_4,K_2]=\mathcal{U}_\d^\dagger H^1[K_2,K_4]\mathcal{U}_\d,
\end{equation}
and, together with the cyclic property of the trace, this implies that
\begin{eqnarray}\label{sd_utiyama_ising}
\frac{\tilde{\pf}_{\sf U}(K_1,K_2,K_3,K_4)}
{\left(4\sinh(2K_1) 
\sinh(2K_3)\right)^{MN}}&=&\tr \left[e^{-H^1[h_1,h_3]}e^{-H^0[K_4,K_2]}
e^{-H^1[h_3,h_1]}e^{-H^0[K_2,K_4]}\right]^N \nonumber\\
&=&\tr \left[e^{-H^1[K_4,K_2]}e^{-H^0[h_3,h_1]}e^{-H^1[K_2,K_4]}
e^{-H^0[h_1,h_3]}\right]^N \nonumber\\
&=&\frac{\tilde{\pf}_{\sf U}(K_1^*,K_2^*,K_3^*,K_4^*)}
{\left(4\sinh(2K_1^*) 
\sinh(2K_3^*)\right)^{MN}}, 
\end{eqnarray}
demonstrating an {\it exact} self-dual mapping  of the Utiyama-Ising model 
for {\it any}
finite lattice.
The dual couplings follow from comparing the first and second lines 
of Equation \eqref{sd_utiyama_ising}
\begin{eqnarray}\label{utiyama_dual_couplings}
h_1^*&\equiv&-\frac{1}{2}\ln\tanh K_1^*=K_4,\ \ \ \ 
h_3^*\equiv-\frac{1}{2}\ln\tanh K_3^*=K_2,\\
K_4^*&=&h_3\equiv-\frac{1}{2}\ln\tanh K_3,\ \ \ \ 
K_2^*=h_1\equiv-\frac{1}{2}\ln\tanh K_1.\nonumber
\end{eqnarray}
This completes the bond-algebraic study of the self-duality properties
of the Utiyama-Ising lattice model \(\tilde{\pf}_{\sf U}(K_1,K_2,K_3,K_4)\).

Equations \eqref{sd_utiyama_ising} and \eqref{utiyama_dual_couplings},
together with Equations \eqref{isquare}, \eqref{itr}, and \eqref{ihex} afford a simple
derivation of the self-duality relation for the square lattice Ising model first derived by 
Kramer and Wannier \cite{KW},
\begin{equation}\label{is}
\pf_{\sf I}(K_1,K_2)=A (K_1,K_1^*) \, \pf_{\sf I}(K_1^*,K_2^*) ,
\end{equation}
and of the duality relation between the hexagonal and triangular lattices referred to
by Onsager \cite{onsager} and written down by Wannier \cite{wannier}
\begin{equation}
\pf_{\sf IH}(K_1,K_2,K_4)=A(K_1,K_1^*,K_3^*) \, \pf_{\sf IT}(K_1^*,K_2^*,K_3^*)
\end{equation}
(with analytic functions $A$, see Section \ref{classical&quantum}). 
The last duality follows from
the fact that if \(K_3\rightarrow 0\), then \(K_4^*\rightarrow \infty\).

It is important to stress that we have derived the self-duality of 
Equation \eqref{sd_utiyama_ising}, using bond algebras, for any finite 
or infinite lattice. One could also derive this self-duality by starting from an 
appropriate quantum Ising chain and use the STL decomposition
(see Reference \cite{bookEPTCP}). However, in this case it is necessary to 
perform the thermodynamic limit in the extra dimension such that 
the conditions of the Trotter theorem are satisfied \cite{schulman}.

\subsection{The classical vector Potts model}
\label{sec8.3}

The \(D=2\)  VP model introduced
in Section \ref{sec4.1.1} \cite{taroni, bookEPTCP}  
has interesting but hard to uncover duality properties. Bond-algebraic methods
provide a powerful approach to 
unveil those duality properties
that rely on the bond-algebraic isomorphism
of Section \ref{sec4.1.1} and the results from Appendix \ref{appF}.

The transfer matrix of the VP model in an $M\times N$ lattice, with partition function written 
in Equation \eqref{classicalVP}, is given by ($\theta_i=2\pi s_i/p$, $s_i=0,1,\cdots,p-1$)
\begin{eqnarray}
&&\langle s'| {T}_{\sf VP}(K_x,K_y)|s\rangle\equiv\\
&&\exp\left[\sum_{i=1}^M\ K_y\cos(\theta_{i}'-\theta_i)+\sum_{i=1}^{M-1}\
K_x\cos(\theta_{i+1}-\theta_i)+ K_x\cos(\theta_M)\right],\nonumber
\end{eqnarray} 
where \(M\) is the number of sites along the horizontal 
direction. The last term represents 
a classical self-dual BC that will allow us exploit the
exact bond-algebraic isomorphism discussed in Section \ref{sec4.1.1}.  
Thus, in the following we will assume open BCs along the 
horizontal direction and periodic BCs along the vertical direction.
If we further
assume that the states \(|s\rangle\equiv\bigotimes_{i=1}^M |s_i\rangle\), \(s_i=
0,\cdots, p-1\) represent the basis diagonalizing the Weyl group algebra matrices \(U_i\)
of Equation \eqref{VsandUs}, then we can write 
\begin{eqnarray}
{T}_{\sf VP}(K_x,K_y)=e^{-H^0[K_x]}\ \prod_{i=1}^M\ e^{-H^1_i[K_y]},
\label{transfermaVPD2}
\end{eqnarray}
where
\begin{eqnarray}
H^0[K_x]=-\frac{K_x}{2}\sum_{i=1}^{M-1} (U_{i+1}^\dagger U_i+ U_{i}^\dagger U_{i+1})
-\frac{K_x}{2}\left(U_M+U_M^\dagger\right),
\label{transfer_VPH0}
\end{eqnarray}
and $H^1_i[K_y]$ is an operator whose matrix elements are given by
\begin{eqnarray}
\langle s_i'|e^{-H^1_i[K_y]}|s_i\rangle=e^{K_y\cos(\theta_{i}'-\theta_i)}.
\label{transfer_VP}
\end{eqnarray}
To determine \(H^1_i[K_y]\), we rewrite Equation \eqref{transfer_VP} in matrix form
\begin{equation}
e^{-H^1_i[K_y]}=\sum_{m=0}^{p-1}\ e^{K_y \cos \theta_m}\ V_i^m ,
\end{equation}
and apply the results of Appendix \ref{appF} to show that \(H^1_i[K_y]\) must 
be of the form 
\begin{equation}\label{non_diag_VP}
H^1_i[K_y]=-\sum_{m=0}^{p-1}\ h_m(K_y)\ V_i^m\ ,
\end{equation}
(recall that \((V^{m})^\dagger=V^{p-m}\) and  \((U^{m})^\dagger=U^{p-m}\)) with
\begin{equation}\label{manyhs}
h_m(K_y)=\frac{1}{p}\sum_{s=0}^{p-1}\ \cos(m \theta_s) \ln\left(\sum_{l=0}^{p-1}
\ e^{K_y\cos \theta_l}\cos(l \theta_s) \right)\ .
\end{equation}
This completes the factoring  of the transfer matrix \({T}_{\sf VP}(K_x,K_y)\) 
of Equation \eqref{transfermaVPD2}.

The isomorphism defined in Equations \eqref{aut_finite_VP1} and 
\eqref{aut_finite_VP2} determines dual forms  
for \(H^0[K_x]\) and \(H^1[K_y]=\sum_i H_i^1[K_y]\),
\begin{eqnarray}
H^{0}[K_x]&\dual&H^{0 D}[K_x]= -\frac{K_x}{2}\sum_{i=1}^{M}\ (V_i+V_i^\dagger),\\
H^{1}[K_y]&\dual&H^{1 D}[K_y]= -\sum_{i=1}^{M-1} \sum_{m=0}^{p-1}\ h_m(K_y)\ (U^\dagger_{i+1}
U_i)^m-\sum_{m=0}^{p-1}\ h_m(K_y)U_M^m\nonumber,
\end{eqnarray}
that translates into an {\it exact} classical duality for the VP model. The 
{\it exact} classical dual reads \(\widetilde{\pf}^D_{\sf VP}=\tr\left(e^{-H^{0 D}[K_x]}
e^{-H^{1 D}[K_y]}\right)^N\), and can be written down longhand
with the help of the results of Appendix \ref{appF}. The important point
to notice is that the Boltzmann weight of the dual model  has the general 
structure
\begin{equation}
e^{\sum_{m=0}^{p-1}\ K^*_{\nu m}\cos (m \theta'-m \theta)},\ \ \ \ 
\nu=x,y,
\end{equation}
with \(s',\ s\) representing the states of a pair of nearest neighbors, and
\(K^*_{x m}\) a function of \(K_y\) alone, while
\(K^*_{y m}\) a function of \(K_x\) alone. 

Clearly, the VP model {\it is not self-dual} for arbitrary \(p\) and
arbitrary couplings. However, the model is approximately self-dual
in the extreme anisotropic limit with \(K_y \gg K_x\) and it is
exactly self-dual for \(p=2,3,4\). We study these aspects of the 
VP model in the next two sections.

\subsubsection{Approximate self-duality in the extreme anisotropic limit}

As explained in Appendix \ref{appF}, in the limit in which \(K_y\)
becomes extremely large \(K_y\rightarrow\infty\), Equation \eqref{non_diag_VP}
simplifies to 
\begin{equation}
H^1_i[K_y]\approx K_y+\frac{\lambda}{2}(V_i+V^\dagger_i),
\ \ \ \ \ \ \frac{\lambda}{2}=e^{K_y(\cos\frac{2\pi}{p} -1)}
\end{equation}
(see Equation \eqref{plarge_K}), so that 
\begin{eqnarray}
\widetilde{\pf}_{\sf VP}(K_x,K_y)&\approx& e^{MNK_y}\tr\left[e^{\frac{\lambda}{2}\sum_i 
(V_i+V^\dagger_i)}e^{-H^0[K_x]}\right]^N\nonumber\\
&\approx&e^{MNK_y}\tr\left[ e^{-H^0[\lambda]}  e^{\frac{K_x}{2}\sum_i 
(V_i+V^\dagger_i)}\right]^N \\
&\approx&e^{MN(K_y-K_y^*)}\widetilde{\pf}_{\sf VP}(K_x^*,K_y^*), \nonumber
\end{eqnarray}
with dual couplings
\begin{equation}
\lambda^*\equiv 2e^{K_y^*(\cos\frac{2\pi}{p} -1)}=K_x,\ \ \ \ 
K_x^*=\lambda \equiv 2 e^{K_y(\cos\frac{2\pi}{p} -1)}.
\end{equation}
We emphasize that this approximate self-duality, in the extreme anisotropic limit,
 is valid for {\it any} value of $p$. We next consider {\it exact} self-dualities for the 
 particular cases $p=2,3,4$.

\subsubsection{The particular cases \(p=2, 3\), and $4$}\label{p234}

Let us start with the simplest $p=2$ Ising case.
If \(p=2\), $U=U^\dagger=\sigma^z$ and \(V=V^\dagger=\sigma^x\). Then, \({T}_{\sf VP}(K_x,K_y)=
e^{-H^0[K_x]}e^{-H^1[K_y]}\)
can be written in terms of 
\begin{eqnarray}
H^0[K_x]&=&-K_x\sum_{i=1}^{M-1}\sigma^z_i\sigma^z_{i+1}-K_x\sigma^z_M , \\
H^1[K_y]&=&-\frac{M}{2}\ln(2\sinh (2K_y))+\frac{1}{2}\ln\tanh K_y \sum_{i=1}^M\ \sigma^x_i,
\end{eqnarray}
which is simply the anisotropic Ising model on a rectangular $M\times N$ lattice, studied
in Section \ref{sec8.2}.

If \(p=3\), then \(V^2=V^\dagger\), and $H^1[K_y]$ becomes 
\begin{equation}
H^1[K_y]=-M h_{0,3}
-\frac{1}{3}\ln\left[\frac{e^{K_y}+2e^{-\frac{1}{2}{K_y}}}{e^{K_y}-e^{-\frac{1}{2}
{K_y}}}\right]\sum_{i=1}^M
\ (V_i+V_i^\dagger),
\end{equation}
with \(h_{0,3}=\frac{1}{3}\ln\left[(e^{K_y}+2e^{-\frac{1}{2}{K_y}})
(e^{K_y}-e^{-\frac{1}{2}{K_y}})^2\right]\), where $h_{m,p}=h_m$ of Equation \eqref{manyhs} 
for a particular $p$.
\(H^0[K_x]\) is identical to Equation \eqref{transfer_VPH0}, with \(U\)s appropriate for
\(p=3\). It follows that \(H^1\dual H^0\) and \(H^0\dual H^1\), rendering 
\(\widetilde{\pf}_{\sf VP}\) self-dual. The classical dual couplings 
follow as usual from comparing 
the transfer matrices of the original and dual models
\begin{equation}
K_x^*=\frac{2}{3}\ln\left[\frac{e^{K_y}+2e^{-\frac{1}{2}{K_y}}}{e^{K_y}-e^{-\frac{1}{2}
{K_y}}}\right],
\ \ \ \ \ \ K_x=\frac{2}{3}\ln\left[\frac{e^{K^*_y}+2e^{-\frac{1}{2}{K^*_y}}}{e^{K^*_y}
-e^{-\frac{1}{2}{K^*_y}}}\right] .
\end{equation}

Finally, consider the case \(p=4\). The general structure of \(H^1[K_y]\) is 
\begin{equation}
H^1[K_y]=-M h_{0,4}-h_{1,4}\sum_{i=1}^M\ (V_i+V_i^\dagger) - h_{2,4}\sum_{i=1}^M\ V_i^2,
\end{equation}
but, as a matter of fact, \(h_{2,4}\) vanishes. A simple application of 
Equation \eqref{manyhs} shows that
\begin{equation}
h_{2,4}=\frac{1}{4}\ln\left[\frac{(2+e^{K_y} +e^{-K_y})(-2+e^{K_y} 
+e^{-K_y})}{(e^{K_y} -e^{-K_y})^2}\right]=0.
\end{equation}
Also, \(h_{1,4}=-(1/2)\ln\tanh(K_y/2)\).
It follows that \(\widetilde{\pf}_{\sf VP}\) is again self-dual, with dual couplings
\begin{equation}
\frac{K^*_x}{2}=-\frac{1}{2}\ln\tanh\frac{K_y}{2}, \ \ \ \ \ \ 
\frac{K_x}{2}=-\frac{1}{2}\ln\tanh\frac{K_y^*}{2} ,
\end{equation}
which are equivalent to the ones for the Ising model. This is not
surprising since the \(p=4\) VP model is equivalent to two decoupled classical Ising models 
\cite{savit}.

\subsection{The classical eight-vertex model}
\label{eightvertexM}

In this section we use bond algebras to study some aspects
of the eight-vertex (8V) model on a square lattice
\cite{sutherland,fan_wu}.
The 8V has an almost self-evident self-duality,
and a duality that clarifies to some extent its
connection to the anisotropic $d=1$ quantum Heisenberg model. 

The 8V model is among the most thoroughly studied exactly solvable models 
in statistical mechanics \cite{baxter}, largely because it shows 
non-universal critical exponents. The first breakthrough in the 
study of the 8V model
came with Sutherland's observation \cite{sutherland}
that its transfer matrix commutes with the quantum Heisenberg
Hamiltonian, provided both models'
couplings are suitably adjusted. It is natural then to ask
whether there is a connection stronger than integrability.
Libero and Drugowich de Felicio \cite{defelicio} 
have shown that the STL (path integral) representation
of (a model dual to) the anisotropic Heisenberg model is dual to the
8V model on suitable regions of those models coupling spaces. The problem
with this remarkable connection is that (strictly speaking) it only
holds true provided some couplings are  
infinitesimally small, and others infinitely large 
(typical of the STL representation of quantum models \cite{suzuki}).
In contrast, our results in this section are exact, and reproduce
those of Reference \cite{defelicio} for appropriately chosen couplings. 

The transfer matrix used by Sutherland \cite{sutherland} in his original
work is not convenient for our purposes.
To associate a simple transfer matrix to the 8V model
we consider its Ashkin-Teller (AT) representation \cite{wu}, 
\begin{eqnarray}
\pf_{\sf 8V/AT}=\sum_{\{\mu_\r,\tau_\r\}}\exp\Big[\sum_\r\sum_{\nu=1,2}
&&(K_{0,\nu}+K_{1,\nu}\ \mu_\r\mu_{\r+\bm{e_\nu}}\\
&&+K_{2,\nu}\ \tau_\r
\tau_{\r+\bm{e_\nu}}+K_{4,\nu}\ \mu_\r\mu_{\r+\bm{e_\nu}}\tau_\r
\tau_{\r+\bm{e_\nu}})
\Big],\nonumber
\end{eqnarray} 
that features two independent classical Ising variables 
\(\mu_\r, \tau_\r=\pm1\) at each site of a square lattice of size \(M\times N\).
The relations connecting the couplings \(K_{i,\nu}\) to the parameters
of the 8V model can be found in Reference \cite{defelicio}. 
The additive constants \(K_{0,1}, K_{0,2}\) are irrelevant to what follows,
so we ignore them, but they can easily be reintroduced if needed. Then 
we can write as usual \(\pf_{\sf 8V/AT}=\tr(T_1T_0)^N\), with  
\begin{eqnarray}
T_0&=&\prod_{i=1}^{M-1}e^{K_{1,1}\ \mu^z_i\mu^z_{i+1}}e^{K_{2,1} \tau^z_i\tau^z_{i+1}}
e^{ K_{4,1}\ \mu^z_i\mu^z_{i+1}\tau^z_i\tau^z_{i+1}},\label{transferAT}\\
T_1&=&\prod_{i=1}^M(e^{K_{1,2}}+e^{-K_{1,2}}\ \mu^x_i)(e^{K_{2,2}}+e^{-K_{2,2}}\ \tau^x_i)
(e^{K_{4,2}}+e^{-K_{4,2}}\ \mu^x_i\tau^x_i).
\end{eqnarray}
Since the \(\mu\)-spins are independent of (commute with) the \(\tau\)-spins,
we see that \(\mathcal{A}_{\sf 8V/AT}=\mathcal{A}_{\sf I}\otimes\mathcal{A}_{\sf I}\),
where \(\mathcal{A}_{\sf I}\) is the bond algebra of the quantum Ising chain studied
in Section \ref{sec3.4}. It follows that we can apply the self-duality mapping
of the Ising chain to the \(\mu\)-bonds while leaving the \(\tau\)-bonds fixed
(or {\it viceversa}), or we can apply the mapping simultaneously to both types of bonds.
The first case defines a duality that we will not
study here any further. The second one defines a self-duality.

\begin{figure}
\centering
\includegraphics[angle=0, width=.8\columnwidth]{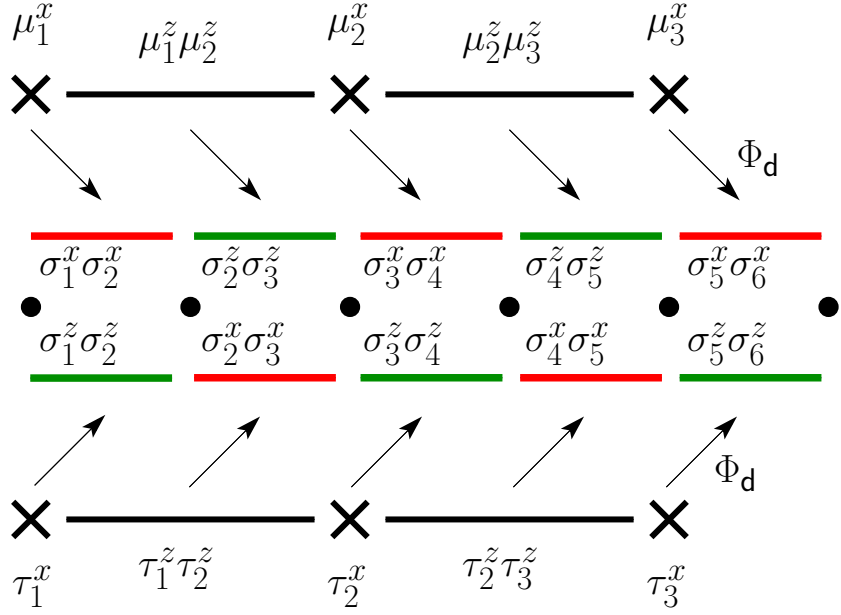}
\caption{The duality mapping \(\Phi_\d\) of Equation \eqref{atellertoh}, 
for \(M=3\) sites. The top and bottom chains represent the two types of independent 
bonds in the AT model, the middle chain represents the independent
bonds in the Heisenberg model (the bond \(\sigma^y_i\sigma^y_{i+1}=
-\sigma^x_i\sigma^x_{i+1}\sigma^z_i\sigma^z_{i+1}\) is not independent).
Notice that both models act on state spaces of the same
dimensionality \(2^{2M}\).}.
\label{attohfig}
\end{figure}

The 8V model is connected through integrability \cite{sutherland}
to the anisotropic Heisenberg model, so we would like to see if we can find
a representation of \(\mathcal{A}_{\sf 8V/AT}\) in terms of the bonds of the Heisenberg
model. A simple bond-algebraic analysis reveals one such representation:
\begin{eqnarray}
\mu^x_i&\dual&\sigma^x_{2i-1}\sigma^x_{2i},\ \ \ \ \ \ \ \ \
\tau^x_i\dual\sigma^z_{2i-1}\sigma^z_{2i}, \ \ \ \ \ \
i=1,\cdots,M, \label{atellertoh}\\
\mu^z_i\mu^z_{i+1}&\dual& \sigma^z_{2i}\sigma^z_{2i+1},\ \ \ \
\tau^z_{i}\tau^z_{i+1}\dual\sigma^x_{2i}\sigma^x_{2i+1},\ \ \ \ \ \ 
i=1,\cdots,M-1.\nonumber
\end{eqnarray}
This duality mapping \(\Phi_\d\) is illustrated in Figure \ref{attohfig}.

It follows that the transfer matrix of Equation \eqref{transferAT}
admits the dual representation
\begin{eqnarray}\label{heisenbergtransfer}
T_0^D&=&\prod_{i=1}^{M-1}e^{K_{1,1}\ \sigma^z_{2i}\sigma^z_{2i+1}}\ 
e^{K_{2,1}\ \sigma^x_{2i}\sigma^x_{2i+1}}\ 
e^{-K_{4,1}\ \sigma^y_{2i}\sigma^y_{2i+1}},\\
T_1^D&=&\prod_{i=1}^M(e^{K_{1,2}}+e^{-K_{1,2}}\ \sigma^x_{2i-1}\sigma^x_{2i})
(e^{K_{2,2}}+e^{-K_{2,2}}\ \sigma^z_{2i-1}\sigma^z_{2i})
(e^{K_{4,2}}-e^{-K_{4,2}}\ \sigma^y_{2i-1}\sigma^y_{2i}) \nonumber \\
&=&C^M\prod_{i=1}^Me^{K_{1,2}^*\ \sigma^x_{2i-1}\sigma^x_{2i}}\ 
e^{K_{2,2}^*\ \sigma^z_{2i-1}\sigma^z_{2i}}\ 
e^{-K_{4,2}^*\sigma^y_{2i-1}\sigma^y_{2i}} , \nonumber
\end{eqnarray}
where 
\begin{equation}
\sinh(2K_{i,2}) \sinh(2K_{i,2}^*)=1,\ \ \ \ \ \ \ \ i=1,2,4,
\end{equation}
and \(C^2=1/(8\sinh(2K_{1,2}^*)\sinh(2K_{2,2}^*)\sinh(2K_{4,2}^*))\), see Equation
\eqref{dualt0dising}.
We can re-write Equation \eqref{heisenbergtransfer} as
\begin{equation}
T_0^D\ =\ e^{H_{\sf even}},\ \ \ \ \ \ \ \ T_1^D\ 
=\ C^M\ e^{H_{\sf odd}},
\end{equation}
where
\begin{equation}
H_{\sf even}=\sum_{i=1}^{M-1}\sum_{\nu=1}^3\ J_\nu^{\sf e}\ \sigma^\nu_{2i}\sigma^\nu_{2i+1},
\ \ \ \ \ \ \ \ 
H_{\sf odd}=\sum_{i=1}^{M}\sum_{\nu=1}^3\ J_\nu^{\sf o}\ \sigma^\nu_{2i-1}\sigma^\nu_{2i},
\end{equation}
with couplings
\(J^{\sf e}_1=K_{2,1},\ J_2^{\sf e}=-K_{4,1},\ 
J_3^{\sf e}=K_{1,1}\), and \(J^{\sf o}_1=K_{1,2}^*,\ 
J_2^{\sf o}=-K_{4,2}^*,\ J_3^{\sf o}=K_{2,2}^*\).
Next we can use the BCH formula to write
\begin{equation}\label{h8vbch}
T_1^DT_0^D\ =\ C^M\ e^{H_{\sf odd}+H_{\sf even}+[H_{\sf odd},\ H_{\sf even}]/2+\ \cdots}.
\end{equation}
The expression \(H_{\sf H}=H_{\sf odd}+ H_{\sf even}\) defines an anisotropic Heisenberg
model in the region of coupling space where \(J_\nu^{\sf e}=J_\nu^{\sf o}\). 
Also, the result of Reference \cite{defelicio} follows by choosing the couplings so that
we can neglect \([H_{\sf odd},\ H_{\sf even}]\) and higher-order terms in Equation 
\eqref{h8vbch}. This completes our discussion of the classical 8V model in its
AT representation. 

In closing, we would like to comment on the bond algebra mapping of Equation 
\eqref{atellertoh} from the point of view of {\it quantum} dualities, 
specially in connection to the problem of generating STL (path integral)
representations of the Heisenberg model. The STL representation of 
quantum lattice models is the starting point for quantum
Monte Carlo techniques \cite{booksuzuki}, and the fact that
the Heisenberg model does not have a simple STL representation was
noticed already in the early Reference \cite{suzuki} (by simple we mean a local or 
quasi-local representation with real and positive Boltzmann weights). In contrast,
the mapping of Equation \eqref{atellertoh} (or more precisely, its 
inverse \(\Phi_\d^{-1}\) produces a dual representation
of the Heisenberg model \(H_{\sf H}=
\sum_{i=1}^{2M-1}\sum_{\nu=1}^3\ J_\nu\ \sigma^\nu_{i}\sigma^\nu_{i+1}\),
\begin{eqnarray}
H_{\sf H}^D&=&(J_1\mu^x_M-J_2\mu^x_M\tau^x_M+J_3\tau^x_M)+\\
&&\sum_{j=1}^{M-1}(J_1(\mu^x_j+\tau^z_j\tau^z_{j+1})-J_2(
\mu^x_j\tau^x_j+\mu^z_j\mu^z_{j+1}\tau^z_j\tau^z_{j+1})
+J_3(\tau^x_j+\mu^z_j\mu^z_{j+1})),\nonumber
\end{eqnarray}
that, as shown in Reference \cite{defelicio},
does have a simple STL representation (e.g., the classical AT
model just studied. For this reason the Hamiltonian \(H_{\sf H}^D\) above is
sometimes known in the literature as the quantum AT model \cite{kadanoffat}).
This fact illustrates one of the important applications of our bond-algebraic 
approach stressed in Reference \cite{con}, that is, that the method allows 
a systematic search for simple STL (or path integral) representations of 
quantum problems.  

Let us present in closing the dual variables that follow from 
Equation \eqref{atellertoh}. After extending the duality 
mapping as
\begin{equation}
\Phi_\d(\tau^z_1)=\sigma^x_1,\ \ \ \ \ \ \ \ \ \ \ \
\Phi_\d(\mu^z_M)=\sigma^z_{2M},
\end{equation}
we can compute (\(j=1,\cdots,M\)) the dual variables
\begin{eqnarray}
\hat{\mu}^x_j&=&\sigma^x_{2j-1}\sigma^x_{2j},\ \ \ \ \ \ \ \ \ \ \ \ 
\hat{\mu}^z_j=\prod_{m=2j}^{2M}\sigma^z_m,\\
\hat{\tau}^x_j&=&\sigma^z_{2j-1}\sigma^z_{2j},\ \ \ \ \ \ \ \ \ \ \ \
\hat{\tau}^z_j=\prod_{m=1}^{2j-1}\sigma^x_{m}.
\end{eqnarray}
where we write \(\hat{\mathcal{O}}\) for the image of \(\mathcal{O}\) under
\(\Phi_\d\), \(\mathcal{O}\dual\hat{\mathcal{O}}\). 
The formal counterparts of these dual variables
for the infinite-size quantum AT model were presented in Reference \cite{shankar}.

\subsection{Classical dualities in \(D=3\) and \(D=4\) dimensions}
\label{sec8.4}


\subsubsection{\(D=3\) Ising/\(\mathbb{Z}_2\) gauge models}

The $D$-dimensional Ising gauge (or \(\mathbb{Z}_2\) gauge) model features Ising 
spin variables $\sigma_{(\r,\mu)}=\pm 1$ residing on the links of a 
$N=N_1\times N_2\times \cdots \times N_D$ hyper-cubic lattice. 
The model was introduced by Wegner \cite{wegner_ising} as an example 
of a system displaying a phase transition without the existence of 
a (local) Landau order parameter. Its partition function is given by
\begin{equation}\label{wegner_model}
\pf_{\sf IG}(K)=\sum_{\{\sigma_{(\r,\mu)}\}}\ \exp\left[K\sum_\r\sum_{\mu>\nu}\ 
\sigma_{(\r,\mu)}\sigma_{(\r+\bm{e_\mu},\nu)}\sigma_{(\r+\bm{e_\nu},\mu)}
\sigma_{(\r,\nu)}\right] .
\end{equation}
In \(D=3\) dimensions  \(\mu,\nu=1,2,3\), and the vertex operator $A_\r$ that flips  
 the sign, \(\sigma \rightarrow -\sigma\), of all six Ising spins 
sharing the vertex $\r$ is a gauge symmetry of the model. 

To set up a transfer matrix for \(\pf_{\sf IG}\)(K), one introduces
a condition that partially fixes the gauge invariance. That condition 
amounts to consider only those spin configurations that satisfy the 
constraint \(\sigma_{(\r,3)}=1\) at every link \((\r,3)\). Since any spin configuration 
that does not satisfy this condition can be obtained from a compliant 
configuration, it follows that
\begin{equation}
\pf_{\sf IG}(K)=\tilde{N}_{\sf G} \, \pf_{\sf IG}'(K),
\end{equation}  
where \(\pf_{\sf IG}'(K)\) is the partition function of Equation \eqref{wegner_model},
with the sum over spin configurations being replaced by the sum over those 
configurations that satisfy the constraint, \(\sum_{\{\sigma_{(\r,\mu)}\}}\rightarrow
\sum_{\{\sigma_{(\r,\mu)}\}}'\), and  $\tilde{N}_{\sf G}$ is a factor that takes into account 
the remaining configurations not included in  \(\pf_{\sf IG}'(K)\).  
 
As long as we compute only expectation values of 
gauge-invariant quantities, we can replace \(\sum_{\{\sigma_{(\r,\mu)}\}}\rightarrow 
\tilde{N}_{\sf G} \, \sum_{\{\sigma_{(\r,\mu)}\}}'\) everywhere, and the factor 
\(\tilde{N}_{\sf G}\) drops out, 
which is a manifestation of the fact that the gauge symmetries of the model provide 
a redundant description.  On the other hand, the partially gauge fixed partition function 
\(\pf_{\sf IG}'(K)\) can be written in terms of the transfer matrix 
\begin{equation}
\frac{T_{\sf IG}(K)}{(2\sinh(2K))^{N_1N_2/2}}=e^{-H^1[K]}\ e^{-H^0[K]},
\end{equation}
with operators
\begin{equation}
H^1[K]\equiv\frac{1}{2}\ln\tanh K \sum_\r\sum_{\mu=1,2}\ \sigma^x_{(\r,\mu)},
\ \ \ \ \ H^0[K]=-K\sum_\r B_{(\r,3)},
\end{equation}
where now \(\r=m^1\i+m^2\j\) denotes the {\it  vertices of a $N_1\times N_2$ square lattice},
representing the planes of constant
\(r^3\) in the original cubic lattice, and \(B_{(\r,3)}\) is the plaquette operator defined 
in Equation \eqref{gauge_plaquette}. For ease of presentation, we assume that \(N_1N_2\) 
is sufficiently large and thus avoid introducing dual boundary corrections, that could 
anyway be computed
by the techniques already developed in previous sections. 
The residual gauge invariance 
present in \(\pf_{\sf IG}'(K)\) translates into gauge symmetries of 
\(H^1[K]\) and \(H^0[K]\) that were discussed at length in Section \ref{sec3.12}, 
together with their duality properties.

It finally follows that the Ising gauge partition function can be written as
\begin{equation}
\pf_{\sf IG}(K)=N_{\sf G} (2\sinh(2K))^{N/2}\tr\left[(e^{-H^1[K]}
e^{-H^0[K]})^{N_3}P_{\sf GI}\right]
\end{equation}
where \(N_{\sf G}\) counts the total gauge redundancy, and \(P_{\sf GI}\) is 
the orthogonal projector onto the space of gauge invariant states. The next
step is to use the projective, gauge reducing duality introduced in Section
\ref{sec3.12} to write (recall that \(P_{\sf GI}=U_\d^\dagger U_\d\) is a projector and 
commutes with $H^0[K]$ and $H^1[K]$)
\begin{eqnarray}\label{classical_wegner}
\frac{\pf_{\sf IG}(K)}{N_{\sf G} (2\sinh(2K))^{N/2}}&=&
\tr\left[(U_\d e^{-H^1[K]}U_\d^\dagger\ 
U_\d e^{-H^0[K]}U_\d^\dagger)^{N_3}\right] \nonumber \\
&=&\tr\left[ (e^{-\frac{1}{2}\ln\tanh K \, \sum_\r\sum_{\mu=1,2}\sigma^z_\r 
\sigma^z_{\r+\bm{e_\mu}}}\ e^{K\sum_\r \sigma^x_\r})^{N_3}\right] \nonumber  \\
&=&\frac{\pf_{\sf I}(K^*)}
{(2\sinh(2K^*))^{N/2}}, 
\end{eqnarray}
with \(K^*=-\frac{1}{2}\ln\tanh K\), and \(\pf_{\sf I}\) the partition function
of the \(D=3\) Ising model. This completes the bond-algebraic proof of the 
duality between the \(D=3\)  Ising and Ising gauge models \cite{wegner_ising}. 

The quantum duality underlying this classical duality was extended
to cover $p$-state VP/\(\mathbb{Z}_p\) gauge models in 
Section \ref{sec5.3}. We can thus generalize the classical duality of this 
section to 
$p>2$. If \(p=2,3,\) or \(4\) (\(p=2\) being the case we
just covered in Equation \eqref{classical_wegner}), then the quantum duality
implies that the corresponding \(D=3\) classical VP/\(\mathbb{Z}_p\) 
gauge models are dual (the partition function for the \(\mathbb{Z}_p\) 
gauge models is presented in the next section, Equation \eqref{class_zpgauge}). 
On the other hand, if \(p\geq 5\), then the quantum duality
translates into an {\it exact} classical duality between models that are 
modified versions of the \(D=3\) classical VP/\(\mathbb{Z}_p\) 
gauge models. These modified classical models can be computed with the
aid of Appendix \ref{appF}, along the lines sketched in Section \ref{sec8.3}.
We next concentrate on the specifics of this  generalization but in $D=4$ dimensions.

\subsubsection{A new family of \(D=4\) self-dual \(\mathbb{Z}_p\) gauge theories}
\label{classical_zpgauge}

In this section we study classical \(D=4\) \(\mathbb{Z}_p\) gauge theories
of the Wilson type \cite{yoneya}, and of a new type that has the advantage
of being exactly self-dual for any \(p\). The two theories coincide for  \(p=2,3,4\),
corresponding to the cases where the \(\mathbb{Z}_p\) partition function/Euclidean
path integral of the Wilson type is exactly self-dual. The following discussion closely 
parallels that of the VP model of Section \ref{sec8.3}, and clarifies
the strong connection between the self-dual structure of both models \cite{con}.

The Wilson-type action of $D$ dimensional \(\mathbb{Z}_p\) gauge theories
is a VP-like generalization of Wegner's Ising ($p=2$)
gauge model of Equation \eqref{wegner_model},
\begin{equation}\label{class_zpgauge}
\pf_{\sf WG}(K)=\sum_{\{\theta_{(\r,\mu)}\}}\ \exp\left[K\sum_\r\sum_{\mu>\nu}\ 
\cos \Theta_{(\r,\mu\nu)} \right],
\end{equation}
where 
\begin{eqnarray}
\Theta_{(\r,\mu\nu)}=
\theta_{(\r,\mu)}+\theta_{(\r+\bm{e_\mu},\nu)}-\theta_{(\r+\bm{e_\nu},\mu)}-
\theta_{(\r,\nu)},  \ \ \ \ \ \mu,\nu=1,\cdots,D ,
\end{eqnarray}
with discrete angles \(\theta_{(\r,\nu)}=2\pi 
s_{(\r,\nu)}/p\), \(s_{(\r,\nu)}=0,\cdots,p-1\), placed on the links of 
a hyper-cubic lattice with vertices \(\r\). {}From now on, we focus on the
\(D=4\) case. The initial interest in \(\pf_{\sf WG}\) \cite{yoneya}
was stimulated by work of 't Hooft on confinement in QCD \cite{thooft}, that stresses
the importance to confinement of the fact that the center  of 
the gauge group \(SU(p)\) is \(\mathbb{Z}_p\) (\(p\) here stands for the number
of colors). {}From this viewpoint, \(p=3\) is especially important, and we will
show that  \(\pf_{\sf WG}\) is self-dual \cite{yoneya}. On the other hand,
for large \(p\), one can think of \(\pf_{\sf WG}\) as an approximation to compact 
QED \cite{horn,frohlich_gauge}. In \(D=4\), compact QED is known to show a 
phase transition between a confinement 
and Coulomb phases \cite{vettorazzo_forcrand, panero}, and it was shown in
Reference \cite{frohlich_gauge} that this feature is shared by \(\pf_{\sf WG}\)
for \(p\) sufficiently large.   

To determine the transfer matrix of \(\pf_{\sf WG}\), $T_{\sf WG}$, we 
need to partially fix the gauge of the model by considering only configurations
that satisfy the constraint \(\theta_{(\r,4)}=0\) (we take \(\mu=4\) to
be the Euclidean time direction). Since any other configuration can obtained
from one satisfying this constraint by a gauge transformation, the restriction
has no physical consequence as long as we compute averages of gauge-invariant
observables only. Under these conditions, we can write 
\begin{equation}
\pf_{\sf WG}(K)=N_{\sf G}\tr [{T_{\sf WG}(K)}]^{N_4},\ \ \ \ \ \ \ \  \mbox{} 
\ \ \ \ T_{\sf WG}(K)= e^{-H^1[K]}e^{-H^0[K]},
\end{equation}
with \(N_{\sf G}\) a counting factor introduced to compensate for the gauge-fixing 
condition, and
\begin{equation}
H^1[K]=-\sum_\r\sum_{\nu=1}^3\sum_{m=0}^{p-1}\ h_m(K)V^m_{(\r,\nu)},
\ \ \ \ \ \ \ H^0[K]= -K\sum_\r\sum_{\nu=1}^3\ B_{(\r,\nu)}. 
\end{equation}
The bonds \(B_{(\r,\nu)}\) were defined in Equation \eqref{3d_Uplaquettes}, and 
the couplings \(h_m(K)\) are {\it identical} to those computed for the 
VP model, Equation \eqref{manyhs} (see also Appendix \ref{appF}).
This exact form of the transfer matrix \(T_{\sf WG}\)  is, 
to the best of our knowledge,
computed here explicitly for the first time for arbitrary \(p\) (In Reference
\cite{yoneya}, Yoneya performs this computation for \(p=2,3,4\), but  
is unable to extend his approach to larger \(p\)).

{}From this point on, the analysis of the duality properties of \(\pf_{\sf WG}\) proceeds
just as that for the VP model, except that the appropriate
bond-algebra mapping is the one of Section \ref{sec6.2}, Equation \eqref{map_gauge_zp}.
It follows, just as for the VP model, that \(\pf_{\sf WG}\) is self-dual for
\(p=2,3,4\) \cite{yoneya} (see Section \ref{p234}), and for $p>4$ dual to a 
different \(\Z_p\) gauge model with Boltzmann weights of the form 
\begin{equation}
\exp\left[K^*_{\mu m}\cos(m\Theta_{(\r,\mu\nu)})\right],
\end{equation}
with \(\mu<\nu=1,2,3,4,\) and \(m=0,\cdots,p-1\). 
The dual couplings \(K^*_{\mu m}\) can be computed in closed form with
the help of the results of Appendix \ref{appF}. This completes our discussion of the
duality properties of Wilson-type \(\Z_p\) gauge theories.

Next we would like to follow a different approach that emphasizes the possibility
of using {\it quantum} models with known duality properties to construct {\it classical}
models with known duality properties as well. Notice that this methodology 
(quantum to classical) 
reverses the 
direction of the path that we have been following in Section  \ref{sec8} (classical
to quantum).
In the light of previous discussion,
we see that the quantum $p$-state gauge model studied in Section \ref{sec6.2}, 
for $p\geq 5$,  
is not
exactly related to the Wilson-type action of Equation \eqref{class_zpgauge}.
On the other hand, that quantum model is exactly self-dual for any \(p\), so we would like 
to determine its related classical \(p\)-state gauge theory.
To this end, we need to identify what classical gauge model \(\pf_{\sf sdG}\),
in an appropriate gauge, has 
\begin{equation}
T_{\sf sdG}=e^{\frac{\kappa}{2}\sum_\r\sum_{\mu=1}^{3}\ (V_{(\r,\mu)}+
V^\dagger_{(\r,\mu)})}\ e^{\frac{K}{2}\sum_\r\sum_{\mu=1}^{3}\ (B_{(\r,\mu)}+
B^\dagger_{(\r,\mu)})}
\end{equation}
for transfer matrix. Then it will follow from the self-duality of Equation 
\eqref{map_gauge_zp} that \(\pf_{\sf sdG}\) is exactly self-dual {\it for any p}.

It is not difficult to compute
the matrix elements of \(T_{\sf sdG}\) (see Appendix \ref{appF}). Then we can reconstruct the 
partition function  
\begin{eqnarray}\label{pstateEM}
\pf_{\sf sdG}&=&\sum_{\{\theta_{(\r,\mu)}\}}\  \exp \left [ \sum_\r K (
\cos \Theta_{(\r,12)}+ \cos \Theta_{(\r,23)}+\cos \Theta_{(\r,31)})  \right ] \times\\
&&\ \ \ \exp \left [\sum_\r
\sum_{m=0}^{p-1}\ \epsilon_m(\kappa) (\cos(m\Theta_{(\r,41)})+\cos(m\Theta_{(\r,42)})
+\cos(m\Theta_{(\r,43)}) ) \right] .\nonumber
\end{eqnarray}
This is the new, exactly self-dual classical gauge theory we were after. Notice that this
self-duality exchanges the Boltzmann weight of the second line of Equation 
\eqref{pstateEM}, that can be associated with the Boltzmann weight of the 
\(\Z_p\) electric field, with the one of the first line, that can be associated
with the Boltzmann weight of the \(\Z_p\) magnetic field. 

Finally, let us compute the coefficients \(\epsilon_m(\kappa)\),
which are an essential ingredient in the definition of \(\pf_{\sf sdG}\).
The starting point is to rewrite 
\(
e^{\frac{\kappa}{2}(V_{(\r,\mu)}+
V^\dagger_{(\r,\mu)})}=\sum_{m=0}^{p-1}e^{u_\kappa(m)}V^m_{(\r,\mu)}\ ,
\)
with (\(\theta_s=2\pi s/p\))
\begin{equation}
u_\kappa(m)=\ln\left[\frac{1}{p}\sum_{s=0}^{p-1}\cos(m\theta_s)e^{\kappa
\cos\theta_s}\right] . 
\end{equation}
Since \(u_\kappa(l)\) is even, 
it can be represented as a discrete cosine series, \(u_\kappa(l)=\sum_{m=0}^{p-1}\ 
\epsilon_m(\kappa) \cos(l\theta_m)\). 
This completes the specification of the \(\epsilon_m(\kappa)\).

\subsection{Dualities for continuum models of classical statistical physics}
\label{sec8.5}

We have so far considered lattice models of classical statistical mechanics. In this 
section we want to show by example that our bond-algebraic approach can indeed 
be used to establish duality transformations in many-body problems defined 
in continuum space-time. 

Consider the Hamiltonian of a chain of coupled quantum harmonic oscillators 
\begin{equation}\label{sd_phonons}
H_{\sf Ph}[1/m, k]=
\sum_i\ ( \frac{p_i^2}{2m}+\frac{1}{2}k(x_{i+1}-x_i)^2)
=\ \mathcal{U}_\d^\dagger\, H_{\sf Ph}[k,1/m]\, \mathcal{U}_\d\ ,
\end{equation}
whose normal modes represent acoustic phonons. Last equality indicates the 
quantum self-duality relation derived in Section \ref{sec4.3} with 
$k=m\omega^2$.

On one hand, standard manipulations which involve the 
STL decomposition used before \cite{bookEPTCP}, 
but  now applied to continuous degrees of freedom \cite{schulman},
lead to 
\begin{eqnarray}
\tr e^{-\beta H_{\sf Ph}[1/m,k]}&=&\mathcal{N}\lim_{N\rightarrow\infty}
\int\prod_{j=1}^N\prod_i dx_{i,j} \\ 
&\times& \exp\left[-\sum_{j=1}^N\sum_i\ \left(
\frac{mN}{2\beta}(x_{i,j+1}-x_{i,j})^2+\frac{k\beta}{2N}
(x_{i+1,j}-x_{i,j})^2\right)\right] \nonumber ,
\end{eqnarray}
with $\cal N$ a normalization factor. This mapping 
relates a chain of coupled 
quantum harmonic oscillators to a classical $D=2$ array of springs. 
On the other hand, it is standard in the path integral
context to interpret the limit 
\(N\rightarrow \infty\) as a sum over classical path 
configurations (in Euclidean time),
that is, as an Euclidean path integral \cite{schulman}
\begin{eqnarray}
\tr e^{-\beta H_{\sf Ph}[1/m,k]}&=&\\
&&\hspace*{-3.0cm}=\int \prod_i \mathcal{D}x_i\ \exp\left[\int_0^\beta d\tau\ 
\sum_i\ \left(\frac{1}{2}m\left(\frac{dx_i}{d\tau}\right)^2+
\frac{1}{2}k(x_{i+1}-x_i)^2\right)\right]\nonumber\\
&&\hspace*{-3.0cm}=\int \prod_i \mathcal{D}x_i\ \exp\left[\int_0^\beta d\tau\ 
\sum_i\ \left(\frac{1}{2}m^*\left(\frac{dx_i}{d\tau}\right)^2+
\frac{1}{2}k^*(x_{i+1}-x_i)^2\right)\right ], \nonumber
\end{eqnarray}
where \(\mathcal{D}x_i\) denotes the Wiener measure on the space of paths.
The last equality results from applying the quantum self-duality of Equation
\eqref{sd_phonons}, and thus constitutes an elementary example of a duality for Feynman
path integrals. The dual couplings \(m^*,\ k^*\) satisfy the relations 
\begin{equation}
km^*=1,\ \ \ \ \ \ \ \ \ \ mk^*=1.
\end{equation} 


\subsection{Classical disorder variables from quantum ones}
\label{appB}

This section exploits the combined application of 
the transfer matrix and bond-algebraic 
techniques to compute classical disorder variables for arbitrary 
models of classical statistical mechanics. 

The notion of {\it disorder variable} 
was introduced in the context of the $D=2$ classical Ising model
 \cite{kadanoff}, and further exploited
in Reference \cite{Kadanoff_Ceva}, as part of 
a scheme, based on the operator
product expansion technique,  to compute its critical exponents. 
It seems 
reasonable to expect that classical disorder variables (defined in terms 
of the classical Kramers-Wannier duality 
\cite{Kadanoff_Ceva}) should be related to quantum dual variables,
but the explicit connection has not been published 
before, most likely because the relation only becomes self-evident
when quantum dualities are recognized as unitary transformations. 
This connection is, however, of great importance  because
there is no simple way to generalize the construction
of classical disorder variables for the Ising model to different models, 
other than the {\it quantum route} that we are going to take next.  
This route depends critically on recognizing quantum dualities
as unitary transformations.

Consider a $D=2$ Ising model on an $M\times N$  lattice with 
cylindrical topology as in Section \ref{sec8.2}
(\(N\) is the number of sites along the periodic BC). 
We can write its partition function in terms of a transfer matrix $T$ as
\begin{equation}
\pf_{\sf I}(K_1,K_2)= \tr T^N,
\end{equation}
and more importantly, we can write two-point correlation functions as
\begin{equation}\label{two_point}
\langle \sigma_{m',n'}\ \sigma_{m,n}\rangle(K_1,K_2)=
\frac{\tr ( T^{(N-n')}\ \sigma^z_{m'}\ T^{(n'-n)}\ \sigma^z_m\ T^n)}{\tr T^N} .
\end{equation}

On the other hand, the transfer matrix \(T\) can be related to a quantum
problem
\begin{eqnarray}
&& \frac{T(K_1,K_2)}{\left(2\sinh(2K_1) \right)^{M/2}}=e^{-H^1[h_1]}e^{-H^0[K_2]},
\end{eqnarray}
where
\begin{eqnarray}
H^1[h_1]=-h_1 \sum_{i=1}^{M} \sigma^x_{i} , \ \ \ \ 
H^0[K_2]=-K_2\sigma^z_{M}-K_2\sum_{i=1}^{M-1} 
\sigma^z_{i}\sigma^z_{i+1}  ,
\end{eqnarray}
with \(h_1=-\frac{1}{2}\ln\tanh K_1\).
Applying the bond-algebraic results of  
Section \ref{sec3.7} one can  show that \(H^1\)
is dual to \(H^0\),
\begin{equation}
H^1[h_1]=\mathcal{U}_\d^\dagger H^0[h_1]\mathcal{U}_\d,\ \ \ \
H^0[K_2]=\mathcal{U}_\d^\dagger H^1[K_2]\mathcal{U}_\d,
\end{equation}
which implies that the two-point correlation 
defined in Equation \eqref{two_point} can be written as
\begin{eqnarray}
\langle \sigma_{m',n'}\ \sigma_{m,n}\rangle(K_1,K_2)&=& \nonumber \\
&=&
\frac{\tr ( \mathcal{U}_\d T^{(N-n')}\mathcal{U}_\d^\dagger\ 
\mathcal{U}_\d \sigma^z_{m'}\mathcal{U}_\d^\dagger\ \mathcal{U}_\d
T^{(n'-n)}\mathcal{U}_\d^\dagger\ \mathcal{U}_\d\sigma^z_m\mathcal{U}_\d^\dagger\ 
\mathcal{U}_\d T^n\mathcal{U}_\d^\dagger)}{\tr 
(\mathcal{U}_\d T^N\mathcal{U}_\d^\dagger)} \nonumber \\
&=&\frac{\tr (\widehat{T}^{(N-n')}\ \mu^z_{m'}\ \widehat{T}^{(n'-n)}\ \mu^z_m\ 
\widehat{T}^{n})}{\tr \widehat{T}^{N}} \nonumber \\
&=& \langle \mu_{m',n'}\ \mu_{m,n}\rangle(K_1^*,K_2^*) ,
\label{kada_ceva_dual}
\end{eqnarray}
with quantum dual variables $\mu^z_m$ defined in Section \ref{sec3.10}, 
and $K^*_1=-\frac{1}{2}\ln\tanh K_2$, $K^*_2=h_1$. 
Since the classical disorder variables of Reference \cite{Kadanoff_Ceva}
are essentially defined through the relation \eqref{kada_ceva_dual}, 
we see that the quantum dual variables \(\mu^z_m\) are indeed
quantum disorder variables themselves. Notice that in contrast
to the classical approach of Reference \cite{Kadanoff_Ceva}, the
quantum approach allow us to compute
\begin{equation}
\langle\mu_{m,n}\rangle(K_1^*,K_2^*)=
\frac{\tr (\widehat{T}^{(N-n)}\ \mu^z_m\ 
\widehat{T}^{n})}{\tr \widehat{T}^{N}},
\end{equation}
a quantity that could not even be defined at the classical level 
(the approach of Reference \cite{Kadanoff_Ceva} can only make sense
of {\it correlators} of disorder variables).

While it would be impossible to extend the techniques
of Reference \cite{Kadanoff_Ceva} to, say, the VP model,
we see now that our bond-algebraic technique afford an straightforward 
solution to the problem of defining {\it classical} disorder 
variables in general: They can be derived from their quantum 
counterparts and the transfer matrix formalism.

\section{Applications of dualities}
\label{sec9}

Thus far, we illustrated, how to derive (hitherto known and also unknown) 
dualities within our bond algebraic approach. 
As we have shown, our approach applies to both quantum and classical
dualities. In the current section, we will discuss applications of dualities.
We will present some spectral consequences of dualities, general 
techniques such as fermionization in arbitrary spatial dimensions, integrability conditions, 
and dimensional reductions.

\subsection{Self-dualities and phase transitions}
\label{sec9.1}

As is well appreciated, one of the most powerful consequences of
dualities are constraints that may be imposed on the phase diagrams
of dual systems. These become particularly
potent in the case of self-dual systems.
Thus, we will now turn to the practical consequences
of self-dualities.  Specifically, we will:

(i) Analyze the relation between self-dualities and the existence/non-existence of phase 
transitions, and the resulting constraints for the spectrum of 
self-dual systems. We discuss extensions to situations wherein more
than one coupling constant is present and in which
a self-dual point is replaced by a self-dual line
or surface.

(ii) Detail some consequences of the constraint 
for derivatives of general quantities at and away
from the self-dual point.

(iii) Briefly discuss the general solution to the constraint discussed above.

\subsubsection{Self-duality and the existence/non-existence of
phase transitions}

A finite temperature phase transition is characterized by a non-analyticity in the
free energy of the system at the transition point (or transition line, etc.). Similarly, a zero temperature
(quantum) phase transition relates a level crossing, or an avoided level crossing, 
and non-analyticities in the ground state energy
of the quantum Hamiltonian at hand.

When present, a self-duality relates the energy levels (and thus the free energy and all related
thermodynamic quantities) at one coupling (or temperature) to those at another
coupling (or temperature). Thus, whenever a phase transition
occurs at  one value of the coupling $\lambda$ it must also occur
at the dual coupling $\lambda^*$. A corollary of the above is that 
(1) if the system exhibits an odd
number of transitions as a function of $\lambda$ then 
one transition must occur at the self-dual point $\lambda_{\sf sd}$.
Similarly, (2) if an even number of transitions are present, then
no phase transition can occur at $\lambda_{\sf sd}$.
A further consequence of self-dualities is that (3) if a system is devoid of
transitions in a particular region $\lambda_a < \lambda <\lambda_b$ then it must
also be devoid of transitions in the related dual region whose
endpoints are given by $\lambda_a^*$ and $\lambda_b^*$ 
(where $\lambda_a^*$ and $\lambda_b^*$ are
the dual counterparts to $\lambda_a$ and $\lambda_b$). Similar remarks can
be made when more than one coupling constant (and/or temperature)
are involved. In such cases self-dual points translate into 
self-dual lines or surfaces and regions devoid of singularities
appear in a higher dimensional parameter space. These statements
are simple yet  proved to be extremely potent
over many decades.

We comment on examples in which cases (1)-(3) above are, respectively, realized:

(1) As discussed in detail in earlier sections, 
the classical $D=2$  Ising model is self-dual \cite{KW}
and displays only a single transition separating the high temperature
disordered phase to the low temperature ordered phase. Thus, 
the $D=2$ Ising model orders at the self-dual inverse 
temperature given by $\beta_{c} = (\ln(1+ \sqrt{2})/2)$.
This value found by Kramers and Wannier matches, as it
must, the critical temperature found by Onsager 
in his exact solution of the  same model.

(2) The $p$-state VP model of Section \ref{sec4.1.1} exhibits, for $p>4$, three phases
and thus its self-dual point does not correspond to a point of non-analyticity \cite{ortiz}. 
(As noted earlier, for $p=2$, the VP model becomes the Ising model
of case (1) above; the same also holds true for $p=4$.) 

(3) By the use of the self-duality of the $D=3$ Ising matter coupled gauge theory, 
it can be shown \cite{nprd} that the confining
phase of this system (weak couplings) is smoothly connected to
its Higgs phase (when all couplings
are large). When the union of the region of 
phase space that is free of transitions
(as proved by the Lee-Yang theorem) 
is taken with its dual counterpart, there
is a region that is free of non-analyticities 
connecting the above two phases \cite{fradkin_shenker}.

The above three consequences, which can be appended by additional
constraints that we will elaborate below, can be applied, mutatis mutandis, 
not only to questions concerning thermodynamic phase transitions
but also to general non-equilibrium phenomena. For instance, we may consider the dynamics
derived from a self-dual Hamiltonian (whether classical 
or quantum). The equations of motion governing the system
dynamics are identical under
the interchange of a coupling constant $\lambda$ with its dual $\lambda^*$.
Consequently, both in the quantum and classical arenas, any transitions associated
with the character of the system dynamics as parameters are changed
must satisfy relations (1)-(3) when the Hamiltonian is self-dual. 

We now comment on the spectral properties of self-dual Hamiltonians. 
Consider a Hamiltonian of the form 
\begin{eqnarray}
H[\lambda] = H_{0} + \lambda H_{1}.
\label{hlh1}
\end{eqnarray}
for which a duality transformation
$\mathcal{U}_{\d} H_{0} \, \mathcal{U}_{\d}^{\dagger} = H_{1}$, $
 \mathcal{U}_{\d} H_{1} \, \mathcal{U}_{\d}^{\dagger} = H_{0}$, 
interchanges the types of bonds present in the two Hamiltonians $H_{0}$ and $H_{1}$. 
Thus, $\mathcal{U}_{d}$ is a unitary operator that
implements a self-duality of the Hamiltonian of Equation \eqref{hlh1}, i.e., 
$\mathcal{U}_{\d} H[\lambda] \, \mathcal{U}_{\d}^{\dagger}=\lambda H[1/\lambda]$ 
with a self-dual point $\lambda_{\sf sd}=1$. This implies that the eigenvalues satisfy
\begin{equation}
E_{n}(\lambda) = \lambda E_{n}(1/\lambda).
\label{evaldual}
\end{equation}

Equation (\ref{evaldual}) 
constitutes the most general constraint of self-duality
on a Hamiltonian of the type of Equation (\ref{hlh1}) \cite{kogut}.
(We allude here to the ``most general'' constraint as 
dualities encompass unitary transformations 
(thus preserving the spectrum) and Equation \eqref{evaldual}
is the sole constraint on the energy eigenvalues $E_{n} (\lambda)$ that arises from the duality.)
It follows that the free energy of the 
quantum system 
at an inverse temperature $\beta$
similarly satisfies $F_q(\beta, \lambda) = \lambda F_q(\beta, 1/\lambda)$. 
By taking derivatives of the free energy, it is seen that 
the (average) internal energy and other general thermodynamic
quantities satisfy identical relations. Relating
values of $\lambda >1$ to those with $\lambda <1$ suggests
a ``halving" of the degrees of freedom. We will, later on, explicitly see
various manifestations of this.

With the identification of the self-duality relation $\lambda \leftrightarrow \lambda^* = 1/\lambda$,
we may now invoke the likes of corollaries (1)-(3) above.
If a level crossing occurs at a point $\lambda$
then a level crossing must occur at the point $\lambda^* = 1/ \lambda$.
That is, if there exist two levels (denoted by $n$ and $m$)
that cross at a particular coupling $\lambda$: $E_{n}(\lambda) = E_{m}(\lambda)$
then it follows from Equation \eqref{evaldual} that $E_{n}(1/\lambda) = E_{m}(1/\lambda)$.
Also, if there exists two quantum phase transitions (two non-analyticities in 
$E_0(\lambda)$, as in the VP model of Section \ref{sec4.1.1}), and one happens at 
the point $\lambda_{c1}$, the second must happen at $\lambda_{c2}=1/\lambda_{c1}$, such that 
$\lambda_{c1} E_0(\lambda_{c2})=E_0(\lambda_{c1})$.
When examining the classical analogue of a zero temperature quantum system 
defined by a Hamiltonian of the form of Equation (\ref{hlh1}), the $\lambda \leftrightarrow 1/\lambda$
duality transformation translates, similar to discussions in earlier sections, 
into a (generally non-trivial) duality transformation
relating the inverse temperatures $\beta \leftrightarrow \beta^*$.
Such a case occurs, as we saw earlier, in the quantum Ising
chain (whose Hamiltonian is of the form of Equation \eqref{hlh1} and
whose classical counterpart is given by the $D=2$ Ising model). 

A relation similar to that of Equation \eqref{evaldual} trivially appears for
a Hamiltonian that is symmetric under the permutations of
more than one type of a pair of bonds. That is,
we may consider $H[\lambda_{1}, \lambda_{2}, \cdots, \lambda_{p}]= 
H_{0} + \lambda_{1} H_{1} + \lambda_{2} H_{2} + \cdots
+ \lambda_{p} H_{p}$ with different unitary operators $\mathcal{U}_{\d}$ that may 
exchange, for instance,  $H_{0}$ with $H_{i >0}$ (i.e., 
$\mathcal{U}_{\d} H_{i} \, \mathcal{U}_{\d}^{\dagger} = H_{0}$
and its inverse). In such a case, the simple extension 
of Equation \eqref{evaldual} in a higher dimensional coupling space 
is 
\begin{eqnarray}
H[\lambda_{1}, \cdots, \lambda_{p}] = \lambda_{i}  
H[\lambda_{1},\cdots, \lambda_{i-1}, 1/\lambda_{i}, \lambda_{i+1}, \cdots, \lambda_{p}].
\end{eqnarray}

We next briefly and explicitly discuss constraints appearing for classical self-dualities
where the free energies of two dual classical systems are the same
only up to an additive non-singular contribution (see Equation \eqref{first_time_statistics}). 
In such
cases the free energy of the classical system $F$ satisfies 
\begin{eqnarray}
\label{simple_f}
F(K) = F(K^*) + f (K,K^*)
\end{eqnarray}
where $f$ is a regular function
of $K$ and $K^*$. 
Differentiation of $F$ relative to temperature yields the average energy.
At the self-dual point $K= K^*=K_{c}$, the energy is given by
$E= \frac{\partial f}{\partial K}|_{K=K_{c}}/
( 1- \frac{\partial K^*}{\partial K}|_{K= K_{c}})$.
For the classical $D=2$ ($N$ sites) Ising model wherein $\exp(-2 K^*)= \tanh K$, 
we may determine the exact energy at the self-dual point, which is equal to  
$E/N= -\sqrt{2}$ \cite{bookEPTCP}.
Similarly, by differentiating Equation \eqref{simple_f} twice relative to $K$, 
we find that
\begin{eqnarray}
T^{2} C_{V}(K) - (T^{*})^{2} C_{V}( K^*) 
\left (\frac{\partial K^{*}}{\partial K}\right )^{2} =
- \left (\frac{\partial^{2} f}{\partial K^{2}}  + E(K^*) 
\frac{\partial^{2} K^*}{\partial K^{2}} \right ),
\end{eqnarray}
where $C_{V}$ is the specific heat at constant volume. 

As is well appreciated \cite{bookEPTCP}, the existence of a self-dual point
implies that whenever it is a critical point
the critical indices 
must be the same on both sides of the transition
(as they always are in any system, self dual or
not) and that the {\em amplitudes} associated
with the self-dual point must be the same
on both sides of the transition. (In 
general critical systems, the amplitudes
need not be the same on both sides of
the critical point.) Any singular contributions in the vicinity
of the critical point must be given by 
the dependence of the free energy $F$
on both sides of the transition point. 
If $K^*$
is linear in $K$ near the critical point, 
then as the derivatives of the same
function $F$ on both sides of the 
transition point will determine the 
behavior of any critical quantity,
by virtue of Equation \eqref{simple_f}, 
the critical behavior must
be the same on both 
sides of the transition. 

\subsubsection{Constraints in the absence of phase transitions}

Equation \eqref{evaldual} leads to constraints on the derivatives of the energy levels and 
all thermodynamic quantities in the absence of phase transitions.
Specifically, by differentiating both sides of Equation \eqref{evaldual}, we find that 
($E^{(j)}_n(x)=\partial^{j} E_n(x)/\partial^j x$)
\begin{eqnarray}
\label{longnd}
E_n^{(1)} (\lambda) &=&  \frac{\lambda E_n(1/\lambda) - E_n^{(1)}(1/\lambda)}{\lambda}, \nonumber
\\ E_n^{(2)} (\lambda) &=& \frac{E_n^{(2)}(1 /\lambda)}{\lambda^{3}}, \nonumber
\\ E_n^{(3)}(\lambda)& =& -  \frac{3 \lambda E_n^{(2)}(1/ \lambda) 
+ E_n^{(3)}(1/\lambda)}{\lambda^{5}}, \nonumber 
\\ E_n^{(4)}(\lambda) &=& \frac{12 \lambda^{2} E_n^{(2)}(1/\lambda) 
+ 8 \lambda E_n^{(3)}(1/\lambda) 
+ E_n^{(4)}(1/\lambda)}{\lambda^{7}}, \cdots . 
\end{eqnarray}
If $E_{n}(\lambda$) is analytic in a domain that includes the 
self-dual point $\lambda_{\sf sd} =1$, then
\begin{eqnarray}
\label{sedl}
E_{n}^{(1)} (1) &=&  \frac{1}{2} E_{n}(1), \nonumber
\\ E^{(3)}(1) &=& - \frac{3}{2} E^{(2)}(1), \nonumber
\\ E^{(5)}(1)  &=& 15 \left (E^{(2)}(1) - \frac{1}{2} E^{(4)}(1) \right ), \cdots .
\end{eqnarray}
The spectrum of the self-dual system captured by Equation \eqref{evaldual} gives
rise to equivalent (``dual'') pairs of equations from even and odd orders in 
$(\lambda-\lambda_{\sf sd})$
about the self-dual point. This manifests the aforementioned ``halving'' of the parameters
characterizing the function. The large degeneracy 
manifest in these equations enables a large number of possible 
solutions to Equation \eqref{evaldual} as we discuss next.

\subsubsection{General self-dual spectra}

The spectra that are analytic at the self-dual 
point form only a small subset of all possible solutions of Equation \eqref{evaldual}.
The self-dual point may mark a transition point between two different phases
(wherein $\{E_{n}(\lambda)\}$ are no longer differentiable to arbitrary order).
Most of the examples that we considered in this article fall into this category.
The self-dual point of the quantum Ising chain ($h=J$ in 
Equation \eqref{infinite_ising_transverse}, i.e., $\lambda = J/h =1$  at the self-dual point)
constitutes a point where a quantum phase transition occurs;
in particular, the gap between the ground state and the first excited
state of the quantum Ising chain scales as $\Delta E = 2 h |1- \lambda|$ \cite{kogut}.

We now discuss the most general possible form of the self-dual 
energies. Identical forms to those presented appear for 
all thermodynamic quantities in self-dual systems. 
Equation \eqref{evaldual} is the sole
condition imposed on the spectrum
from self-duality. 
It is easy to see, by direct substitution, that if $w_{n}(\lambda)$
is a solution to Equation \eqref{evaldual} then so is $w_{n}(\lambda) + 
\lambda w_{n} (1/\lambda)$.
Conversely, for {\em any} function $w_{n}(\lambda)$, the combination 
$w_{n}(\lambda) + \lambda w_{n} (1/\lambda)$ satisfies Equation \eqref{evaldual}.
Thus, Equation \eqref{evaldual} is 
satisfied if and only if 
\begin{eqnarray}
E_{n}(\lambda) = w_{n}(\lambda) +  \lambda w_{n}(1/\lambda),
\label{ensd}
\end{eqnarray} 
with $w_{n}$ representing arbitrary functions. 
This general solution suggests a halving
of the degrees of freedom allowed by the function
(the function formed by the sum in Equation \eqref{ensd} must be ``even'' under the interchange of 
$\lambda$ with $1/\lambda$) and trivially allows for a rich variety of forms. 
 Depending on the form of
the functions $w_{n}$ in Equation \eqref{ensd}, 
the spectra $E_{n}(\lambda)$ can be either analytic or non-analytic at 
$\lambda = \lambda_{\sf sd} =1$. Similarly, these functions 
enable transition points $\lambda^*$
where level crossing occurs (in particular, those where the ground state changes character
as the gap between the ground state and the lowest excited state vanishes) to 
be such that the energy variations to a given order are continuous or discontinuous.   

An equivalent alternate solution to Equation (\ref{evaldual}) in the 
non-analytic case is simple.  
Consider any set of functions $\{E_{n}(\lambda)\}$ 
defined on the interval $[\lambda_{\sf sd}, \infty)$, 
and from this set define the functions on the remaining segment $\lambda \in (0,\lambda_{\sf sd}]$.
This is similar to a standard complex inversion transformation 
(applied now only on the half real line 
$\lambda \ge 0$) with the additional scale factor of $\lambda$ on the 
right-hand side of Equation \eqref{evaldual}.
We can make these functions continuous and {\em non-differentiable} at $\lambda=\lambda_{\sf sd}=1$
by defining functions on $[\lambda_{\sf sd}, \infty)$ such that the function has an
ill-defined derivative on reflection (Equation (\ref{evaldual})) as the left 
hand and right hand derivatives (if they exist)
do not match.
For instance, if for a particular level (say, $n=0$) the eigenvalue behaves as
 $E_{0}(\lambda) = \sqrt{\lambda-1}$
for $ \lambda \ge 1$ then we will need to set  $E_{0}(\lambda) = \sqrt{\lambda(1-\lambda)}$ for 
all $0 < \lambda  \le 1$ in order to satisfy Equation (\ref{evaldual}). 
Similar constructs can be implemented for all levels $n$.
The same may also be done for higher order derivatives
(for, e.g., a divergent second order derivative at $\lambda=1^{+}$).

\subsection{Correlation functions}
\label{sec9.2}

The unitary character of the duality transformation emphasized in this article
(Equation \eqref{d_as_u}) further allows for a related, very
simple but powerful, relation concerning the correlation lengths 
(and times). To establish this relation, we can employ the 
transfer matrix formalism.  We consider a classical system in which along
one spatial (or temporal) direction, the system has length $N_{1}$
and a corresponding transfer matrix $T$. As detailed in Section \ref{classical&quantum} and Section
\ref{sec8},  
the transfer matrix in the classical problem may be related to a quantum Hamiltonian
(and viceversa). If two quantum Hamiltonians $\tilde{H}(\tilde{\phi})$ and $H(\phi)$ 
are dual to each other (and thus share the same spectrum), the time evolution of  
the dual fields $\tilde{\phi}$ and $\phi$ are identical (the corresponding eigenstates
of the two Hamiltonians evolve with identical frequencies). The same holds true
for the imaginary time evolution of two sets of fields. In particular, 
the gap $\Delta E$ between
the ground state and the first excited state (which determines the asymptotic long time
correlations of the system) is the same in both systems. The same, of course, holds true
not only for the long time limit but for any time separation. 

Within the quantum to classical correspondence \cite{bookEPTCP},
an imaginary time axis in the quantum problem is replaced by
an additional spatial axis in a static equilibrium thermodynamic
classical problem. As detailed in earlier sections,
 the eigenvalues $\{ E_{n} \}$ of the quantum Hamiltonian $H$ 
 are replaced
by the eigenvalues $\{\lambda_{n} \}$ of the classical transfer matrix $T$.
Commonly, the transfer matrix eigenvalues can be numbered in list of descending absolute
values $|\lambda_{0}| \ge |\lambda_{1}| \ge |\lambda_{2}| \ge \cdots$.
For finite size transfer matrices (as these commonly arise in $D=1$ classical systems,
or 0+1 space-time dimensional quantum systems), the Perron-Frobenius theorem
guarantees that the largest eigenvalue ($\lambda_{0}$) is non-degenerate. 
In higher dimensional systems (i.e., classical $D>1$ dimensional 
systems), the eigenvalues can become degenerate (and indeed
do become degenerate) at critical transitions.
In terms of the transfer matrix
eigenvalues $\{\lambda_{n} \}_{n=0}^{\dim{T}-1}$
(with $\dim{T}$ being the dimension of the transfer
matrix $T$),
 the partition function reads 
 \begin{eqnarray}
 \label{transfer_Z}
{\cal Z} = \sum_{n=0}^{\dim{T}-1} \lambda_{n}^{N}.
 \end{eqnarray}
For $D=1$ systems, the transfer matrix can have a finite number
of eigenvalues while in higher dimensions the number of eigenvalues 
(the size of the transfer matrix) is infinite, in the thermodynamic limit
in which the system is of infinite extent along all spatial directions.
The correlation function is completely determined by the eigenvalues
and eigenvectors of the transfer matrix. In particular, the inverse correlation length
determining the correlations at large distances
is given by
\begin{eqnarray}
\xi^{-1} = - \ln (\lambda_{1}/\lambda_{0}).
\label{transfer-xi}
\end{eqnarray}
(Within the corresponding quantum problem in imaginary time, 
an analogous relation relates the correlation time with the 
gap between the first excited state and the ground state; the transfer
matrix eigenvalues of the classical problem are related to exponentials of
the energies in the quantum problem.)
As emphasized in this article, a duality between two systems implies the 
equivalence of the spectrum of the two theories (up to an overall constant
factor). The partition functions of dual systems are equal to one
another.  In what follows, we will denote the $n$th transfer matrix eigenvalues of
systems (systems (1) and (2)) by $\lambda_{(1);n}$ and $\lambda_{(2);n}$.
{}From Equation \eqref{classd_from_q} we have
\begin{eqnarray}
{\cal Z}_{1} = \sum_{n=1}^{\dim{T}-1} \lambda_{(1);n}^{N} = A \  {\cal Z}_{2} =
A \sum_{n=0}^{\dim{T}-1} \lambda_{(2);n}^{N}  .
\end{eqnarray}
Amongst other things, this implies that the correlation lengths
of the two systems are the same, as by Equation (\ref{transfer-xi})
\begin{eqnarray}
\xi_{(1)}^{-1} = -  \ln(\lambda_{(1);1}/\lambda_{(1);0})
= - \ln(\lambda_{(2);1}/\lambda_{(2);0}) = \xi_{(2)}^{-1}.
\end{eqnarray}
Analogous relations follow not only for the asymptotic large distance correlation length
but rather for all correlation lengths set by $[- \ln (\lambda_{n}/\lambda_{0})]$. 

\subsection{Fermionization as a duality}
\label{sec9.3}

As is well known, the quantum {\it exchange} statistics of elementary
degrees of freedom (whether fermionic, bosonic, or spin) can, quite generally, 
be readily transformed. For instance, spin $S=1/2$ $SU(2)$ operators can be mapped onto 
spinless fermions by a transformation known as the Jordan-Wigner transformation 
\cite{JW}, and  generalizations to higher spin, or even arbitrary algebras 
exist \cite{GJW}.  The generalized Jordan-Wigner mappings represent ingenious constructs 
based on non-local isomorphic mappings between the degrees of freedom in question \cite{GJW}.  Since these 
dictionaries are independent of any Hamiltonian (or action) they generically fail 
to preserve locality by introducing strings in the interactions, in particular in spatial dimensions $d>1$. 
Amongst many other benefits, these mappings readily allows for 
exact solutions of many \(d=1\) dimensional models, including the XY 
and  transverse field quantum Ising models.  This is so as the transformations map
spin quadratic forms onto non-interacting fermionic terms (i.e., quadratic forms), 
that may be exactly solved by algebraic means. 

By contrast, dualities  are {\it model specific transformations}
that always preserve locality (in the Hamiltonians), and are designed to take
the most advantage of each model's peculiarities. To see how transmutation of 
statistics is possible, notice for example that bond algebras of fermionic systems 
feature only (sums of) 
{\it even products} of the elementary fermionic degrees of freedom. 
Since such even products behave very differently from the fermions
themselves, they could in principle be mimicked by bonds of models
featuring other types of elementary degrees of freedom. It follows from the formalism of
Section \ref{sec3.5}, in the absence of gauge symmetries, 
 that the {\it dual variables} that arise from a {\it 
duality connecting models with different statistics} will afford a 
dictionary 
mediating the two statistics. Dualities can also help improve the efficiency of 
standard fermionization techniques in dimension \(d\geq 2\), 
because different dual representations
of one and the same model can behave very differently under generalized
Jordan-Wigner transformations. 

The outline of this section is as follows:
{(i)} First, we will analyze
the Jordan-Wigner transformation
through the prism of bond algebras
and illustrate that it is not at
all necessary to consider non-local
transformations \cite{anfossi_montorsi}. Rather, as alluded to
above, we may focus on the bond algebras
of local bonds (interactions) to illustrate
a mapping from one local Hamiltonian to another. The 
Jordan-Wigner dictionaries arise as dual variable mappings. 
{(ii)} We will then show that the generalization of
a higher dimensional Jordan-Wigner transformation
that maps nearest-neighbor bilinear interactions
of different underlying statistics
to one another is strictly impossible 
on lattices that have closed  loops. In other words, 
there is no local fermionization mapping, that preserves 
the number of degrees of freedom, in dimensions 
$d \geq 2$. There are, of course, four-body (plaquette) interactions 
where a local mapping is still possible. 
{(iii)} Finally, we will show that while
mappings involving local nearest-neighbor
bilinears, and preserving the number 
of degrees of freedom, are impossible, mappings 
involving gauge-reducing dualities 
are indeed possible. We illustrate
this point by fermionizing the \(d=2\) Ising model via its duality to the 
\(\mathbb{Z}_2\) Ising gauge model. Nonetheless, 
dualities that connect models with different statistics 
are constrained  by  quantum-mechanical considerations
{\it that depend on the space-time dimension \(D\)}, and they become
increasingly challenging as \(D\) increases. We will show
an example of this in a family of models that
can be fermionized in \(d=1\) dimensions, but 
cannot be fermionized in higher dimensions.

We would like to mention that $\it bosonization$, a process which not only 
effects exchange statistics transmutation to canonical bosons but also modifies 
the {\it exclusion} statistics \cite{GJW}, could also be in principle 
interpreted as a duality mapping  but it is a less precisely defined 
mathematical transformation which requires some vacuum regularization process. 

\subsubsection{Transmutation of statistics
via bond algebras: Dual variables and the Jordan-Wigner transformation}
\label{stba}

In this section we want to derive the standard Jordan-Wigner mapping as 
a duality transformation.  As
emphasized throughout this article, practically what matters is
not the algebra of the elementary degrees of freedom but rather the algebra 
of the bonds that appear in the Hamiltonian. Consider the simple model of 
non-interacting (spinless) fermions in $d=1$, characterized by the 
$N$ sites tight-binding Hamiltonian
\begin{eqnarray}
\label{Hferm}
H_{\sf fermion} = \lambda \sum_{i=1}^{N-1} 
(c^\dagger_{i}c^{\;}_{i+1}+c_{i+1}^{\dagger} 
c^{\;}_{i}),
\end{eqnarray}
where $c^{\;}_{i}$ and $c_{i}^{\dagger}$ annihilate or create a
spinless fermion at site $i$, and 
\begin{equation}\label{car}
\{c^{\;}_i,\ c^{\;}_j\}=0,\ \ \ \ \ \ \{c^{\;}_i,\ c^\dagger_j\}=\delta_{i,j}
\end{equation}
(\(\{A,B\}=AB+BA\)).  The bonds $c_{i+1}^{\dagger}c^{\;}_{i}$ 
(and their Hermitian conjugates) are 
nilpotent, \((c_{i+1}^{\dagger}c^{\;}_{i})^2=0\).
However, unlike the elementary fermionic degrees of freedom,
bonds of the type 
\(c_{i+1}^{\dagger}c^{\;}_{i}\) (with $1<i<N-1$) each fail
to commute with only three bonds;
\begin{equation}\label{bonds_fermions}
c_{i+1}^{\dagger}c^{\;}_{i}\ \ \ \ \mbox{fails to commute with}\ \ \ \
c^\dagger_ic^{\;}_{i-1},\ \ c^\dagger_{i}c^{\;}_{i+1},\ \ c^\dagger_{i+2}c^{\;}_{i+1}.
\end{equation}
Boundary bonds fail to commute with only two bonds.
(That is, $c_{2}^{\dagger} c^{\;}_{1}$ does not commute with $c^{\dagger}_{1} c^{\;}_{2}$ or
$c_{3}^{\dagger} c^{\;}_{2}$ and, similarly, $c_{n}^{\dagger} c^{\;}_{n-1}$ does not
commute with $c^{\;}_{n} c_{n-1}^{\dagger}$ nor $c^{\dagger}_{n-1} c^{\;}_{n-2}$.)

A model of spin \(S=1/2\) degrees of freedom  that,  we show next, is 
closely related to $H_{\sf fermion}$ is the isotropic XY model
\begin{eqnarray}
\label{Hspin}
H_{\sf XY} = \lambda
\sum_{i=1}^{N-1} (\sigma_{i}^{+}\sigma_{i+1}^{-}+ \sigma_{i+1}^{+}\sigma_{i}^{-}),
\end{eqnarray}
with  $\sigma^{\pm} = (\sigma^{x} \pm i \sigma^{y})/2$ which satisfy the Pauli algebra
\begin{eqnarray}
[\sigma_{i}^{+},\ \sigma_{j}^{-}] =  \sigma_{i}^{z}\ \delta_{i,j}, 
~~~~~[\sigma_{j}^{z},\ \sigma_{i}^{\pm}] =  \pm  2\sigma_{i}^{\pm}\ \delta_{i,j}.
\end{eqnarray} 
As for the spinless fermions, each bond is nilpotent, 
\((\sigma_{i+1}^{+}\sigma_{i}^{-})^2=0\). Similarly,  bonds that share no common site commute, while 
(for $1<i<N-1$)
\begin{equation}\label{bonds_spinons}
\sigma_{i+1}^{+}\sigma_{i}^{-}\ \ \ \ \mbox{fails to commute with}\ \ \ \
\sigma^+_{i}\sigma^-_{i-1},\ \sigma^+_{i}\sigma^-_{i+1},\ \ 
\sigma^+_{i+2}\sigma^-_{i+1} .
\end{equation}
Similar to the fermionic case, the boundary bonds (those with $i=1$ and $i=(N-1)$ above)
fail to commute with only two bonds.

A direct comparison of Equations \eqref{bonds_fermions} and \eqref{bonds_spinons},
together with the fact that both types of bonds are nilpotent, suggests that
the mapping
\begin{equation}\label{fermasd}
c_{i+1}^{\dagger}c^{\;}_{i}\dual \sigma_{i+1}^{+} \sigma_{i}^{-},
\end{equation}
(and the corresponding Hermitian-conjugate relation)
may define a duality isomorphism. The easiest way to check the
assertion is to 
assume that this is indeed the case, and use \(\Phi_\d\) as 
explained in Section \ref{sec3.5} to compute dual 
variables. Then, when written in terms of dual variables, it is easy
to check that \(\Phi_\d\) defines a bond algebra isomorphism. 
{\it These dual variables turn out to define the Jordan-Wigner 
transformation.}

The starting point consists in writing   \(c^\dagger_i\) in terms of bonds.
To this end, we need to {\it extend the bond algebra} by adding two more
generators, \(c^{\;}_1\) and \(c_1^\dagger\). We can then compute
\(c^{\;}_i,\ i=2,3,\cdots,N\), recursively from the relation
\begin{equation}\label{toiterate}
[c^{\;}_{i-1},\ c_{i-1}^\dagger c^{\;}_i]=c^{\;}_i,
\end{equation} 
so that \(c_i\) can be written as a nested multiple commutator of bonds.
The next step is to extend the action of \(\Phi_\d\) to the new generators.
We propose
\begin{equation}\label{mapping_at_boundary}
c^\dagger_1\dual\sigma^+_1, \ \ \ \ c^{\;}_1\dual\sigma^-_1, 
\end{equation}
to be consistent with Equation \eqref{fermasd}, and it follows that
we have to extend the bond algebra of the XY model as well
by adding the two generators \(\sigma^+_1,\ \sigma^-_1\).
Now we can proceed to compute the dual variables. 

Equations \eqref{fermasd},
and \eqref{mapping_at_boundary} turn the recursion relation Equation \eqref{toiterate}
into 
\begin{eqnarray}
c^{\;}_1&\dual& \hat{c}^{\;}_1=\sigma^-_1 \\
~[c^{\;}_{i-1},\ c_{i-1}^\dagger c^{\;}_i]=c^{\;}_i &\dual& 
[\hat{c}^{\;}_{i-1},\ \sigma^+_{i-1}\sigma^-_i]=\hat{c}^{\;}_i,\nonumber
\end{eqnarray}
that can be solved to yield
\begin{equation}
c^{\;}_i\dual\hat{c}^{\;}_i=\prod_{j=1}^{i-1}(-\sigma^z_j) \, \sigma^-_i , \ \ 
c^{\dagger}_i\dual\hat{c}^{\dagger}_i=\prod_{j=1}^{i-1}(-\sigma^z_j) \, \sigma^+_i
\label{JWisod}
\end{equation}
(for instance, \(\hat{c}^{\;}_2=[\hat{c}^{\;}_{1},\ 
\sigma^+_{1}\sigma^-_2]=[\sigma^-_1,\ 
\sigma^+_{1}\sigma^-_2]=-\sigma^z_1\sigma^-_2\)). 
{\it This is nothing else than the Jordan-Wigner transformation}. 
In particular, the fermion number operator $n_i$ transforms as 
\begin{equation}
n_i=c^\dagger_i c^{\;}_i\dual\hat{c}^{\dagger}_i\hat{c}^{\;}_i=
\frac{\mathbb{1}+\sigma^z_i}{2} . 
\end{equation}
The fact that the dual variables \({\hat{c}}^\dagger_i,\ {\hat{c}}^{\;}_i\) satisfy the 
fermionic algebra of Equation \eqref{car} confirms that the mapping 
\(\Phi_\d\) of Equations \eqref{fermasd} and \eqref{mapping_at_boundary}  
defines an isomorphism of (extended) bond algebras.
We see that {\it fermionization is a process local in the bonds}, 
and that the non-local structure of the Jordan-Wigner transformation 
has the same origin as that of other dual variables.  Notice that one can define the 
inverse duality isomorphism 
\begin{equation}\label{fermasd}
\sigma^+_1\idual c^\dagger_1,\ \ \ \
\sigma_{i+1}^{+} \sigma_{i}^{-} \idual c_{i+1}^{\dagger}c^{\;}_{i},
\end{equation}
by inspection of Equation \eqref{JWisod}, and express  spin variables in terms of 
fermions
\begin{equation}\label{invJWisod}
\sigma^-_i\idual\hat{\sigma}^-_i=\prod_{j=1}^{i-1}(1-2n_j) \, c^{\;}_i=
e^{i\pi\sum_{j=1}^{i-1}n_j} \ c^{\;}_i ,
\end{equation}
with the corresponding Hermitian-conjugate relation.

There is some flexibility in the definition of the Jordan-Wigner transformation
that reflects  two independent facts. First, we could
have extended the bond algebra by adding the boundary bonds
\(c^{\;}_N,\ c^\dagger_N\) (and \(\sigma^-_N,\ \sigma^+_N\)), with the resulting 
string operator running from site \(i+1\) to site \(N\) in Equations \eqref{JWisod} and
\eqref{invJWisod}. Second, the boundary term mapping 
 is only defined up to a phase term $\eta$. We could have set
\(c^\dagger_1\dual\eta \, \sigma^+_1\), where \(\eta\eta^*=1\), with the corresponding dual 
variables displaying this overall phase. 
Moreover, one can extend these ideas to establish a duality-based 
derivation of generalized Jordan-Wigner transformations \cite{GJW}. 

\subsubsection{Non-locality and fermionization in $d \geq 2$ dimensions}

We next ask whether nearest-neighbor spin and fermion models
can be related on arbitrary lattices. On general lattices on which 
{\it closed loops} may be drawn, the bond algebras become richer and, in general, prohibit
such a mapping between the simplest choice of elementary bonds involving two
nearest-neighbor spins and two nearest-neighbor fermions.  
The natural extension of the mapping
of Equation \eqref{fermasd} reads
\begin{equation}\label{fermasd_highd}
c^\dagger_\r c^{\;}_{\r'} \qdual \sigma^+_\r \sigma^-_{\r'},
\end{equation}  
where the vectors \(\r, \r'\) now label the endpoints of links on 
{\em an arbitrary $d$-dimensional lattice}.
As we will show below, Equation \eqref{fermasd_highd} generally
fails to define a bond algebra isomorphism  whenever the lattice contains 
at least two closed loops \(C,\ C'\) that share a link.
To see this, let \(\r_1,\r_2,\cdots, \r_N\equiv\r_1\) 
label the sites in \(C\), and \(\r_1',\r_2',\cdots, \r_M'\equiv\r_1'\) the 
sites in \(C'\), listed in order of appearance 
(for some orientation of the loops), and let
\begin{equation}
\r_1=\r_1',\ \ \ \ \ \ \ \ \ \r_{N-1}=\r_{M-1}' ,
\end{equation}
denote the endpoints of the shared link. Then, on one hand, we can compute
the following nested anti-commutator along \(C\),
\begin{eqnarray}\label{xy_loop_indep}
S_{\r_1,\r_{N-1}} &\equiv &
\{\{\{\cdots\{\{\sigma^+_{\r_1}\sigma^-_{\r_2},\ \sigma^+_{\r_2} \sigma^-_{\r_3}\},\ 
\sigma^+_{\r_3} \sigma^-_{\r_4}\}\cdots\},\ \sigma^+_{\r_{N-2}} \sigma^-_{\r_{N-1}}\},\ 
\sigma^+_{\r_{N-1}} \sigma^-_{\r_1}\}  \nonumber \\
&=& \frac{1}{2}(1-\sigma^z_{\r_1}\sigma^z_{\r_{N-1}}),
\end{eqnarray}
and the outcome is in fact {\it independent of the loop} (the same computation along 
\(C'\) would have returned the same result).
In contrast, the corresponding nested anti-commutator in terms of fermionic bonds
does depend on the loop,
\begin{eqnarray}
F_C &\equiv& \{\{\{\cdots\{\{c_{\r_1}^\dagger c^{\;}_{\r_2},\ c_{\r_2}^\dagger c^{\;}_{\r_3}\},\ 
c_{\r_3}^\dagger c^{\;}_{\r_4}\}\cdots\},\ c^\dagger_{\r_{N-2}} c^{\;}_{\r_{N-1}}\},\ 
c^\dagger_{\r_{N-1}} c^{\;}_{\r_1}\} \nonumber\\
&=& (n_{\r_{N-1}}-n_{\r_1})^2\prod_{m=2}^{N-2}(2n_{\r_m}-1)
\end{eqnarray}
with \(n_\r=c^\dagger_\r c^{\;}_\r\). 
If we compute the same quantity along \(C'\), the result reads
\begin{equation}
F_{C'}=(n_{\r_{M-1}'}-n_{\r_1'})^2\prod_{m=2}^{M-2}(2n_{\r_m'}-1)=
(n_{\r_{N-1}}-n_{\r_1})^2\prod_{m=2}^{M-2}(2n_{\r_m'}-1) ,
\end{equation}
so that \(F_C\neq F_{C'}\). On the other hand, it follows from Equations
\eqref{fermasd_highd} and \eqref{xy_loop_indep} that
\begin{equation}
F_C, \ F_{C'}\qdual S_{\r_1,\r_{N-1}}.
\end{equation}

This shows that the mapping of Equation \eqref{fermasd_highd} cannot
define a bond algebra isomorphism (a fermionization of the 
XY model in $d\geq 2$), since it defines a many-to-one mapping. It is a good illustration
of the type of problems that arise in attempts at fermionizing spin models
in more than one dimension, while preserving locality \cite{highdJW} and the
dimension of the state space \cite{verstraete_cirac}.  
Thus, our circle of ideas has closed on itself. We see how a common (and slightly imprecise) lore, relating
the existence of closed loops with the impossibility of  transmutation of statistics 
in a direct physical way, is mathematically realized within the bond
algebraic approach in the above case.
 
One cannot exclude the existence of
other mappings between systems having longer range interactions
and/or being of a simpler form in other representations. For example, 
examples in which more complicated spin interactions
(i.e., not those involving $S=1/2$ spins on nearest-neighbor sites)
may admit a transmutation of statistics. 
Each proposed isomorphism that may
replace Equation \eqref{fermasd_highd} can be verified
(and more generally excluded) by examining the bond algebras of the 
candidate dual systems. This is the subject of the next section.  In particular, 
we will describe a different kind of fermionization scheme that exploits 
gauge-reducing dualities, thus relating two systems with {\it different number of 
degrees of freedom}.

\subsubsection{Dualities and fermionization in 
\(d\geq2\) dimensions}

The result of the last section is a rigorous manifestation of a 
general obstruction to mapping nearest-neighbor
spin Hamiltonians to local fermionic models, and {\it viceversa}, 
in more than one spatial dimension \cite{highdJW}, 
while preserving the dimension of the (Hilbert) state space.  
There are some interesting exceptions, however, 
where the obstructions to higher dimensional fermionization
have been overcome. 
These include the fermionization of (i) the POC model \cite{chen} of Section \ref{sec5.1},
(ii) the $d=2$ dimensional Kitaev's honeycomb spin $S=1/2$ model \cite{Kitaev2006,pachosannals} 
(and extensions thereof) of Section \ref{sec3.6}, 
and (iii) some general related models discussed in Reference \cite{galitski}. 
A mapping between 
{\em local} fermionic models and local spin models is important in practice.
Some treatments  \cite{verstraete_cirac,galitski}
invoke the need to add auxiliary fermions (spins) to map any local fermionic (spin)
model to a {\it sector of a local spin (fermionic)} model, that acts in general on a larger state 
space. 

Bond-algebraic dualities, that are always local transformations,
afford practical alternative (and more general)
approaches to the problem of fermionization in higher dimensions, that may 
or may not involve a change in the number of degrees of freedom. First there
is the basic approach of looking directly for a bond-algebraic duality
that connects the fermion/spin model of interest to a dual spin/fermion model,
as we did at the beginning of this section. At times it is more 
convenient to look for a duality of the model of interest to a dual model
that features the same type of degrees of freedom, but displays interaction terms
that are amenable to standard fermion/spin or spin/fermion transformations.
In either case, a change in the number of degrees of freedom may occur naturally
as a consequence of a gauge-reducing duality. This affords a natural picture
where local fermionization becomes possible, perhaps at the cost of introducing gauge
symmetries. We next illustrate this idea in \(d=2\) dimensions.

The quantum Ising model in \(d=2\) (see Equation \eqref{anyd_ising}) has a large, flexible
bond algebra generated by 
\begin{equation}
\sigma^z_\r\sigma^z_{\r+\i},\ \ \ \ \sigma^z_\r\sigma^z_{\r+\j},
\ \ \ \ \sigma^x_\r.
\end{equation}
This bond algebra contains the bonds of most other \(d=2\) dimensional
spin models of interest (we already
exploited this fact in Section \ref{sec3.6} to find new dualities
for the Heisenberg model). It follows that if we can fermionize the Ising model,
we can translate that fermionization to many other spin models, including the Heisenberg model, 
but it is well 
known that any attempt to rewrite it in terms of fermionic operators (while
preserving the dimension of the state space) returns a {\it non-local} fermionic
Hamiltonian (this is to be expected in the light of the discussion of the
previous section). As explained in Section \ref{sec3.12},
the \(d=2\) dimensional Ising model is dual to the \(d=2\) dimensional
\(\Z_2\) gauge theory defined in Equation \eqref{ising_gauge}. Seen in reverse,
the mapping of Equation \eqref{gauge_to_ising} that establishes this duality,
\begin{eqnarray}
\sigma^z_{\r-\j}\sigma^z_\r \dual \sigma^x_{(\r,1)},\ \ \ \ \ \
\sigma^z_{\r-\i}\sigma^z_\r \dual\sigma^x_{(\r,2)},\ \ \ \ \ \
\sigma^x_\r \dual B_{(\r,3)},
\label{gauge_to_ising_in_reverse}
\end{eqnarray}
can be understood as a prescription to turn the Ising model (and any other
model whose bonds can be written in terms of Ising bonds) into a model
with gauge symmetries, that is identical to the Ising model 
{\it if projected onto the sector of gauge invariant states}.
{}From a different perspective, the duality of Equation 
\eqref{gauge_to_ising_in_reverse}
introduces {\it in a natural way} one extra auxiliary spin (in the language
of References \cite{galitski,verstraete_cirac}) for each spin in 
the original model.

The advantage of this approach is that {\it the dual \(\Z_2\) gauge theory can be 
fermionized} straightforwardly
by a Jordan-Wigner transformation, to read
\begin{eqnarray}\label{fermionic_gauge}
H_{\sf G}&=&\sum_\r\ \left[\bar{n}_{(\r,1)}+\bar{n}_{(\r,2)} \right.\\
&+&\lambda(c^\dagger_{(\r,1)}-c^{\;}_{(\r,1)})
\left.(c^\dagger_{(\r+\i,2)}+c^{\;}_{(\r+\i,2)}) (c^\dagger_{(\r+\j,1)}+c^{\;}_{(\r+\j,1)})
(c^\dagger_{(\r,2)}-c^{\;}_{(\r,2)})\right],\nonumber
\end{eqnarray}
where we have introduced \(\bar{n}_{(\r,\nu)}=1-2c^\dagger_{(\r,\nu)}c^{\;}_{(\r,\nu)}\),
that has the simple properties 
\begin{equation}
\bar{n}^2_{(\r,\nu)}=1,\ \ \ \ \ \ \bar{n}_{(\r,\nu)}
c_{(\r',\mu)}\bar{n}_{(\r,\nu)}=(1-2\delta_{\r,\r'}\delta_{\nu,\mu})
c_{(\r',\mu)}.
\end{equation}
In terms of fermions, the gauge symmetries (constraints) read
\begin{equation}
G_\r=\bar{n}_{(\r,1)}\bar{n}_{(\r,2)}\bar{n}_{(\r-\i,1)}\bar{n}_{(\r-\j,2)}.
\end{equation}
We can combine this result with the results of Section \ref{sec3.6} to
obtain a fermionization of the \(d=2\) quantum Heisenberg model.

Let us close this section with two additional examples of fermionization. 
The fermionization 
of Kitaev's honeycomb model \cite{Kitaev2006,pachosannals}, discussed
briefly near the end of Section \ref{sec3.6}, can  be 
achieved by (i) bond-algebraic techniques \cite{3detc}, while taking note of local
symmetries or (ii) via a special projective method from an extended Hilbert space
(Kitaev's original solution \cite{Kitaev2006}) or, alternatively, via a brute force high dimensional
Jordan-Wigner transformation that leads to a local Hamiltonian due to the presence of
the local symmetries of the model \cite{nussinov_chen}.  Another new example is afforded
by the XM model of Section \ref{sec4.4}, that can be fermionized to read
\begin{eqnarray}
H_{\sf XM}&=&-\sum_\r\ \left [ h\ \bar{n}_\r \right .\\
&+&\left . J (c^{\;}_{\r-\i}+c^\dagger_{\r-\i})
(c^{\;}_{\r-\i+\j}-c^\dagger_{\r-\i+\j})
(c^{\;}_{\r}+c^\dagger_{\r})(c^{\;}_{\r+\j}-c^\dagger_{\r+\j}) \right ].\nonumber
\end{eqnarray}

\subsection{Self-dualities and quantum integrability}
\label{sec9.4}

Two self-dual models   known to any physicist,  
electromagnetism without sources and the quantum Ising chain, 
happen to be integrable as well.  Since the presence of an 
exact self-duality
is a rather unusual property in itself, one is naturally tempted to
conjecture a connection between quantum integrability and self-duality. 

We have seen by now more than enough examples of 
self-dual models that are (most likely) {\it not}  integrable
to grant that the previous argument is without force. That is not,
however, quite the end of the story. As it turns out, one of the
few known criteria for quantum integrability, the Dolan-Grady relations
\cite{dolan_grady}, 
is somewhat connected to self-duality, in that those relations
may in principle be more easily satisfied by self-dual models. 
In fact, of the very few (all one-dimensional) models known to satisfy
these relations, the most relevant one is the quantum Ising chain.
Closer scrutiny, however, seems to make very clear that the Dolan-Grady
relations are essentially unrelated to self-duality. Let us explain
this more clearly. 

Assume that a Hamiltonian can be partitioned
into two pieces $A$ and $B$, 
\begin{eqnarray}
H=A+\lambda B,
\end{eqnarray}
such that \(A\) and
\(B\) satisfy the {\it Dolan-Grady relations}
\begin{eqnarray}\label{dg}
[A,[A,[A,B]]]=c[A,B],\ \ \ \ \ \ [B,[B,[B,A]]]=c[B,A],
\end{eqnarray}
with $c$ some c-number.
It was shown in \cite{dolan_grady}, in a remarkable {\it 
tour-de-force}, that the constraints \eqref{dg} alone
suffice to guarantee that \(H\) is a member of a
family with an infinite number of conserved charges (which can furthermore be
explicitly written down in terms of \(A\) and \(B\)). Clearly,
if \(H\) is self-dual under an exchange of \(A\) and \(B\), 
so that \(A=\mathcal{U}_\d B \, \mathcal{U}_\d^\dagger\), then
either Dolan-Grady relation implies the other. 

In this sense, relations \eqref{dg}
may be in principle more easily satisfied by self-dual models.
But other than that, there is no reason to believe that 
self-duality is related to quantum integrability in any deeper way.
First and foremost, the infinite set of conserved charges exist
whenever \eqref{dg} holds, independently of the presence
or absence of self-duality \cite{baseilhac}.  
Moreover, a chain of coupled harmonic oscillators (\(d=1\)
phonons) afford an example of a model  which is both exactly 
solvable  and self-dual, and yet it does not satisfy neither
relation in \eqref{dg}, which shows that the latter do not constitute
a necessary condition for integrability either. It is important to 
notice also that the Ising chain has {\it two} 
finite, self-dual renditions,
\begin{eqnarray}
A&=&\sum_{i=1}^{N-1}\sigma^z_i\sigma^z_{i+1}+\sigma^z_N,\ \ \ \ \ \
B=\sum_{i=1}^{N}\sigma^x_i;\nonumber\\
A&=&\sum_{i=1}^{N-1}\sigma^z_i\sigma^z_{i+1},\ \ \ \ \ \ \ \ \ \ \ \ \ \,
B=\sum_{i=1}^{N-1}\sigma^x_i, 
\end{eqnarray}
{\it but only the latter satisfies the Dolan-Grady relations}. 

It seems safe to conclude that self-duality and 
quantum integrability are quite independent properties, a fact
that highlights once again
the importance of self-dualities as non-perturbative
probes of strongly-coupled models.   

\subsection{Duality, topological quantum order, and dimensional reduction}
\label{sec_tqo_dim}

In recent years there has been much interest in topological quantum 
order  \cite{wenbook, toric_code,tqo}. 
In topologically ordered systems, the state of the system cannot be characterized by 
local measurements but rather by topological quantities
\cite{toric_code,wenbook}.  One of the main hopes further driving interest
in those systems has been that systems characterized
by non-local order will be immune to local perturbations and thus quantum 
information may be protected for sufficiently long times. 
One of the most important properties of low temperature topological quantum
order is indeed its robustness to local perturbations \cite{toric_code,nsf}. 
As demonstrated in \cite{tqo}, several models harboring that order
reduce, {\em via a bond algebraic duality}, to one-dimensional Ising chains with short-range interaction. 
This suggested the possibility of short autocorrelation
times in these particular models, 
and  several new ones that have been devised and investigated since. 
One of the consequences of this dimensional reduction borne by bond-algebraic dualities 
is that these systems possess memory times that are finite  (i.e., system size 
independent) at all temperatures. This is so as the memory times $\tau$ gleaned from 
(time $t$) autocorrelation functions such
as $\langle Z(0) Z(t) \rangle \sim \exp(-|t|/\tau)$ with $Z(t)$ a quantum bit (qubit) operator are reduced,
via a bond algebraic duality, to the autocorrelation function of a 
local quantity in a lower dimensional system. As the dual lower dimensional system 
remains ergodic at all non-zero temperatures (and thus harbors a finite autocorrelation time),
the memory time in the higher dimensional topologically ordered system
always remains finite as well. This phenomenon is known as {\it thermal fragility} \cite{3detc}.  

There have been further suggestions that 
measurement of  non-local entanglement may detect (and, possibly, even quantify) 
topological order 
\cite{tee1,tee2}. These ideas seem exciting and some  results
have recently appeared in one dimensional systems \cite{1dtqo}. We would like to point out, however,  
that in general, there is no single measure of entanglement that uniquely characterizes 
the quantum state of a system as exemplified by discussions of  {\it generalized entanglement} \cite{barnum1}. 
In systems that exhibit conventional  ``Landau orders'', global symmetries link degenerate states to
one another (and, in particular, link the degenerate ground states to one another).
Let us label any such orthonormal ground states by 
$\{ |g_{\alpha} \rangle \}$.
It was suggested in a series of works \cite{tqo,tqo1} that what  
differentiates such topologically
ordered states from conventional states is the existence of ``$d$-dimensional gauge-like 
symmetries'' \cite{tqo,batista_nussinov}. We briefly comment on particular aspects of these \cite{tqo,tqo1}. 
Symmetry operators ($T_{\alpha \beta}$) may 
connect topologically ordered ground states to one another,
\begin{eqnarray}
T_{\alpha \beta} | g_{\alpha} \rangle = | g_{\beta} \rangle.
\end{eqnarray}
The operators $\{T_{\alpha \beta} \}$ realize a group 
that characterizes 
classes of topologically ordered states. (Moreover, these symmetries ensure that
topological order exists at non-zero temperatures \cite{tqo,tqo1}.) In the jargon 
of generalized entanglement \cite{barnum1}, 
the topologically ordered states are entangled relative to local observables.
Measures of entanglement such as topological entanglement entropy
\cite{tee1,tee2} are, on their own, insufficient  for the 
determination of topological quantum order \cite{tqo,tqo1}. The group theoretical
classification of  $d$-dimensional gauge-like symmetries, however, does enable a natural framework
for the analysis of systems with topological order. 
One would like, of course,  to have a generalized order parameter that may ascertain the existence
of phase transitions in topologically ordered systems. 

Dualities are of paramount importance in such systems as they enable the construction 
of precisely such order parameters when they exist. In those cases its determination 
is based
on the following maxim: 
``Given a Hamiltonian (or corresponding classical action) that displays a phase transition,
there exists a dual Hamiltonian (or action) for which the phase transition is made
evident by the existence of an order parameter. That is, there exists a duality
that maps systems with topological quantum orders into systems with standard
(Landau type) order parameters that signal the breaking of a global symmetry when transitions
are present. In the original language, the dual variable (the order parameter
of the dual theory) may be non-local. As illustrated in \cite{tqo}, the correlations between
local quantities in one such dual basis can become, in the original basis, 
non-local correlation functions (that lead to ``string'' or ``brane'' orders). 
In prominent examples of topologically ordered systems such as Kitaev's toric
code model \cite{toric_code} and three-dimensional extensions \cite{3detc}
such a duality can be constructed as discussed 
earlier (see also \cite{3detc,tqo,tqo1}). The system
is no longer entangled in the dual basis.

To make the discussion concrete, we will first consider the XXYYZZ
model \cite{chamonmodel,sergey}, then briefly comment on another 
spin $S=1/2$ model on a honeycomb lattice \cite{ly},  and finally 
discuss the three-dimensional Kitaev's TC model. Details concerning
the bond algebraic mappings leading to the results given below will
be presented in a forthcoming publication \cite{noc}. 

The XXYYZZ model is a spin $S=1/2$ system on a face centered cubic (FCC) lattice. 
As is well known, an FCC lattice can be viewed as comprised of all of the odd (or even) sites
of a cubic lattice. By even sites, we allude here to sites for which the 
sum of the $x$, $y$, and $z$ coordinates (in units of 
the lattice constant which we set to unity) is even. The basis
vectors along the Cartesian directions are $\hat{e}_{x,y,z}$. 
The Hamiltonian \cite{chamonmodel, sergey} is of the form 
\begin{eqnarray}
H = -J \sum_{u \in {\sf even}} h_{u},
\label{cham}
\end{eqnarray}
with interaction terms
given by  
\begin{eqnarray}
\label{huxxyyzz}
h_{u} = S^{x}_{u-\hat{e}_{x}} S^{x}_{u+\hat{e}_{x}} S^{y}_{u-\hat{e}_{y}} S^{y}_{u+\hat{e}_{y}}
S^{z}_{u-\hat{e}_{z}} S^{z}_{u+\hat{e}_{z}}.
\end{eqnarray}
That is, the interaction terms are given by  the product of spins over 
all sites surrounding an even sublattice
site with the product being of the form ``XXYYZZ'', wherein the component of the spin appearing
in the product is determined by its relative location relative to the center of the octahedron formed by the
six sites.  
By a bond algebraic mapping \cite{noc} {\em the $d=3$ dimensional XXYYZZ model can be 
mapped into four decoupled ($D=1$) Ising spin
chains}. That is, the bond algebra satisfied by the interaction terms
$h_{u}$ of Equation \eqref{huxxyyzz} is identical to that of the bonds in four decoupled 
classical Ising chains. Thus, a duality transformation implements
an exact dimensional reduction.  A consequence of this mapping is that
the system remains ergodic at all temperatures $T>0$.

We next briefly comment on the honeycomb lattice model of Reference \cite{ly}.
In our language, that Hamiltonian is composed of a sum of two types of mutually commuting
bond operators for each minimal hexagonal plaquette. The square of each of the bond operators, 
as well as the product of all bond operators (of each of the two types) over the entire lattice is 
constrained to be one.
Replicating the analysis of \cite{con,3detc,tqo,tqo1}, this system
can be exactly mapped onto two decoupled classical  Ising chains. As in the system 
of \cite{chamonmodel}, this model exhibits finite, system size independent, 
autocorrelation times at all temperatures. 

Finally, we  focus on  the $d=3$ Kitaev's  TC  model defined in Reference 
\cite{3detc},  which corresponds to Equation \eqref{KTCMd3} with $J_x=J_z$ and $h_x=h_z=0$. 
 As discussed in \cite{3detc,tqo}, the finite temperature partition function of this 
 system is given by the product of the partition functions
of a $D=1$ Ising chain and a $D=3$ Ising gauge model:
${\cal Z}_{{\sf Kitaev}, d=3}= {\cal Z}_{{\sf I},  D=1} \times {\cal Z}_{{\sf IG}, D=3}$.
Specifically, $[A_{\r}, B_{(\r,\nu\mu)}]=0$ and the bond algebras formed by the two sets 
of operators are decoupled. Moreover, the bond algebra formed by the vertex operators $A_{\r}$ alone
is isomorphic to that of bonds in a classical nearest-neighbor Ising chain. The bond algebra formed by
the operators $B_{(\r,\nu\mu)}$ alone is identical to that of bonds (plaquette terms) in the classical $D=3$ 
Ising gauge theory (which is dual to the $D=3$ dimensional Ising model). 

We now discuss how order in this topologically ordered system can be ascertained by non-local measures. 
Below the critical temperature of the Ising gauge theory (which is dual to that of the Ising model),
the system exhibits non-trivial order as ascertained by the non-local asymptotic character of the Wilson
loop $W_{R} = \prod_{(\r,\nu\mu) \in R} B_{(\r,\nu\mu)}$ where the region $R$  is taken to be 
arbitrarily large.
In the limit of large $R$, in the ordered low temperature phase, $W_{R}$ scales as
$W_{R}  \sim \exp(-c_{1} |\partial R|)$ where $|\partial R|$ is the perimeter of the region
$R$ with $c_{1}$ a positive constant. (This asymptotic behavior of $W_{R}$ 
is known as a ``perimeter law''.)
By contrast, at high temperatures, $W_{R}$ satisfies for large $R$ an ``area law"
and scales as $\exp(-c_{2} |R|)$, 
where $|R|$ denotes the number of plaquettes in $R$ (the area of $R$) and $c_{2}$
is also a positive constant.  
These non-local measures delineating the high and low temperature phases
of the Ising gauge theory formed  by the bonds $B_{(\r,\nu\mu)}$
(fleshed out by the Wilson loops, or an equivalent 
kink operator formed in the quantum version of the model) 
can be mapped onto local measures in the dual theory (a local on-site magnetization
in the $D=3$ classical Ising model or the $d=2$ quantum version, respectively).

%

\appendix

\section{Duality by Fourier transformation}
\label{appA}

As explained in Section \ref{classical&quantum}, the bond-algebraic
approach to dualities covers both quantum and classical dualities
in a unified fashion \cite{con}. Here we summarize the
authors elaboration of the standard approach to classical dualities 
\cite{wegner,wu_wang,savit,malyshev}, for ease of reference
and comparison to our new approach. We also take the opportunity
to point out the difficulties in extending the standard approach
to produce non-Abelian dualities, 
difficulties that may perhaps be overcome eventually with the help
of bond-algebraic techniques (see Section \ref{classical&quantum}). 
With these goals in mind, we keep
the group structure of the models we are going to consider fairly
general, 
but restricted  to nearest-neighbors
interactions on a hyper-cubic lattice {\it with periodic} BCs. 
Basic ideas work the same 
in any lattice, and can be generalized to include 
gauge interactions \cite{savit, malyshev}.

Standard techniques for classical dualities can only be applied
to partition functions that meet certain requirements. First, it should
be possible to represent the elementary degrees of freedom  \(g_\r\) 
on each site \(\r\) of a  lattice by the elements  
of some group \(G\),  \(g_\r\in G\) (the work of Wilson on lattice field theory 
\cite{wilson}, see Appendix \ref{appE}, is an example of such a setting). 
Groups can be Abelian or non-Abelian; examples of Abelian groups are 
$\mathbb{Z}_p, \mathbb{Z},  U(1), \mathbb{R}$, while $SU(2)$ or 
$SO(5)$ constitute non-Abelian instances.  
Second, the partition function should be of the form 
\begin{equation}\label{general_group_pf}
\pf=\sum_{\{g_\r\}}\prod_{\r}\prod_{\nu=1,\cdots,D}\ 
\exp\left[-u(g_{\r+\bm{e_\nu}}g_\r^{-1})\right],
\end{equation}
where the sum over configurations \(\sum_{\{g_\r\}}\) could represent
a multiple integral, depending on the group. (Notice that we
have absorbed the temperature factor in the definition of the 
interaction energy \(u(g_{\r+\bm{e_\nu}}g_\r^{-1})\).)  The
crucial point is that the interaction depends on
the state \(g_\r\) and its neighbor \(g_{\r+\bm{e_\nu}}\) through
the combination \(g_{\r+\bm{e_\nu}}g_\r^{-1}\). It is 
standard practice for Abelian groups to write \(g_{\r+\bm{e_\nu}}-g_\r\) instead of 
\(g_{\r+\bm{e_\nu}}g_\r^{-1}\).  For example, in the  
$U(1)$ case \(g_{\r+\bm{e_\nu}}-g_\r^{-1}\)
stands for \(\theta_{\r+\bm{e_\nu}}-\theta_\r\), and the actual
interaction is \(e^{i\theta_{\r+\bm{e_\nu}}}e^{-i\theta_\r}\).
We will often write \(u(g_{\r+\bm{e_\nu}}-g_\r)\) when  
the group underlying the partition function of Equation 
\eqref{general_group_pf} is Abelian, 
and  \(u(g_{\r+\bm{e_\nu}}g_\r^{-1})\)
if we want to discuss Abelian and non-Abelian groups on an equal footing.

Next we review basic facts about Fourier analysis on
groups \cite{terras}. A key idea is to use a set of distinguished functions 
\(\chi_{\alpha}:G\rightarrow\mathbb{C}\) (a  
generalization of plane waves \(e^{ikx}\) to arbitrary groups) to write down
Fourier-like expansions. The distinguished functions are the 
characters of the group \(G\) whose irreducible representations
we label with the letter \(\rho\). One can always write the character expansion
\begin{equation}
u(g)=\sum_{\rho} \hat{u}(\rho)\chi_{\rho}(g) ,
\end{equation}
{\it provided \(u:G\rightarrow\mathbb{C}\) is a class function}, \(u(g_1g_2)=u(g_2g_1)\),
and where the coefficients \(\hat{u}(\rho)\) are unique. (If \(G\) is
Abelian every function can be expanded in terms of characters.) Physical interactions
fall within this category  since
the potential \(u(g_{\r+\bm{e_\nu}}g_\r^{-1})\) is always a 
{\it symmetric} (\(u(g^{-1})=u(g)\)) class function, so that 
\begin{equation}
u(g_{\r+\bm{e_\nu}}g_\r^{-1})=u(g_\r g_{\r+\bm{e_\nu}}^{-1}) ,
\end{equation}
which for Abelian groups takes the more familiar form
\(u(g_{\r+\bm{e_\nu}}-g_\r)=u(g_\r-g_{\r+\bm{e_\nu}})\).
It then follows that the Boltzmann weights can expanded as
\begin{equation}\label{ch_exp}
e^{-u(g_{\r+\bm{e_\nu}}g_\r^{-1})}=\sum_{\rho}e^{-u^D(\rho)}
\chi_\rho(g_{\r+\bm{e_\nu}}g_\r^{-1}).
\end{equation}  
The {\it dual Boltzmann weights}  \(e^{-u^D(\rho)}\) are not necessarily real and 
positive.

We want next  to rewrite the partition function in terms of the dual
Boltzmann weights.  To this end, we need to associate
a {\it new} degree of freedom \(\rho_{(\r,\nu)}\), 
to the {\it links} of the lattice, that is labelled by the irreducible representations
\(\rho\) of the group \(G\) that represents the degrees of freedom \(g_\r \in G\) on
the sites. We can then use Equation \eqref{ch_exp} to rewrite Equation 
\eqref{general_group_pf} as
\begin{equation}\label{spinfoam}
\pf=\sum_{\{\rho_{(\r,\nu)}\}}
\ I(\{\rho_{(\r,\nu)}\})\ e^{-\sum_{\r,\nu}\ u^D(\rho_{(\r,\nu)})},
\end{equation}
where
\begin{equation}\label{bigI}
I(\{\rho_{(\r,\nu)}\})=\sum_{\{g_\r\}}\prod_{\r}\prod_{\nu=1,\cdots,D}\chi_{\rho_{(\r,\nu)}}
(g_{\r+\bm{e_\nu}}g_\r^{-1}).
\end{equation}
The fundamental difference between Abelian and non-Abelian 
dualities manifests in the behavior of \(I(\{\rho_{(\r,\nu)}\})\).

Assume first that \(G\) is Abelian (we now write 
\(g_1+g_2\) instead of \(g_1g_2\)). In this case 
the irreducible representations can be labelled 
by the elements of another {\it Abelian} group \(\hat{G}\) (the Pontryagin
dual of \(G\) \cite{morris}), in such a way that
if \(\hat{k}_1,\hat{k}_2\in \hat{G}\) label two irreducible representations
of $G$ with characters \(\chi_{\hat{k}_1}\) and \(\chi_{\hat{k}_2}\), then
\begin{eqnarray}\label{char1}
\chi_{\hat{k}_1}(g)\chi_{\hat{k}_2}(g)&=&\chi_{\hat{k}_1+\hat{k}_2}(g),
\ \ \ \ \ \ \chi_{-\hat{k}}(g)=\chi^*_{\hat{k}}(g) , \nonumber \\
\chi_{\hat{k}}(g_1)\chi_{\hat{k}}(g_2)&=&\chi_{\hat{k}}(g_1+g_2),
\ \ \   \chi_{\hat{k}}(-g)=\chi^*_{\hat{k}}(g), 
\end{eqnarray}
with \({}^*\) indicating complex conjugation.
{\it These are essential relations for
deriving classical dualities}. 
Let us list the Pontryagin
duals of the groups most often used
\begin{equation}
\widehat{\Z_p}=\Z_p,\ \ \ \ \widehat{\Z}=U(1),\ \ \ \
\widehat{U(1)}=\Z,\ \ \ \ \widehat{\mathbb{R}}=\mathbb{R},
\end{equation}
with characters given by
\begin{eqnarray}
\chi_{m}(n)&=&e^{i\frac{2\pi mn}{p}},\ \ \ \ \ \ \ \ \ \ \ \ \ \ \ \ \ \
n,m\in \Z_p, \nonumber\\
\chi_{e^{i\theta}}(m) \equiv \chi_\theta(m)&=&e^{im\theta},
\ \ \ \ \ \ \ \ \ \ \ \ \ \ \ \ \ \ \ \  
m\in\Z, \ \ \theta\in[0,2\pi),  \nonumber \\
\chi_m(e^{i\theta})\equiv \chi_m(\theta)&=&e^{im\theta},
\ \ \ \ \ \ \ \ \ \ \ \ \ \ \ \ \ \ \ \ 
\theta\in[0,2\pi),\ \ m\in\Z,  \nonumber \\
\chi_k(x)&=&e^{ikx},\ \ \ \ \ \ \ \ \ \ \ \ \ \ \ \ \ \ \ \ \,
x,k\in\mathbb{R}.
\end{eqnarray}
In general, characters of Abelian groups satisfy the completeness
relations 
\begin{equation}\label{completeness}
\sum_{g\in{G}}\ \chi_{(\hat{k}-\hat{k}')}(g)=A_{G} \, \delta_{\hat{G}}(\hat{k}-\hat{k}'),
\ \ \ \ \ \ 
\sum_{\hat{k}\in\hat{G}}\ \chi_{\hat{k}}(g-g')=A_{\hat{G}}\, \delta_G(g-g'),
\end{equation}
where  \(\delta\) is the Kronecker or Dirac delta depending on the group, and $A$
is a normalization constant that is also group dependent.

\begin{figure}[h]
\begin{center}
\includegraphics[width=0.55\columnwidth]{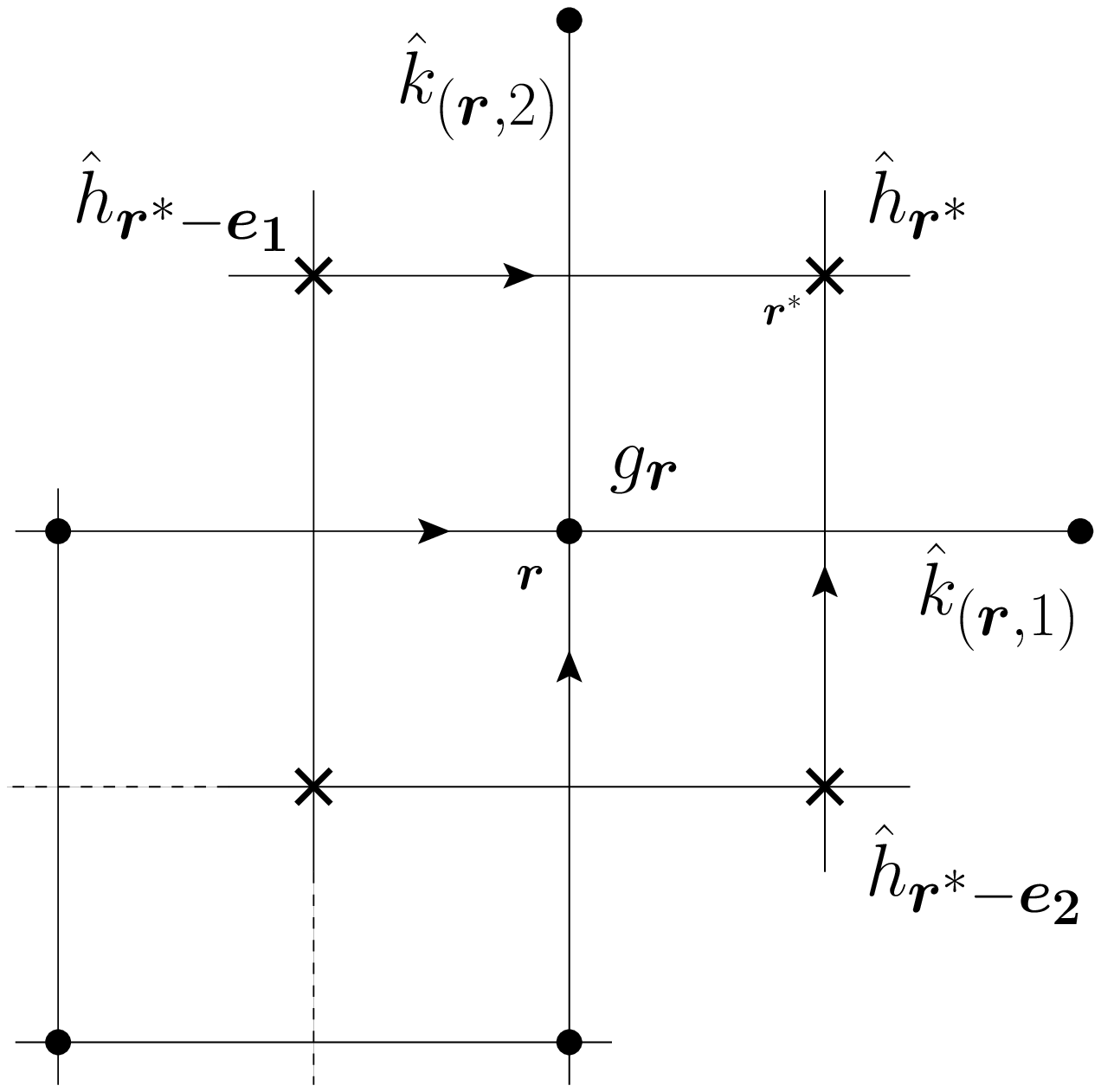}
\end{center}
\caption{The partition function \(\pf\) of 
Equation \eqref{general_group_pf} features degrees of freedom \(g_\r\)
placed on the sites (marked with crosses) of one of the square lattices.
The other dual square lattice (with sites marked by heavy dots)
supports the degrees of freedom \(\hat{h}_{\r^*}\)
of the dual partition function \(\pf^D\) of Equation \eqref{dual_z}.
The arrows specify the standard orientation
on both lattices, determined by the basis vectors \(\i,\j\).
The configurations of link variables \(\hat{k}_{(\r,\nu)}\) that satisfy the 
divergenceless condition of equation \eqref{Iabelian} can be parametrized in terms of 
configuration of dual variables \(\hat{h}_{\r^*}\), as in Equation 
\eqref{divergenceless_configs}, that reflects the choice of orientation for links.}
\label{dualsimplicials}
\end{figure}
We next exploit these properties 
to simplify the expression for \(I(\{\hat{k}_{(\r,\nu)}\})\), 
 Equation \eqref{bigI}, in the $D=2$ dimensional case 
\begin{eqnarray}
I(\{\hat{k}_{(\r,\nu)}\})&=&\sum_{\{g_\r\}}\prod_{\r}\prod_{\nu=1,2}
\chi_{\hat{k}_{(\r,\nu)}}
(g_{\r+\bm{e_\nu}}-g_\r) \nonumber\\
&=&\prod_\r (\sum_{\{g_\r\}}\ \chi_{-\hat{k}_{(\r,1)}-\hat{k}_{(\r,2)}+
\hat{k}_{(\r-\bm{e_1},1)}+\hat{k}_{(\r-\bm{e_2},2)}}(g_\r) ) \nonumber \\
&=&\prod_\r A_{G}\ \delta_{\hat{G}} (-\hat{k}_{(\r,1)}-\hat{k}_{(\r,2)}+ 
\hat{k}_{(\r-\bm{e_1},1)}+\hat{k}_{(\r-\bm{e_2},2)} ), 
\label{Iabelian}
\end{eqnarray}
where we have used Equations \eqref{char1},  and \eqref{completeness}. 
The partition function for an $N$ sites lattice is
\begin{equation}\label{middle_step}
\pf=A_{G}^{N}\sum_{\{\hat{k}_{(\r,\nu)}\}} \prod_\r e^{-u^D(\hat{k}_{(\r,\nu)})}
\delta_{\hat{G}} (-\hat{k}_{(\r,1)}-\hat{k}_{(\r,2)}+ 
\hat{k}_{(\r-\bm{e_1},1)}+\hat{k}_{(\r-\bm{e_2},2)}),
\end{equation}
and one needs to resolve the constraints embodied in the delta function. 
This is where the concept of a {\it dual lattice} (formally defined at the end of this 
appendix) enters the scene. 
Since every site \(\r\) is surrounded
by a square of the dual lattice (see Figure \ref{dualsimplicials}), one can 
{\it parametrize} those configurations
of \(\hat{k}_{(\r,\nu)}\)s {\it that satisfy the constraints} 
in terms of degrees of freedom on the
sites \(\r^*\) of the dual lattice, 
\begin{equation}\label{divergenceless_configs}
\hat{k}_{(\r,1)}=\hat{h}_{\r^*}-\hat{h}_{\r^*-\j} ,\ \ \ \ \ \ 
\hat{k}_{(\r,2)}=\hat{h}_{\r^*-\i}-\hat{h}_{\r^*},
\end{equation}
so that 
\begin{equation}\label{dual_z}
\pf=A_{G}^{N}\sum_{\{\hat{h}_{\r^*}\}} \exp\left[-\sum_\r\sum_{\nu=1,2}\ 
u^D(\hat{h}_{\r^*+\bm{e_\nu}}-\hat{h}_{\r^*})\right]\equiv A_{G}^{N}\ \pf^D.
\end{equation}
The role of the dimension $D$ manifests in the resolution of the constraints. 
For instance, in  a \(D=3\) dimensional cubic lattice, 
we need to parametrize \(\hat{k}_{(\r,\nu)}\) in terms of a combination of 
{\it four} degrees of freedom \(\hat{h}_{\r^*}\) on the corners of the 
square plaquette in the dual (cubic) lattice that is pierced by the link \((\r,\nu)\). 
This is why in \(D=3\), the duals of models with nearest-neighbor interactions
show four-body interactions and gauge symmetries \cite{wegner_ising}. 
The reader can find worked out
examples in the reviews listed at the beginning of this section, and in the 
book \cite{bookEPTCP}. 

In the non-Abelian case  the relation
\(\chi_{\hat{k}}(g_1-g_2)=\chi_{\hat{k}}(g_1)\chi^*_{\hat{k}}(g_2)\) 
{\it does not hold} in general, since
\begin{equation}
\chi_\rho(g_1g_2^{-1})=\tr D_\rho(g_1) D^\dagger_\rho(g_1)\neq 
\chi_\rho(g_1)\chi^*_\rho(g_2),
\end{equation}
(\(D_\rho\) are the operators of the representation \(\rho\)) 
unless \(\rho\) happens to be a scalar representation. This prevents
the key simplification in Equation \eqref{Iabelian}
 that allows to perform the sums \(\sum_{\{g_\r\}}\)
to obtain only \(\delta\) constraints. More specifically, the sums \(\sum_{\{g_\r\}}\)
can still be performed in the non-Abelian case, but the resulting 
\(I(\{\rho_{(\r,\nu)}\})\) has a complicated structure that {\it prevents reinterpreting
the dual representation of \(\pf\) of Equation \eqref{spinfoam} as a dual
partition function}.
{\it It is in this sense that the standard approach to classical dualities based on 
Fourier analysis fails for non-Abelian models}. It has been argued recently \cite{oeckl}
that the dual form of Equation \eqref{spinfoam} can be understood as a spin foam
model in the non-Abelian case.

Let us finally recall the formal definition of dual lattice, borrowed
from simplicial topology \cite{bredon}. A
$D$-dimensional lattice (or more precisely, simplicial complex)
$\Lambda$ is characterized by
a set of  vertices (lattice sites, $D=0$  objects), edges (links,
$D=1$  objects), elementary faces ($D=2$), and so on, up to the elementary
$D$-dimensional volumes, and incidence relations. 
The dual lattice $\Lambda^*$ is obtained by
placing a vertex at the center of each $D$-dimensional volume, and connecting any
two dual  vertices that share an $(D-1)$-dimensional face with a link piercing
that face. Similarly, the construction of the dual lattice assigns to
each $(D-2)$-dimensional  simplex in the initial lattice one, and only one,
\((D=2)\)-dimensional face in
the dual lattice, and so on. Hence a lattice duality maps $D=s$-dimensional
elementary simplices of the lattice to $(D-s)$-dimensional simplices of its dual, 
in such a way that whenever two $D=s$-dimensional simplices share an $(D=s-1)$-dimensional
face, the corresponding $(D-s)$-dimensional dual simplices are connected by an 
$(D-s+1)$-dimensional simplex of the dual lattice that {\it pierces} that connecting 
$D=(s-1)$-dimensional simplex (face). 
For example, a $D=2$ triangular  lattice is dual to an hexagonal
lattice,  and a square lattice is dual to itself (i.e., it is {\it
self-dual}). It is well known,  though not necessarily obvious, that a
lattice duality transformation is its own inverse. That is, the lattice
$\Lambda^*$ dual to a lattice $\Lambda^{**}$, which is itself a dual
lattice, is just the original lattice $\Lambda$, i.e., 
$\Lambda^{**}=\Lambda$. 

\section{Conditions for spectral equivalence}\label{sec3.1}

In this appendix we analyze general conditions any two physical 
systems must satisfy in order to share  identical energy spectra. 

It is a basic postulate of quantum mechanics that
observables should  be represented by Hermitian operators \cite{dirac,PT}.
Hermiticity guarantees that observables always admit an spectral decomposition,
and, in particular, if the spectral decomposition of a Hamiltonian is known,
\begin{eqnarray}
H=\int dE\ E \, P_E ,
\label{PEG}
\end{eqnarray}
where \(P_E=P_E^2=P_E^\dagger\)  is the 
projector onto the Hilbert space sector of energy $E$.   
Then both the thermodynamics and  quantum dynamics it determines
are known as well. This follows because
thermodynamic quantities can be computed from the (canonical) partition function
\begin{equation}
{\cal Z}(\beta)= \int dE~ e^{-\beta E} \rho(E),
\label{Zr}
\end{equation}
where $\rho(E)={\sf Tr}\ P_E$ is the density of states at energy $E$;
and the dynamics of any observable is determined by Heisenberg's equation of motion,
\begin{equation}
\frac{d{{\cal O}_t}}{dt}=\frac{i}{\hbar}[H,{\cal O}_t],
\end{equation}
that can be solved explicitly as 
\begin{equation}
{\cal O}_t=\int dE dE'\ P_{E'}{\cal O}_0P_E\ e^{\frac{i}{\hbar}(E'-E)t}.
\end{equation}


Suppose now that the Hamiltonian \(H\) is defined on a Hilbert space
${\cal{H}}$ of finite dimension ${\sf dim}({\cal{H}})$. Then the 
energy levels \(E_i\) are determined as the roots of the eigenvalue (or secular)
equation 
\begin{equation}
\det (H-E)=0.
\end{equation}
On the other hand, the secular equation can be written solely in terms
of the powers $ {\sf
Tr}(H^{k})$, $k=1,2, \cdots,  {\sf dim}({\cal{H}})$, 
\begin{eqnarray}
0 &=&  \prod_{i=1}^{{\sf dim}({\cal H})} (E-E_{i}) \nonumber \\
&=&  E^{{\sf dim}({\cal{H}})}- ({\sf Tr}(H)) E^{ {\sf
dim}({\cal{H}})-1}  + \frac{1}{2}[({\sf Tr}(H))^{2} - {\sf Tr}(H^{2})]
E^{ {\sf dim}({\cal{H}})-2} \nonumber \\ 
&-& \frac{1}{6} [({\sf Tr}(H))^{3} - 3 {\sf Tr}(H) {\sf Tr}(H^{2})   +2
{\sf Tr}(H^{3})]  E^{ {\sf dim}({\cal{H}})-3}  \!\! +\! \frac{1}{24} [({\sf
Tr}(H))^{4} \! + \! 3 ({\sf Tr}(H^{2}))^{2}   \nonumber \\ 
&+& 8 ({\sf Tr}(H)) ({\sf Tr}(H^{3})) - 6
{\sf Tr}(H^{2}) ({\sf Tr}(H))^{2} - 6 {\sf Tr}(H^{4})]  E^{ {\sf
dim}({\cal{H}})-4} + \cdots~.  
\label{spec_eq}
\end{eqnarray}
We  see that the energy levels are uniquely determined by 
the traces $\{{\sf Tr}(H^{k})\}_{k=1}^{{\sf dim}{\cal{H}}}$.
It follows that two Hamiltonians \(H_1\), \(H_2\) 
can have identical energy levels only if 
\begin{equation}
\tr(H^{k}_1)=\tr(H_2^k),\ \ \ \ \ \ k=0,1,\cdots,{\sf dim}{\cal{H}}. 
\end{equation}
{}From an algebraic perspective, one can argue that
energy levels are determined 
by the algebra \(\{\mathbb{1},H, H^2,\cdots\}\) generated by \(H\) and \(\mathbb{1}\) (see
Reference \cite{algebras}, Theorem 3.30, for a precise statement of this idea).

Next we review well known relations concerning the one-to-one
correspondence between the density of states $\rho(E)$ and the partition
function ${\cal Z}(\beta)$. 
Equation (\ref{Zr}) expresses the partition function ${\cal Z}(\beta)$ 
as a Laplace transform of the density of states $\rho(E)$. The inverse
Laplace transform
\begin{eqnarray}
\rho(E) = \frac{1}{2 \pi i} \lim_{W \to \infty} \int_{\gamma - i
W}^{\gamma + iW}  e^{\beta E}~  {\cal Z}(\beta)~  dE,
\label{Laplace}
\end{eqnarray}
(with the real number $\gamma$ chosen such that it is  greater than the
real part of all the zeroes of ${\cal Z}(\beta)$) uniquely defines the
density of states $\rho(E)$. Thus, if two systems share the same
partition function ${\cal Z}(\beta)$ then they will have the same
density of states.  In systems with bounded Hamiltonians $H$ defined on
arbitrarily  large yet finite size lattices $\Lambda$ (or arbitrarily
large yet finite continuum volumes), the partition function is analytic
for all $\beta$ and the expansion 
\begin{equation}
{\cal Z}(\beta) = {\sf Tr}\ e^{-\beta H} = \sum_{n=0}^{\infty}
\frac{(-1)^{n}}{n!}  \beta^{n} {\sf Tr}({H}^{n})
\end{equation}
converges for all $\beta$. It follows that if for two Hamiltonians
$H_{1}$ and $H_{2}$, describing two different systems, we have that
\begin{equation}
{\sf Tr}({H}_{1}^{n}) =  {\cal N} ~ {\sf Tr}({H}_{2}^{n})
\label{equiva}
\end{equation}
for all natural numbers $n$ with ${\cal N}$ a constant, then  $H_{1}$
and $H_{2}$ lead to identical partition functions (up to an overall
multiplicative constant), thus encapsulating identical physics.  (If both
Hamiltonians act on Hilbert spaces of the same dimensionality ${\sf
dim}{\cal{H}}$ then, from Equation (\ref{spec_eq}),  it will suffice to
verify Equation \eqref{equiva} for $n=1, 2, \cdots, {\sf
dim}{\cal{H}}$).

\section{Lattice quantum field theory}
\label{appE}

We review the basics of lattice 
quantum field theory (LQFT) \cite{wilson_kogut, LQFT} 
to clarify the connection between the lattice models studied in this paper 
and QFT. 

LQFT regularizes field theories by approximating
fields taking values in the continuum space-time $x^\mu=(x^0,x^1,
\cdots , x^d)$ ($x^0$ is the time axis) with fields 
defined on a hyper-cubic space lattice of 
spacing \(a\), i.e., $\r=(am^1,\cdots,am^d)$ 
with \(m^i\) an integer.  We thus sample fields $\phi$ by 
retaining only their values at lattice sites $\r$
\begin{equation}
\phi(x^0,x^1,\cdots,x^d)\ \longrightarrow\ \phi_\r(x^0),
\end{equation}
or links, if we are dealing with a vector field, or plaquettes, etc.
Suppose next that the field's dynamics is specified by the
action ($\mathcal{L}$ is the Lagrangian density)
\begin{equation}
S=\int dx^0d^dx\ \mathcal{L}(\phi,\partial_\mu\phi,\cdots).
\end{equation}
Then we can specify the dynamics of the lattice field by 
a suitable discretization of the action $S$,
  \begin{equation}
S=\int dx^0d^dx\ \mathcal{L}\ \longrightarrow\ \int dx^0
\ \sum_\r\ a^d\mathcal{L}_\r=S_{\sf L}.
\end{equation}
The so called ultraviolet regulator \(a\) reduces the number of degrees
of freedom from an uncountable infinity in the continuum to a countable one
in the lattice, and it is formally removed by taking the limit
\(a\rightarrow 0\). 

The lattice action \(S_{\sf L}\) describes  a mechanical
model that can be quantized according to 
well understood schemes of quantization. The resulting quantum theory, 
known as lattice Hamiltonian approach, 
represents a lattice approximation to the QFT. It was used
repeatedly in Section \ref{sec6},  for instance Sections \ref{sec6.1} and 
\ref{sec6.2}. Let us consider here another example relevant to this paper, 
a simple non-linear sigma model in \(D=d+1=2\) dimensions, 
\begin{equation}
S_{\sf XY}=\frac{1}{2}\int d^2x\ \left(\partial_0\phi^* 
\partial_0\phi- \lambda
\partial_1\phi^*\partial_1\phi-\eta(\phi^*\phi-1)\right),
\end{equation}
where \(\eta\) is a Lagrange multiplier that forces 
the complex scalar field \(\phi\)  to take values on the unit circle. 
The corresponding classical lattice field theory 
\begin{eqnarray}
S_{\sf LXY}=\int dx^0\ a\sum_i\ (\frac{1}{2}(\partial_0\theta_i)^2- 
\frac{\lambda}{a^2}(1-\cos(\theta_{i+1}-\theta_i))),
\end{eqnarray}
where we have parametrized the field as \(\phi_i=e^{i\theta_i}\), 
describes a set of planar rigid rotators that 
has to be quantized according to 
Dirac's quantization scheme \cite{das}. 
The resulting quantum model is described by the Hamiltonian 
($L_i=-i\partial/\partial\theta_i$)
\begin{equation}
H_{\sf LXY}=\frac{1}{a}\sum_i\  (\frac{1}{2} L_i^2+
\lambda (1-\cos(\theta_{i+1}-\theta_i))  ) ,
\end{equation}
which is nothing but the XY model of
Section \ref{sec5.1}.

This scheme to regularize QFTs 
can be exploited directly \cite{kogut_susskind,fradkin_susskind}. 
However, the standard approach to study non-perturbative 
properties amounts to combining this lattice regularization
with Feynman's path integral method. {\it The resulting formalism discretizes both
space and time}, recasting quantum field theoretic problems in 
the language of classical statistical mechanics, and is ideally suited
for numerical simulations. We describe it next. 

Let us start with the simple case of a
single particle in an external potential $V(x)$.
Consider Feynman's expression for transition amplitudes (Feynman's path integral \cite{schulman}),
\begin{eqnarray}
\langle x_f,t_f\vert x_i,t_i\rangle &=& \langle x_f\vert e^{-iH(t_f-t_i)}\vert
x_i\rangle \label{transition_amplitude}\\
&=&\lim_{N\rightarrow \infty}
\left(\frac{m}{2\pi i\delta t}\right)^{\frac{N}{2}}
\int dx_1\cdots dx_{N-1}\
e^{i\delta t\sum_{j=0}^{N-1}\left(\frac{1}{2}m (x_{j+1}-x_j)^2
\delta t^{-2}-V(x_j)\right)}, \nonumber
\end{eqnarray} 
where \(x_0\equiv x_i\), \(x_N\equiv x_f\), 
\(\delta t\equiv(t_f-t_i)/N\),
and 
\begin{equation}
H=-\frac{1}{2m}\frac{d^2}{dx^2}+V(x)
\label{Hamilt_part1d}
\end{equation}
the Hamiltonian 
operator of a particle moving in one dimension.
Equation \eqref{transition_amplitude} reduces the evaluation of 
transition amplitudes to the evaluation of a multiple integral, by means
of an approximation that {\it discretizes time} (notice for future 
reference, that the argument
of the exponential is the (discrete form of the) classical action of 
the particle). 

Our goal, however, is to use Feynman's formula to study ground
state properties. So consider the quantum partition function 
${\cal Z}(\beta)$ of Equation \eqref{Zr}. We can
expect that at very low temperatures ${\cal Z}(\beta)$ 
will be dominated by the ground energy level,
\begin{equation}\
\lim_{\beta\rightarrow \infty} {\cal Z}(\beta) \approx e^{-\beta E_\Omega} \rho(E_\Omega),
\end{equation}
or more precisely,
\begin{equation}\label{dominated}
E_\Omega =\lim_{\beta\rightarrow \infty}\ -\frac{1}{\beta}\ln {\cal Z}(\beta).
\end{equation}

On the other hand, the partition function corresponding to the Hamiltonian \eqref{Hamilt_part1d}
\begin{eqnarray}
{\cal Z}(\beta)&=&\tr e^{-\beta H}=\int dx\ \langle x\vert e^{-\beta H}\vert x\rangle =
\int dx\ \langle x, -i\beta\vert x,0\rangle\\
&=&\lim_{N\rightarrow \infty}
\left(\frac{m N}{2\pi \beta}\right)^{\frac{N}{2}}
\int dx_1\cdots dx_{N}\
e^{-\frac{\beta}{N}\sum_{j=0}^{N-1}\left(\frac{1}{2}m (x_{j+1}-x_j)^2
(\beta/N)^{-2}+V(x_j)\right)}, \nonumber
\end{eqnarray}
where now \(x_0\equiv x_N\), and the last integral expression can be 
interpreted as a {\it classical} partition function for \(N\) oscillators.

Let us apply these ideas to a bosonic scalar field $\phi$ in \(D=2\) dimensions,
\begin{equation}\label{si_b}
S_{\sf B}=\int d^2x (\frac{1}{2}\eta^{\mu\nu}\partial_\mu\phi
\partial_\nu\phi-V(\phi)),
\end{equation}
that for special forms of \(V(\phi)\) was studied in Section 
\ref{sec6.1}. We start by discretizing the model in space
and quantize it canonically. We then apply the formulas just 
derived for one particle 
\begin{eqnarray}
\pf_{\sf B}(\beta)&=&\left(\frac{ N}{2\pi \beta}\right)^{\frac{NM}{2}}
\int\prod_{i=0}^{M-1}d\phi_i(N-1)\cdots d\phi_i(0)
\nonumber \\
&\times&\ e^{-\frac{\beta}{N}\sum_{j=0}^{N-1}
\sum_{i=1}^{M}a\left(\frac{N^2}{2\beta^2}(\phi_i(j+1)-\phi_i(j))^2+
\frac{1}{2a^2}(\phi_{i+1}(j)-\phi_i(j))^2+V(\phi_i(j))\right)},
\end{eqnarray}
from which we can compute, for instance, quantum ground state properties
(\(M=L/a\), where \(L\) is the length in the space direction).
It is standard practice to fine-tune the spacing in the ``time'' direction
\(\beta/N\) equal to the lattice spacing \(a\), {\it so that \(aN=\beta\)}.
Then we can write 
\begin{eqnarray}\label{homogeneous_a}
\!\!\!\!\pf_{\sf B}(a)=\left(\frac{ 1}{2\pi a}\right)^{\frac{NM}{2}}
\int\prod_{r^0=0}^{N-1}\prod_{r^1=0}^{M-1}d\phi_\r\ e^{-a^{2}
\sum_\r\left(\frac{1}{2a^2}\sum_{\nu=0,1}\ (\phi_{\r+\bm{e_\nu}}-\phi_\r)^2+
V(\phi_\r)\right)}\ ,
\end{eqnarray}
where now \(\r=(ar^0,ar^1)\) includes the ``time'' coordinate.


Another important example is that of gauge fields. The problem of finding a 
convenient discretization for the Yang-Mills
action 
\begin{equation}
S_{\sf YM}=-\frac{1}{4g^2}\int d^0xd^dx\ \tr F_{\mu \nu}F^{\mu \nu},
\end{equation}
was solved by Wilson \cite{wilson}. Wilson's action reads
\begin{equation}\
S_{\sf W}=-\frac{1}{2g^2}\sum_\r\sum_{\mu<\nu}\ \left(\Re\
\tr\left(U_{(\r,\nu)}U_{(\r+\bm{e_\nu},\mu)}U^\dagger_{(\r+\bm{e_\mu},\nu)}
U^\dagger_{(\r,\mu)} \right)-N\right),
\end{equation}
where the \(U\)s, defined on the links of the lattice, are elements of the 
gauge group (assumed for simplicity to be some unitary or special unitary group
of dimension \(N\)), and \(\Re\) stands for the {\it real part}.  If we 
take the gauge group to be \(U(1)\) 
\begin{eqnarray}
U_{(\r,\nu)}=e^{i\theta_{(\r,\nu)}},
\end{eqnarray}
and Wilson's action reads
\begin{equation}
S_{\sf W}=-\frac{1}{2g^2}\sum_\r\sum_{\mu<\nu}\ \left(
\cos(\theta_{(\r,\nu)}+\theta_{(\r+\bm{e_\nu},\mu)}-
\theta_{(\r+\bm{e_\mu},\nu)}-\theta_{(\r,\mu)})-1\right),
\end{equation}
which is the standard action for {\it compact} QED. Notice that if
one expands the cosine to lowest order one obtains
\begin{equation}
S_{\sf W}=-\frac{1}{4g^2}\sum_\r\sum_{\mu<\nu}\ \left(
\theta_{(\r,\nu)}+\theta_{(\r+\bm{e_\nu},\mu)}-
\theta_{(\r+\bm{e_\mu},\nu)}-\theta_{(\r,\mu)}\right)^2,
\end{equation}
which is the standard discretization of Maxwell's
QED action.

In closing, let us notice that the quantum Hamiltonian 
for a lattice field theory can be recovered in principle from  
\(\pf\propto\int e^{-S_{\sf LE}}\), by using the transfer matrix
formalism, see Section \ref{classical&quantum}.

\section{Bond-algebraic dualities in finite-size systems}
\label{appG}

{}From a practical point of view, our approach to dualities
is specially attractive because it permits to study dualities
and self-dualities in {\it finite-size systems}, both classical and quantum. In this appendix, we 
illustrate through example the power of the bond-algebraic 
approach. One can simply check the duality mapping in a 
finite system and then safely extrapolate to the thermodynamic limit. 
The role  boundary terms play in determining bulk properties  is marginal. 

\begin{figure}[h]
\begin{center}
\includegraphics[width=0.7\columnwidth]{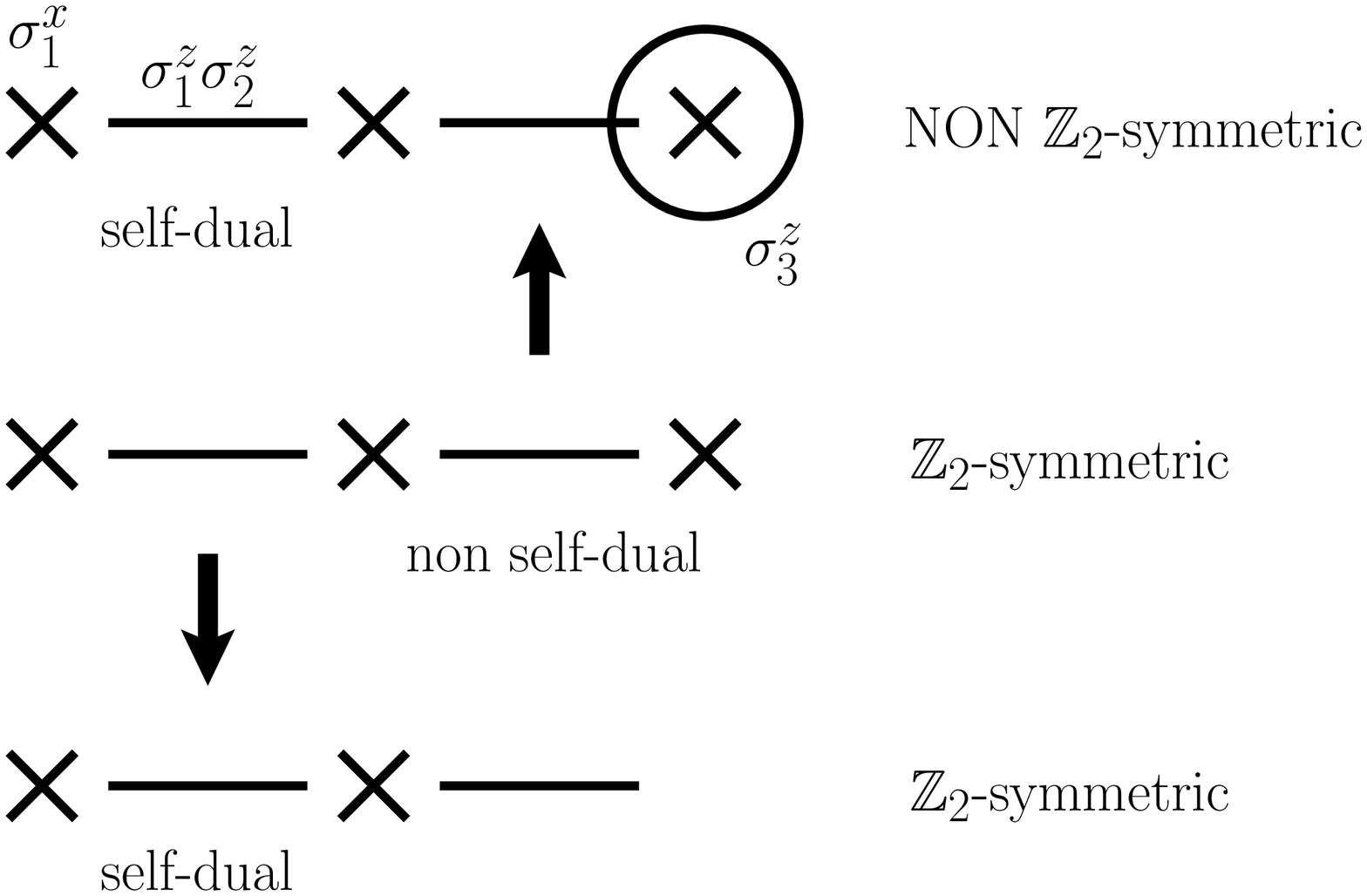}
\end{center}
\caption{Here the crosses represent the bonds \(\sigma^x_i,\ i=1,2,3\), the 
links represent the bonds \(\sigma^z_i\sigma^z_{i+1}\), and 
the circle on the right end of the top chain represents the bond 
\(\sigma^z_3\). A finite, open quantum Ising chain (illustrated by the chain
in the middle section of the figure) is not self-dual, but admits
two different self-dual BCs (illustrated by the top and
bottom chains) that differ in their effect
on the \(\Z_2\) symmetry.}
\label{shorty_ising}
\end{figure}

Let us start with the Ising model. We would like to study 
both open and periodic BCs, so we need at least three
spins. Then we have the Hamiltonian 
\begin{equation}\
H_{\sf I}=h(\sigma^x_1+\sigma^x_2+\sigma^x_3)+
J(\sigma^z_1\sigma^z_2+\sigma^z_2\sigma^z_3)
\label{finiteIsingquantumch}
\end{equation}
with open BCs, that is {\it not} self-dual 
(its energy levels are not symmetric in \(J\) and \(h\)), 
and commutes with \(Q=\sigma^x_1
\sigma^x_2\sigma^x_3\). To restore self-duality, we can
either {\it remove},
\begin{equation}
H_{\sf sd}=H_{\sf I}-h\sigma^x_3\
\end{equation}
or {\it add}
\begin{equation}\label{ising_three_spins}
H_{\sf sd}=H_{\sf I}+J\sigma^z_3\
\end{equation}
a bond (see Figure \ref{shorty_ising}). These options are not equivalent, 
because the first one preserves the \(\mathbb{Z}_2\) symmetry,
while the second does not.

The picture is quite different if we consider {\it periodic} 
BCs
\begin{equation}
H_{\sf I}=h(\sigma^x_1+\sigma^x_2+\sigma^x_3)+
J(\sigma^z_1\sigma^z_2+\sigma^z_2\sigma^z_3+\sigma^z_3\sigma^z_1).
\end{equation}
Again, the model is {\it not} self-dual, but it is \(\mathbb{Z}_2\)-symmetric
and translationally invariant. Remarkably, we can restore self-duality
just by ``decorating" a bond, preserving  all the 
symmetries of the model
\begin{equation}
H_{\sf sd}=h(\sigma^x_1+\sigma^x_2+\sigma^x_3)+
J(\sigma^z_1\sigma^z_2+\sigma^z_2\sigma^z_3+\sigma^z_3Q\sigma^z_1).
\end{equation}
This Hamiltonian is local in terms of fermionic degrees of freedom (i.e.,
after a Jordan-Wigner transformation \cite{GJW}) and its spectrum 
is self-dual. 

The bonds of the \(O(3)\)-symmetric Heisenberg model
\begin{equation}
H_{\sf H}=J_x(\sigma^x_1\sigma^x_2+\sigma^x_2\sigma^x_3)
+J_y(\sigma^y_1\sigma^y_2+\sigma^y_2\sigma^y_3)+
+J_z(\sigma^z_1\sigma^z_2+\sigma^z_2\sigma^z_3),
\end{equation}
can be written in terms of the bonds of the self-dual Ising model
of Equation \eqref{ising_three_spins}. Hence we can use the
Ising model self-duality mapping  to generate a duality for the
Heisenberg model. The dual Hamiltonian reads
\begin{equation}
H_{\sf H}^D=J_x(\sigma^z_2+\sigma^z_3\sigma^z_1)-J_y(\sigma^x_3\sigma^z_2
+\sigma^z_3\sigma^x_2\sigma^z_1)+J_z(\sigma^x_3+\sigma^x_2),
\end{equation}
and a simple calculation confirms that it shares the energy levels
of \(H_{\sf H}\). Notice though that the non-Abelian
symmetry of \(H_{\sf H}\) is hidden in \(H_{\sf H}^D\).

Next we consider gauge-reducing dualities as described 
in Section \ref{sec3.12}. 
Dualities that eliminate gauge symmetries connect models
with state spaces of different dimensions, and so
the unitaries \(U_\d\) that represent them are projective,
meaning that they either map a state to another state with the
same norm, or to zero. In other words, unlike symmetries
and ordinary dualities that are represented by square matrices,
gauge-reducing dualities are represented by {\it rectangular} 
matrices.

Let us consider an elementary example, the \(d=2\)-dimensional \(\mathbb{Z}_2\) 
gauge model of Section \ref{sec3.12} restricted to just {\it two}
lattice sites, 
\begin{equation}
\bar{H}_{\sf G}=J\sigma^z_1\sigma^z_2+h(\sigma^x_1+\sigma^x_2),
\end{equation} 
so that the model features just one gauge symmetry, 
\(G=\sigma^x_1\sigma^x_2\). As discussed in Section \ref{sec3.12}, 
gauge symmetries are actually {\it constraints}. The state space of
\(\bar{H}_{\sf G}\) above has dimension,
\(2^2\), but contains states that are not physical, because
they are not invariant under the gauge transformation \(G\).

The fact that the state space of a gauge model contains non-physical states
has an impact on its {\it energy levels}. In general, three things could happen 
to any specific level: i) 
Its energy eigenstates are all gauge-invariant and so physical; ii) 
some eigenstates are gauge-invariant 
and some are not, so that the gauge constraints reduce the level's degeneracy;
iii) {\it none} of its energy eigenstates
are gauge-invariant, and so the energy level
is not physical (not realizable in experiments). Notice on the other hand
that the energy eigenstates cannot be all gauge invariant, so either ii) and/or
iii) must actually occur for any gauge system. \(\bar{H}_{\sf G}\) in particular shows
a combination of i) and iii), as one can check explicitly. On the other hand,
it is more convenient to use dualities to solve the gauge constraints.

The point is that we can find a different representation 
of the bond algebra of \(\bar{H}_{\sf G}\) that affords a dual Hamiltonian 
\(\bar{H}^{D}_{\sf G}\) that captures only the gauge-invariant physics
of \(\bar{H}_{\sf G}\).
Since there is only one gauge symmetry, the space of gauge invariant states
has dimension \(2^2/2=2\), so that we should be looking for a duality
to a model with a two-dimensional state space. With this in mind,
we find right away 
\begin{equation}\label{tiny_gauge}
\sigma^z_1\sigma^z_2\dual \sigma^x,\ \ \ \ \sigma^x_1\dual\sigma^z,
\ \ \ \ \sigma^x_2\dual\sigma^z,
\end{equation}
so that \(\Phi_\d(G)=\sigma^x\sigma^x=\mathbb{1}\), and the dual Hamiltonian 
reads 
\begin{equation}
\bar{H}^{D}_{\sf G}=J\sigma^x+2h\sigma^z.
\end{equation}
It is easy to check that \(\bar{H}_{\sf G}\) displays four non-degenerate 
energy levels, two of them identical to the two levels of  
\(\bar{H}^{D}_{\sf G}\). It follows (and
this can be checked explicitly) that the energy eigenstates of \(\bar{H}_{\sf G}\)
that belong to the levels not present in \(\bar{H}^{D}_{\sf G}\) are not gauge-invariant.

The projective unitary that implements the mapping of Equation 
\eqref{tiny_gauge} is a \(2\times4\) matrix that can be constructed as follows. 
The bond \(\sigma^z_1\sigma^z_2\) and the symmetry \(\sigma^x_1\sigma^x_2\)  
taken together form a complete commuting set, with 
simultaneous eigenstates
\begin{eqnarray}
\sqrt{2}\ \vert 1,\pm 1\rangle=
\begin{bmatrix}
1\\
0
\end{bmatrix}
\otimes
\begin{bmatrix}
1\\
0
\end{bmatrix}
\pm
\begin{bmatrix}
0\\
1
\end{bmatrix}
\otimes
\begin{bmatrix}
0\\
1
\end{bmatrix},
\\\
\sqrt{2}\ \vert -1,\pm 1\rangle=
\begin{bmatrix}
1\\
0
\end{bmatrix}
\otimes
\begin{bmatrix}
0\\
1
\end{bmatrix}
\pm
\begin{bmatrix}
0\\
1
\end{bmatrix}
\otimes
\begin{bmatrix}
1\\
0
\end{bmatrix}.\
\end{eqnarray}
On the other hand, \(\sigma^z_1\sigma^z_2\)
is itself dual to a complete commuting set for the target space,
\(\sigma^z_1\sigma^z_2\dual \sigma^z\). Thus \(U_\d\) must map
the \(\pm\) eigenstates of \(\sigma^z_1\sigma^z_2\) to the 
corresponding \(\pm\) eigenstates of \(\sigma^z\), and simultaneously
map the \(-\) eigenstates of  \(\sigma^x_1\sigma^x_2\) to the zero
vector (this is so because \(\sigma^x_1\sigma^x_2\) must act as the 
identity in the space of the dual model, and the identity has only 
the eigenvalue \(1\)). This defines  \(U_\d\) uniquely
\begin{equation}\
U_\d=\vert1\rangle\langle1,1\vert+\vert-1\rangle\langle-1,1\vert=
\frac{1}{\sqrt{2}}
\begin{bmatrix}
1& 0& 0& 1\\
0& 1& 1& 0
\end{bmatrix}. 
\end{equation}
Now it is easy to check that 
\begin{equation}\
U_\d\sigma^x_1U_\d^\dagger=U_\d\sigma^x_2U_\d^\dagger=\sigma^x.
\end{equation}
Furthermore,
\begin{equation}\
U_\d U_\d^\dagger=
\begin{bmatrix}
1& 0\\
0& 1
\end{bmatrix},
\ \ \ \ \ \
U_\d^\dagger U_\d=\frac{1}{2}
\begin{bmatrix}
1& 0& 0& 1\\
0& 1& 1& 0\\
0& 1& 1& 0\\
1& 0& 0& 1
\end{bmatrix},
\end{equation}
that is, both \(U_\d U_\d^\dagger\) and \(U_\d^\dagger U_\d\) are 
{\it projectors}, \(\left(U_\d U_\d^\dagger\right)^2=U_\d U_\d^\dagger\), 
\(\left(U_\d^\dagger U_\d\right)^2=U_\d^\dagger U_\d\). This is the 
trademark of a partial isometry. Notice also that {\it 
\(U_\d^\dagger U_\d\) is the projector onto the space of 
gauge invariant states} (that is, in this example, the space of states invariant
under \(G=\sigma^x_1\sigma^x_2\)).

\begin{figure}[h]
\begin{center}
\includegraphics[width=0.4\columnwidth]{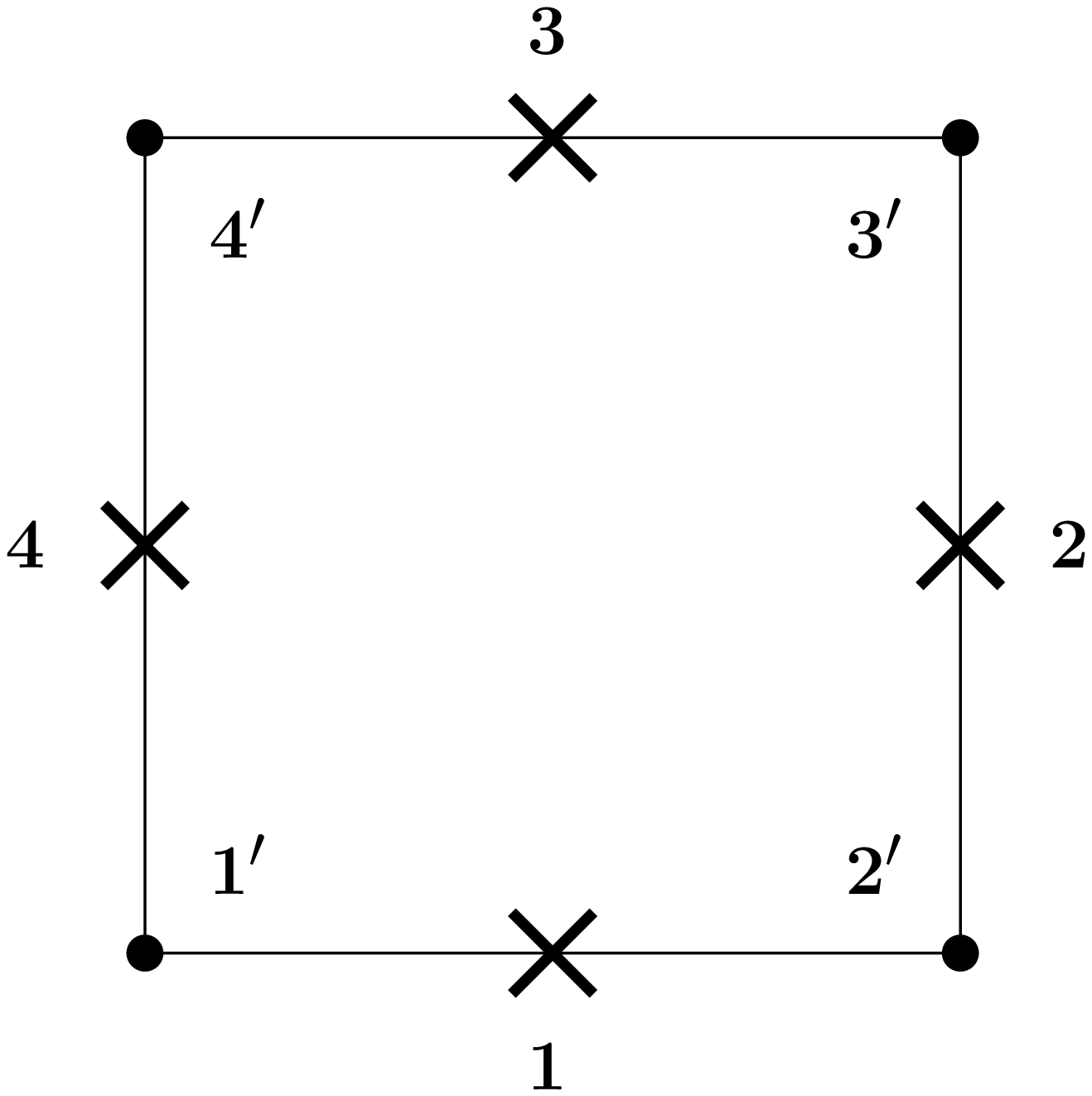}
\end{center}
\caption{The labeling of sites and links used to define several models
in this appendix. Depending
on the model, we place (or not) a quantum or classical spin on the 
sites/links of this square.}
\label{sq_gauge}
\end{figure}

Now that we understand the most elementary example in detail, let
us consider the slightly more challenging one, the $\mathbb{Z}_2$ Higgs field 
of Section \ref{sec5.5}, restricted to four lattice sites,
\begin{eqnarray}
H_{\sf AH}&=&H_{\sf G}+ 
\kappa(\sigma^z_{1'}\sigma^z_1\sigma^z_{2'}+\sigma^z_{2'}\sigma^z_2\sigma^z_{3'}
+\sigma^z_{3'}\sigma^z_3\sigma^z_{4'}+
\sigma^z_{4'}\sigma^z_4\sigma^z_{1'})\nonumber\\
&+&\lambda(\sigma^x_{1'}+\sigma^x_{2'}+
\sigma^x_{3'}+\sigma^x_{4'}),
\end{eqnarray}
with $H_{\sf G}=J\sigma^z_1\sigma^z_2\sigma^z_3\sigma^z_4+h(\sigma^x_1+\sigma^x_2
+\sigma^x_3+\sigma^x_4)$, and  gauge symmetries
\begin{equation}\label{small_gauge_ah}
\sigma^x_1\sigma^x_{2'}\sigma^x_2,\ \ \ \ 
\sigma^x_2\sigma^x_{3'}\sigma^x_3,\ \ \ \
\sigma^x_3\sigma^x_{4'}\sigma^x_4,\ \ \ \ \sigma^x_4
\sigma^x_{1'}\sigma^x_1.
\end{equation}
Here \(1',2',3'\) and \(4'\) denote the vertices of 
the square with sides \(1,2,3,4\), see Figure \ref{sq_gauge}. 

In the presence of matter fields, all four 
gauge symmetries of Equation \eqref{small_gauge_ah} are independent. 
Hence the space of gauge invariant states has dimension \(2^8/2^4=2^4\).
This means that the dual model without gauge symmetries must feature only
four spins. This suggest the bond algebra mapping,
\begin{eqnarray}
&&\sigma^z_{1'}\sigma^z_1\sigma^z_{2'}\dual\sigma^z_1,\ \ \ \ \ \ 
\sigma^z_{2'}\sigma^z_2\sigma^z_{3'}\dual\sigma^z_2,\nonumber\\
&&\sigma^z_{3'}\sigma^z_3\sigma^z_{4'}\dual\sigma^z_3,\ \ \ \ \ \ 
\sigma^z_{4'}\sigma^z_4\sigma^z_{1'}\dual\sigma^z_4,\\
&&\sigma^x_{1'}\dual\sigma^x_4\sigma^x_1,\ \
\sigma^x_{2'}\dual\sigma^x_1\sigma^x_2,\ \
\sigma^x_{3'}\dual\sigma^x_2\sigma^x_3,\ \  
\sigma^x_{4'}\dual\sigma^x_3\sigma^x_4,\nonumber
\end{eqnarray}
that leaves the two bonds corresponding to the gauge field fixed,
and maps the gauge symmetries of Equation \eqref{small_gauge_ah} to
\(\mathbb{1}\). The dual Hamiltonian \(H_{\sf AH}\dual H^D_{\sf AH}\)
reads
\begin{eqnarray}
 H^D_{\sf AH}=H_{\sf G}+
\kappa(\sigma^z_1+\sigma^z_2+\sigma^z_3+\sigma^z_4)
+
\lambda(\sigma^x_4\sigma^x_1+\sigma^x_1\sigma^x_2+
\sigma^x_2\sigma^x_3+\sigma^x_3\sigma^x_4),
\end{eqnarray}
that as expected features only four spins and shows no local (nor global) 
symmetries. One can check explicitly that
every energy level of \(H^D_{\sf AH}\)
is also present in the spectrum of \(H_{\sf AH}\), but \(H_{\sf AH}\) has
other levels in addition to the physical, gauge-invariant, ones.

Next we consider the exact self-duality of the classical Ising model on the 
Utiyama lattice for 
just four classical spins. The model was described in
Section \ref{sec8.2}, Equations \eqref{pf_uti}, \eqref{transfer_uti}, and 
\eqref{hs_uti}, for a \(2M\times 2N\) lattice. If we 
set \(M=N=1\), we get the partition function (and its dual from Equation 
\eqref{sd_utiyama_ising})
\begin{eqnarray}\label{small_uti}
\tilde{\pf}_{\sf U}&=&\sum_{\{\sigma_{i'}\}}\ e^{K_4(\sigma_{1'} \sigma_{2'}+
\sigma_{3'})+K_2(\sigma_{3'}\sigma_{4'}+\sigma_{2'})+(K_1+K_3)(\sigma_{1'}
\sigma_{4'}+\sigma_{2'}\sigma_{3'})} \\
&=&A\sum_{\{\sigma_{i'}\}}\ e^{K^*_4(\sigma_{1'} \sigma_{2'}+
\sigma_{3'})+K^*_2(\sigma_{3'}\sigma_{4'}+\sigma_{2'})+(K^*_1+K^*_3)
(\sigma_{1'}\sigma_{4'}+\sigma_{2'}\sigma_{3'})}=A \, \tilde{\pf}^D_{\sf U} \nonumber,
\end{eqnarray}
where $A=\sinh(2K_1)\sinh(2K_3)/ \sinh(2K_1^*)\sinh(2K_3^*)$, 
with  sites \(i'=1,\cdots,4\) shown in Figure \ref{sq_gauge},  and the 
dual couplings \(K^*_{i},\ i=1,\cdots,4\) can be read from Equation 
\eqref{utiyama_dual_couplings}. With all these elements in place, 
one can check
the identity relation of Equation \eqref{small_uti}.
 Notice the essential role played by the new classical
self-dual BCs embodied in the terms \(K_2\sigma_{2'}\)
and \(K_4\sigma_{3'}\).  

\section{Classical Poisson dualities}
\label{poisson}

Appendix \ref{appA} describes an algorithm to compute
dual forms of classical partition functions that is typically
connected to bond-algebraic dualities for models
that admit a transfer matrix formulation (see Sections \ref{classical&quantum}
and \ref{sec8}). In this section we describe a quite different algorithm 
whose connection to bond algebras is unknown. The resulting type of
duality, that we call Poisson duality, exploits the identity
\begin{equation}\label{poissonn}
\sum_m\delta(x-m)=\sum_m e^{2\pi i mx} ,
\end{equation}
that follows from the Poisson summation formula. Poisson dualities
work specifically for models that have integer-valued degrees of freedom.
We have managed to generalize Poisson dualities to general Abelian groups through
the corresponding generalized Poisson formulas, but we will not report these results
here. 

Let \(\{m_\r\}\) denote
integer-valued configurations (\(m_\r\in\mathbb{Z}\)), 
and \(\mathcal{E}\{m_\r\}\) the cost function of the 
configuration, so that the partition function of interest reads
\begin{equation}\label{intpartf}
\pf=\sum_{\{m_\r\}}e^{-\mathcal{E}\{m_\r\}} .
\end{equation}
We can use Equation \eqref{poissonn} to rewrite \(\pf\) as
\begin{eqnarray}
\pf&=&\sum_{\{m_\r\}}\int \left[\prod_\r dx_\r\ 
\delta(x_\r-m_\r)\right]\ e^{-\mathcal{E}\{x_\r\}}\\
&=&\sum_{\{m_\r\}}\int\prod_\r dx_\r\ e^{2\pi i
\sum_\r m_\r x_\r}e^{-\mathcal{E}\{x_\r\}}.\nonumber
\end{eqnarray}
Defining the dual cost functional \(\mathcal{E}^D\) through the
Fourier transform
\begin{equation}\label{dual_energy}
e^{-\mathcal{E}^D\{y_\r\}}=\int\prod_\r dx_\r\ e^{2\pi i
\sum_\r y_\r x_\r}e^{-\mathcal{E}\{x_\r\}},\ \ \ \ y_\r\in\mathbb{R},
\end{equation}
we have the duality relation
\begin{equation}
\pf^D\equiv \sum_{\{m_\r\}}e^{-\mathcal{E}^D\{m_\r\}}=\pf.
\end{equation}

The usefulness of this duality is dictated by the difficulty
to  compute the Fourier transform of Equation \eqref{dual_energy}.
For Gaussian models, like the $D=2$ SoS model,
\begin{equation}\label{energy_sos}
\mathcal{E}_{\sf SS}\{m_\r\}=-\sum_{\r}\sum_{\nu=1,2}\ K_\nu (m_{\r+\bm{e_\nu}}-m_\r)^2 ,
\end{equation}
\(\mathcal{E}^D\) can be computed in closed form, and the dual
model represents a lattice Coulomb gas \cite{bookEPTCP}. But in general,
we need to resort to approximations and/or numerics. 


\section{Exponential of operators in the Weyl group algebra}
\label{appF} 

This appendix presents closed-form expressions for the exponentials and
logarithms of operators in the Weyl group algebra defined in Section \ref{sec4.1}. 
As discussed in 
Section \ref{sec8.3}, these expressions allow us to establish
{\it exact} connections between a large class of classical and 
quantum $p$-states models, and  to our knowledge are not available in the literature. 

As mentioned in Section \ref{sec4.1.1}, 
the operator \(V\) generates an algebra of circulant matrices 
\cite{aldrovandi}. It follows that 
\begin{equation}
e^{\sum_{m=0}^{p-1}\ a_mV^m}=\sum_{m=0}^{p-1}\ b_mV^m.
\end{equation}
Our goal is to find closed-form expressions for the coefficients \(a_m\) in terms
of \(b_m\) (useful for performing classical-to-quantum mappings), 
and for the coefficients \(b_m\) in terms of \(a_m\) (useful for performing
quantum-to-classical mappings). To achieve this we have to recall
that the discrete Fourier transform \(F\)
puts \(V\) in diagonal form, \(FVF^\dagger=U\) (see Equation \eqref{wga_aut}), 
and that \(\tr (U^{m\dagger}U^n)/p=\delta_{m,n}\). It follows that
\begin{equation}
b_m=\frac{1}{p}\tr \left[U^{p-m}\ e^{\sum_{l=0}^{p-1}\ a_lU^l}
\right],\ \ \ \ \ \
a_m=\frac{1}{p}\tr \left[U^{p-m}\ln\left(\sum_{l=0}^{p-1}\ b_lU^l\right)
\right]\ .
\end{equation}
In physical applications, the \(a_m\) are Hermitian-symmetric,
\(a_{p-m}=a_m^*\) (to guarantee that \(\sum_{m=0}^{p-1}\ a_mV^m\) 
is a Hermitian operator), and the \(b_m\) are real and positive. Thus it
is convenient to assume that both set of coefficients satisfy \(a_{p-m}=a_m\),
\(b_{p-m}=b_m\), and the relations between them simplify to 
\begin{eqnarray}
b_m&=&\frac{1}{p}\sum_{s=0}^{p-1}\ \cos (\frac{2\pi ms}{p}) \ e^{\sum_{l=0}^{p-1}\ 
a_l\cos(\frac{2\pi l s}{p})},\label{cos_bs_and_as}\\
a_m&=&\frac{1}{p}\sum_{s=0}^{p-1}\ \cos(\frac{2\pi m s}{p})\ln\left(\sum_{l=0}^{p-1}\ 
b_l\cos(\frac{2\pi l s}{p})\right).
\end{eqnarray}
These are the expressions that are most useful in physical applications. 

Suppose next that \(b_m=e^{K\, u(m)}\), where \(K\) is a positive constant, and
\(u(m)\) is a real function of \(m=0,\cdots,p-1\) (for example,
\(u(m)=\cos(\frac{2\pi m}{p})\) for the classical VP model of Section \ref{sec8.3}).
We would like to study the behavior of the \(a_m\) to next-to-leading order
in \(K\), in the limit that \(K\) grows very large
(this could happen at low temperature, or in the context of a STL
decomposition \cite{bookEPTCP}, see Section \ref{sec8}). Notice
that in this limit
\begin{equation}\label{approx_cs}
\sum_{l=0}^{p-1}\ b_l\cos(\frac{2\pi l s}{p})\approx 
e^{K u(0)}\left(1+2e^{K (u(1)-u(0))}\cos(\frac{2\pi s}{p})\right)
\end{equation}
to next-to-leading order, assuming that the inequalities
\begin{equation}
0>(u(1)-u(0))>(u(2)-u(0))>\cdots
\end{equation}
hold. The factor two in Equation \eqref{approx_cs} is due to 
the symmetry \(u(p-l)=u(l)\).
Replacing expansion \eqref{approx_cs} into Equation \eqref{cos_bs_and_as} leads to
\begin{equation}\label{plarge_K}
a_m\approx K u(0)\delta_{m,0}+e^{K(u(1)-u(0))}(\delta_{m,1}+\delta_{m,p-1}),
\ \ \ \ \ K\rightarrow\infty,
\end{equation}
where we have used  \(\ln(1+x)\approx x\) and the 
identity
\begin{equation}
\frac{1}{p}\sum_{l=0}^{p-1}\ \cos(\frac{2\pi l s}{p})\cos(\frac{2\pi l m}{p})=
\frac{1}{2}\left(\delta_{s,m}+\delta_{s,p-m}\right).
\end{equation}

\noindent
{\bf Acknowledgments}

This research was supported in part by the National Science Foundation 
under Grant No. NSF PHY05-51164. We thank J. Cardy, 
D. J. Gross, A. Kostelecky, H. Nishimori, E. Knill, and L. Oxman for many 
fruitful discussions.


\begin{thebibliography}{10}
\bibitem{remark_HS}
We note, in passing, that albeit being notably different, 
this transformation has some similarity to the so-called
Holstein-Primakoff transformation between spin generators
and the harmonic oscillator creation and annihilation operators.
\end{thebibliography}
\end{document}